\RequirePackage[l2tabu]{nag}		
\pdfoutput=1
\documentclass[a4paper,11pt,leqno,openbib]{memoir} 
\usepackage{datetime}
\usepackage{ifpdf}
\ifpdf
\fi
\ifdraftdoc 
	\usepackage{draftwatermark}				
	\SetWatermarkScale{0.3}
	\SetWatermarkText{\bf Draft: \today}
\fi
\newsubfloat{figure}
\newsubfloat{table}
\settrimmedsize{297mm}{210mm}{*}
\settypeblocksize{634pt}{448.13pt}{*} 
\setulmargins{4cm}{*}{*} 
\setlrmargins{*}{*}{1.5} 
\setmarginnotes{17pt}{51pt}{\onelineskip} 
\setheadfoot{\onelineskip}{2\onelineskip} 
\setheaderspaces{*}{2\onelineskip}{*} 
\checkandfixthelayout
\usepackage{pdfpages}

\usepackage{fouriernc}
\usepackage[T1]{fontenc}

\OnehalfSpacing 
\setsecnumdepth{subsection} 
\maxsecnumdepth{subsubsection}
\usepackage{multirow}
\usepackage{hhline}
\usepackage{url}
\usepackage{graphicx}					
\usepackage{epstopdf}
\usepackage{amssymb}
\usepackage{deluxetable}
\usepackage{caption}
\usepackage{rotating}
\usepackage{memhfixc}
\def\aj{Astronomical Journal}
\def\apj{Astrophysical Journal}
\def\apjl{Astrophysical Journal Letters}
\def\mnras{Monthly Notices of the Royal Astronomical Society}
\def\physrep{Physics Reports}

\def\aap{Astronomy \& Astrophysics}
\def\nat{Nature}
\def\pasp{Astronomical Society of the Pacific, Publications}
\def\apjs{The Astrophysical Journal Supplement Series}
\def\Msun{\mathrm{M_{\odot}}}
\def\msun{$M_{\odot}$}
\def\Lsun{\mathrm{L_{\odot}}}
\def\Rsun{\mathrm{R_{\odot}}}
\def\arcsec{"}
\def\arcmin{'}
\def\sax{SAX~J1748.9-2021\,}
\def\com{COM-SAX~J1748.9-2021\,}

\definecolor{alma}{RGB}{175,39,47}
\usepackage{calc,soul,fourier}
\makeatletter 
\newlength\dlf@normtxtw 
\newsavebox{\feline@chapter} 
\newcommand\feline@chapter@marker[1][4cm]{%
	\sbox\feline@chapter{%
		\resizebox{!}{#1}{\fboxsep=1pt%
			\colorbox{alma}{\color{white}\thechapter}%
		}}%
		\rotatebox{90}{%
			\resizebox{%
				\heightof{\usebox{\feline@chapter}}+\depthof{\usebox{\feline@chapter}}}%
			{!}{\scshape\so\@chapapp}}\quad%
		\raisebox{\depthof{\usebox{\feline@chapter}}}{\usebox{\feline@chapter}}%
} 
\newcommand\feline@chm[1][4cm]{%
	\sbox\feline@chapter{\feline@chapter@marker[#1]}%
	\makebox[0pt][c]{
		\makebox[1cm][r]{\usebox\feline@chapter}%
	}}
\makechapterstyle{daleifmodif}{

	\renewcommand\printchapternum{\null\hfill\feline@chm[2.5cm]\par}

} 
\makeatother 
\chapterstyle{daleifmodif}
\makepagestyle{myvf} 
\makeoddfoot{myvf}{}{\thepage}{} 
\makeevenfoot{myvf}{}{\thepage}{} 
\makeheadrule{myvf}{\textwidth}{\normalrulethickness} 
\makeevenhead{myvf}{\small\textsc{\leftmark}}{}{} 
\makeoddhead{myvf}{}{}{\small\textsc{\rightmark}}
\newcommand{\clearemptydoublepage}{\newpage{\thispagestyle{empty}\cleardoublepage}}
\makeindex
\usepackage{import}

\usepackage{lipsum}					
\usepackage{amsfonts} 					
\usepackage[centertags]{amsmath}			
\usepackage{stmaryrd}					
\usepackage{amssymb}					
\usepackage{amsthm}					
\usepackage{newlfont}					
\usepackage{layouts}					
\usepackage{longtable,rotating}			
\usepackage[applemac]{inputenc}			
\usepackage{wasysym}					
\usepackage{mathrsfs}					
\usepackage{float}						
\usepackage{verbatim}					
\usepackage{upgreek }					
\usepackage{latexsym}					
\usepackage[round,authoryear]{natbib}		

\usepackage{url}						
\usepackage{etex}						
\usepackage{fixltx2e}					
\usepackage[spanish,english]{babel}		
\usepackage{color}                    				
\usepackage[colorlinks=true,
		     allcolors=black]{hyperref}              
\usepackage{memhfixc}					
\usepackage{enumerate}					
\usepackage{footnote}					
\usepackage{microtype}					
\usepackage{rotfloat}					
\usepackage{alltt}						
\usepackage[version=0.96]{pgf}			
\usepackage{tikz}						
\usetikzlibrary{arrows,shapes,snakes,
		       automata,backgrounds,
		       petri,topaths}				
\newcommand{\pgftextcircled}[1]{                                                                    
    \setbox0=\hbox{#1}%
    \dimen0\wd0%
    \divide\dimen0 by 2%
    \begin{tikzpicture}[baseline=(a.base)]%
        \useasboundingbox (-\the\dimen0,0pt) rectangle (\the\dimen0,1pt);
        \node[circle,draw,outer sep=0pt,inner sep=0.1ex] (a) {#1};
    \end{tikzpicture}
}
\newcommand{\blackged}{\hfill$\blacksquare$}
\newcommand{\whiteged}{\hfill$\square$}
\newcounter{proofcount}

\let\oldsqrt\sqrt
\def\sqrt{\mathpalette\DHLhksqrt}
\def\DHLhksqrt#1#2{%
\setbox0=\hbox{$#1\oldsqrt{#2\,}$}\dimen0=\ht0
\advance\dimen0-0.2\ht0
\setbox2=\hbox{\vrule height\ht0 depth -\dimen0}%
{\box0\lower0.4pt\box2}}
\newcommand{\mycaption}[2][\@empty]{
	\captionnamefont{\scshape} 
	\changecaptionwidth
	\captionwidth{0.9\linewidth}
	\captiondelim{.\:} 
	\indentcaption{0.75cm}
	\captionstyle[\centering]{}
	\setlength{\belowcaptionskip}{10pt}
	\ifx \@empty#1 \caption{#2}\else \caption[#1]{#2}
}
\newcommand{\mysubcaption}[2][\@empty]{
	\subcaptionsize{\small}
	\hangsubcaption
	\subcaptionlabelfont{\rmfamily}
	\sidecapstyle{\raggedright}
	\setlength{\belowcaptionskip}{10pt}
	\ifx \@empty#1 \subcaption{#2}\else \subcaption[#1]{#2}
}
\usepackage{lettrine}
\newcommand{\initial}[1]{%
	\lettrine[lines=3,lhang=0.33,nindent=0em]{
		\color{alma}
     		{\textsc{#1}}}{}}

\begin{document}
%
%
%
%
%
\frontmatter
\pagenumbering{roman}
\includepdf{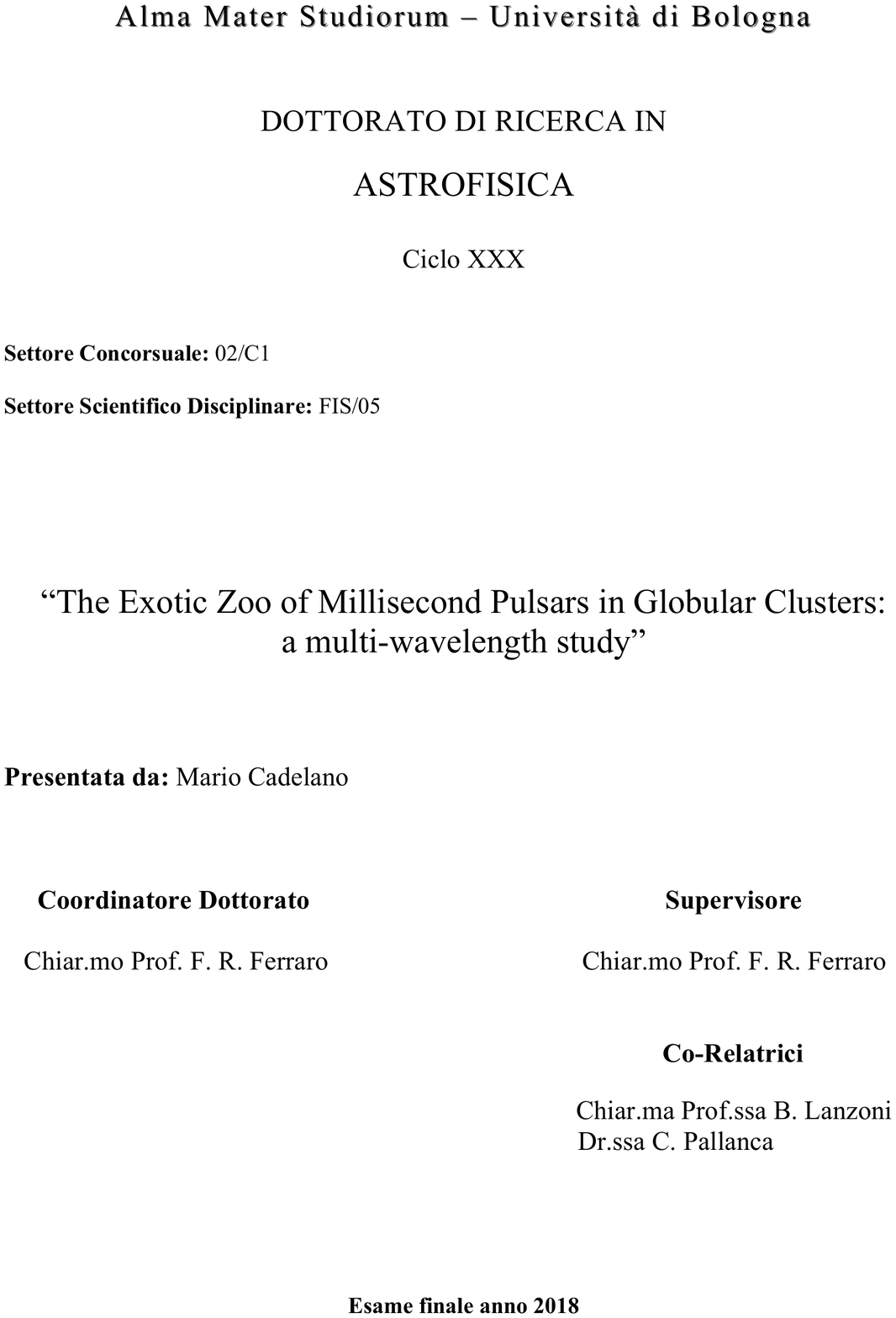}
\clearemptydoublepage

\includepdf[pages={1-}]{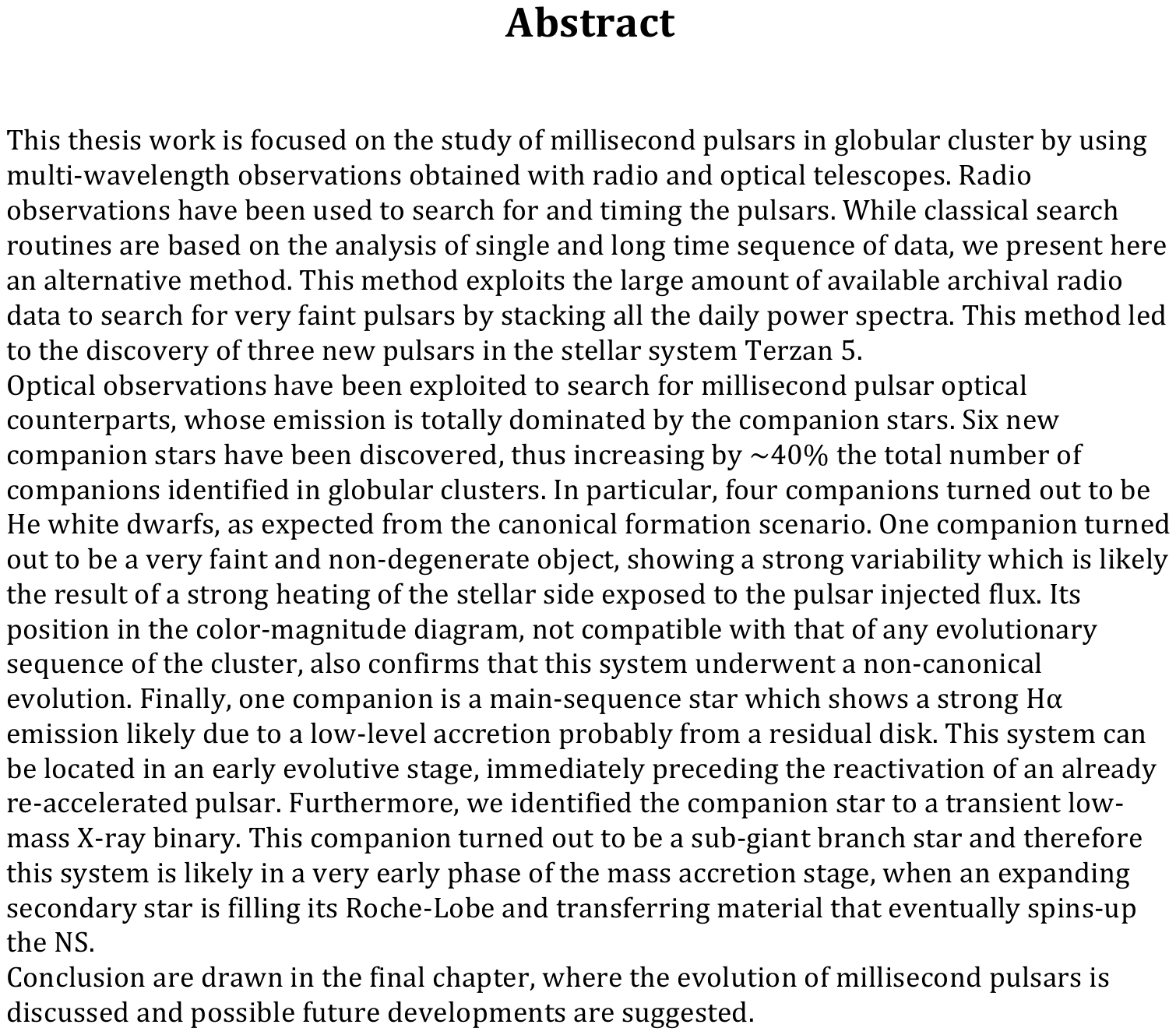}

\clearpage

\clearemptydoublepage

%
%
%

\thispagestyle{empty}
$$
\vspace{6cm}
$$
\begin{flushright}
{\it To my Walden...}

\cleardoublepage
\thispagestyle{empty}
$$
\vspace{6cm}
$$
{\it When we become the enfolders of those orbs, \\and the pleasure and knowledge of every thing in them,\\ shall we be fill'd and satisfied then? \\ \vspace{0.2cm} No, we but level that lift to pass and continue beyond.}\\
\vspace{1cm}
- Walt Whitman, {\it Leaves of Grass} -
\end{flushright}

\clearpage
\clearemptydoublepage
%
%
\clearemptydoublepage
\renewcommand{\contentsname}{Table of Contents}
\maxtocdepth{subsection}
\tableofcontents*
\addtocontents{toc}{\par\nobreak \mbox{}\hfill{\bf Page}\par\nobreak}
\clearemptydoublepage
\listoftables
\addtocontents{lot}{\par\nobreak\textbf{{\scshape Table} \hfill Page}\par\nobreak}
\clearemptydoublepage
\listoffigures
\addtocontents{lof}{\par\nobreak\textbf{{\scshape Figure} \hfill Page}\par\nobreak}
\clearemptydoublepage
%
%
\mainmatter
\chapter*{Introduction}
\addcontentsline{toc}{chapter}{Introduction} 
\sectionmark{Introduction}
\chaptermark{Introduction}
\markboth{Introduction}{Introduction}

\initial{P}ulsars (PSRs) are rapidly spinning and highly magnetized neutron stars (NSs), mostly known as emitters of pulsating radiation in the radio bands. Their characteristic pulsated emission is the result of a collimated beam of radiation located in correspondence of the magnetic poles, misaligned with respect to the spinning axis. Therefore, the observer receives a pulsed signal every time the emission beams sweep past the line of sight, once per rotation. This is the so-called ``lighthouse effect''.

The first PSR was discovered about 50 years ago by Jocelyn Bell Burnell, who observed an intense source of pulsation in the sky with a stable period of 1.33 s \citep{hewish68}, later confirmed to be a rapidly spinning NS by Franco Pacini and Thomas Gold \citep{gold68}. Both of them independently proposed that such a rotating and magnetized star, made mostly of neutrons and formed after the supernova explosion of a high-mass star, could emit a radiation similar to that observed by Jocelyn Bell Burnell. \\

To date, $\sim2600$ PSRs have been discovered throughout he Galaxy\footnote{\url{http://www.atnf.csiro.au/people/pulsar/psrcat/}}. The vast majority of them are isolated systems spinning at periods of about 1 s. However, a second important class, populated by $\sim400$ objects, shows very short spin periods (around 5 ms) and are usually called ``millisecond pulsars'' (MSPs). They are old NSs, reaccelerated by mass and angular momentum accretion from an evolving companion star and therefore they are commonly observed in binary systems with an exhausted and deeply peeled secondary star, usually a white dwarf (WD) with a He core \citep[e.g.,][]{stairs04}.

Being highly collisional systems, globular clusters (GCs) are the ideal habitat for the formation of MSPs \citep{verbunt87,hut91,ransom05a,hessels07}. Indeed, about 40\% of MSPs are found in these systems, despite the Galaxy is $\sim10^3$ times more massive than the entire GC system. This can be explained taking into account that MSPs in GCs are the outcome not only of the standard evolution of primordial binary systems, but also of non primordial binaries created through stellar dynamical encounters in the dense internal region of GCs.  Before this thesis work, 146 MSPs were known in 28 GCs \footnote{\url{http://www.naic.edu/~pfreire/GCpsr.html}}. However, population synthesis models suggest that the next generation of radio telescopes will unveil few hundreds or even few thousands of new MSPs within the 150 Galactic GCs. Such a large population will hopefully include holy grails of PSR astronomy such as MSP-MSP binaries or even MSP-black hole binaries, expected to be formed exclusively in the extreme environments of GC cores through dynamical interactions \citep{freire04,verbunt14}. \\

MSPs in GCs can provide a wealth of science spacing in many different fields, from stellar evolution to accretion physics, from dynamics to general relativity tests. In the following we briefly introduce some of the science that can be performed by studying these objects. Some of this aspects will be discussed with more details throughout the thesis:

\begin{itemize}

\item Thanks to the stability of their signal over time, MSPs are extremely precise clocks. Deviations from their precision can be caused by their acceleration induced by the host GC potential field. Thus such deviations can be exploited to study the structure of the cluster itself, its physical characteristics, dynamical status, possibly probing or excluding the presence of an intermediate-mass black hole in the center \citep[e.g.,][]{prager17}. Moreover, the measurements of MSP proper motions allow to constrain the orbit of the host cluster within the Galaxy, shedding light on the formation and evolution of these stellar systems and the Milky Way itself across their $\sim12$ Gyr lifetime.

\item The study of MSPs in GCs can be used to improve out understanding of the dynamical evolution of these dense stellar systems. For instance, the identification of a large population of MSPs is crucial to constrain the rate of ejection and the NS retention fraction in the early stage of GC evolution. Furthermore, different GCs show differences in the properties of their MSP population. For example, the numerous population of MSPs in 47 Tucanae shows properties basically identical to those of the Galactic field population, while M15 shows a population composed exclusively of exotic systems, such as isolated MSPs and double NS systems. This is tightly related to the cluster dynamical status which can therefore be constrained identifying as many MSPs as possible in many different GCs \citep{verbunt14}.

\item Since MSPs are mostly located in binary systems, their companion stars, identifiable through optical observations, provide a complementary view on the properties of these systems. Indeed, according to the canonical formation scenario, MSP companions are expected to be WDs with a He core. However, several deviations from this scenario can occur. These deviations are usually imprinted on the characteristics of the companion stars and thus the identification of the optical counterparts can help shedding light on the MSP evolutionary paths \citep{ferraro01a,fer03_msp,pallanca10,pallanca14a,mucciarelli13}. The identification and characterization of MSP optical counterparts is also a powerful tool to study stellar evolution under extreme conditions (due to the MSP energetic emission), as well as the accretion mechanism and the physical processes occurring in the intra-binary space, where the interplay between the PSR wind of relativistic particles and material lost from the outer envelope of the companion star becomes dominant \citep[e.g.,][]{bogdanov05}.

\item MSPs can be used to test general relativity on small scales thanks to independent measurements (through PSR timing) of different relativistic orbital properties of the binaries, so-called ``post-Keplerian parameters'' \citep[e.g.,][]{burgay03,kramer06,breton08}. The number of systems clearly showing such effects is enhanced in GCs thanks to perturbations of the binaries due to close encounters with other stars. Binary MSPs also provided in the past the first indirect evidence of gravitational waves emission through the measurement of the orbital period decay of a double NS system \citep{hulse75}. Finally, mass measurements through PSR timing and spectroscopy of the companion star could in principle be used to measure NS masses with high precision, thus possibly setting constraints on the equation of state of ultra-dense matter, which is a still open question in fundamental physics \citep{freire08_m5,demorest10}.

\end{itemize}

This thesis presents a multi-wavelength study of MSPs in GCs performed through radio and optical observations. These observations have been exploited through different approaches aimed to discovering new and extremely faint objects, obtain long-term highly precise timing solution and identify the optical counterparts to systems experiencing different evolutionary stages. The thesis is organized as follows:

\begin{itemize}

\item In Chapter 1 we provide a theoretical background on PSR properties, emission physics and evolution to the stage of MSPs. The properties of the different families of MSPs are then presented. Finally, we show why GCs are the ideal habitat for the formation of these systems and what kind of science has been (and can be) done by studying their population of MSPs.

\item In Chapter 2 we present the main radio and optical methodologies used to study MSPs in GCs, from the techniques used to search for them and timing their signals in the radio bands, to the identification and characterization of their optical counterparts through optical observations.

\item In Chapter 3 we report on the discovery of three new MSPs in the stellar system Terzan~5. These have been identified not though classical search routines, but using a different approach which exploits the large archival observations of this cluster, which has been routinely monitored in the past years.


\item In Chapter 4 we present the optical identification and characterization of the companion star to the black-widow system M71A. Black-widows are a sub-class of MSPs showing eclipses of the radio signal. They largely deviate from the canonical formation scenario and the companion star is expected to be very different from the ``typical'' He WD.

\item In Chapter 5 we report on the identification and characterization of the optical counterparts to four MSPs in the GC 47 Tucanae. All the discovered companion stars are He WD, as expected from the standard formation scenario. Their properties are analyzed and discussed.

\item In Chapter 6 we focus our attention on the optical counterpart to an accreting MSP system in the GC NGC 6440. Accreting MSPs are a sub-class of transient low-mass X-ray binaries showing pulsations from an already recycled NS, which is however still inactive as a radio MSP. The optical counterpart gives us the chance to probe the properties of the companion star in a system experiencing an evolutionary stage immediately preceding that of radio MSPs.

\item In Chapter 7 we report on the identification of the optical counterpart to EXO~1745-248, a low-mass X-ray binary in the stellar system Terzan~5. This binary is characterized by recurrent outburst episodes triggered by active mass accretion from an evolving secondary star to the NS. The characterization of such a system allows to get insights on the very early stages of NS binary evolution, where the re-acceleration of a slow and inactive pulsar is still on-going.

\item Finally, in the Conclusion chapter, we summarize the results obtained in this work, we draw our final conclusions and future prospects on the study of PSRs in GCs.

\end{itemize}

\chapter{From Neutron Stars to Millisecond Pulsars}
\label{CapIntro}
\begin{flushright}
\textit{}
\end{flushright}

\vspace{1cm}

\initial{T}his chapter is aimed at providing a general theoretical background for a proper study of the population of millisecond pulsars in globular clusters. We will present a general view on the neutron star properties and on the pulsar phenomenology, focusing on the physical processes that characterize their peculiar signals. Then, we will show what is and how forms a millisecond pulsar, presenting the number of different evolutionary scenarios that create the exotic zoo of these objects. We will conclude showing why globular clusters are the ideal site for the formation of millisecond pulsars and what kind of science can be performed with a multi-wavelength approach.

\clearpage

\section{Neutron Stars}

Neutron stars (NSs) are rapidly spinning and magnetized compact stars composed mostly of neutrons. They are the core remnant of the gravitational collapse of massive stars ($M>11 \; \Msun$) which eject their envelopes through supernova explosions. 

The primary energy source of a star is given by thermonuclear fusion reactions, which can sustain the stellar structure until the exhaustion of the thermonuclear fuel (i.e. H, He and heavier elements till, at most, the iron group elements like Fe and Ni). Once this is exhausted, the only stellar energy source left is the gravitational one. This phenomenon has different outcomes depending on the original mass of the star:
\begin{itemize}
\item Stars with masses smaller than $\sim11 \; \Msun$ end their life as white dwarfs (WDs): compact ($\mathrm{\sim \, R_{\oplus}}$) and hot ($\mathrm{\sim10^{4-5} \; K}$) stars sustained by the pressure of degenerate electrons and mostly made of C and O or, eventually, of O, Ne and Mg.
\item Stars with masses between $\sim11 \; \Msun$ and $\sim 25 \; \Msun$ become NSs: extremely compact objects ($\sim 10$ km), sustained by the pressure of degenerate neutrons.
\item For stars with masses larger than $\sim25 \; \Msun$, the gravitational collapse leads directly to the formation of a black hole.
\end{itemize}
While the processes leading to the formation of a WD are relatively steady and smooth events, the formation of a NS is preceded by the violent emission of most of the star gravitational energy through an explosion known as ``core-collapse supernova''. During this process, the outer envelopes of the star are blown away, while only a small inner core is left, formed from the collapse of the previous star nucleus, mostly composed of iron group elements and degenerate electrons (vaguely resembling a WD composed of Fe nuclei). This nucleus is subject to free fall collapse since its mass is expected to exceed the Chandrasekhar limit\footnote{It is the mass limit below which a system supported by the pressure of degenerate electrons (such as a WD) is stable. Above this value, the pressure can no longer sustain the gravitational collapse.} of $\sim1.4 \; \Msun$ and a remnant composed mostly of neutrons is formed. In fact, during this phase, the density in the stellar core is large enough that neutronization of matter is favored through $\mathrm{\beta-decay}$ reactions:
\begin{equation}
\label{betadecay}
p^+ + e^- \rightarrow n + \nu_e
\end{equation}
This process results in the formation of a tremendous amount of neutrons. As a rough estimate, the ratio between the stellar nucleus mass (close to the Chandrasekhar limit of $\sim1.4 \; \Msun$) and the neutron mass ($\sim 1.7\times10^{-24}$ g) is of the order of $10^{57}$, thus easily explaining why these remnants are called ``neutron stars".\\

NSs are extremely compact stars. Assuming that in a NS the speed of sound is smaller than the speed of light, it has been demonstrated \citep{lattimer90} that the minimum radius is about 1.5 times its Schwarzschild radius (${R_S}$):
\begin{equation}
{R_{min}\simeq1.5 \; R_S=\frac{3GM}{c^2}=6.2 \; \mathrm{km} \; \left(\frac{M}{1.4 \; \Msun}\right)}
\end{equation}
where $G$ is the gravitational constant, $c$ the speed of light and $M$ the NS mass. On the other hand, a maximum radius can be obtained assuming that the rotating star is stable against centrifugal forces. For a NS with rotational period $P$, we find:
\begin{equation}
{R_{max} \simeq \left( \frac{GMP^2}{4\pi^2}\right)^{1/3} = 16.8 \; \mathrm{km} \; \left(\frac{M}{1.4 \; \Msun}\right)^{1/3} \left( \frac{P}{\mathrm{ms}}\right)^{2/3}}
\end{equation}

The fastest known rotating NS \citep[PSR J1748$-$2446ad;][]{hessels06} spins with a period of $\sim1.39$ ms, implying a maximum radius of $\sim21$ km for a typical mass of $1.4 \; \Msun$. Assuming that the NS is born from the collapse of a WD-like structure with a radius $\mathrm{\sim R_{\oplus}}$, it follows that its final radius can be up to $\sim500$ times smaller than that of the progenitor one. Indeed, most models predict NSs to have a  radius of $\mathrm{10 \; km \; - 20 \; km}$ \citep{lattimer01}. Although such measurements are observationally challenging, high-energy data are confirming this range of values \citep[e.g.,][]{guillot13,guver13}.
This can also explain why NSs are observed as rapidly rotating stars: under the (weak) assumption that during the gravitational collapse there is no mass and/or angular momentum loss in the stellar core and that the NS is a sphere of uniform density, applying the conservation of angular momentum law before and after the collapse we find:
\begin{equation}
{I_i \Omega_i = I_f \Omega_f \; \; \; \Longrightarrow \; \; \;  \Omega_f = \Omega_i \frac{R_i^2}{R_f^2}}
\end{equation}
where ${I=\frac{2}{5}MR^2}$ and ${\Omega=2\pi/P}$ are the moment of inertia and angular velocity of the NS (or of the stellar nucleus, before the collapse), respectively. Setting ${R_i =500\; R_f}$, we find that ${P_f = 10^{-6}P_i}$. Considering a typical ${P_i \approx 10^3}$~s, it results that ${P_f \approx  10^{-3}}$ s.\\

Shortly after their formation, NSs are extremely hot objects with temperature as high as $10^{11}$~K. However, very efficient cooling mechanisms allow the star to reach, in a time-scale of $10^5$~years, a typical temperature of about $10^6$ K \citep[see][and reference therein]{potekhin15}. Therefore the black-body emission of these objects is peaked at soft X-ray wavelengths, even though, as explained in the following, most of the NS radiation is emitted through non-thermal mechanisms. \\

A possible alternative NS formation mechanism is the so-called ``Accretion Induced Collapse": if a massive WD in a binary system is accreting material from a standard evolving star, its mass can eventually exceed the Chandrasekhar limit. It follows a gravitational collapse that likely creates a NS \citep[e.g.][]{freire14}.
 
\subsection{The Internal Structure of a Neutron Star}
\label{interiors}

The NS structure is able to contrast the gravitational collapse thanks to the pressure granted by degenerate neutrons. As for any degenerate system, the stellar structure can be described by an equation of state, that provides a link between the density and the pressure of the star at different stellar radii.

Given the extreme physical properties of NSs, testing the equation of state in Earth laboratories is prohibitive. Indeed, for a  typical NS with a mass of $1.4 \; \Msun$ and a radius of about 10 km, the average density is $\mathrm{\sim6.7\times10^{14} \ g \ cm^{-3}}$, about the same order of magnitude of  the density of an atomic nucleus. While the equation state of matter with density large as the nuclear density is known, beyond this value relativistic field theories have not been able, so far, to constrain this equation, due to our lack of knowledge of the properties of the strong force in such an extreme regime.  Observational constraints can be obtained through the simultaneous measurement of NS radii and masses, which is, however, quite challenging \citep[see, e.g.,][]{lattimer01,guillot14}.

A description of the different kinds of equation of state proposed so far by theoretical physicists is beyond the goals of this thesis. Briefly, in most of them a NS with a very short interval of radii (10.5 km $-$12 km) is expected to be found in the mass range from $0.5 \; \Msun$ to $2.5 \; \Msun$ \citep{lyne12}. The bulk of the NS population is known to have a mass around $1.35 \; \Msun$ \citep[][]{stairs04}, corresponding to radii in the interval of 10.5 km and 11.2 km.

The maximum mass allowed to a system supported by the pressure of degenerate electrons is well known as the Chandrasekhar limit. The same value for a system supported by the pressure of degenerate neutrons is however unknown, as a consequence of our ignorance about the equation of state. A possible value, the so-called "Tolman-Oppenheimer-Volkov" limit, is of about $2.5 \; \Msun$ \citep{lattimer01}. However, none of the NSs with a precise mass measurement reach such a value, although few candidates are known \citep{freire08_m5,demorest10}, especially among the class of recycled NSs (see Section~\ref{msp}). \\

\begin{figure}
\centering
\includegraphics[width=6cm]{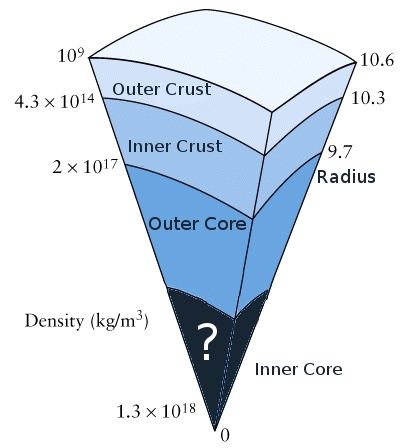}
\caption[Schematic representation of the NS internal structure.]{Schematic representation of the NS internal structure. Credits: A. Patruno.}
\label{NSstructure}
\end{figure}

A schematic representation of the NS structure is reported in Figure~\ref{NSstructure}, where it can be seen that the star density varies by about 9 orders of magnitude between the surface ($\mathrm{\rho\sim10^9 \ g \ cm^{-3}}$)  and the central regions ($\mathrm{\rho\sim10^{18} \ g \ cm^{-3}}$). The most external regions of the crust are composed of a thin atmosphere of atomic nuclei and free electrons, whose typical thickness is of few meters. Below that, the matter already reach the typical density of a WD and it is composed of a solid lattice, mostly of Fe nuclei.
Descending into the star interiors, the energetic of the system favors the neutronization of matter:  the production of isotopes with a large number of neutrons (e.g. $\mathrm{Ni^{62}}$, $\mathrm{Ni^{64}}$, $\mathrm{Ni^{66}}$), through $\beta$-decay reactions like in \ref{betadecay}. Although such isotopes would be unstable in normal condition and neutrons would decade into protons, this cannot happen in a complete degenerate structure. Below the crust, where densities are very close to the nuclear ones, the matter is expected to be in an superfluid state, where heavy atomic nuclei are embedded in a sea of neutrons, with some protons and electrons. Finally, in the very central regions, where the composition of the matter is still extremely unclear, and the density overcomes the nuclear one, elementary particles can behave in unpredictable ways. Some models predict that the matter can be in exotic states, where the extreme densities force the disruption of neutrons into mesons and kaons, eventually also creating a solid inner core. Other models predict an interior composed of free quarks or a Bose-Einstein condensate. In any case, the real structure of NS interiors is still to be defined \citep[see, e.g.,][]{shapiro83}.

\section{Phenomenology of Pulsars}
NSs are commonly observed as radio pulsars (PSRs), which are highly magnetized and rotating objects characterized by a peculiar emission of pulsated and stable radiation at radio wavelengths. As we will see, most of the PSR properties can be naturally explained in a scenario where charged particles, located in the NS surface, are accelerated in a highly magnetized and rotating structure. 
\subsection{Pulse Properties}
\begin{figure}
\centering
\includegraphics[width=10cm]{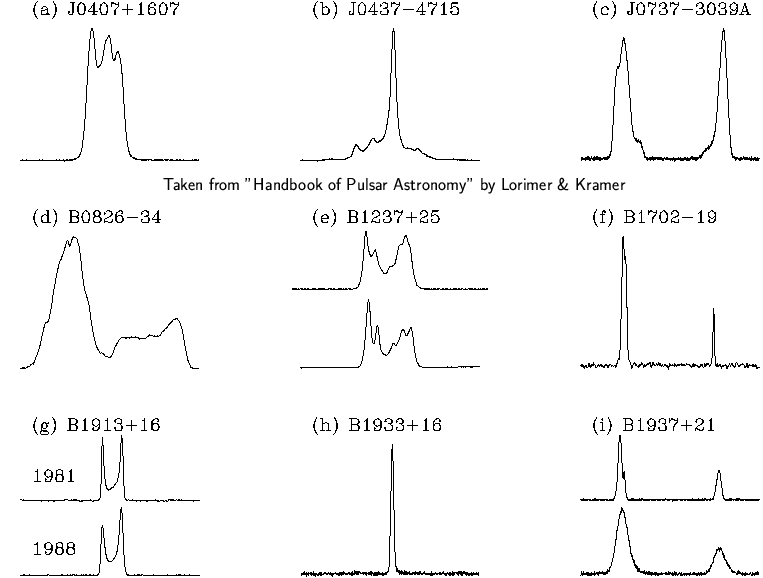}
\caption[Integrated pulse profiles for a sample of PSRs.]{Integrated pulse profiles for a sample of PSRs.}
\label{pulseprofiles}
\end{figure}

The typical pulsed emission is the result of the so-called {\it lighthouse effect}, where an emission beam (or eventually two), located at the magnetic pole (or poles), sweep past our line of sight once per NS rotation. PSR spin periods cover a broad range of values: the slowest one is PSR J18414$-$0456 s with a spin period of 11.79 s \citep{dib14}, while the fastest one is PSR J1748$-$2446ad \citep{hessels06}, spinning at about 1.39 ms. Each PSR is characterized by a distinct pulse profile, which is usually very stable at the same observing frequency and therefore  can be thought as a fingerprint of the PSR itself. Being very weak radio sources (typical flux densities are of the order of $10^{-5} -1$ Jy), the pulse profile is obtained by summing together hundreds or thousands of pulses (this process is called ``folding'') creating an ``integrated pulse profile'' like those reported in Figure~\ref{pulseprofiles}. The shape of each pulse profile is due to the size and geometry of the emission beam, as well as the angle between the line of sight and the beam. The main pulse profile is sometimes followed by a secondary pulse called ``interpulse'', separated by 180$^{\circ}$ from the main pulse and likely due to a secondary emission beam originate from the opposite magnetic pole.

The radio PSR emission is characterized by a strong dependence of the emitted flux density on the observing frequency. This dependence can be described by a simple power-law ($S_\nu \propto \nu^{\alpha}$) with a steep spectral index $0\gtrsim\alpha\gtrsim-4$ \citep{maron00}. 

\subsection{Derivation of the Pulsar Physical Parameters}
\label{PSRparam}
The emission of a PSR does not come for free, but at the expenses of a large reservoir of kinetic rotational energy. This is the reason why PSR periods are observed to increase with time. The spin-down rate (${\dot P = dP/dt}$) is related to the rotational kinetic energy as follows \citep[see][for more details]{handbook}:
\begin{equation}
\label{erot}
{\dot E \equiv - \frac{dE_{rot}}{dt} = -\frac{d(I\Omega^2/2)}{dt}=-I\Omega \dot \Omega = 4\pi^2I \dot P P^{-3}    }
\end{equation}
where ${\dot E}$ is the ``spin-down luminosity'' and represents the kinetic energy output emitted by the NS. Assuming a typical moment of inertia  ${I=10^{45}}$ g cm$^{2}$, we find:
\begin{equation}
{\dot E \simeq 3.95\times 10^{31} \; \mathrm{erg \; s^{-1}} \;  \left( \frac{\dot P}{10^{-15}}\right) \left( \frac{P}{\mathrm{s}}\right)^{-3}  }
\end{equation}
Only a very small fraction of this spin-down luminosity is actually converted into radio emission. Most of it is converted into high energy radiation and relativistic wind of particles that heavily interact with the interstellar medium (the so-called ``PSR wind nebulae'') or, if present, with a companion star \citep{kirk09}. \\

As we will see in the following, PSRs have strong dipole magnetic fields. A rotating magnetic dipole with a moment |{\bf m}| emits electromagnetic waves at its rotation frequency. The radiation power is:
\begin{equation}
\label{edipole}
{ \dot E_{dipole} = \frac{2}{3c^3}|{\mathbf m}|^2 \Omega^4 \sin^2\alpha }
\end{equation}
where $\alpha$ is the angle between the magnetic and the spin axis. Equating equation~(\ref{erot}) to equation~(\ref{edipole}), we find:
\begin{equation}
\label{omegadot}
{\dot \Omega = - \left( \frac{2|{\mathbf m}|^2 \sin^2\alpha}{3Ic^3}\right)\Omega^3 }
\end{equation}
which shows the expected spin period evolution. By expressing this equation using a simple power-law and in terms of the rotational frequency $\nu$, this relation becomes:
\begin{equation}
\label{nudot}
{ \dot \nu = -K \nu^n}
\end{equation}
where $n$ is called ``breaking index'' and $K$ is assumed to be a constant. For a pure magnetic dipole breaking, the index is $n=3$. However, other dissipation mechanisms, such as the PSR wind, may carry away part of the rotational kinetic energy. The measured values of $n$ varies from 1.4 to 2.9 \citep{kaspi02}, confirming that the pure magnetic dipole breaking is just a poor assumption. \\

Rearranging equation~(\ref{nudot}) in terms of the spin period derivative, a first order differential equation is obtained. Integrating this equation, assuming a constant $K$ and $n\neq1$, a simple relation for the PSR age can be obtained:
\begin{equation}
{ T = \frac{P}{(n-1)\dot P} \left[ 1 - \left( \frac{P_0}{P} \right)^{n-1} \right] }
\end{equation}
where ${P_0}$ is the PSR birth spin period. Assuming ${P_0<<P}$ and ${n=3}$, this equation simplifies in the so-called ``characteristic age'':
\begin{equation}
\label{tage}
{\tau_c \equiv \frac{P}{2\dot P} \simeq 15.8 \; \mathrm{Myr} \; \left( \frac{P}{\mathrm{s}} \right) \left( \frac{\dot P}{10^{-15}} \right)^{-1} }
\end{equation}
Such an age is easily measurable through observations. However, given the assumptions, its value is not a very reliable measurement of the true PSR age. Thus it should be treated carefully and only as a rough estimate. Rearranging the two latter equations, we find an equation for the spin period at birth:
\begin{equation}
{P_0 = P \left[ 1 - \left(\frac{n-1}{2}\right) \frac{T}{\tau_c} \right]^{\left( \frac{1}{n-1} \right)}}
\end{equation}
Therefore, if the breaking index is measured and the true age of the PSR is independently determined (e.g. by measuring the age of an associated supernova remnant), the PSR spin period at birth can be inferred. The range of possible PSR birth periods is still unclear. Some estimates suggest a broad range of values from tens to hundreds of milliseconds \citep[e.g.][]{migliazzo02,kramer03}.\\

PSRs are highly magnetized systems. Although direct measurements of the magnetic field strength are not possible, an estimate of its intensity can be obtained from equation~(\ref{omegadot}), since the magnetic field strength is related to the magnetic moment by the relation ${B\approx|{\mathbf m}|/r^3}$. It follows that the magnetic field strength at the NS surface (${B_S}$) is:
\begin{equation}
{ B_S = \sqrt{{\frac{3Ic^3}{8\pi^2R^6 \sin^2\alpha} P \dot P}  }  }
\end{equation}
Using the same typical $I$ as before, a typical radius ${R=10}$ km and assuming $\alpha=90^{\circ}$, we find:
\begin{equation}
\label{bsurface}
{ B_S = 3.2\times10^{19} \; G \; \sqrt{{P \dot P}} \simeq 10^{12} \; \mathrm{G} \left(\frac{\dot P}{10^{-15}}\right)^{1/2} \left(\frac{P}{\mathrm{s}}\right)^{1/2}   }
\end{equation}
As in the case of the characteristic age, given all the assumptions, this equation does not provide a precise measure of the true surface magnetic field, but only a rough estimate. As we shall see in Section~\ref{evolution}, this characteristic surface magnetic field can cover a broad range of values from $\sim10^8$ G to $\sim10^{15}$ G.

\section{Pulsar Magnetosphere and Emission Mechanism}
\begin{figure}
\centering
\includegraphics[width=10cm]{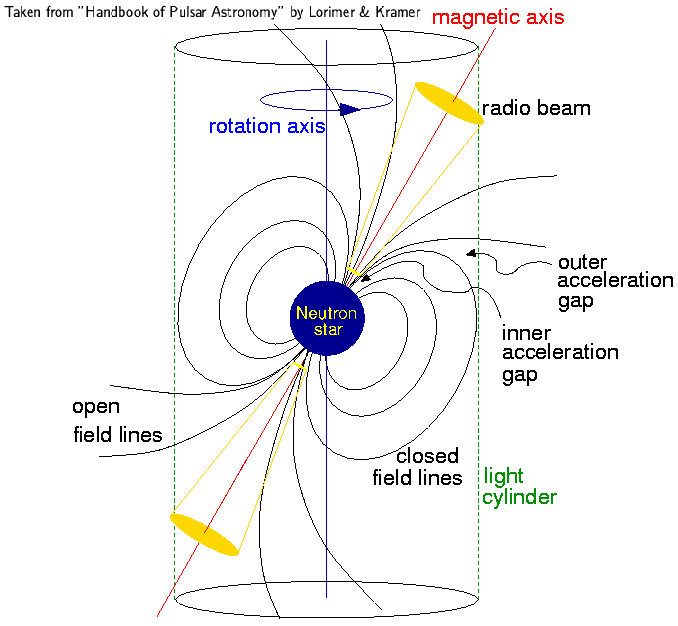}
\caption[Schematic representation of a NS and its magnetosphere]{Schematic representation of a rotating NS and its surrounding magnetosphere}
\label{magnetosphere1}
\end{figure}

\begin{figure}
\centering
\includegraphics[width=10cm]{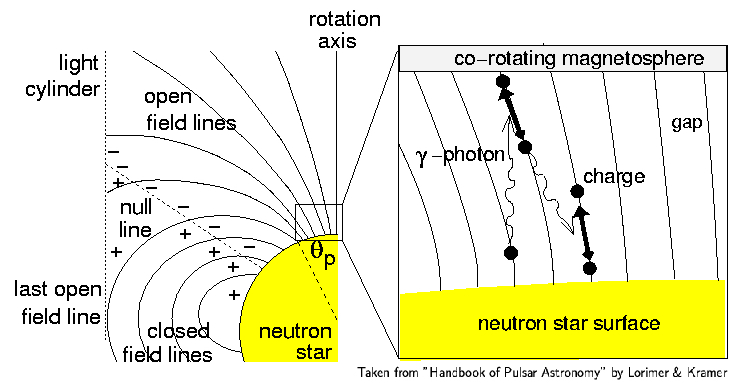}
\caption[PSR magnetosphere in the Goldreich-Julian model.]{PSR magnetosphere according to the Goldreich-Julian model.}
\label{magnetosphere2}
\end{figure}
In Section~\ref{interiors} we saw that the NS surface and atmosphere are composed of charged particles. These are subject to Lorentz forces much stronger than the gravitational ones. In a rotating and magnetized system such as a NS, this leads to the formation of a strong electric field able to extract plasma from the surface, which surrounds the NS following the magnetic field lines and creating the ``PSR magnetosphere''. A simple, yet very educative model of this phenomenon has been presented by \citet{goldreich69} shortly after the discovery of the first PSR. This model assumes that the spin and magnetic axes of the PSR are aligned and shows that in any point of the sphere, the magnetic field {\bf B} induces an electric field ${\mathbf{E}=(\boldsymbol{\Omega} \times \mathbf{r}) \times \mathbf{B}}$. In a perfectly conductive sphere this will be balanced by a distribution of charges that create an electric field able to grant, at any point ${\mathbf{r}}$, a force-free state. This can be mathematically expressed as follows:
\begin{equation}
\label{forcefree}
{ \mathbf{E} + \frac{1}{c} \left(\boldsymbol{\Omega}\times \mathbf{r} \right) \times \mathbf{B} = 0}
\end{equation}
Under the assumption that there is vacuum outside the sphere, the surface charges induce a quadrupole moment that corresponds to a surface electric field:
\begin{equation}
{ E_{||} = \left(  \frac{ \mathbf{E} \cdot \mathbf{B} }{B}  \right)_{r=R} = - \frac{\Omega B_S R}{c} \cos^3{\theta}   }
\end{equation}
This field induces an electric force on charged particles that is stronger than the gravitational pull. As a consequence, charges are easily extracted from the surface to create a surrounding dense plasma. The charge distribution around the NS is arranged in such a way that the electric field ${E_{||}}$ is shielded and thus the force-free state (see equation~\ref{forcefree}) is maintained also outside the star. This plasma outside the NS is subjected to the same electromagnetic field as the NS interior and thus the charged particles are forced to rigidly co-rotate with the star. Co-rotation can however occur up to a distance where the plasma is rotating at a speed equal to the speed of light. This co-rotation limit defines a radius known as ``light cylinder radius'' and is given by:
\begin{equation}
{ R_{LC} = \frac{c}{\Omega}=\frac{cP}{2\pi}  \simeq 4.77\times10^{4} \; \mathrm{km} \; \left( \frac{P}{\mathrm{s}} \right)   }
\end{equation}
The existence of such a limiting radius divides the magnetosphere into two distinct regions as shown in Figures~\ref{magnetosphere1} and \ref{magnetosphere2}: the ``closed field lines region'' where the magnetic field lines close within the light cylinder and the ``open field lines region'' where the magnetic field lines cannot close. 

The presence of the open field line region provides the physical ground for the NS to emit the pulsated radiation. Indeed, the radio beam, responsible of the pulsed signal, is confined within this region. The exact physical mechanism that produce the PSR emission is still poorly understood. However, the generally accepted model identifies the emitting region as a cone-shaped beam centered on the magnetic poles, within a so-called ``inner acceleration gap'' (see Figures~\ref{magnetosphere1} and \ref{magnetosphere2}). Here, the plasma extracted from the surface by a residual electric field, is accelerated along the open field lines, thus emitting photons in a direction tangential to that of the open field line at the point of emission. This mechanism is usually called ``curvature emission'' \citep{komesaroff70} and the very high brightness temperatures measured imply that this mechanism is a coherent process, at least at radio wavelengths. Briefly, the charges extracted from the NS surface are accelerated, reaching relativistic energies ($\gamma\lesssim10^7$), along the magnetic field lines where they emit $\gamma$-ray photons by curvature emission or inverse Compton scattering with lower-energy photons. These photons, due to the presence of the strong magnetic field, can split into an electron-positron pair if the initial energy is larger than twice the electron rest mass. In such a way, a new generation of particle is created and it will emit high energy photons and create pairs again. The net result is the creation of an avalanche of secondary pair plasma. This phenomenon results in the multiplication by a factor up to $10^4$ of the initial plasma density and it is believed that the acceleration of this secondary plasma produces the commonly observed radio emission \citep[see][and reference therein, for more details]{handbook}. As a consequence, the opening angle of the radio beam is a function of both the width of the open field line region and the height where the emission occurs. However, it is worth reminding that the observed pulse shape is also a function of the structure of the emitting regions within the cone and of the portion of the emission cone crossed by the observer line of sight.

For the sake of completeness, we specify that the PSR pulsated emission has been observed also at higher energies, such as optical, X-ray and $\gamma$-ray wavelengths. The emission mechanism in these bands is quite different from the radio one and likely located, as well as in the inner acceleration gap, also in an outer acceleration gap: a region in the outer magnetosphere close to the light cylinder (see Figure~\ref{magnetosphere1}). High energy photons from these outer gaps are probably created through synchrotron and curvature emission (in a lower $B$ regime, with respect to the radio emission case), while high energy photons from the inner gap are likely due to inverse Compton scattering.

\section{Pulsar Evolution}
\label{evolution}

\begin{figure}
\centering
\includegraphics[width=10cm]{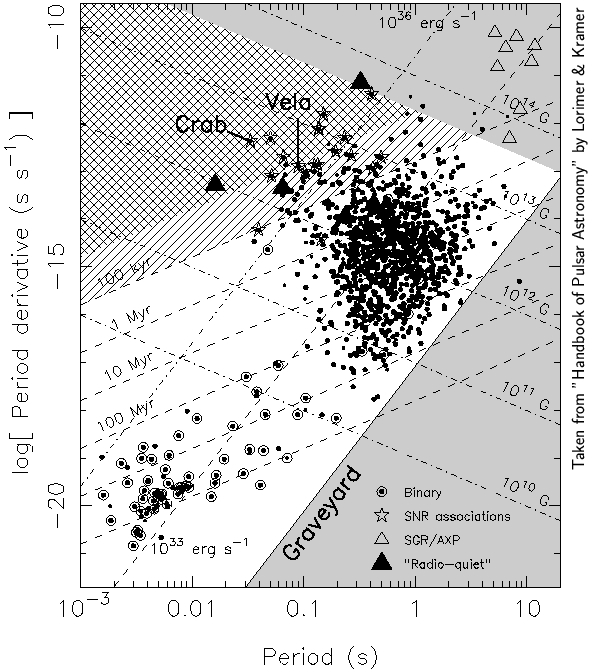}
\caption[The $P-\dot{P}$ for a large sample of radio PSRs.]{The $P-\dot{P}$ for a large sample of radio PSRs. Binary systems are highlighted with circles. PSRs associated to supernova remnants are plotted  as empty stars, magnetars as empty triangles and radio quiet PSRs as filled triangles. Lines at constant magnetic field, characteristic age and spin-down luminosity are also plotted. The gray shaded area is the region called ``graveyard'' where the PSR is expected to be no more observable.}
\label{ppdot}
\end{figure}

The evolution of a PSR can be described through a spin period vs spin period derivative ($P-\dot{P}$) diagram, reported in Figure~\ref{ppdot} \citep{handbook}. From the analysis of the plot, at least two distinct families of PSRs can be distinguished. The bulk of the PSR population is located at ``long'' spin periods, centered around 0.5 s, and spin-down rates of about $10^{-15}$: these are classical isolated NSs, spinning down after being formed in a supernova event. They are characterized by typical magnetic field of $10^{11-13}$ G and ages of about $10^{6-7}$ yr.
A second family can be identified in those objects with extremely short spin periods, around 5 ms, and low spin-down rates around $10^{-20}$. Given their typical spin periods, these objects are called ``millisecond pulsars" (MSPs). From equation~(\ref{tage}) is it clear that this population is composed of PSR older than the bulk of the population, being older  than $10^8$ yr and, on the basis of equation~(\ref{bsurface}), having low magnetic field strength of about $10^{8-9}$ G. Furthermore, the vast majority of them is located in binary systems. Indeed, as we will see in the following, their formation requires the presence of a companion star. \\

According to the most plausible evolutionary scenarios  \citep{handbook}, PSRs are born with small spin periods around tens of milliseconds and large spin-down rates ($\lesssim10^{-15}$), thus located in the upper left section of the $P-\dot{P}$ diagram. While they slow down, losing rotational kinetic energy,  they move toward the bulk population with period around 0.5 s. This evolution is likely accompanied by a decay of the magnetic field strength and thus a decrease of the spin-down rate. The slow down process keeps moving the PSR toward the rightest regions of the diagram where they eventually cross the so-called ``death line" and enter the ``PSR graveyard". This is a region where the combination of spin periods and spin-down rates is not suitable to provide an efficient emission mechanism and therefore the NS is no more able to emit as a PSR.
 
The bulk of the PSR population is located close to the death line. This is because the evolutionary time scales from the birth region to this region of the diagram is quite fast, the typical timescale being of about $10^{5-6}$ yr. Indeed, young PSRs have short spin periods and strong magnetic fields that allow an efficient energy emission, and thus a rapid slow down. When they reach spin periods larger than $\sim0.5$ s, this is not true anymore and the evolution in the $P-\dot{P}$ diagram is slowed down. After $10^7$ yr the emission eventually becomes too weak to be detectable and the PSR moves toward and beyond the death line.\\

Across the death line, we should be no more able to see the PSR emission. However, if the dead NS is located in a binary system, accretion phenomena can transfer mass and angular momentum from the companion to the NS. This can spin-up the NS till periods of the order of milliseconds, where the emission mechanism can be reactivated and the NS can be visible again in the radio band as a MSP. The same mechanisms are also thought to be responsible of the burying of the magnetic field \citep{bisno74}: this is why MSPs are observed in the lower left region of the diagram, where objects with very low spin-down rates (and thus weak magnetic field) are located.\\

In the $P-\dot{P}$ diagram it can be also distinguised a third family of objects, composed by PSRs with long spin periods but very high spin-down rates (of $\sim10^{-11}$) corresponding to strong magnetic field strengths (around $10^{14-15}$ G). These objects are called ``magnetars" or in some cases ``Anomalous X-ray Pulsars'' or ``Soft Gamma-ray Repeaters'' and they are characterized by high energy emission \citep{kaspi17}. A description of this family of PSRs is beyond the goals of this thesis.

\section{Millisecond Pulsars}
\label{msp}
As we saw in the previous section, MSPs represent an old and evolved population of NSs, resulting from mass accretion from a companion star onto a dead NS.

The first MSP, 4C21.53, was discovered in 1982 by \citet{backer82} and \citet{alpar82}. It was long thought that this source could host a NS as primary star. However, no pulsations were discovered till the search was moved to very short spin periods, in a region where no PSR was known. In this way, it was discovered a 1.6 ms pulsation from this source, confirming its PSR nature, but with an unprecedented short spin period. To date, $\sim400$ MSPs are known\footnote{See \url{https://apatruno.wordpress.com/about/millisecond-pulsar-catalogue/}}.\\

\begin{figure}[h]
\centering
\includegraphics[width=10cm]{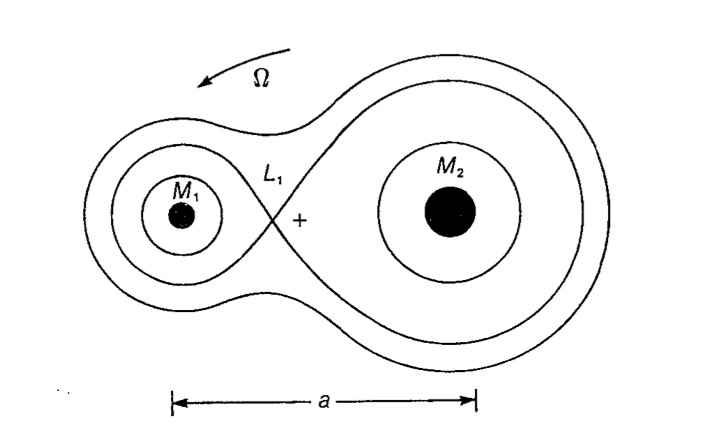}
\caption[Surfaces of equal gravitational potential in a binary system.]{Surfaces of equal gravitational potential in a binary system, as seen from the equatorial plane. The two drop-like structures intersecting at the Lagrange point $L1$ are called Roche Lobes of the stars with masses ${M_1}$ and ${M_2}$, rotating around the center of gravity with angular velocity $\Omega$.}
\label{rochelobe}
\end{figure}

The formation of a MSP begins in a binary system where the massive star explodes as a supernova, creating a NS. This phenomenon can release an amount of energy large enough to unbound the binary. However, it can be shown, starting from the virial theorem and assuming spherical symmetry, that the binary system survives the explosion if the total amount of mass ejected during the supernova event is less than half the total mass of the binary system. After the explosion, if the binary system survives, it likely is on an eccentric orbit and the NS is visible as a classical PSR for about $10^{6-7}$ yr.
The accretion phenomenon usually starts when the companion star evolves from the main sequence stage and becomes a red giant star. If in this stage the companion fills its Roche Lobe\footnote{The Roche Lobe is defined as the region, around a star in a binary system, within which orbiting material is gravitationally bound to that star.} (Figure~\ref{rochelobe}), transfer of material from the surface of the companion toward the NS can occur through the stellar wind or, more commonly, through the Lagrange point $L1$ (see Figure~\ref{rochelobe}). The mass transfer onto the NS implies also the transfer of angular momentum, thus resulting in the acceleration of the compact object. A limiting spin period is reached thanks to the equilibrium between the magnetic pressure of the accreting NS and the ram pressure of the in-falling matter \citep{batta91}. Its value is usually of few milliseconds and hence a MSP is formed. This phenomenon allows us to observe old NSs, whose emission should not be observable, having passed the death line. This is why MSPs are also known as "recycled PSRs" and the accretion/reacceleration stages as "recycling scenario''.\\

Aside from the dominant radio pulsed emission, MSPs are also strong high energy emitters, both in the X-ray and in the $\gamma$-ray regimes \citep{rutledge04,bogdanov11,xing16}. The high energy emission can be due both to thermal and non thermal processes, like the emission of photons in the surroundings of the MSP magnetosphere or to intra-binary shocks between the relativistic wind of particles emitted by the MSP and the material lost by the companion star.
On the other hand, the optical emission of MSPs is dominated by the companion stars \citep[e.g.][]{ferraro01a,fer03_msp,stappers01,breton13} and thus optical observations are the only window to characterize the binary system properties from the secondary star point of view and to study the physics and the evolution of stars under the extreme conditions due to the presence of a NS. 

\subsection{Canonical Formation Scenarios}

Canonical formation scenarios are quite branched and include many kind of MSPs. A simple but illustrative classification can be done on the basis of the companion mass. This quantity is not always easily measurable. In some cases its value can be constrained by the identification of the companion star in the optical bands while in the case of relativistic binary systems, PSR timing can provide precise mass measurements of both the primary and the secondary stars. \\
\begin{figure}
\centering
\includegraphics[width=13cm]{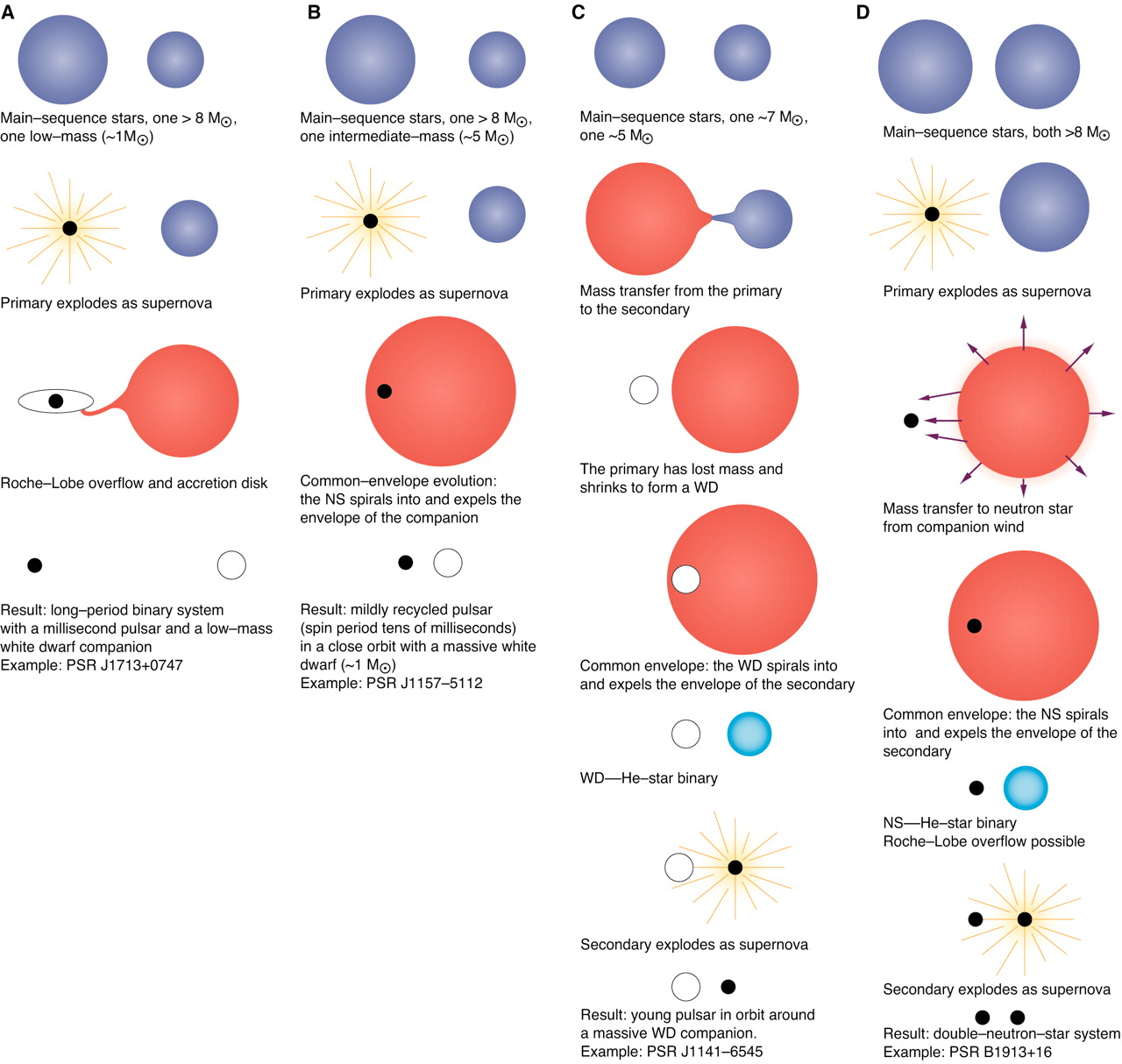}
\caption[MSPs formation scenarios.]{Schematic representation of  MSP canonical formation scenarios, taken from \citet{stairs04}. Panel A shows the evolutions of MSPs with low-mass companion stars, panel B and C the evolution of MSPs with intermediate-mass companions and finally panel D the evolution of MSPs with high-mass companion stars. }
\label{stairs}
\end{figure}
A schematic representation of the different possible evolutionary scenarios is presented in Figure~\ref{stairs} (see \citealt{stairs04}). As can be seen, following the formation of the primary PSR and the subsequent accretion stages, the secondary object becomes an inactive system from the thermonuclear reactions point of view, usually a WD. We can distinguish three main MSP families on the basis of the companion mass:
\begin{itemize}
\item MSPs with low-mass companion stars: it is the most common class and the companion stars are usually WD with a He core and mass $<0.5 \; \Msun$.
\item MSPs with intermediate-mass companion stars. They usually have massive C-O WD companions with masses $\sim1 \; \Msun$.
\item MSPs with high-mass companion stars, where the latter has a mass larger than $1 \; \Msun$, typically another NS.
\end{itemize}

\subsubsection*{MSPs with low-mass companion stars}
The vast majority of binary MSPs belongs to this class. They are formed from binary systems with a low-mass ($\sim1 \; \Msun$) secondary star. This scenario is depicted in section A of Figure~\ref{stairs}. Once the NS is formed and the secondary is evolved to the red giant phase, mass accretion can occur via Roche-Lobe overflow, through the formation of an accretion disk. 

The accretion stages are commonly observed in systems called ``Low-Mass X-ray Binaries'' (LMXBs)\footnote{For the sake of completeness, it is worth mentioning that it also exist a sub-class of LMXB with a stellar mass black hole as primary star with different properties with respect to those with a NS primary.}, whose intense X-ray radiation is due to the accretion disk \citep{tauris06}. These systems have been long thought to be the progenitor of MSPs and this connection has been recently confirmed by the discovery of a new class of binary systems, the so-called ``transitional MSP'' (tMSPs), which we will describe in Section~\ref{accreting}. The ``low-mass" prefix is used to distinguish them from the class of ``High-Mass X-ray Binaries", with more massive companion stars, that are thought to be the progenitors of double PSR systems (\citealt{stairs04}, scenario D of Figure~\ref{stairs}).  

This phase can last as long as $\sim1$ Gyr and thus the final result is a rapidly spinning MSP with spin periods usually shorter than 10 ms, and a companion star whose envelope has been completely stripped out, thus a low-mass WD ($\sim 0.1 \; \Msun-0.2 \; \Msun$) with a He core or, more rarely, with a C-O core and an outer envelope of H and He.

\subsubsection*{MSPs with intermediate-mass companion stars}

They are the outcome of the evolution of binary systems where the secondary star has a mass in the range $3 \; \Msun-6 \; \Msun$. The evolutionary scheme is reported in sections B and C of Figure~\ref{stairs}. In the former case, the primary star creates a NS and when the secondary evolves to be a red giant, a common envelope phase allows the NS to accrete material from the secondary and thus to spin the NS up. At the end, we find a mildly recycled MSP (spin periods of tens of ms) with a massive WD with a C-O core \citep[see, e.g.,][]{lorimer06,pallanca13b}. In the latter scenario, both the stars have masses lower than $\sim8 \; \Msun$. The more massive one firstly becomes a red giant and starts to transfer mass to the secondary star via Roche-Lobe overflow. The primary becomes a massive WD with a C-O or O-Mg core. When the secondary evolves, a common envelope phase allows mass transfer from the secondary to the primary. Finally the secondary star, which has accreted matter in the early stages and lost its envelope in the later ones, explodes as a supernova and a young PSR is formed \citep[see, e.g.,][]{kaspi00}.

\subsubsection*{MSPs with high-mass companion stars}

It is the scenario depicted in section D of Figure~\ref{stairs}. Briefly, two main sequence stars, both with masses larger than $8 \; \Msun$, evolve. When the first NS is formed and the second one reaches the red giant phase, mass transfer onto the NS can occur via the stellar wind lost by the secondary star. After a common envelope phase, where the accretion can still occur, the secondary star explodes as a supernova and a second NS is formed. These systems are usually characterized by large spin periods ($\geq20$ ms) and eccentric orbits ($0.1\lesssim e \lesssim 0.9$). Double NS systems are known, like PSR B1913+16 or B2127+11C \citep{jacoby06,weisberg10}, but only in one case the pulsations have been detected from both the NSs \citep{burgay03}. All of them are highly relativistic systems and thus suitable to be used as probes of general relativity \citep[e.g.][]{hulse75}.

\subsection{The Spiders: Black-Widows and Redbacks}
\begin{figure}[b]
\centering
\includegraphics[width=10cm]{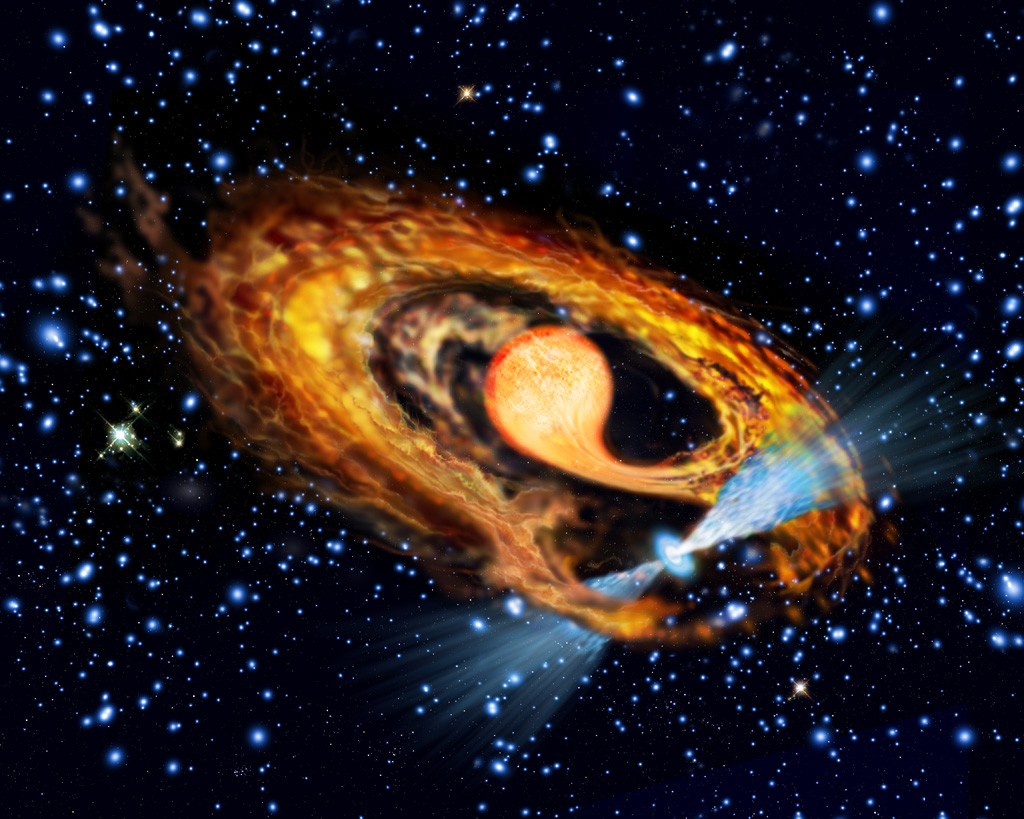}
\caption{An artistic representation of a spider MSP system. Credits: Ferraro F. R., ESO.}
\label{artisticRB}
\end{figure}

There is another class of MSPs with low-mass companion star, whose properties are not predicted by the canonical evolutionary scenario described before. These systems are named ``spiders'' and classified in two species: Black-Widows (BW) and Redbacks \citep[RB, see, e.g.,][]{fst88,ruderman89,ray12,roberts13}. They are also called ``eclipsing MSPs'' since they are characterized by periodical eclipses of the radio signal lasting for a significant fraction of the orbit, usually during the PSR superior conjunction. This phenomenon cannot be due to the occultation of the MSP by the companion star, but to absorption or scattering between the radio photons and ionized material located between the emitter and the observer. This material is likely produced from the ablation of the companion star surface by the relativistic wind of particles and high energy radiation emitted by the re-activated PSR (and so the name of the two murderous spiders). An artistic representation is reported in Figure~\ref{artisticRB}.\\

\begin{figure}
\centering
\includegraphics[width=10cm]{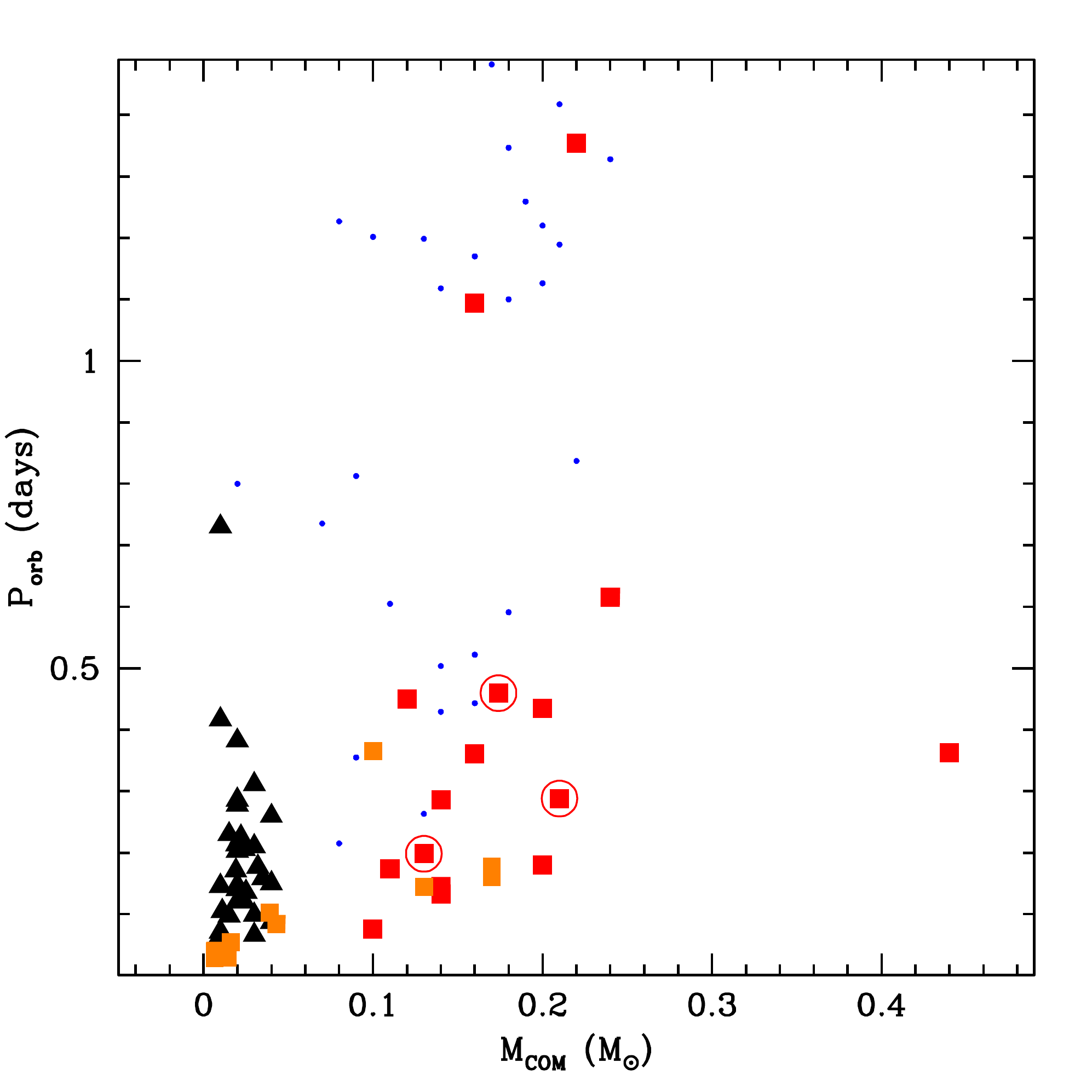}
\caption[Orbital period as a function of the companion mass for a sample of MSPs.]{Orbital period as a function of the companion mass for a sample of MSPs. Blue points mark the position of a small sample of canonical MSPs with He WD companion stars. Black triangles, red and orange squares mark the positions of BWs, RBs and AMXPs, respectively. Finally, the three open circles around the red squares identify the three tMSPs.}
\label{porbm2}
\end{figure}

Among a population of 402 MSPs, 22 RBs and 40 BWs are known so far\footnote{see \url{https://apatruno.wordpress.com/about/millisecond-pulsar-catalogue/}}. The distinction between the BW and the RB classes is made on the basis of the companion star mass: while RB companions can have masses in the range $0.1 \; \Msun-0.5 \; \Msun$, BWs have extremely low-mass companion stars, of the order of $\leq0.05 \; \Msun$ (see Figure~\ref{porbm2}). Both the kinds of companions are expected to be non-degenerate or, at least, semi-degenerate stars, as observationally confirmed by their optical identification \citep[e.g.][]{stappers01,pallanca10,breton13}. Indeed, in the case of a WD companion, the surface gravity would be too strong and the projected radius too small to produce such a mass loss from its atmosphere. The typical orbital periods of spider MSPs can cover a range from few hours to a day (Figure~\ref{porbm2}) and so they always present very compact orbits with no measurable eccentricities, as a result of the strong tidal interaction between the two components of binary system. A strong tidal locking also forces the companion star to rotate around his rotational axis with the same period of the orbit and thus, as in the Earth-Moon case, the same region of the star is constantly exposed to the MSP.  Interestingly, spider MSPs are X-ray and $\gamma$-ray emitters \citep[e.g.][]{bogdanov05,ray12}. The bulk of the high energy radiation is due to non thermal emission from an intra-binary shock formed between the PSR wind and the material outflowing from the companion star \citep{roberts14}.
 
 There are few cases of MSPs with an expected extremely low-mass companion that show no radio eclipses. The typical mass function of their companion star is around $10^{-5}\; \Msun$, while that of BW is usually $\geq3\cdot10^{-4} \; \Msun$. Since the mass function depends on the orbital inclination angle as ${\sin^3(i)}$, it is reasonable that these systems are BWs seen at low inclination angle \citep{freire05}. Indeed, the eclipsing material is expected to be able to intercept the radio beam and our line of sight only at high inclination angles. \\

The physical mechanisms leading to the formation of these systems are still debated. Simulations by \citet{chen13} show that RBs and BWs are the outcome of different evolutionary paths, where the PSR irradiation efficiency is the discriminant factor. At odds with these results, the simulations by \citet{benvenuto14} show that the evolution of RBs is bifurcated, with some of them evolving into BWs and the others producing canonical MSPs with a He WD companion. Possibly, the progressive evaporation of BW companions could lead to their total disruption and thus to the formation of isolated MSPs.

\subsection{Accreting and Transitional Millisecond Pulsars}
\label{accreting}
The bridge between LMXBs and MSPs is found in a class of peculiar binaries called ``Accreting Millisecond X-ray Pulsars'' (AMXPs). These are a subgroup of transient LMXBs that show, during outbursts, X-ray pulsations from a rapidly rotating NS. During these outbursts, the matter lost from the non-degenerate companion star via Roche-Lobe overflow is channeled down from a truncated accretion disk onto the NS magnetic poles, producing X-ray pulsations at the rotational frequencies ($\nu\geq100$ Hz)  of a weakly magnetized (${B\sim10^{8-9}}$ G) NS \citep[see][and references therein]{patruno12}.  

AMXPs emission is powered by the accretion itself, at odds with that of classical MSPs, powered by the NS rotation. Indeed, the presence of active accretion halts the radio emission mechanism (and indeed pulsations have never been detected in these systems, despite deep radio searches; \citealt{patruno09}) and the pulsations are visible thanks to the heating of the inner acceleration gap due to the presence of accretion flow. Such a flow is however a non steady phenomenon and indeed, during quiescence, X-ray pulsations have never been observed. All AMXPs are transient systems, running through cycles of outburst and quiescence with recurrence times different among systems and with durations from few days to several years. The spin frequencies of these systems confirm that the re-acceleration of the dead NS already took place during previous stages. 

The total number of AMXPs systems currently known is 20  \citep{patruno12,sanna17,sanna17b,strohmayer17}. All of them are compact or even ultracompact binaries with orbital periods usually much shorter than 1 day and companion stars with masses usually smaller than $1 \; \Msun$, in a similar fashion to that of spider MSPs (see Figure~\ref{porbm2}). Given the range covered by the orbital periods, these systems are unlikely to be the progenitors of canonical MSPs, while a tight connection with the spiders is obvious given their orbits and the properties of their companion stars.\\

The discovery of AMXPs has been an important validation for the MSP recycling scenario, probing that accretion of matter during the LMXB phase can indeed spin-up a slowly rotating NS. The final confirmation of this scenario arrived in 2013, almost 50 years after the discovery of the first binary PSR. \citet{papitto13} found that the transient X-ray source IGR J18245$-$2452 in the globular cluster (GC) M28 had an accretion outburst during March 2013, when it reached a luminosity larger than $10^{36}$ erg s$^{-1}$. They discovered that during this outburst the NS was emitting X-ray pulsations with a period of 3.9 ms. These have been used to obtain a timing solution that showed that the spin and orbital properties of this object were identical to that of PSR J1824$-$2452I, a RB system already known in the cluster. No radio pulsations have been detected during the outburst phase, while they reappeared less than two weeks after the end of the outburst, thus probing that this system was swinging between a pure accretion powered state, where it behaves like an AMXP, and a pure rotational powered state, where it behaves like an eclipsing radio MSP. These systems have been named ``transitional MSPs'' (tMSPs) and to date, only two more are known \citep{archibald09,demartino15}. All of them during the rotation powered state are classified as RBs (see Figure~\ref{porbm2}). The possibility that also BWs can swing to an accretion powered state is still open.

\section{Globular Clusters: the Ideal Millisecond Pulsar Factories}

GCs are gravitationally bound systems composed of up to $10^6$ stars and orbiting the Galaxy halo and bulge. The gravitational forces due to the ensemble of stars results in systems with spherical symmetry and high central densities. They are very old systems with typical ages of $\sim12$ Gyr and metallicities $\mathrm{-2.0\lesssim[Fe/H]\lesssim-1.0}$, but up to solar values for the systems in the Galactic bulge (e.g. Terzan~5). 

The central high densities are high enough that GC are the only place in the universe where dynamical interactions between stars occur in a timescale much shorter than the stellar ages. Therefore, the stars within the cluster, during their lifetime, are subjected to a large number of dynamical interactions such as exchange interactions, fly-byes, collisions, creation and disruption of binary systems. This results in the production of a large number of exotic objects such as MSPs, LMXBs, blue straggler stars and cataclysmic variables.\\

Although the absolute number of MSPs in the Galactic field is larger than that in GCs (252 vs 149), their number per unit mass is larger in the latter. In other words, $\sim40\%$ of MSPs are found in GCs, despite the Galaxy is $\gtrsim10^{3}$ times more massive than the entire GC system. This is easily explainable in a scenario in which MSPs in the field are the outcome of the evolution of primordial binary systems, while those in GCs can be the result of the evolution of both primordial systems and new binary systems formed later through dynamical interactions.
\begin{figure}
\centering
\includegraphics[width=10cm]{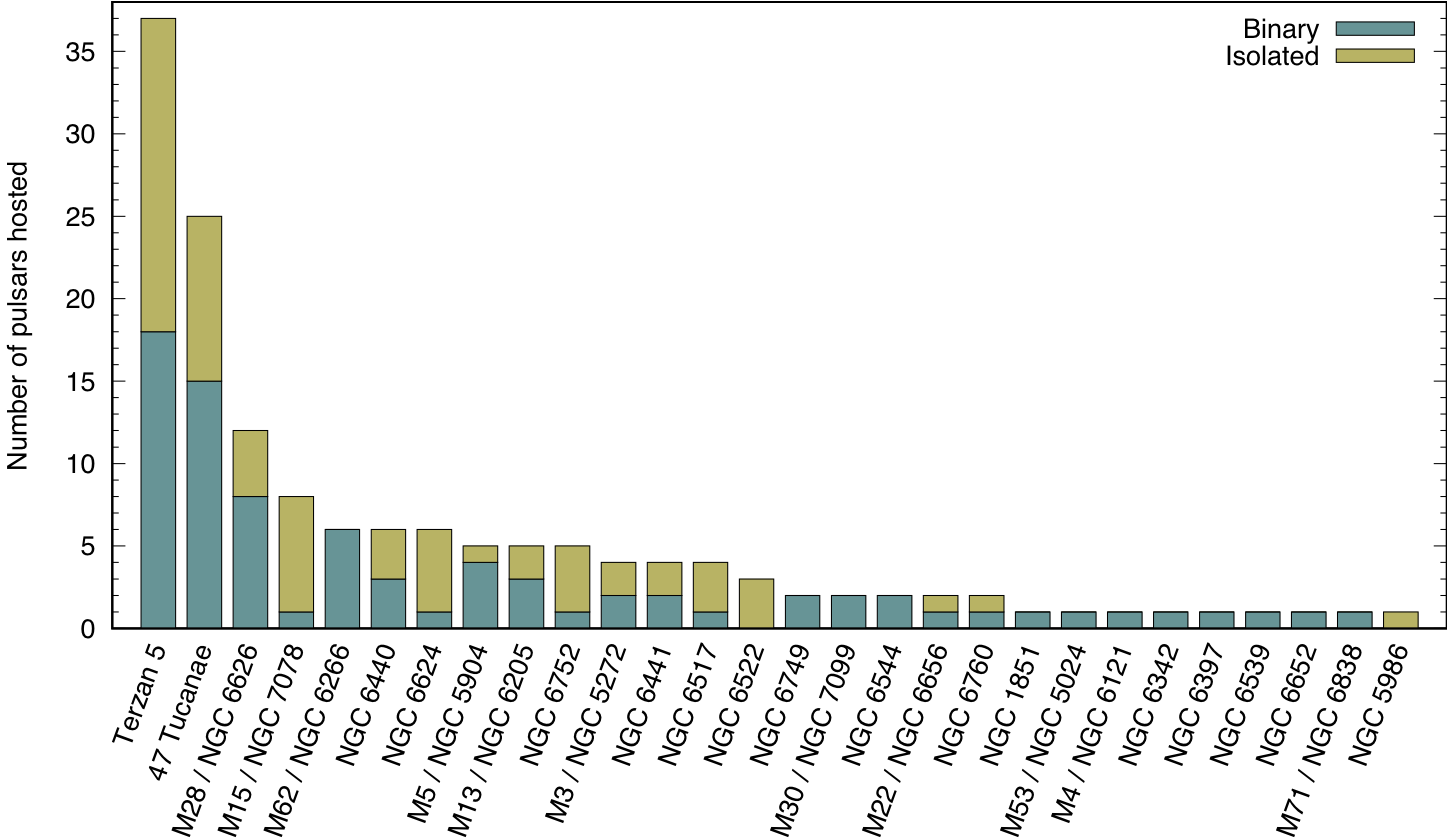}
\caption[Number of MSPs identified in each GCs.]{Number of MSPs identified in each GCs. The results shown in Chapter~\ref{cap_t5} are already included here. Courtesy of A. Ridolfi.}
\label{GCpsr}
\end{figure}
\begin{figure}
\centering
\includegraphics[width=7.cm]{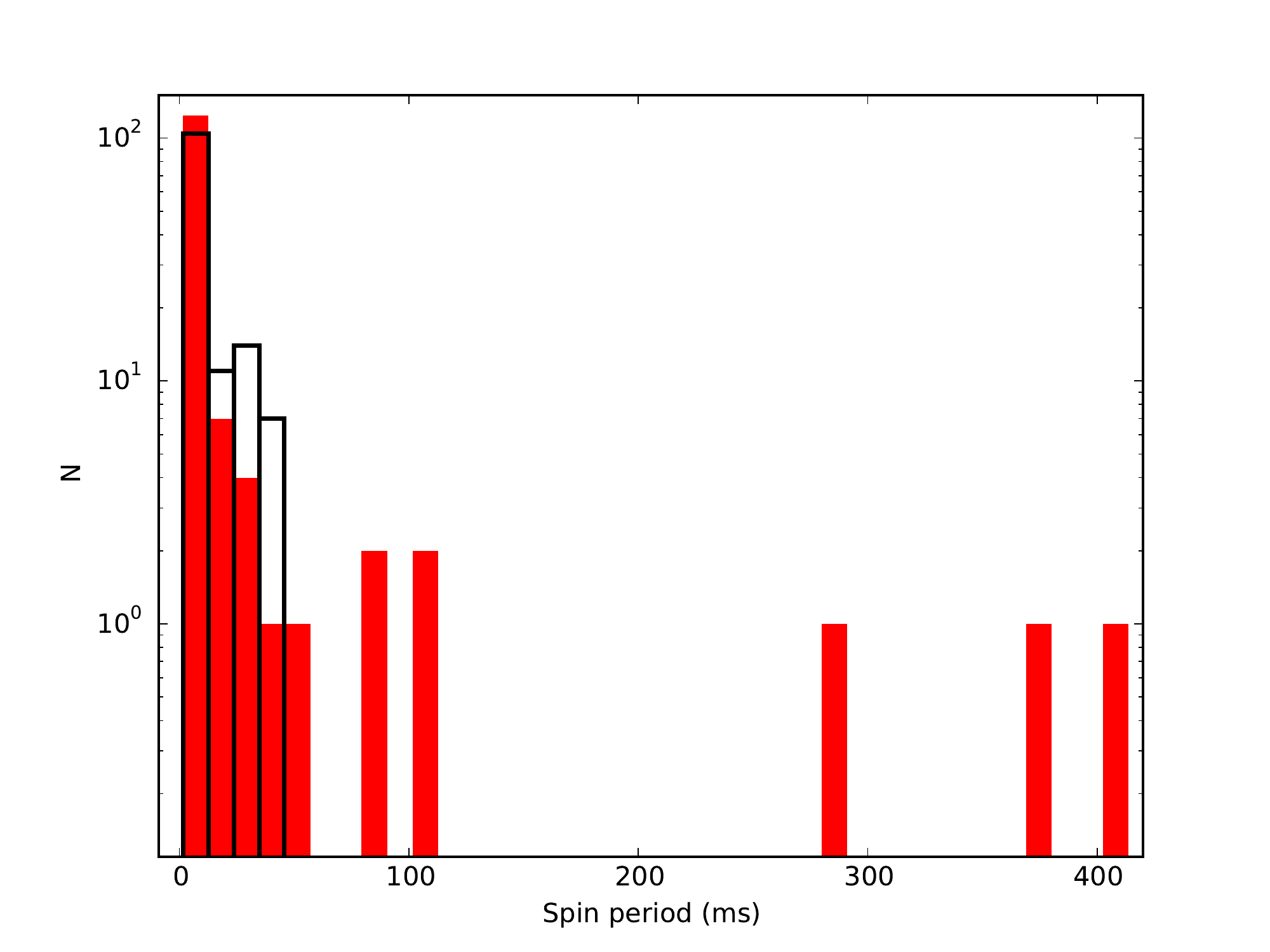}
\includegraphics[width=7.cm]{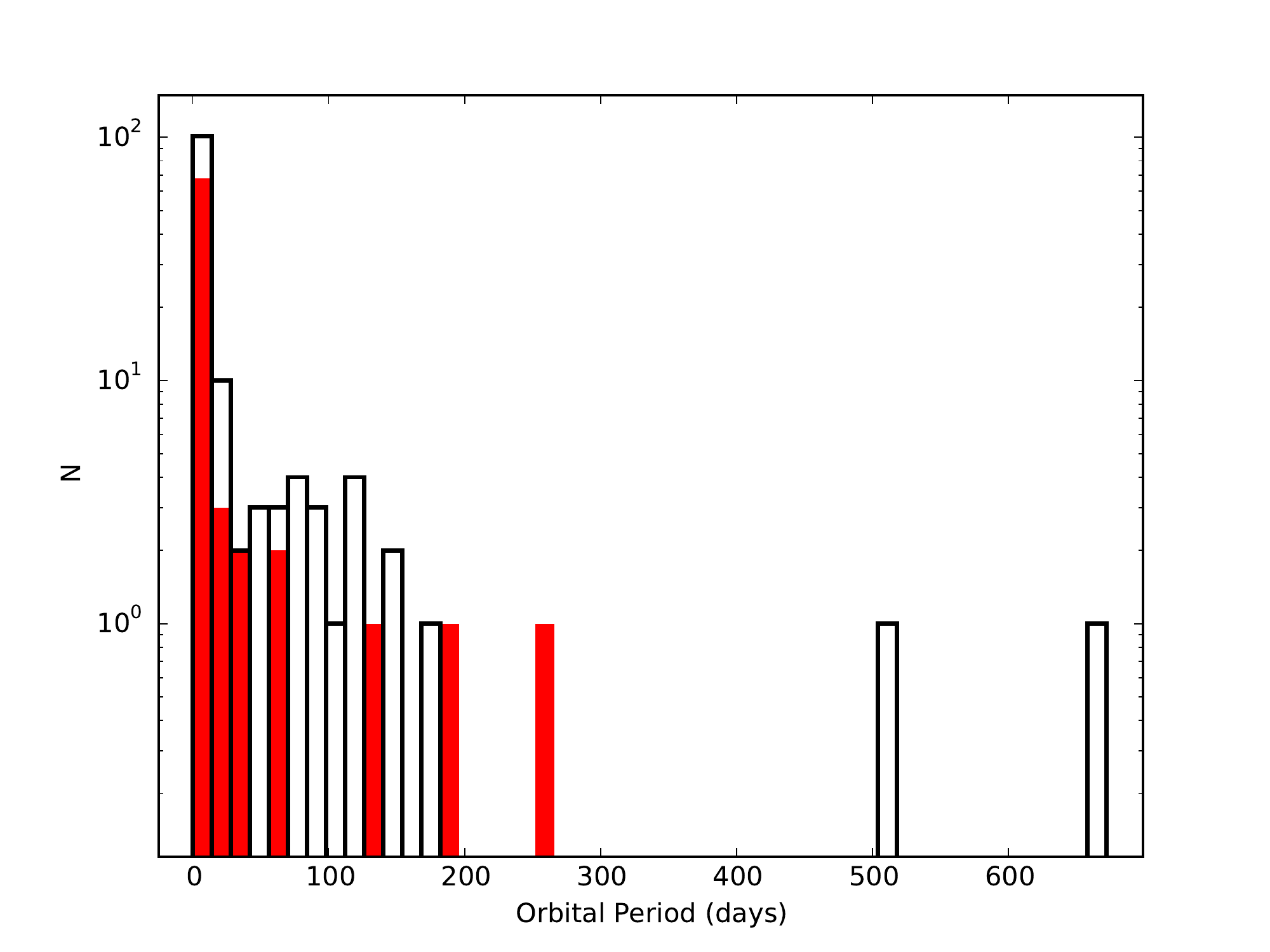}
\includegraphics[width=7.cm]{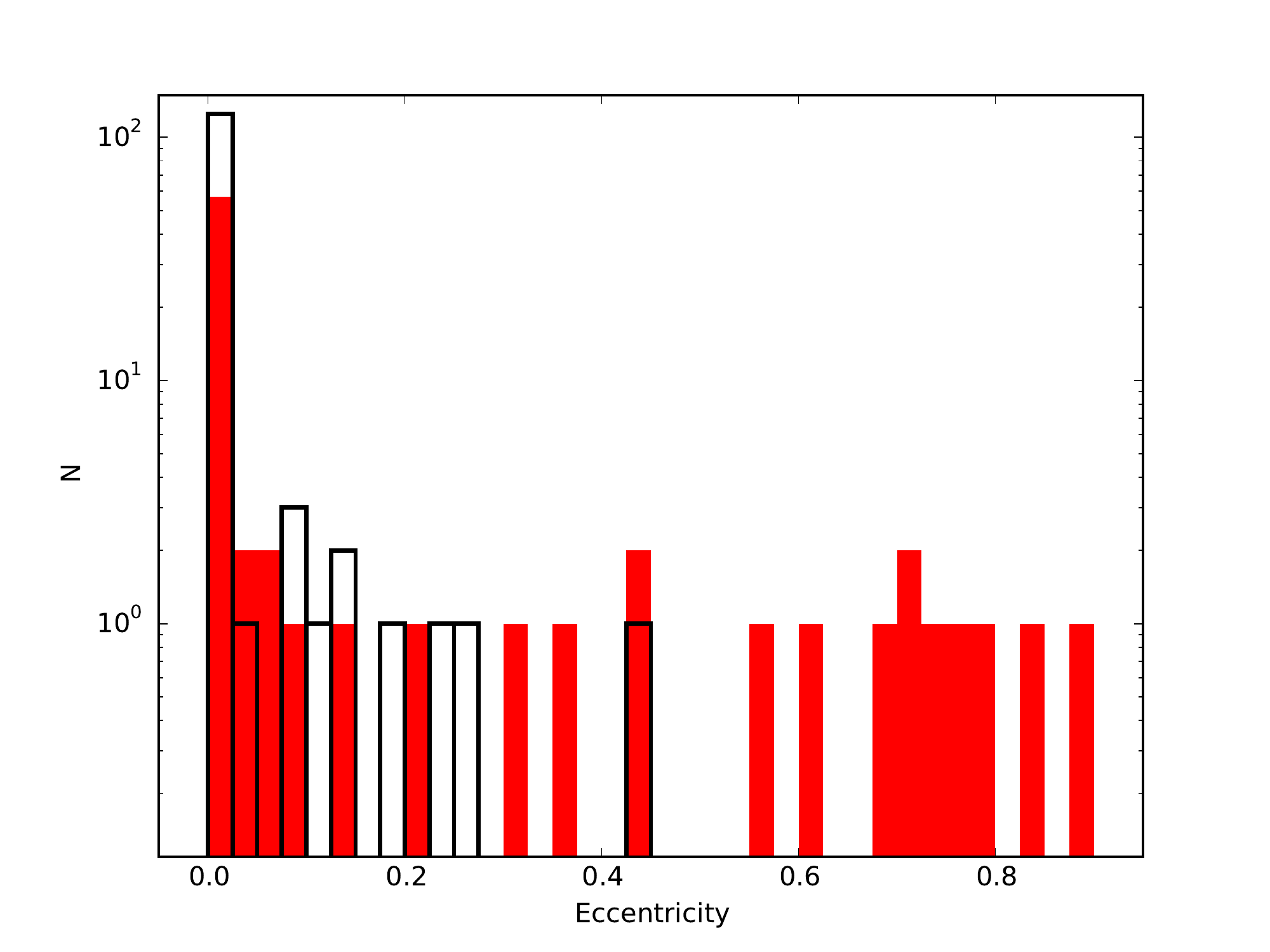}
\caption[Distribution of MSP spin periods, orbital periods and eccentricities.]{{\it Top left panel:} the red bars show the histogram of the spin periods of MSPs in GCs, while the empty black bars that of MSPs in the Galactic field. {\it Top right panel:} same as before but for the distribution of orbital periods. {\it Bottom panel:} same as before but for the distribution of eccentricities.}
\label{histoGC}
\end{figure}

Figure~\ref{GCpsr} shows the number of MSPs in each cluster. As can be seen, half of the population is located in only three GCs: Terzan~5, 47~Tucanae and M28. The first system alone hosts $\sim25\%$ of the entire population, thus suggesting an environment particularly suitable for creating new binaries. Most of these MSPs are likely the outcome the canonical evolutionary scenario, although the population properties are quite different than that in the Galactic field. As can be seen in Figure~\ref{histoGC}, the distribution of spin periods in GCs shows a distribution similar to that of Galactic field MSPs, but with a significant number of cases of long spin period objects, usually interpreted as mildly recycled systems, where the accretion has been halted by interactions with other stars. The distribution of orbital periods shows that GCs host more compact binaries than the Galactic field, as the result of energy exchanges with flying-by stars. Finally, the distribution of eccentricities shows that GCs host a  significant number of eccentric binaries. According to the canonical scenario, the accretion stages lead to the formation of MSPs in almost perfectly circularized orbits. The trend observed in GCs suggests that dynamical interactions often perturb binaries, forcing them on eccentric orbits. 

\subsection{Dynamical Interactions}
\begin{figure}
\centering
\includegraphics[width=8.5cm]{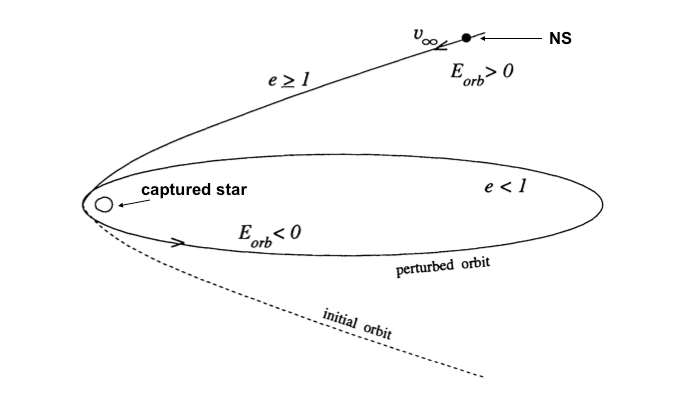}
\includegraphics[width=5.5cm]{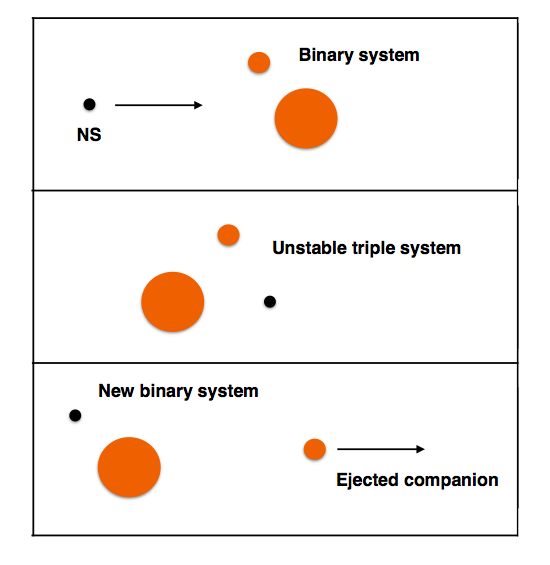}
\caption[Schematic representation of dynamical interactions in GCs.]{{\it Left panel:} A schematic representation of a tidal capture event. Taken from \citet{mardling96}. {\it Right panel:} Schematic representation of an exchange interactions between an isolated NS and a preexisting binary system.}
\label{dyna}
\end{figure}

We will briefly review here three main dynamical interactions that are likely to occur in the central regions of GCs and able to produce exotic binaries such as LMXBs and consequently MSPs.
\begin{itemize}
\item {\bf Tidal capture:} it occurs when an isolated NS, flying-by close to a main sequence star, menage to create a gravitational bond with it and thus a new binary system \citep[e.g.][see left panel of Figure~\ref{dyna}]{clark75} that can evolve through the scenario A depicted in Figure~\ref{stairs}. Simulations by \citet{ivanova08} suggest that only a small fraction ($\sim2\%$) of NSs is able to create a binary system in this way.
\item {\bf Direct collision:} it obviously occurs when a NS directly collide with another cluster star, eventually creating a new binary system. In a GC with 11 Gyr old stars, about $\sim5\%$ of NSs are expected to form a binary system through this mechanism \citep{ivanova08}.
\item{\bf Exchange interaction:} it is an interaction between a star and a preexisting binary system, where the less massive component of the binary is kicked out \citep[e.g.][]{hut83}. In the context of MSP evolution, this can happen in two different ways. In the first way, an isolated NS interacts with a binary system and, after the creation of an unstable triple system, the less massive companion of the triple is kicked out (see the right panel of Figure~\ref{dyna}). The remaining binary will eventually go through the LMXB phase. In the second way, an isolated star interacts with a binary system containing a NS, possibly already recycled, and its companion star, most commonly a low-mass He WD. If the perturber is more massive than the current companion star, the latter is expelled and the NS will obtain a new companion star, thus starting, or eventually re-starting, the recycling phase. Simulations by \citet{ivanova08} suggest that a large number of NSs (around $\sim20\%-25\%$) is able to create a binary system through this mechanism; this value can increase up to $\sim40\%-50\%$ in the very inner region of dense GCs.
\end{itemize}
Finally, there are other mechanisms, such as fly-byes \citep{hut83}, that can shrink or, at the contrary, force the binary systems on large eccentric orbits. Moreover, dynamical interactions between an isolated star and a preexisting binary system can also destroy the latter, without forming a new binary. This can explain the large fraction of isolated MSPs found in GCs.\\

\begin{figure}
\centering
\includegraphics[width=10cm]{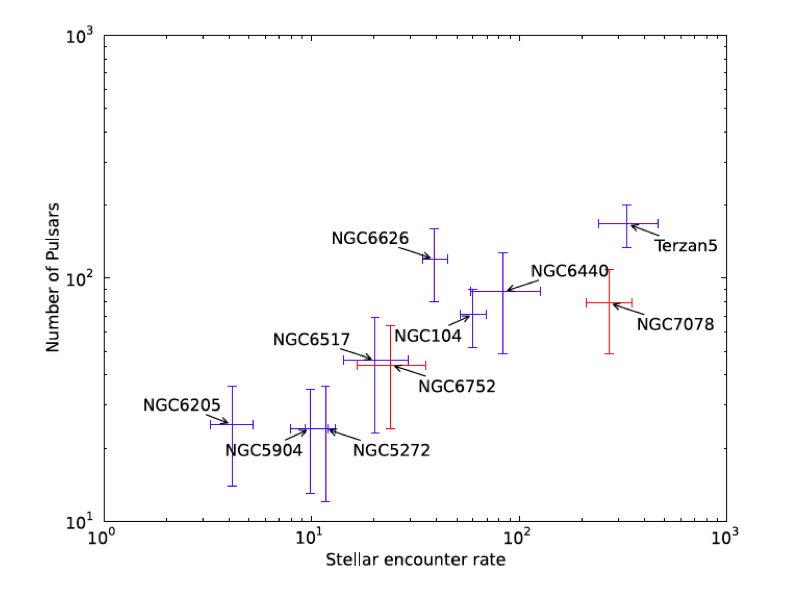}
\caption[Number of MSPs in GCs as a function of the stellar encounter rate.]{Number of MSPs expected to be hosted in GCs as a function of the stellar encounter rate. Taken from \citet{bahramian13}.}
\label{gamma}
\end{figure}

An indicator of the rate of stellar interactions in the core of GCs is called ``collisional parameter'' or ``stellar encounter rate'' ($\Gamma$; \citealt{verbunt87}) and can be expressed as:
\begin{equation}
{\Gamma \propto \int{n_c \; n \; A \; v \; dV} \propto \rho^{1.5}_0 r_c^2}
\end{equation}
where $n$ and ${n_c}$ are the stellar density and the compact object density, respectively, $A$ is the cross-section of the encounter, $v$ is the relative velocity between the two interacting objects, $V$ is the GC volume, $\rho_0$ and ${r_c}$ are the GC central density and core radius, respectively. As can be seen in Figure~\ref{gamma}, the number of MSPs in GCs, predicted by population synthesis models, clearly correlates with the stellar encounter rate.

The bulk of the MSP population in GCs is composed of binary systems. However, some clusters, such as M15, show a larger number of isolated MSPs or more exotic systems, such as double NS binaries. In order to take into account these differences among GCs, \citet{verbunt14} proposed a new indicator, quantifying the interaction rate per binary: $\mathrm{\gamma \propto \rho_0 / \sigma}$, where $\sigma$ is the velocity dispersion of stars in the cluster core. In GCs with larger values of $\gamma$, a larger number of interactions between single stars and binary systems should occur. Therefore these GCs are expected to host a larger number of exotic systems, such as isolated MSPs, eccentric binary MSPs or mildly recycled MSPs. The Galactic GCs with the highest value of $\gamma$ are NGC 6652 and NGC 6624, while M3 and M13 have the lowest values.

\section{Millisecond Pulsars in Globular Clusters: State of the Art}
\label{msp_gc}
In the last decades, the study of MSPs in GCs at different wavelengths has provided a wealth of science \citep[see][and references therein]{hessels15} spanning from fundamental physics to dynamics and stellar evolution. GCs host the most peculiar MSP systems, such as the fastest spinning one \citep{hessels06}, potentially the most massive NSs known \citep{freire08,freire08_m5}, at least one double NS system \citep{jacoby06}, a triple system composed of a MSP orbiting a WD and a planet, as well as various eccentric binaries and other non canonical systems. Before this thesis work, 146 MSPs were known in GCs. However, population synthesis show that the next generation of radio telescopes will be able to unveil from few hundreds to few thousands of new MSPs \citep[e.g.][]{turk13}. This large population will hopefully also include unique systems such as MSP-MSP binaries or even MSP-black hole binaries, which are expected to formed in the extreme environments of GC cores through dynamical interactions. \\   

Fundamental physics can be probed, for example, setting constraints on the equation of state of NS or by testing general relativity or alternative gravitation theories on small scales \citep{hessels06,freire08_m5,demorest10}. The discovery of new very fast spinning MSPs, or even sub-MSP systems, will strongly constrain the maximum radius and hence the NS equation of state \citep{lattimer01,hessels06}. Moreover, MSPs in GCs can in principle experience multiple recycling episodes if companion swappings occur. These should lead to the creation of even more fast spinning and also more massive NSs, thus also allowing  to probe the maximum mass for such systems. Indeed, the mass of NSs is relatively easy to measure in GCs due to the presence of more eccentric systems with respect to the Galactic field. These are likely to show orbital relativistic effects such as the periastron precession (see Chapter~\ref{binaries}) that allow precise mass measurements \citep[e.g.][]{jacoby06,freire08_m5}. \\

The cumulative population of MSPs in a cluster can provide a wealth of information about the cluster structure, potential well, proper motion and also the dynamical status. This is feasible mainly because the spin period derivative of a MSP in a GC is not only due to its intrinsic spin-down rate, but also to its acceleration in the cluster potential field. Indeed, a PSR with intrinsic spin period ${P_0}$, moving with respect to the observer, is observed with a period:
\begin{equation}
{ P = \left[ 1 + \left( \mathbf{V_p} - \mathbf{V_{bary}} \right) \cdot \frac{\mathbf{n}}{c} \right] P_0}
\end{equation}
where ${\mathbf{V_p}}$, ${\mathbf{V_{bary}}}$ and ${\mathbf{n}}$ are the PSR velocity, the velocity of the solar system barycenter and the unit vector along the line of sight, respectively. Following \citet{phinney93} it can be shown that the derivatives of this equation give an equation for the spin period derivative expressed as the measured acceleration ${a_{meas}}$:
\begin{equation}
\label{acc_meas}
{ \frac{a_{meas}}{c} = \left( \frac{\dot P}{P}\right)_{meas} = \left( \frac{\dot P}{P}\right)_{int} + \frac{a_{GC}}{c} +  \frac{a_{add}}{c}    }
\end{equation}
where ${\left( \frac{\dot P}{P}\right)_{int}}$ is the ratio between the intrinsic ${\dot P}$ and $P$, ${a_{GC}}$ is the acceleration due to the GC gravitational field and ${a_{add}}$ is any additional acceleration such as that due to the Galaxy potential field. Therefore, measuring the spin period derivative of a MSP population is a powerful tool to study the cluster potential in a very unique and direct way, possibly probing or excluding the presence of an intermediate-mass black hole in the center. For example, \citet{prager17} used the ensemble of MSPs in Terzan~5 to constrain the physical characteristics of this system, such as the structural parameters, the core density and the total mass, also excluding the presence of a central intermediate-mass black hole more massive than $\sim10^4 \; \Msun$. All their results are completely independent with respect to the optical derived quantities, though fully consistent.

Furthermore, measurements of an ensemble of MSP proper motions provide a direct and precise measurements of the GC motions in the sky. This can be used to constrain the orbits of clusters in the Galactic field, thus providing insight into their evolution. Precise and long-term timing of the MSPs in 47 Tucanae also revealed, for the very first time, the presence of an ionized intracluster medium \citep{freire01a}. The cumulative population can also be used to study the complex interplay between dynamics and stellar evolution, since binary NSs in GC cores are believed to play a major role in delaying the core collapse phase of the system. Moreover, different clusters appear to host different kinds of MSP populations in terms of spin period distribution, location within the cluster, fraction of binaries versus isolated \citep{verbunt14}. All this is tightly linked to the cluster dynamical status and therefore MSP studies combined with that of other exotic systems, such as blue straggler stars \citep{ferraro12}, can provide major improvements in this field.\\

Evolutionary scenarios and accretion physics can be probed with multi-wavelength studies of these systems. In particular, the identification of the optical counterpart to binary systems is of utmost importance since it opens the possibility of evaluating the frequency and timescales of dynamical interactions in dense stellar systems, to explore the impact of dynamics on MSP evolutionary paths, and to investigate stellar evolution under extreme conditions \citep[see, e.g.,][]{fer03_dyna,fer03_mass,sabbi03b,sabbi03a,mucciarelli13,ferraro15}. Indeed the properties of MSP companion stars are extremely different from those of standard stars, due to the effects of the mass transfer phase and of the gravitational field, intense high energy emission and relativistic wind of the NS. In the case of WD companions, it is possible to estimate the masses and cooling ages of the systems through direct comparison of their properties with stellar evolutionary models \citep{edmonds01,fer03_msp,sigurdsson03}. Moreover, spectroscopic follow-ups allow to measure the radial velocities of the companion star, thus providing the mass ratio of the binary. This, together with the radio timing properties of the system, can lead to precise mass measurements of both the components of the binary \citep{fer03_mass,cocozza06,antoniadis13}. These measurements benefit from the known GC distance and optical extinction, thus reducing the uncertainties on the estimated quantities with respect to the case of MSPs in the Galactic field. Furthermore, the derived WD cooling ages are a more appropriate measurement of the system age with respect to the characteristic PSR ages derived from radio timing (see Section~\ref{PSRparam} and \citealt{tauris12a,tauris12b}). The correct determination of MSP ages is an important tool for studying the spin evolution and constraining the physics of the recycling phases (see, e.g., \citealt{vankerkwijk05} and the references therein).

Also X-ray counterparts allow to study the physics of the MSP intra-binary space and of the accretion physics. They also provide a way to put constraints on the NS radii \citep[e.g.][]{bogdanov05,bogdanov11,guillot13}. Interestingly, the {\it Fermi} $\gamma$-ray satellite revealed emission from a sample of GCs \citep{abdo10,zhang16} that host (or are expected to host) a large population of MSPs.\\

\begin{figure}
\centering
\includegraphics[width=10cm]{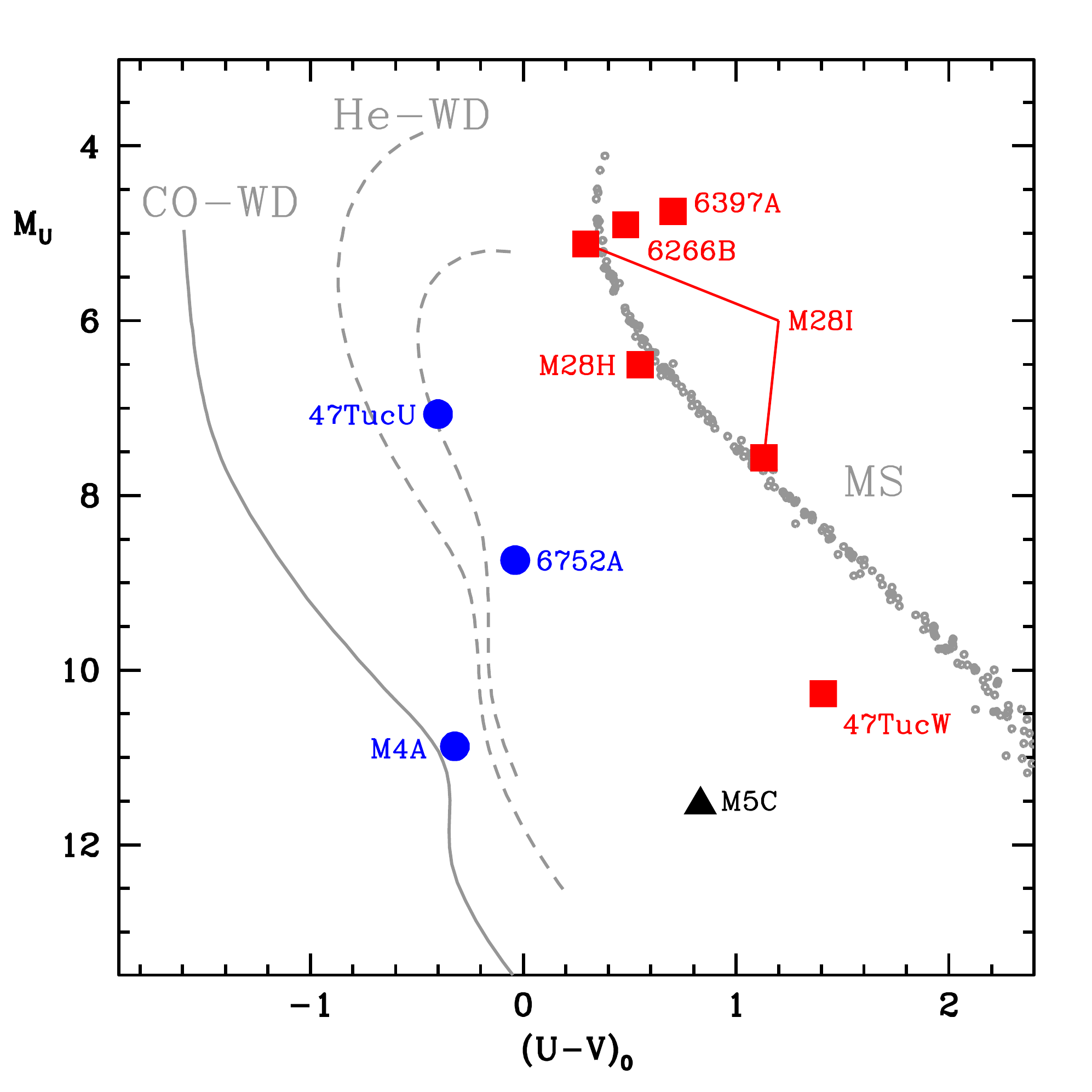}
\caption[Position of the MSP optical counterparts in an absolute CMD.]{Position of the MSP optical counterparts in an absolute CMD. Blue dots mark  the position of  canonical MSPs, while red squares and black triangles the position of RBs and BWs, respectively.}
\label{figcom1}
\end{figure}

Despite the importance of MSP optical counterparts, their identification in crowded stellar systems like GCs is extremely challenging. Only nine companions were identified in GCs before this thesis work, especially thanks to the high resolution performances of the {\it Hubble Space Telescope} cameras. We will briefly review their properties in the following, while the positions of the nine companion stars in an absolute color-magnitude diagram (CMD) are reported in Figure~\ref{figcom1}.
\begin{itemize}
\item {\bf Canonical systems:} the companion to PSR J0024$-$7204U (COM-47TucU\footnote{All the MSPs in GCs are usually named as the cluster name followed by a letter.}, \citealt{edmonds01}), to PSR J1740$-$5340A (COM-6752A, \citealt{fer03_msp}) and to PSR B1620$-$26 (COM-M4A, \citealt{sigurdsson03}) are He WD systems. Therefore they likely are the result of the canonical evolutionary scenario\footnote{Please note that M4A is a triple system with a Jupiter-mass third object. Thus it certainly experienced a deviation from the canonical scenario. This is, so far, the only planet known in a GC.}. All of them are WD with low-masses around $0.15 \; \Msun - 0.4 \; \Msun$ and cooling ages in a range from 0.5 Gyr to 3 Gyr.
\item {\bf Redback systems:} five companions belong to the RB class. The companion to PSR J1740$-$5340A (COM-6397A, \citealt{ferraro01a}) is a bright and tidally distorted star located in the red side of the CMD at about the turn-off level. It has a mass of $0.2 \; \Msun-0.3 \; \Msun$, and \citet{mucciarelli13} spectroscopically confirmed that it likely lost up to $80\%$ of its original mass as a consequence of accretion and ablation due to the PSR wind. \\ Similarly, the companions to PSR J1701$-$3006B and PSR J1824$-$2452H (COM-6266B and COM-M28H, \citealt{cocozza08,pallanca10}) are tidally distorted star located close to the cluster main sequence. The companion to PSR J0024$-$7204W (COM-47TucW, \citealt{edmonds02}) is a faint star and its light curve shows that it is subjected to a very strong heating by the PSR injected flux. Finally, the companion to the tMSP J1824$-$2452I (COM-M28I, \citealt{pallanca13}) is another main sequence-like star if observed during the rotation powered state, while it becomes brighter and bluer during the accretion powered state. Also a strong ${H_\alpha}$ emission has been detected during its accretion powered state, which completely disappears during quiescence.
\item {\bf Black-Widow systems:} only the optical counterpart to the BW PSR J1518+0202C (COM-M5C, \citealt{pallanca14a}) was identified before this thesis work. It is an extremely weak and variable star showing the signatures of a very strong heating of the stellar side facing the MSP. It is located in an anomalous region between the main sequence and the WD cooling sequence, where no stars in standard evolutionary stages are expected.
\end{itemize}

\chapter{Searching, Timing and Optical Identification of Millisecond Pulsars in Globular Clusters}

\vspace{1cm}

\initial{I}n this chapter we summarize the main methodologies used to study pulsars at radio and optical wavelengths. First of all, the main issues arising in classical pulsar searches are presented, along with the techniques routinely used to overcome them in order to discover these faint and exotic objects. Then we present the timing analysis technique and how it can be used to infer a wealth of physical properties of isolated and binary pulsars both. Finally, we show how to identify and characterize the optical counterparts to millisecond pulsars in globular clusters, by using photometric and spectroscopic techniques.

\clearpage

\section{Searching for Pulsars}
\label{searching}
Pulsars (PSRs) are weak radio sources whose periodic signal is usually embedded in a sea of noise and terrestrial interferences. These signals, on their way from the source to the observer, interact with the interstellar medium through a number of physical processes such as dispersion, scattering, scintillation and Faraday rotation. This, eventually together with the PSR motion within a binary system, make basically impossible the identification of new PSRs by simply analyzing the total power registered by the radio telescope\footnote{Although the first PSRs were discovered in this way.}. In the following, the main steps of the typical PSR searches are described.

\subsection{Dispersion and Dedispersion of the Signal}

\begin{figure}
\centering
\includegraphics[width=7cm]{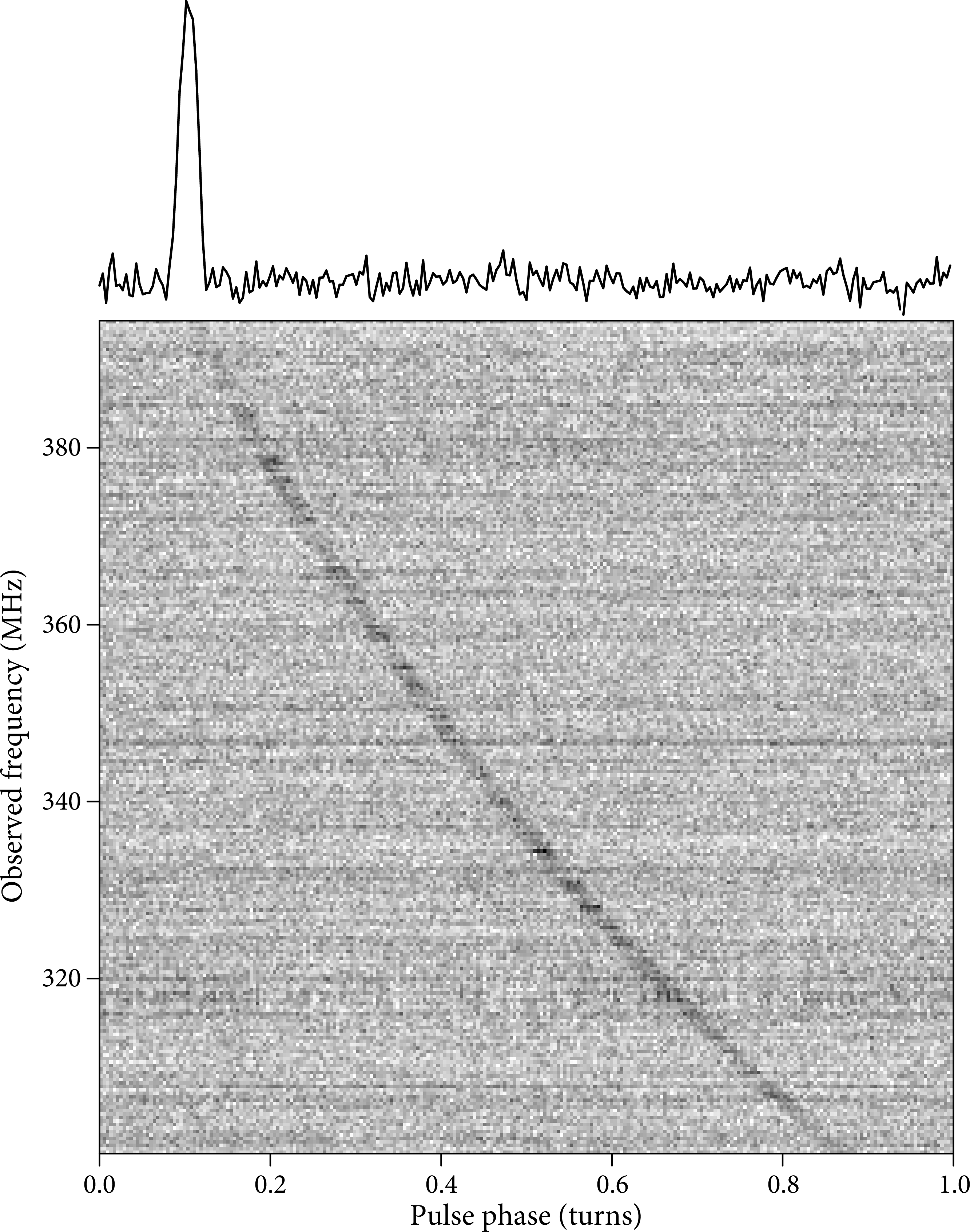}
\caption[The effect of the dispersion of the PSR signal.]{The effect of the dispersion of the PSR signal. The gray scale pattern show the uncorrected dispersed signal of PSR J1400$+$50, as a function of the observing frequency. The top panel report the corrected pulse profile. Figure credit: NRAO.}
\label{dispersion}
\end{figure}

The first main problem to deal with when searching for PSRs is the dispersion of their signal: frequency dependent time of arrivals of the PSR electromagnetic waves due to their travel through a cold and ionized interstellar medium (see Figure~\ref{dispersion}). In order to compensate the intrinsic faint emission of PSRs, observations covering large frequency bandwidth are needed, thus heavily introducing such a dispersive effect that needs to be corrected, otherwise the signal would be broadened and the signal to noise ratio (S/N) heavily reduced. \\

To understand the basic physics behind this phenomenon, consider that the group velocity of the propagating wave, emitted by a PSR, is ${v_g=c\mu}$ where $\mu<1$ is the frequency dependent refracting index of the ionized interstellar medium. As a consequence, the PSR signal, while traveling a distance $d$ from the source to Earth, experiences the following delay with respect of a signal of infinite frequency:
\begin{equation}
{ t = \left( \int_{0}^{d}{\frac{dl}{v_g}} \right) - \frac{d}{c}  \equiv A\times \frac{DM}{f^2}}
\end{equation}
where $f$ is the wave (observing) frequency and the quantity DM is called ``dispersion measure'' and it is defined as:
\begin{equation}
{ DM = \int_0^d n_e dl}
\end{equation}
where $n_e$ is the electron density and $A$ is the ``dispersion constant'' ${A \equiv \frac{e^2}{2\pi m_e c}}$, where $e$ and ${m_e}$ are the electron charge and mass, respectively. It follows that the delay between two observing frequencies is:
\begin{equation}
\label{deltat}
{ \Delta t \simeq 4.15 \times 10^6 \; \mathrm{ms} \; \times DM \times \left(f_1^{-2} - f_2^{-2} \right)}
\end{equation}
We will briefly show in the following the two main methods used to remove such an effect from PSR data.

\subsubsection*{Incoherent Dedispersion}
It is the simplest way to compensate the effect of the signal dispersion. It consists of splitting the observing frequency bandwidth into a large number of independent frequency channels and then applying to each one of them an appropriate delay measurable by equation~(\ref{deltat}), in which ${f_1}$ is a reference frequency channel (usually the central observing frequency) and ${f_2}$ the central frequency of channel to be corrected.

It is obvious that the main issue of incoherent dedispersion is the finite bandwidth of the single channels, that cannot be made arbitrary small,. Therefore each channel inherently retain a small dispersive delay. Such as dispersive delay across a single frequency channel strongly depends on the observing frequency. Therefore, broader channel bandwidths can be used at higher frequency observations. 

\subsubsection*{Coherent Dedispersion}

A much more efficient method, known as coherent dedispersion, allows to completely remove the dispersive effect on the signal by using the phase information of the incoming voltage ${v(t)}$ induced in the telescope by the incoming electromagnetic wave. Once this is measured, it is possible to recover the intrinsic complex voltage ${v_{int}(t)}$ as originated by the PSR \citep[see][and references therein]{handbook}.

Briefly, this process is based on the fact that the modification of the electromagnetic wave by the interstellar medium can be described as the result of a ``phase'' filter, or ``transfer function $H$'', which can be easily analyzed in the Fourier domain. For a signal observed at a central frequency ${f_0}$ and bandwidth ${\Delta f}$, the Fourier transforms ${V(t)}$ and ${V_{int}(t)}$ of ${v(t)}$ and ${v_{int}(t)}$, respectively, can be expressed as:
\begin{equation}
{V(f_0 + f)= V_{int}(f_0 + f)H(f_0 + f)}
\end{equation}
These are non-zero only for ${|f|<\Delta f /2}$ and thus:
\begin{equation}
{  v(t) = \int_{f_0 - \Delta f /2}^{f_0 + \Delta f /2} V(f) e^{i2\pi f t}df}
\end{equation}
\begin{equation}
{  v_{int}(t) = \int_{f_0 - \Delta f /2}^{f_0 + \Delta f /2} V_{int}(f) e^{i2\pi f t}df}
\end{equation}
It can be shown that the transfer function $H$ can be written as:
\begin{equation}
{ H(f_0 + f) = e^{-ik(f_0+f)d} = e^{+i\frac{2\pi A}{(f+f_0)f_0^2}DM f^2}    }
\end{equation}
where $k$ is the wavenumber. By applying the inverse function of $H$ to the sampled voltage data, the voltage as originally emitted by the PSR, and thus completely dedispersed, is recovered.

\subsection{Standard Search Procedures}

The standard search procedures start from the raw data that have to be, first of all, dedispersed in a range of DMs in order to create a number of time series, one per DM value. The step interval between different DM trials should be chosen as a tradeoff, being not too large, so that the PSR signal at a true DM lying in the middle of two trials is not excessively broadened to be missed, and being not too small, to avoid to waste computational time. To better quantify this, consider a simple top-hat pulse of intrinsic width ${W_{int}}$, the observed width ${W_{eff}}$ at an incorrect DM value that differs by $\Delta$DM from the true value is:
\begin{equation}
{ W_{eff} = \sqrt{{W_{int}^2 + (k_{DM} \times |\Delta DM| \times \Delta f/ f^3)^2}}}
\end{equation}
where $k_{DM}$ is a constant. It can be shown that the S/N is both a function of the PSR period and observed width \citep[see Section~A1.22 of][]{handbook}:
\begin{equation}
{S/N \propto \sqrt{{\frac{P-W_{eff}}{W_{eff}}}}}
\end{equation}
On the basis of S/N required for the analysis, an appropriate DM step interval has to be chosen. This can be as large as few or tens of DM units for very long period PSRs, to few tenths for very short period PSR, such as millisecond pulsars (MSPs) .\\

For very long integration time data, such as those needed to study MSPs in globular clusters (GCs), the effect of the rotation of the Earth around the Sun has to be taken into account by applying a barycentric correction. The standard approach to this aim is to refer the topocentric collected data to that of the solar system barycentre. To do this, the starting time of the observation has to be shifted appropriately to match the arrival time of the first sample at the solar system barycentre. Then, the arrival time of each following samples, has to be corrected taking into account the relative motion between the observer and the solar system barycentre.\\

\begin{figure}[b]
\centering
\includegraphics[width=12cm]{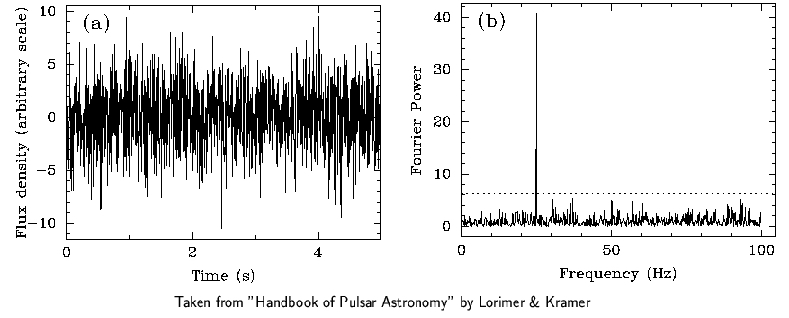}
\caption[Sinusoidal signal as view in the time and Fourier space.]{{\it Panel a):} A noisy time series containing a purely sinusoidal 25 Hz signal. {\it Panel b):} The power spectrum of the Fourier transform of the same time series. The dashed horizontal line is the power threshold above which a peak is considered a PSR candidate instead of a noise fluctuation.}
\label{fourier}
\end{figure}

Once the dedispersed and barycentered time series are obtained, they can be investigate to search for periodic signals. One of the most common routines used to do this is the Fourier transform of the time series and analysis of the resulting power spectrum. However, the continuous form of the Fourier transform cannot be applied, since the time series are a set of independently sampled data points. A discrete Fourier transform is then used and the resulting power spectrum is composed of bins with width equal to the inverse of the total integration time of the analyzed observation. To illustrate the effectiveness of Fourier transforms, in Figure~\ref{fourier} a sinusoidal signal is plotted both in a noisy time series and in its Fourier power spectrum. As can be seen, the signal is hopelessly confused within the noise in the former, while it appears as a strong peak in the Fourier space. 
In real life, however, PSRs are not perfectly sinusoidal signals. Indeed they can have very complex pulse profiles with different duty cycles. This means that the power spectrum powers will be distributed between the fundamental frequency and a significant number of harmonics. Since these contain a significant fraction of the PSR power, methods to perform incoherent harmonic summing have been successfully implemented \citep[see, e.g.,][]{ransom02}. \\

The following and final step is to select, in the power spectra obtained at different DMs, all the peaks above a chosen threshold. The resulting list of candidates will likely contain a number of false detections, radio frequency interferences (RFI) and, hopefully some new PSRs. The latter are expected to appear multiple times in the list, at different DMs, with larger S/N in the proximity of the PSR true DM. All the candidates can be folded, modulo the best period and DM known, starting from the raw data, in order to confirm or reject their PSR nature. This is a procedure in which, given all the known PSR ephemerides (the spin period, position and DM, at least), the corresponding rotational phase of each sample of the observation is calculated, in order to obtain, at the end, an integrated pulse profile, coherently summing all the pulse available in the observation. This represent an average emission of the neutron star (NS) as a function of the rotational phase. The folded data of a real PSR candidate is expected to present a signal across the whole bandwidth and possibly (but not necessarily) across the whole observation, with a peak at the true DM, period and eventually period derivative.

\subsubsection*{Pulsars in Binary Systems}

The method described in the previous section is not suitable to the identification of PSRs in binary systems, especially those with short period orbits (i.e. much shorter than the observation time). Indeed, the motion of the PSR in the binary system results in a Doppler shift of the spin frequency that spreads the signal over a number of adjacent spectral bins in the power spectra, thus significantly diluting the S/N.

\begin{figure}[!h]
\centering
\includegraphics[width=10cm]{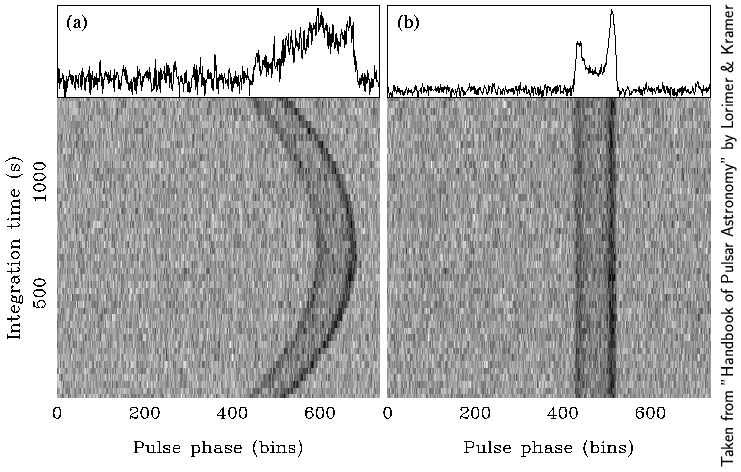}
\caption[Pulse profiles with and without correction for the orbital motion.]{{\it Panel a):} Folded pulse profile of PSR B1913$+$16 without correction for the apparent changes of the spin period.  {\it Panel b):} The same folded data, but assuming a constant acceleration to compensate the motion of the PSR in the binary.}
\label{accel}
\end{figure}

Several algorithms have been built to compensate, at least partially, this problem. The most famous one is a time domain technique that allows  to recover the PSR signal by resampling the time series to the rest frame of an inertial observer with respect to the PSR. Therefore, a standard periodicity search on this modified time series should be able to easily detect the PSR signal. This method is particularly efficient if the binary system orbital parameters are known, since the PSR velocity with respect to the observer as a function of time is known. However, in the most common blind searches, assuming a Keplerian model for such a velocity would require a search in a five-dimensional space and thus a prohibitive computational time. The simplest way to overcome this problem is to assume that, within the observations, the orbital acceleration of the PSR is constant and therefore its velocity scales linearly with the observing time. These are called ``acceleration searches'', and they are usually carried out on time series corrected assuming different values of orbital accelerations. The effectiveness of this method in recovering the PSR signal is shown in Figure~\ref{accel}. The main limitation of this method is that for long integration times the assumption of constant acceleration becomes very poor, as the PSR go through a large fraction of its orbit.

There are also a number of alternative methods, especially frequency-domain techniques. In the case of GCs, the so called ``phase-modulation searches'' \citep{ransom03} have been routinely and successfully implemented. Having typical integration times of several hours, GCs observations usually sample many PSR orbital periods. The phase-modulation exploit the fact that the signal in a power spectrum covering several orbits has a characteristic shape imprinted by the constantly changing PSR spin frequency and it can be described as a family of Bessel functions forming sidebands around the true spin period of the PSR. Once this sidebands are localized, it is usually simple to recover the orbital period, semi-major axis and epoch of ascending node of the binary from sideband spacing, width and phases, respectively.

\subsection{Flux Density Estimates}

It can be demonstrated (see appendix A1.4 of \citealt{handbook}) that the S/N of a top-hat pulse of period $P$, width $W$ and peak amplitude ${T_{peak}}$ above a system noise temperature ${T_{sys}}$ can be written, starting from the radiometer equation \citep{dicke46}, as follows:
\begin{equation}
\label{radiometer}
{ S/N = \sqrt{{n_p t_{obs} \Delta f}} \left(\frac{T_{peak}}{T_{sys}} \right) \frac{\sqrt{{W(P-W)}}}{P}    }
\end{equation}
where ${t_{obs}}$ is the total observing time, ${\Delta f}$ the observing bandwidth and ${n_p}=1$ or  ${n_p}=2$ for single polarization observations and two orthogonal summed polarization observations, respectively. To express this value in terms of flux density, we know that the peak flux density of the pulsar is ${S_{peak}=T_{peak}/G}$, where $G$ is the telescope gain expressed in units of Jy$^{-1}$. Since the mean flux density across the whole period is simply the peak flux density times the duty cycle ($W/P$), we get:
\begin{equation}
{ S_{mean} = S_{peak} \left( \frac{W}{P} \right) = \frac{T_{peak}W}{GP}}
\end{equation}
Substituting in equation~(\ref{radiometer}) the expression for ${T_{peak}}$ obtained in the latter formula, we obtain an equation that can be used to roughly measure the PSR flux density:
\begin{equation}
{ S_{mean} = S/N \frac{T_{sys}}{G\sqrt{{n_p t_{obs} \Delta f}}} \sqrt{{\frac{W}{P-W}}}    }
\end{equation}

In the context of PSR searches, this equation can be used to estimate the limiting detectable flux density of an observation, corresponding to a minimum signal to noise ratio ($\mathrm{S/N_{min}}$):
\begin{equation}
{ S_{min} =  \beta(S/N)_{min} \frac{T_{sys}}{G\sqrt{{n_p t_{obs} \Delta f}}} \sqrt{{\frac{W}{P-W}}}  }
\end{equation}
where $\beta\gtrsim1$ is a correction factor introduced to account for system imperfections in the digitization of the signal.

\section{Pulsar Timing}

Once a new PSR is discovered, follow-up observations are usually obtained to monitor the pulse time of arrivals (TOAs), allowing to study the PSR physical properties, such as its spin, astrometric and orbital parameters. 

The pulse TOA is defined as the arrival time of closest pulse at the mid-point of the observation. To obtain such a measure, being PSRs weak radio sources, it is necessary to fold the observation, thus adding together a significant number of pulses. The folded profile is then matched to a reference one, usually the highest S/N detection of the PSR or a noise-free template that can be obtained stacking together a proper amount of folded observations and then fitting the resulting pulse profile as a sum of Gaussian components. The TOA uncertainty (${\sigma_{TOA}}$) can be defined as the ratio between the pulse width over its S/N. Using the radiometer equation, it can be shown that:
\begin{equation}
{\sigma_{TOA} \simeq \frac{W}{S/N} \propto \frac{S_{sys}P \delta^{2/3}}{S_{mean}\sqrt{\mathrm{t_{obs}\Delta f}}}}
\end{equation}
where ${S_{sys}}$ is the system equivalent temperature (high sensitivity radio telescopes have low ${S_{sys}}$) and $\delta$ is the PSR duty cycle. Therefore, the best measurements are obtained observing  short period, narrow pulse and (obviously) bright PSRs with sensitive and wide-band systems. The folding procedure, summing together a large amount of pulses, allows to reduce the TOA uncertainties. For MSPs, few minutes of observation allow to sum together few thousand pulses. This, together with the intrinsic stability of their signal, explains why MSPs provide better timing performances than classical PSRs.

Before starting the TOA analysis, it is necessary to transform all the measured topocentric TOAs into the solar barycentre system, which can be approximated as an inertial frame. This transformation also corrects for relativistic time delays due to the presence of masses in the solar system. Moreover, this procedure allows to combine TOAs obtained at different times in different observatories. All the TOAs are also corrected for a pulse arrival at infinite high frequency, in order to remove any dispersive effect from the data.\\

\subsection{Isolated Pulsars}

To study the properties of PSRs, i.e. to study changes in the pulse arrival times as a function of time, it is necessary to describe the PSR rotations in a reference frame co-moving with the spinning object. The most common way to do it is to use a Taylor expansion of the PSR spin frequency $\nu$:
\begin{equation}
{\nu(t) = \nu_0 + \dot \nu_0 (t-t_0) + \frac{1}{2} \ddot \nu_0(t-t_0)^2 + ...}
\end{equation}
where ${t_0}$ is a reference epoch and $\nu_0$ the spin period at ${t=t_0}$. Alternatively, the same equation can be written in terms of number of pulses received (or expected) from a reference time ${t_0}$:
\begin{equation}
{N = N_0 + \nu_0(t-t_0) + \frac{1}{2}\dot \nu_0 (t-t_0)^2 + \frac{1}{6} \ddot \nu_0(t-t_0)^3 + ...}
\end{equation}
The timing procedure consists of a least square fitting between the measured TOAs and a multi parametric model that is expected to be able to unambiguously account for every single rotation of the PSR between different observations. Basically, the procedure tries to match the measured arrival times to the number of pulses, minimizing the following expression:
\begin{equation}
{ \chi^2 = \sum_i \left(\frac{N(t_i)-n_i}{\sigma_i}\right)^2}
\end{equation}
where ${n_i}$ is the integer number of pulses nearest to ${N(t_i)}$ and ${\sigma_i}$ the TOA uncertainty in units of pulse turns. At the beginning of the procedure, when only few TOAs, covering a time baseline of few days, are usually available, only the spin period is fitted. When longer TOA time baseline are available, also the spin frequency derivatives and the astrometric position can be fitted. The latter is obtained by measuring the annual variation of TOAs due to the Earth orbit around the Sun. Incomplete or incorrect timing models causes systematic trends in the post-fit residuals, while they are expected to be Gaussianly distributed around zero for a good quality fit. The uncertainties on the fitted parameters decreases with follow up observations, thus with longer time baselines and also with observations obtained at different frequencies, particularly useful to better estimate the DM.

\subsection{Binary Pulsars}
\label{binaries}
In case of binary systems, the orbit of the PSR around the center of mass results in periodical modulation of the pulse TOAs. These are a consequence of the variation of the observed spin period due to the Doppler effect as the PSR moves through its orbit with different velocities along the observer line of sight. For non relativistic systems, Kepler's law can be used to fully describe the orbital properties of these systems. Only five parameters need to be fitted in the timing solution, in order to refer the TOAs to the binary barycentre: the orbital period (${P_b}$), the projected semi-major axis (${x \equiv a_p \sin(i) /c}$, where a is the semi-major axis and i the orbital inclination angle), the orbital eccentricity ($e$), the longitude of the periastron ($\omega$) and the epoch of periastron passage (${T_0}$). These parameters fully describe the orbital properties of the system and they can be also used to constrain the secondary mass  through the mass function, that depends on the orbital period and projected semi-major axis:
\begin{equation}
{f(m_p,m_c) = \frac{(m_c \sin(i))^3}{(m_p + m_c)^2} = \frac{4 \pi}{G} \frac{(a_p \sin(i))^3}{P_b^2} = \frac{4 \pi^2}{T_\odot}\frac{x^3}{P_b^2}}
\end{equation}
where ${m_p}$ and ${m_c}$ are the PSR and companion mass, respectively, and ${T_\odot = (GM_\odot/c^3) \approx 4.92 \; \mu s}$. Assuming a standard NS mass of $1.4 \; \Msun$ and an orbital inclination of $90^{\circ}$, a lower limit on the companion mass can be obtained.\\

In the case of a relativistic binary system, the Keplerian description is no more able to fully reproduce the pulse TOAs and so a number of relativistic effects, due to the strong gravitational field and high orbital velocities, have to be included in the timing fit. These are usually parametrized through the so-called ``post-Keplerian'' parameters that for point masses with negligible spin contribution are only a function of the masses of the two orbiting bodies and of the standard Keplerian parameters. Therefore, measuring two post-Keplerian parameters allows to determine the masses of both the PSR and companion star in the frame of a given theory such as general relativity or eventually alternative gravitation theories. The measurement of a third parameter allows consistency checks of the assumed theory, and so, typically, validation of the general relativity theory. In general relativity, the relativistic orbit of a binary can be described through five main post-Keplerian parameters \citep[see][and references therein]{handbook}:
\begin{equation}
{ \dot \omega = 3T_\odot^{2/3} \left( \frac{P_b}{2\pi}\right)^{-5/3} \frac{(m_p + m_c)^{2/3}}{1-e^2}}
\end{equation}
that describes the advance of periastron in units of rad s$^{-1}$, easily measurable for eccentric orbits. A second parameter is the so-called ``gamma parameter'', that expresses, in units of seconds, the modification of arrival times due both to the gravitational redshift induced by the companion star and time dilation as the PSR moves at different speeds in an elliptical orbit:
\begin{equation}
{ \gamma = T_\odot^{2/3} \left( \frac{P_b}{2\pi}\right)^{1/3} e \frac{m_c(m_p+2m_c)}{(m_p+m_c)^{4/3}}}
\end{equation}
Two other important parameters are the ``range'' and ``shape'' of the Shapiro delay:
\begin{equation}
{r = T_\odot m_c}
\end{equation}
\begin{equation}
{s = \sin(i) = T_\odot^{-1/3} \left( \frac{P_b}{2\pi}\right)^{-2/3} x \frac{(m_p+m_c)^{2/3}}{m_c}}
\end{equation}
which is a delay caused by the gravitational field of the companion star and measurable only for nearly edge-on orbits. Finally, the last parameter is the derivative of the orbital period due to the emission of gravitational waves:
\begin{equation}
{ \dot P_b = -\frac{192\pi}{5}T_\odot^{5/3} \left( \frac{P_b}{2\pi}\right)^{-5/3} f(e)\frac{m_p m_c}{(m_p+m_c)^{1/3}}}
\end{equation}
where $f(e)$ is a simple function of the eccentricity.

Orbital properties of binary PSRs, both relativistic and non, can show periodical or secular changes due to different effects, such as emission of gravitational waves, mass loss, tidal dissipation or interaction with the companion stars. In the case of binaries in GCs, these changes can be the result of a changing Doppler effect due to a change in the distance between the solar system barycentre and the PSR, as the latter orbit in the GC potential field. In particular, this effect contaminates the observed values of the orbital period derivative, which by consequence are not a direct measurement of the intrinsic variation of the orbit. The cluster acceleration affects ${\dot P_b}$ the same way it affects ${\dot P}$ (Section~\ref{msp_gc}). Thus measuring both the derivatives can help to disentangle the intrinsic values from those induced by the cluster.

\section{Identification of Optical Counterparts in Globular Clusters}
\label{opticalcom}
In the case of a binary system, the optical counterpart can be identified and it provides a complementary approach to study the properties of these systems from the secondary star point of view. Indeed, as already explained in Chapter~\ref{msp_gc}, the optical emission is dominated by the companion star while that due to the NS emission, both thermal and non thermal, is totally negligible.

The analysis of the optical counterparts is articulated in two main step. The first one is the photometry of stellar system of the GC in order to identify the counterpart in a position coincident with that of the radio PSR, and then analyze its color-magnitude diagram (CMD) position and possible stellar variability. The second step is the spectroscopic follow-up of the counterpart, useful to mass measurements and to study the companion star chemical composition.

\subsection{Photometry of the Companion Stars}
\label{photometry}
Being NSs the most massive stars in GCs, they are usually located in their very central regions as a consequence of dynamical friction effects. Therefore, in order to properly resolve the extreme crowding conditions of these regions, high angular resolution observations are required. Most of the high resolution images used in this thesis have been obtained through the Wide Field Camera 3 (WFC3) and the Advanced Camera for Survey (ACS), both mounted on the {\it Hubble Space Telescope} (HST). Both the detectors provide pixel scales of the order of $\mathrm{0.04\arcsec / pixel}$ and a field of view (FOV) ranging from $160\arcsec \times 160\arcsec$ for the UVIS channel of the WFC3 to $202\arcsec \times 202\arcsec$ for the Wide Field Camera of the ACS. Their performances allow to resolve the crowded stellar population of GC core regions. Indeed, the typical full width half maximum of point sources in these instruments is of about $0.06\arcsec - 0.08\arcsec$. \\

Multi-epoch and multi-filter observations are usually required in order to perform a secure identification and complete characterization of the companion stars. In fact, these are necessary in order to create CMDs, color-color diagrams and properly sampled light curves.

In the following we describe the typical procedure used to perform the photometric analysis of GC data \citep[see, e.g.][]{dalessandro08a,dalessandro08b}. First of all, it is necessary to model the point-spread function (PSF) of each image of the data-set. This can be done by using a sample of $\sim200$ bright but not saturated stars as reference objects. The PSF structure is determined using an iterative procedure and the final model is chosen on the basis of a $\chi^2$ test. The best fit model is usually provided by a Moffat function \citep{moffat69}. The next step is to perform a deep source detection analysis, usually setting a $\sim3\sigma$ to $\sim5\sigma$ detection limit, where $\sigma$ is the standard deviation of the background counts. Once a list of stellar objects is obtained, a PSF-fitting can be performed for each star of each image in order to measure their apparent magnitudes. Classical aperture photometry techniques are not recommended here due to the extreme crowding conditions of GC central regions. The main advantage of using the PSF-fitting method is that, if a proper PSF model has been derived, it is possible to measure the magnitudes of each star avoiding the contamination from the surrounding, eventually brighter, stars and accurately measuring the background level in order to assign a proper photometric uncertainty to the measurements. 

In the resulting catalog of stars obtained after the PSF-fitting, we usually choose to include only stars that have been detected at least in half the images of each photometric filter. Then, for each star, the magnitudes estimated in different images are homogenized, and their weighted mean and standard deviation are finally adopted as the star mean magnitude and its related photometric error \citep[see][]{ferraro91,ferraro92}. However, in order to perform variability studies, for each source we also keep the homogenized magnitudes measured in each frame in all the available filters. 

The catalogs of stars obtained following this procedure contain instrumental magnitudes. Their calibration can be performed by using the following simple equation \citep{holtzman95}:
\begin{equation}
{m_{cal} = m_{instr} + 2.5 \log(t_{exp}) + ZP + AC}
\end{equation}
where ${m_{cal}}$ and ${m_{instr}}$ are the calibrated and instrumental magnitudes, ${t_{exp}}$ is the exposure time of a given observation, $ZP$ is the zero point needed to report the magnitude to a particular photometric system (such as VEGAMAG for HST observations; \citealt{sirianni05}) and $AC$ is the aperture correction that quantifies the stellar flux lost during the PSF-fitting procedure due to the truncation of the PSF wings (unavoidable for computational reasons). Alternatively, if a catalog of photometric standards is available, the calibration can be easily performed by cross-correlations of the catalogs.\\

The first necessary condition for a star to be considered as a possible optical counterpart is to be located in a position compatible with that of the MSP, determined with high precision ($\lesssim0.1\arcsec$) through the radio timing. Therefore, before starting the analysis of the candidates, it is mandatory to obtain a high-accuracy astrometric solution for the stars in the cluster. This is usually done by cross-correlating  the instrumental positions of stars  with the absolute ones ($\alpha$ and $\delta$) reported in the most updated catalogs of astrometric standard stars (such as 2MASS, UCAC4; \citealt{cutri03,zacharias13}). 

\subsubsection*{CMD Position}

\begin{figure}[h]
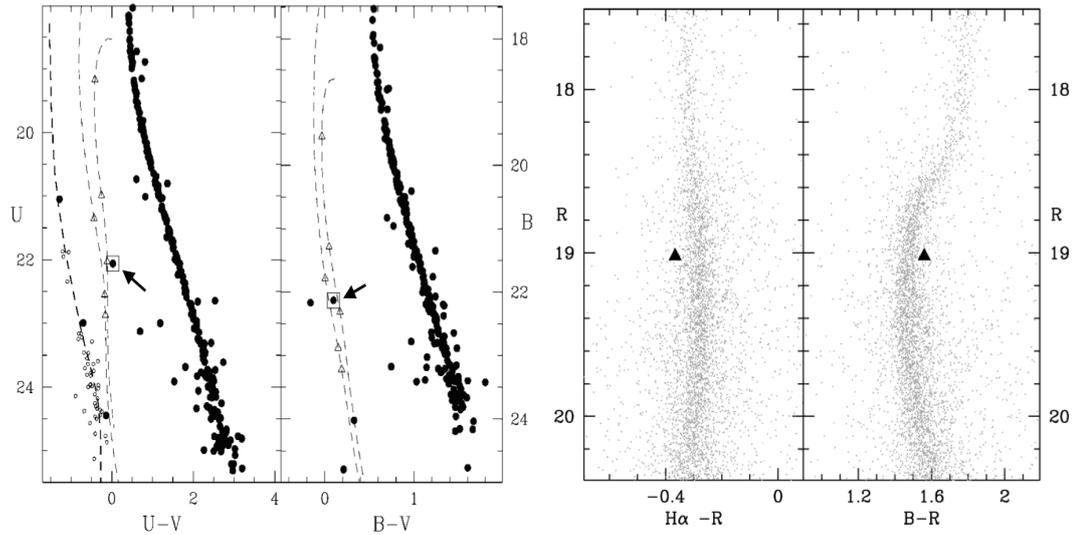

\centering
\includegraphics[width=7.cm,height=7.1cm]{cmdcom1.png}
\includegraphics[width=7.cm,height=7.1cm]{cmdcom2.png}
\caption[CMD position of two MSP companion stars.]{{\it Left panel:} CMD position of the He WD companion to PSR J1911$-$5958A in NGC 6752 \citep{fer03_msp}. {\it Right panel:} CMD position of the companion star to the redback system PSR J1701$-$3006B in NGC 6266 \citep{cocozza08}.}
\label{cmdpos}
\end{figure}

An anomalous position of a candidate star in the CMD is a valid indicator that it could be the MSP companion star. Indeed, since these stars have been going through to an anomalous evolution and could be subjected to perturbations due to the NS emission and gravitational field, their position in the CMD is not usually expected to be compatible with that of the classical stars of the cluster.
From the analysis of the CMD position it is possible to infer the degenerate or non-degenerate nature of the companion star (see Figure~\ref{cmdpos}). In fact, a typical He white dwarf (WD) companion will be  located along the known WD cooling sequences, while non-degenerate objects will be usually more closely located to the main sequence. Moreover, comparing the position of the companion star with that predicted by stellar evolutionary models, it is possible to estimate its mass, radius, age and bolometric luminosity. These results are quite solid for the case of He WD companions, while in the case of non-degenerate companions, the position in the CMD could be heavily faked due to a perturbed, Roche Lobe filling companion. Measuring the mass in such a way usually lead to overestimated values \citep[e.g.][]{pallanca10,mucciarelli13}.
\subsubsection*{Light Curve Variability}

The strongest confirmation on the companion nature of the candidate counterpart is given by stellar variability with a periodicity compatible with that of the binary orbital period. Indeed, the interaction between the two objects likely results in an observable stellar flux variability, especially in the case of high inclination angle orbital planes. 

The analysis of the companion star light curve is an useful tool to study the physical properties of the companion and of the intra-binary system, i.e. how the companion star and the MSP closely interacts and how this alters the properties and evolution of the secondary star, otherwise comparable to that of the other stars in the cluster.

In the following chapters we will adopt the formalism according to which the reference time used to fold the optical measurements is the PSR ascending node. Therefore orbital phases equal to zero and 0.5 correspond to the PSR ascending and descending node, respectively, while orbital phases equal to 0.25 and 0.75 correspond to the PSR superior and inferior conjunction, respectively.

\begin{figure}[t] 
\centering
\includegraphics[width=7cm]{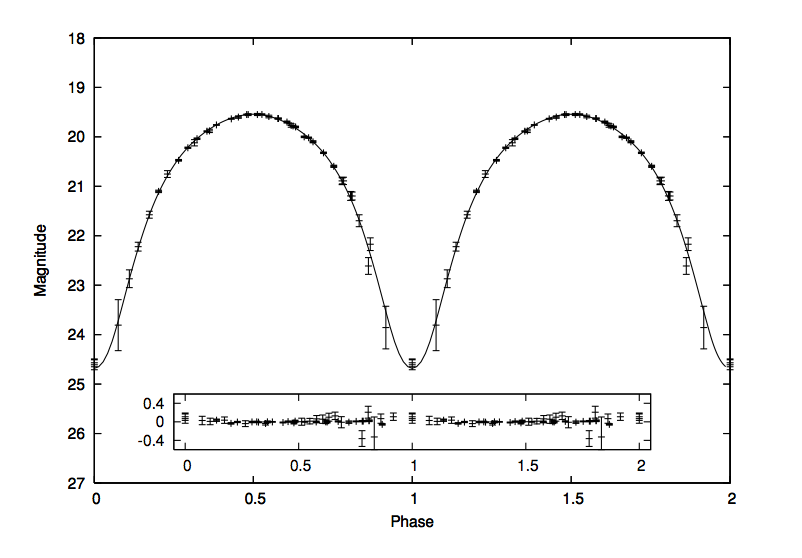}
\includegraphics[width=7cm,height=5cm]{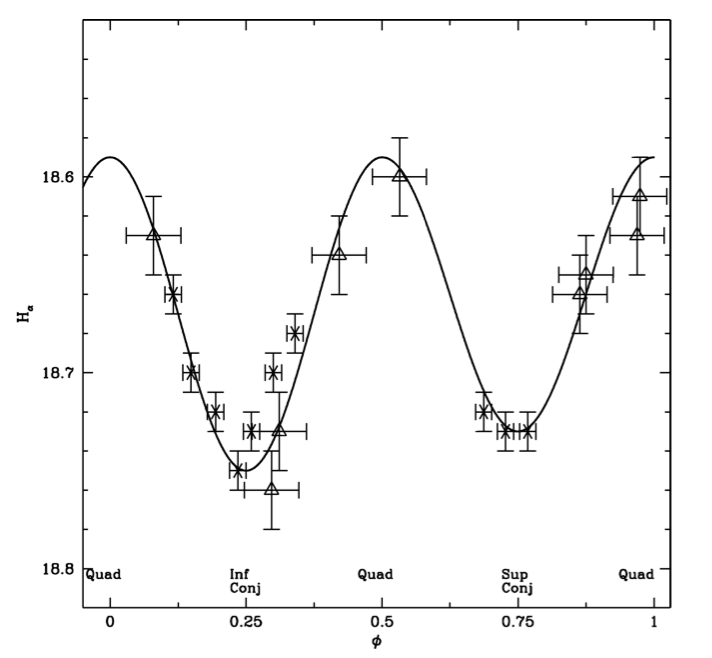}
   \caption[Light curves of the companion stars to PSR B1957+20 and PSR J1701-3006B]{{\it Left panel:} PSR 1957+20 light curve of the companion star from \citet{reynolds07} as an example of single minimum and single maximum structured light curves. {\it Right panel:} Light curve of PSR J1701-3006B from \citet{cocozza08} as an example of double minima and double maxima curves.}
   \label{curvexample}
\end{figure}

The typical light curve of MSP companions can be classified in two main categories. The first one includes light curves with a single minimum and a single maximum structure, usually in correspondence of the PSR superior and inferior conjunctions. This structure is usually interpreted as the result of the heating of the companion side exposed to the PSR injected flux. Indeed, during the PSR inferior conjunction, the stellar side exposed to the PSR is observed, while it is at least partially hidden by the back side of the star during the PSR superior conjunction. The luminosity difference between the minimum and the maximum is a function of the physics of irradiation, properties of the companion star, as well as the orbital inclination angle at which the binary is seen. An example of this curve is reported on the left panel of Figure~\ref{curvexample}. 
\begin{figure}[t] 
   \centering
   \includegraphics[width=12cm]{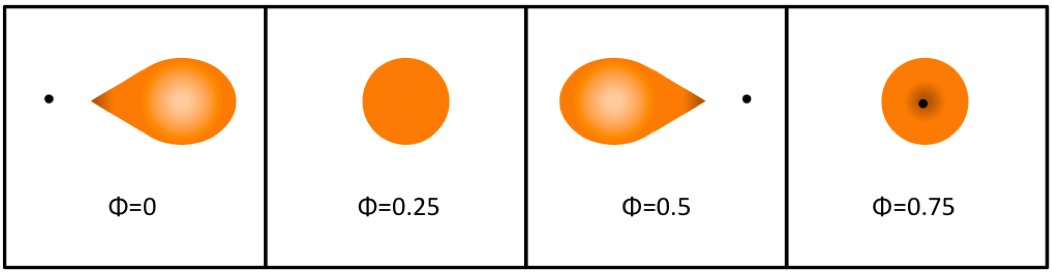} 
   \caption[Representation of the tidal distortion of a companion star.]{Schematic representation of the effects of tidal distortion on MSP companion stars. The system is seen perfectly edge-on. The black dot corresponds to the MSP, while the companion is the orange drop-like structure. Brighter colors correspond to higher temperatures. Credit: \citet{pallanca14b}.}
   \label{drop}
\end{figure}

The second category includes those light curves that show two maxima and two minima, the formers in correspondence of the PSR ascending nodes and the latter in correspondence of the two conjunctions. Thus the companion appears brighter when its lateral sides are observed, while it appears fainter when the back side and the side facing the PSR are observed. This light curve structure is interpreted as the result of the tidal distortion of the companion star that is filling its Roche Lobe and therefore its surface is drop-like structured (Figure~\ref{drop}). Tidal distortions make surface gravity to change in different points of the stellar surface, reaching minimum values in the region facing the PSR and on the opposite side. This results in a stellar variability since the emitted flux ($F$) is proportional to the stellar surface temperature as ${F\propto T^4}$ and to the surface gravity as ${F\propto g}$, and thus ${T\propto g^{1/4}}$ \citep{vonzeipel:fluxgravity}. The maximum values of the surface gravity, and thus of the temperature, are located at the stellar lateral side, visible at the orbital nodes, while the minimum values are located at the front and back ends of the drop-like structure, and thus during the conjunctions. Asymmetric minima can be the result of the different surface gravities between the stellar front and back side, as well as heating of the stellar side facing the MSP. An example of this light curve is reported in the right panel of Figure~\ref{curvexample}.

\subsection{Spectroscopy of the Companion Stars}

We will briefly describe here those studies that can be performed with spectroscopic observations of the companion stars. We will not enter in the details of the procedures, since they are beyond the goal of this chapter and thesis.

\begin{figure}[t] 
   \centering
   \includegraphics[width=7.5cm]{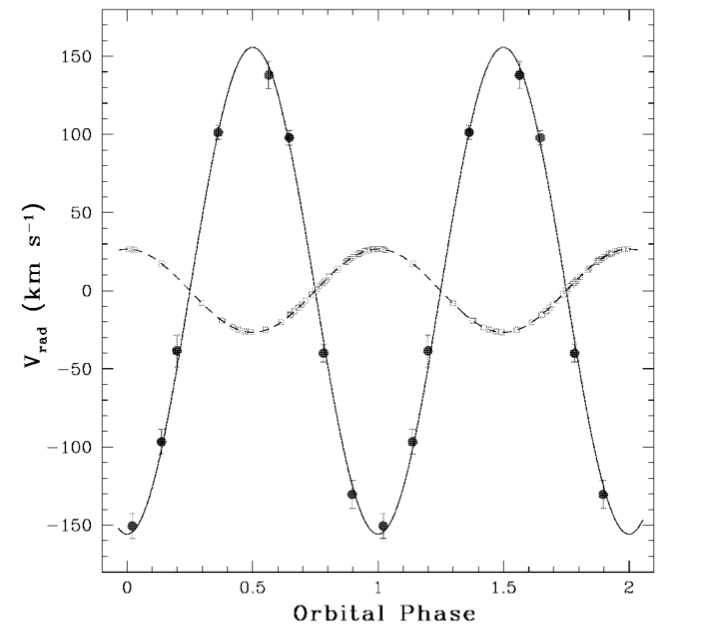} 
      \includegraphics[width=7.5cm]{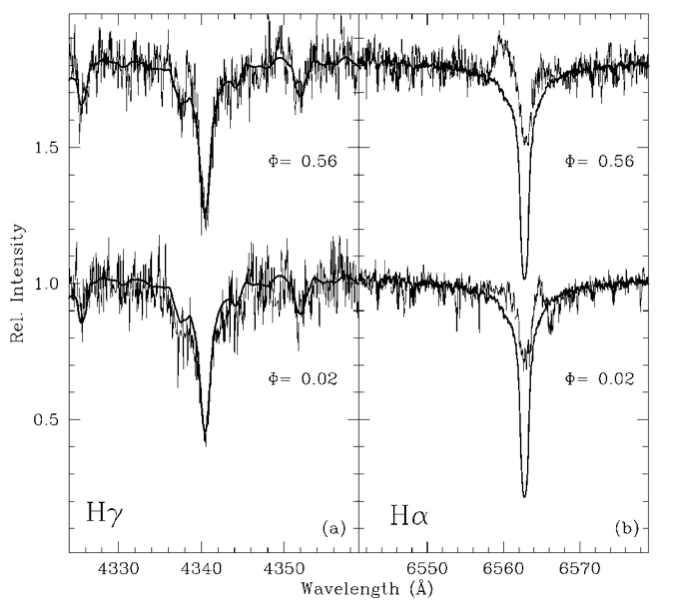} 
   \caption[Radial velocity curve and optical spectra of the companion to PSR J1740-5340A]{{\it Left panel:} Radial velocity curve of the companion to PSR J1740-5340A (black points) and of the PSR itself (open squares). Taken from \citet{fer03_mass}. {\it Right Panel:} Optical spectra of the companion star to PSR J1740$-$5340A \citep{sabbi03a} at two different orbital phases and for two different Balmer spectral lines. The solid black line is a reference spectrum for a classical sub-giant star of the cluster.}
   \label{radialvel}
\end{figure}

Following the discovery and the photometric analysis of the companion star, a spectroscopic follow-up can be performed if the star is bright enough to provide high quality spectra. Such a follow-up is useful to study both the kinematic and chemistry of the system.
The kinematic analysis is aimed at measuring the radial velocities of the companion star during the orbit. This can be done measuring the Doppler shifts of the spectral lines with respect to their rest position at different orbital phases. The radial velocity curve will have the same periodicity of the orbit and an average velocity corresponding to the line of sight velocity of the binary moving within the cluster (see left panel of Figure~\ref{radialvel}). This, together with the radial velocity curve of the MSP, obtained through radio timing, allows to measure the mass ratio between the two components of the binary system. The ratio can be used together with other mass estimates, such as the total mass, the mass function, and eventual inclination angle estimates (e.g. from the light curve analysis) to precisely constrain the mass of the two objects.

The chemical analysis is instead aimed at measuring the chemical properties of the stellar surface \citep[e.g.][]{sabbi03b,sabbi03a}. Of particular interest are the Balmer series line, since the presence of emission, instead of absorption, or of absorption with phase-dependent structures, allow to study the presence of ionized material in the intra-binary and surrounding space (see right panel of Figure~\ref{radialvel}). If the companion star is a WD, the comparison of the observed spectra with simulated ones allows to infer the physical properties of the WD (such as temperature, surface gravity, radius and mass). Moreover, since these stars have been subjected to the removal of the external envelope as a consequence of the accretion and eventually of subsequent ablation, the observed chemical abundances give us a direct view on the regions where the stellar chemical composition has been modified by thermonuclear reactions, thus shedding light on the original properties of the star \citep{mucciarelli13}.

\chapter{Discovery of Three New Millisecond Pulsars in Terzan 5}
\label{cap_t5}
\begin{flushright}
\textit{Mainly based on \citealt{cadelano18}, ApJ, 855:125C}
\end{flushright}
\vspace{1cm}

\initial{W}e report on the discovery of three new millisecond pulsars (namely J1748$-$2446aj,
J1748$-$2446ak and J1748$-$2446al) in the inner regions of the dense stellar system
Terzan 5. These pulsars have been discovered thanks to a method, alternative to the classical search routines, that exploited the large set of
archival observations of Terzan~5 acquired with the Green Bank Telescope over the last 5 years (from
2010 to 2015). This technique allowed the analysis of stacked power spectra obtained by combining 
$\sim$ 215 hours of observation.
J1748$-$2446aj has a spin period of $\sim2.96$ ms, J1748$-$2446ak of
$\sim1.89$ ms (thus it is the fourth fastest pulsar in the cluster) and J1748$-$2446al
of $\sim5.95$ ms. All the three millisecond pulsars are isolated and currently we have timing solutions only for J1748$-$2446aj and J1748$-$2446ak. For these two
systems, we evaluated the contribution to the measured spin-down rate of the
acceleration due to the cluster potential field, thus constraining the intrinsic
spin-down rates, which are in agreement with those typically measured
for millisecond pulsars in globular clusters. Our results increase to 37 the number
of pulsars known in Terzan~5, which now hosts $\sim25\%$ of the entire pulsar population
identified, so far, in globular clusters.\\

\clearpage

\section{Introduction}\label{intro}

To date, the exotic millisecond pulsar (MSP) zoo in globular clusters (GCs) comprehend 146 pulsars (PSRs) in 28 GCs\footnote{see
http://www.naic.edu/$\sim$pfreire/GCpsr.html}. However, population synthesis simulations clearly showed that a very large population of several
thousand MSPs is still to be unveiled \citep{bagchi11,chennamangalam13,turk13,hessels15}. The discovery of several GCs emitting also in the $\gamma$-ray corroborates this results, since such an emission can be exclusively due to a population of MSPs  \citep{abdo10,zhang16}.

The number of MSPs identified in GCs per year decreased abruptly after 2011, 
thus showing the limit in performance and sensitivity reached by the current generation of
radio telescopes. However, for more than one decade, GCs have been routinely observed
at radio wavelength in order to obtain long-term timing solutions of the identified
MPSs. This resulted in the production of a large archive of tens of observations of
the same region of the sky. The work presented here is aimed at showing the large
possibilities of finding new PSRs by exploiting these huge archival data sets. At odds
with traditional PSR searches, which are based on the analysis of a single and
long time sequence of data (see Chapter~\ref{searching}), we present here a method to search for PSRs by
incoherently stacking the power spectra obtained from all the available observations. A
similar procedure has been already successfully implemented { in M15 by \citet{anderson93}}, in Terzan 5 by \citet{sulman05}, leading to the discovery of three isolated MSPs (namely J1748$-$2446af, J1748$-$2446ag and J1748$-$2446ah), and in 47 Tucanae by \citet{pan16}. { All these PSRs are isolated. Indeed, due to the Doppler shifts of the spin frequency induced by the PSR motion in binary systems, this method is only efficient to discover faint isolated PSRs or long period binary PSRs, the latter being unlikely to survive in GCs.}

Among the Galactic GCs, Terzan 5 turned out to be the most amazing MSP factory. In
fact, 34 MSPs have been identified so far in this system \citep[][Ransom et al. 2017,
in preparation]{ransom05a,hessels06,prager17}, which is about $\sim23\%$ of the total number of MSPs
identified in GCs. Terzan 5 is, indeed, one of the most intriguing stellar system in
the Galaxy. \citet{ferraro09, ferraro16} found out that this system is probably
not a genuine GC, but
more likely the pristine remnant of a building block of the Galactic bulge, which was
originally much more massive than today \citep{lanzoni10}. 

Here we present  the identification of three new MSPs in Terzan 5. In Section~\ref{ter5_analysis} we present 
the dataset and the
stacking procedures that led us to the discovery of the new MSPs. In
Section~\ref{ter5_results} we describe the main properties of these new systems and their
timing solutions. Finally, in Section~\ref{ter5_intrinsic} we constrain their acceleration and some of their physical
parameters. 

\section{Observations and Data Analysis}
\label{ter5_analysis}

\subsection{Dataset and initial data reduction}
The work presented here has been performed by using 33 archival observations of
Terzan~5 obtained with the 100-m Robert C. Byrd Green Bank Telescope (GBT) from
August 2010 to October 2015. Observations were acquired at 1.5 GHz and 2.0 GHz using
800 MHz of bandwidth, although radio frequency interference (RFI)  excision
reduced the effective bandwidth to $\sim600$ MHz. Only one observation was obtained
at 820 MHz using 200 MHz of bandwidth.  Observation lengths vary from a minimum of 1.5 hours
to 7.5 hours, the latter typical for the majority of the observations. The total
observation length, resulting from the stack of all these observations, is of about 9
days ($\sim 215$ hours).

{ The data recorded by GUPPI were Full Stokes with 10.24 $\mu$s sampling and 512 channels, each coherently dedispersed in hardware to a dispersion measure (DM) of $\mathrm{238 \ pc \ cm^{-3}}$, which is close to the cluster average. The total intensity (i.e. sum of two orthogonal polarizations) was extracted from those data and downsampled to 40.96 $\mu$s resolution for incoherent dedispersion into 23 DM trials}.

The data have been processed using the {\tt PRESTO} software suite \citep{ransom02}.
We obtained 22 time series per observation ranging from a DM of
$\mathrm{ 233 \ pc \ cm^{-3}}$ to $\mathrm{ 244 \ pc \ cm^{-3}}$ and spaced by $\mathrm{ 0.5 \ pc \
cm^{-3}}$, plus an additional time series at a control DM of $\mathrm{ 100 \ pc \ cm^{-3}}$. The time series have been transformed to the barycenter of the solar system using {\tt TEMPO}. For each sample of the time series of each observation, we subtracted the mean of all
the channels (i.e. we subtracted the DM=$\mathrm{ 0 \ pc \ cm^{-3}}$ time series) and excised
some interference by removing samples with values higher than $4\sigma$, where
$\sigma$ is the standard deviation of all the sample values. Since we aim at stacking
together the power spectra of observations of different lengths, we manually added
samples with null values to the time series of shorter length, thus
obtaining time series of length equal to that of the longest one. We then applied a
fast Fourier transform to all the time series { and squared the complex amplitudes} to obtain the power spectra. Finally,
in the power spectra, we ignored all the spectral bins expected to contain the powers
of all the known Terzan~5 PSRs and their harmonics (also accounting for the shifts due to the binary PSRs orbital motions). We also excised  the most relevant RFI.\\

\subsection{Stacking search procedures}
First of all, we normalized all the available power spectra dividing the spectral powers by the local median value. Then, we summed the 33 individual daily power spectra into a stacked power spectrum for each of the 22 DM trials {and the control DM}. { These final stacked power spectra are nearly chi-squared distributed with 66 degrees of freedom.} In
all these stacked spectra, we performed an harmonic sum, by summing to each power bin the powers of
the corresponding harmonic, { from the second up to the eighth}. This has been done to further enhance the
spectral powers of the still unidentified PSRs. At the end of this, we had 22 stacked power
spectra, plus the one at the control DM.  In order further remove periodic interference that can be
still persistent in high DM spectra, we subtracted from each stacked power spectrum the control
stacked spectrum obtained at DM=$\mathrm{ 100 \ pc \ cm^{-3}}$. In this way, a large fraction of RFI,
present in both the control and the science power spectra, are removed, leaving the signal of the
still undiscovered PSRs virtually unaffected. 

\begin{figure}
\centering
\includegraphics[width=10cm]{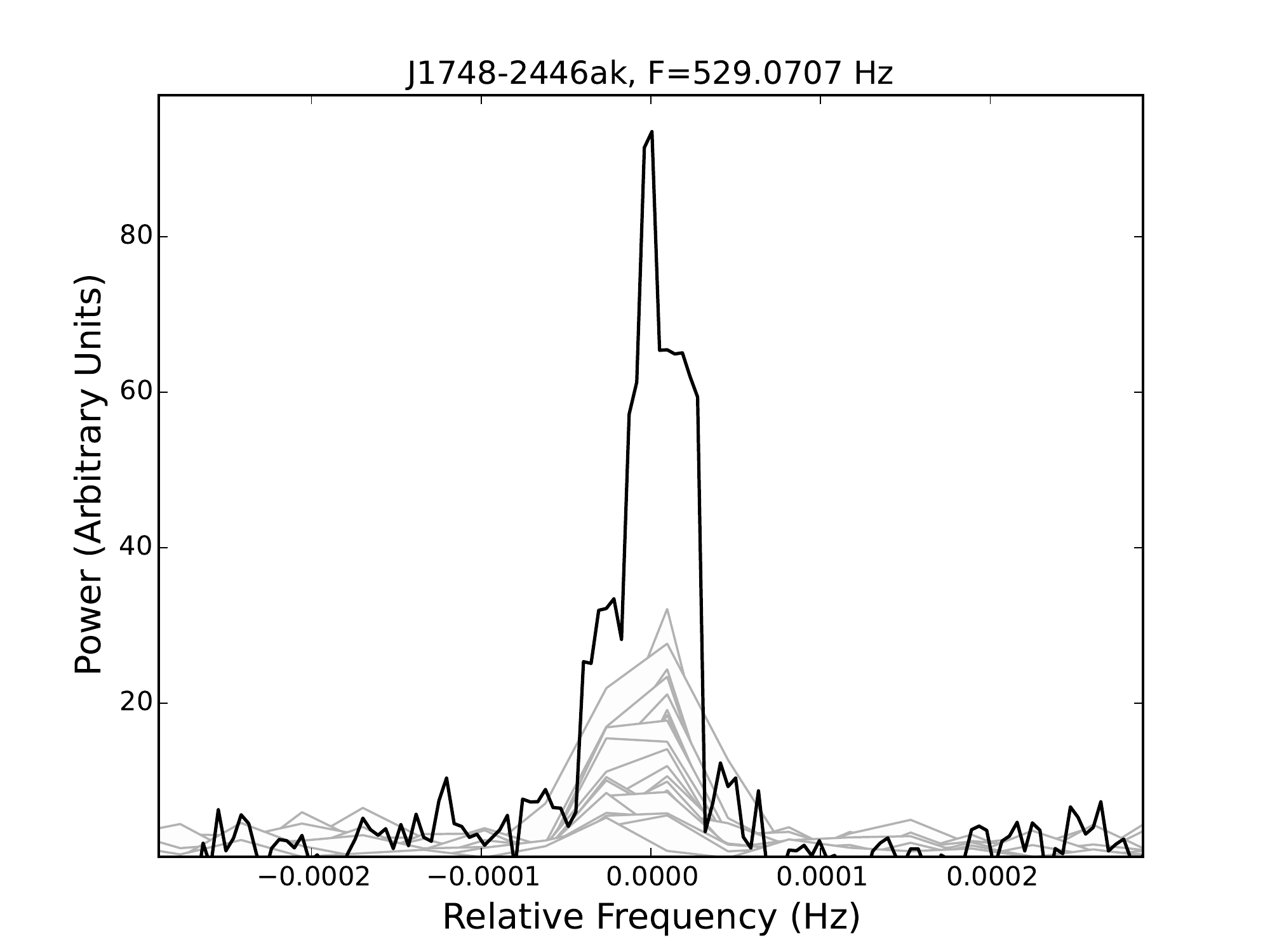}
\caption[Power spectrum around the region of the newly discovered Ter5ak.]{Power spectrum around the region of the newly discovered PSR J1748$-$2446ak (Ter5ak). We plotted in black the stacked { and harmonic summed} power spectrum obtained from $\sim215$ hours of archival observations, while in shaded gray we plotted the power spectra obtained from the single daily observations. {The power in the stacked spectrum is spread over more bins than for single observations due to the effect of the harmonic sum}.}
\label{spec}
\end{figure}

In order to illustrate the effectiveness of the method, in Figure~\ref{spec} we compare
 the stacked power spectrum for one of the newly discovered MSPs, with that
obtained from a single observation where this MSP has a high
signal to noise ratio (S/N). It can be clearly seen how the stacking procedure greatly enhances the
spectral powers of such a faint object.

\begin{figure}[t]
\centering
\includegraphics[width=7.8cm]{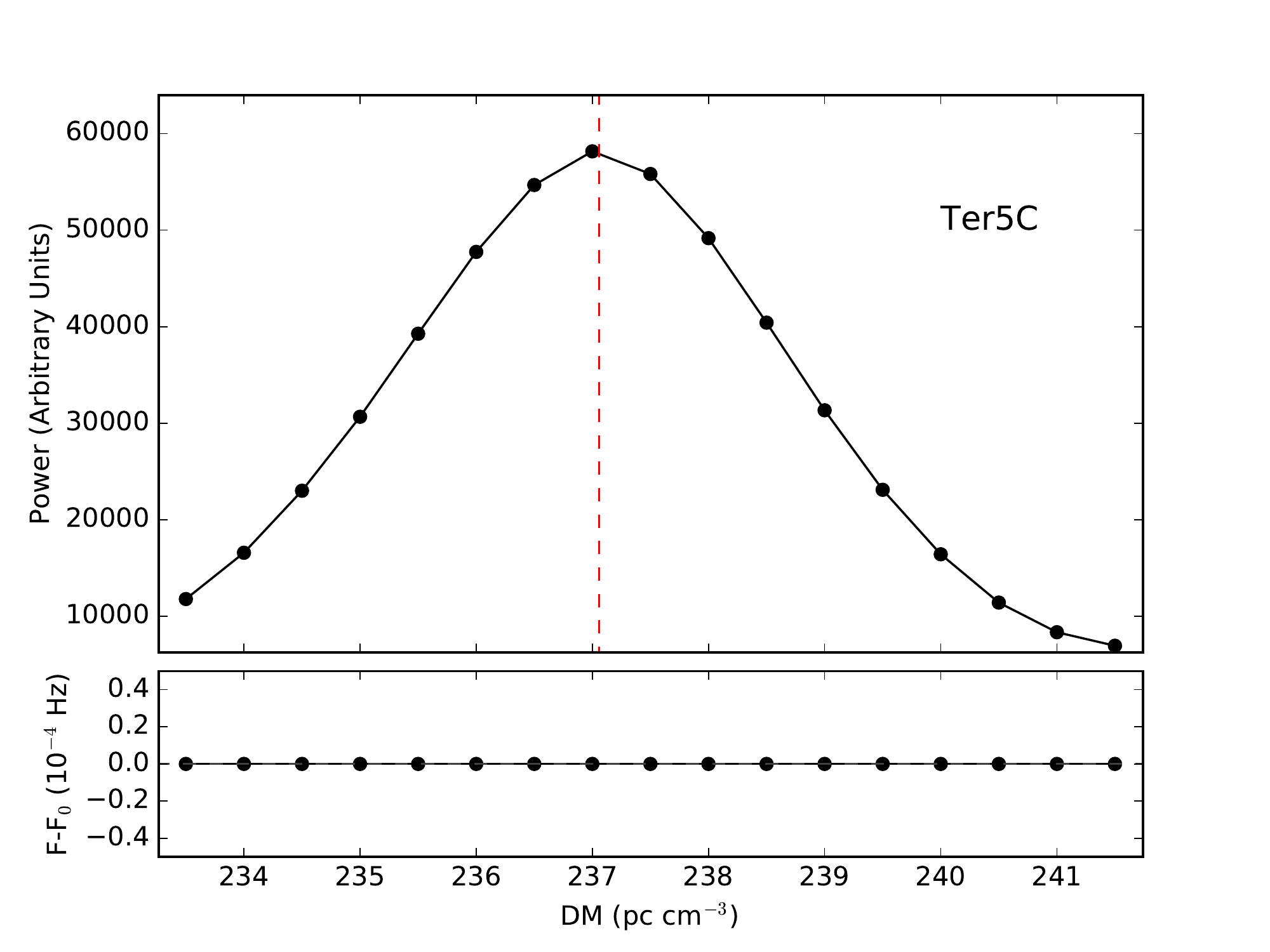}
\includegraphics[width=7.8cm]{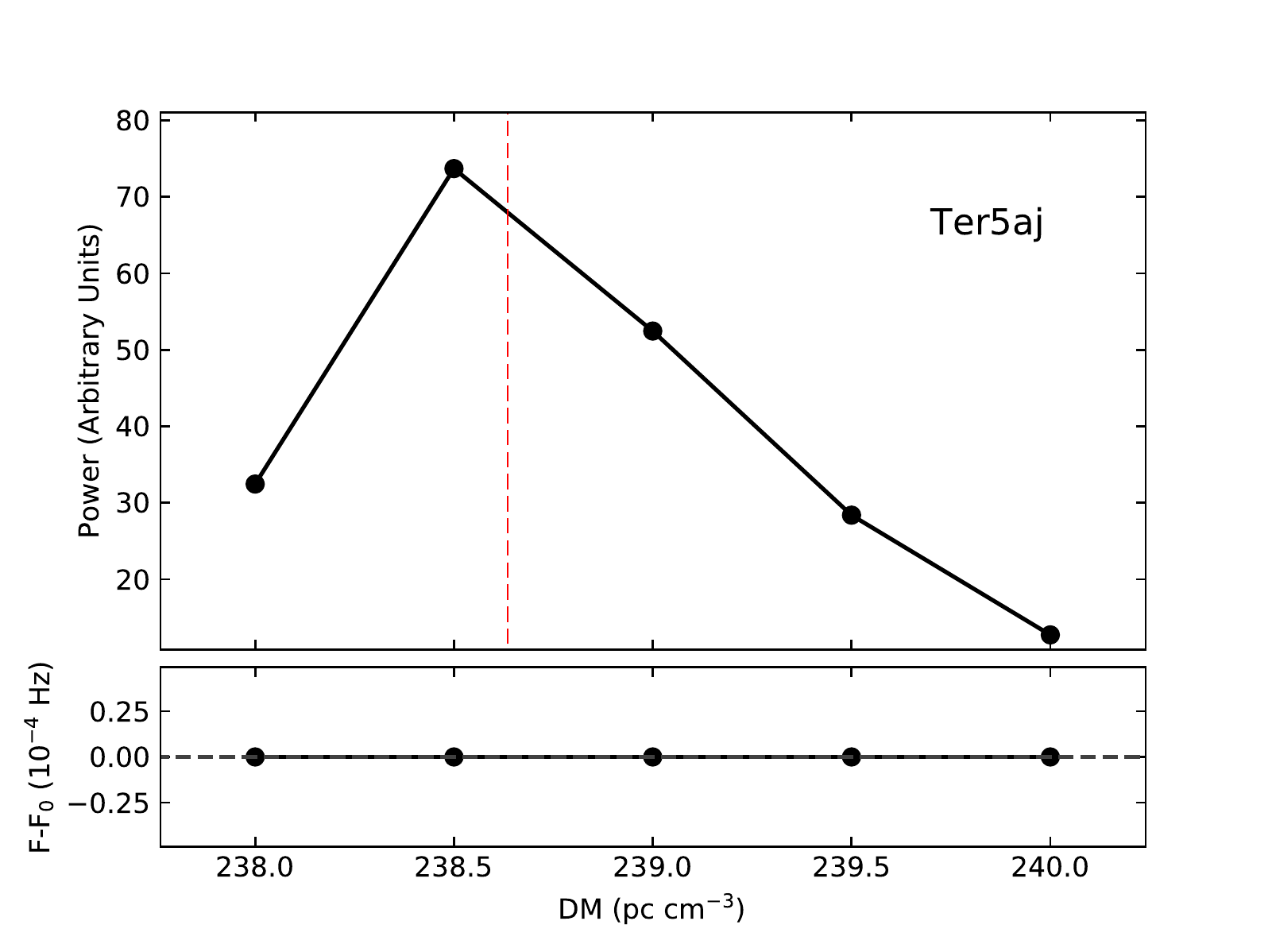}
\caption[Spectral powers as a function of DM for Ter5C and Ter5aj.]{{\it Left Panel:} Spectral powers of a candidate PSR, corresponding to J1748$-$2446C (Ter5C), obtained from the stacked power spectra. The powers are plotted as a function of the DM as obtained from the KD-tree algorithm (see text). The red dashed vertical line is the MSP true DM as derived from its timing solution. A the bottom  of the top panel the frequency difference across the different DMs in which the candidate has been found are plotted. In this case, the candidate has the same frequency in all the DMs, as expected from a real PSR. {\it Right Panel:} same as in the left panel, but for the newly discovered PSR Ter5aj.}
\label{psrC}
\end{figure}

The next step is to select, in the stacked power spectra, the periodic signals that likely originate
from a real PSR instead of from RFI. To do this, we selected in the power spectra all the peaks
above $4.5\sigma$ ($\sigma$ is the standard deviation of the spectral powers) and saved them into a
candidate file. We ended up with 22 candidates files, one per DM, each one containing $\sim13000$
candidate. Therefore the total number of candidates is about $3\times10^{5}$.  We then applied a
KD-tree algorithm to perform a selection of all these candidates. In fact, in the DM-frequency phase
space, PSR candidates are expected to be closely segregated around their DM and spin frequency.
On the other hand, RFI can appear in a large range of DMs, also with slightly different spin frequencies.
We used the KD-tree routine included in the python scipy package.
Briefly, we built a 2D tree in a DM-frequency space. Then, for each candidate, we searched for the
1000 closest neighbors and selected only those with a spin frequency compatible (within a tolerance
of $10^{-5}$ Hz) with that of the candidate. Since the PSR signal is expected to be observed at
contiguous DM values with an almost Gaussian distribution of powers around the true DM value, we
selected as good candidates only those whose closest neighbors are found at contiguous DMs and with
a maximum in the spectral power vs DM space. As an example, we show in Figure~\ref{psrC} the output
of the KD-tree procedure for the case of the known PSR J1748$-$2446C (Ter5C). As can be seen, the
algorithm found, for a candidate corresponding to Ter5C, neighbors at contiguous DMs, with a peak in
the spectral powers very close to the MSP's true DM value. Moreover, the spin frequency does not show
any variation at different DMs, as expected from a real PSR. Applying this procedure to our
candidates, { we discarded $\sim97\%$ of them} and ended up with only $\sim100$ possible PSR
candidates. These have been individually analyzed, folding the single observations at the candidate
frequency and DM corresponding to that of the maximum spectral power, allowing also a search in spin
period, spin period derivative and DM in order to maximize the S/N.

\section{Results}
\label{ter5_results}


The method described in the previous section allowed us to discover three previously unknown MSPs in
Terzan~5: J1748$-$2446aj, J1748$-$2446ak and J1748$-$2446al (hereafter, Ter5aj, Ter5ak and
Ter5al, respectively). We plot in Figure~\ref{profili} the average pulse profiles and the signals as
a function of time for these three new MSPs { in the brightest individual days. Moreover, we have been able to blindly re-detect all the other isolated MSPs known in the cluster\footnote{{ To do this and to obtain the top panel of Figure~\ref{psrC}, we re run the whole procedure without excising the signal of the known PSRs.}}}.\\

The stacking technique described in this work is of particular effectiveness in a high DM system like Terzan 5. Indeed, for typical 1.4 GHz observations, the high DM and very small scintillation bandwidth allow us to average, in the stacked power spectra, over many scintles, making negligible any effect due to diffractive scintillation. The only variability appreciable in the different terms of the power spectra sum is due to refractive scintillation, which can affect the flux densities of PSRs by typically up to a factor of $\sim2$. On the other hand, in a low DM cluster such as, for example, 47 Tucanae, diffractive scintillation can change the measured flux densities by more than an order of magnitude. Therefore few of the power spectrum sums will have high values, while most of them will have much lower values, thus diluting the S/N of the final sum.

\subsection{Ter5aj}

\begin{figure*}
\centering
\includegraphics[width=12cm]{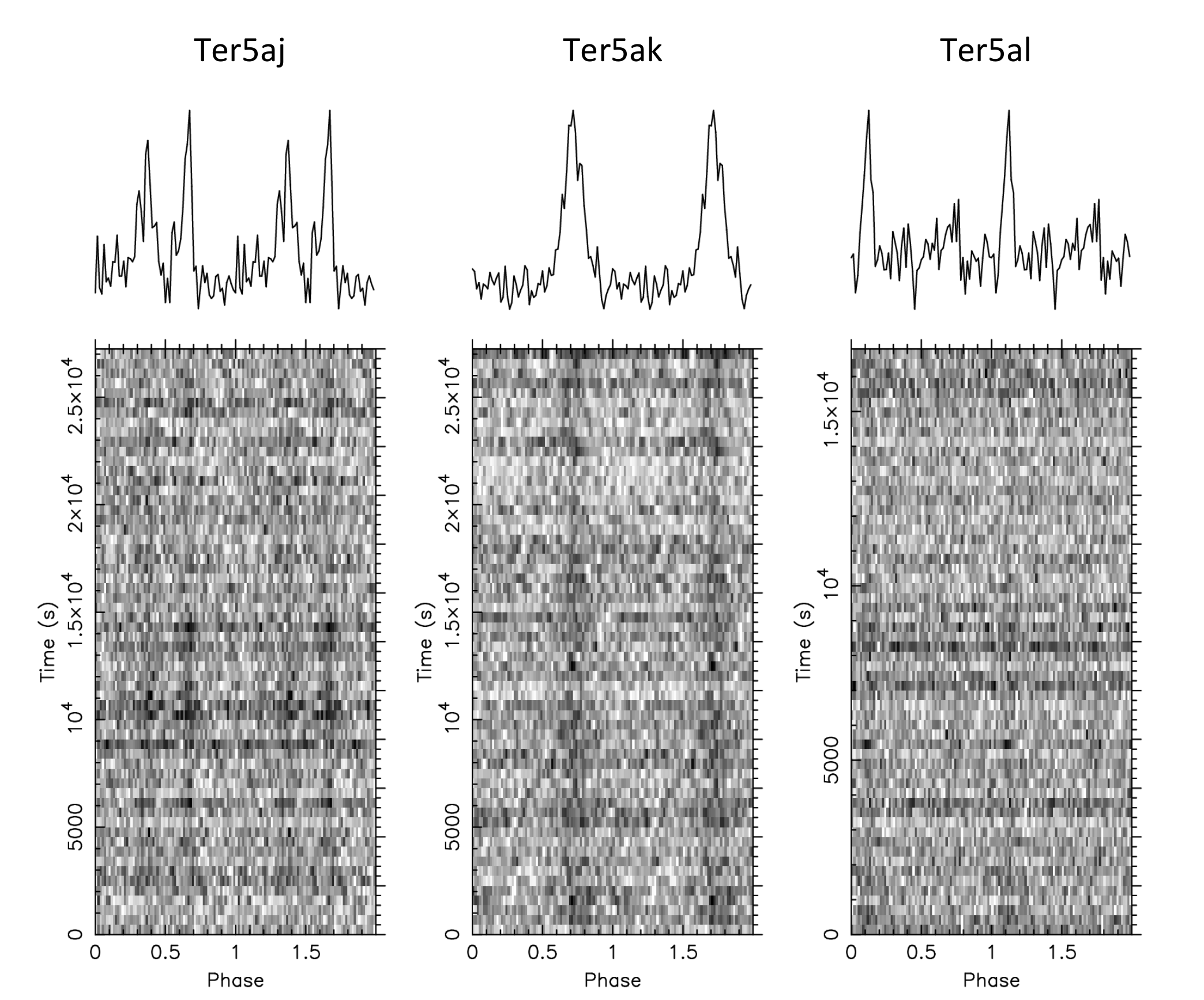}
\caption[Averaged pulse profiles of the best detections of Ter5aj, Ter5ak and Ter5al.]{{\it Top panels:} Averaged pulse profiles of the best detections of Ter5aj (on the left), Ter5ak (in the middle) and Ter5al (on the right). {\it Bottom panel:} Intensity of the signal (gray scale) as a function of the rotational phase and time for each MSP.}
\label{profili}
\end{figure*}

\begin{figure}
\centering
\includegraphics[width=10cm]{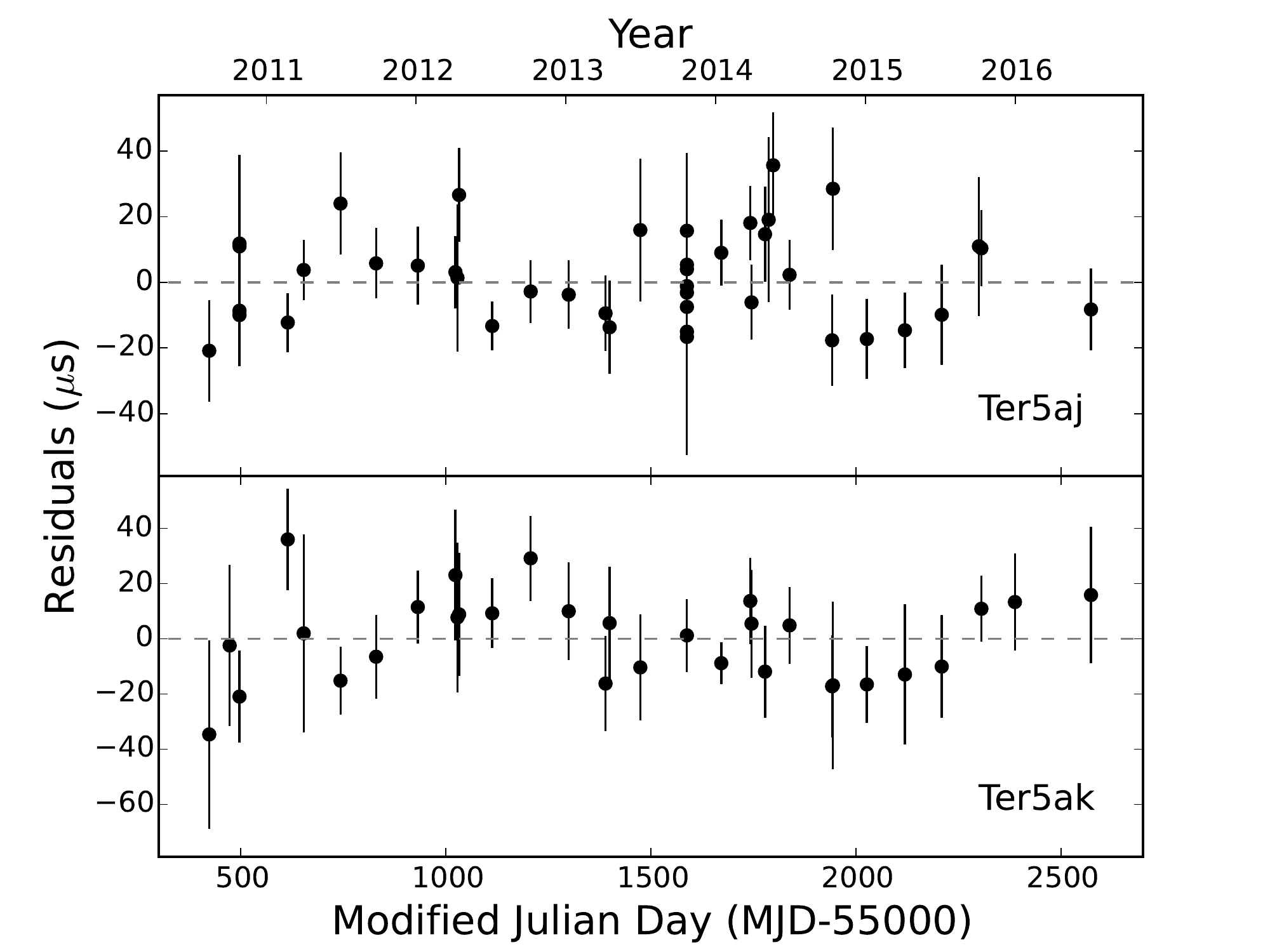}
\caption[Timing residuals for Ter5aj and Ter5ak.]{Timing residuals for Ter5aj (top panel) and Ter5ak (bottom panel).}
\label{resid}
\end{figure}

Ter5aj or J1748$-$2446aj has been discovered with a maximum spectral power at $\mathrm{ DM=238.50 \ pc \
cm^{-3}}$ (see Figure~\ref{psrC}). Folding the single observations, we have been able to clearly identify it and
confirm its PSR nature in all the 33 GUPPI observations. As can be seen from
the left panel of Figure~\ref{profili} (see also the top panel of Figure~\ref{sumprof}), Ter5aj presents a double peaked pulse shape,
where the two peaks are separated by $\sim0.3$ in phase. We extracted the pulse times
of arrival (TOAs) with the {\tt get$_{-}$TOAs} routine within {\tt PRESTO}, using  a 
double Gaussian template, created by fitting the pulses obtained in the
observations where this object has the highest S/N (see left panel of Figure~\ref{profili}). We phase connected all the $\sim6$ years of data using
standard procedures with {\tt TEMPO}\footnote{\url{http://tempo.sourceforge.net}}.  The
timing solution is tabulated in Table~\ref{ajtiming}, the post-fit timing residuals are
reported in the top panel of Figure~\ref{resid} {and the averaged pulse profile, obtained by summing all the daily detections, is reported in the top panel of Figure~\ref{sumprof}}.

\begin{figure}
\centering
\includegraphics[width=10cm]{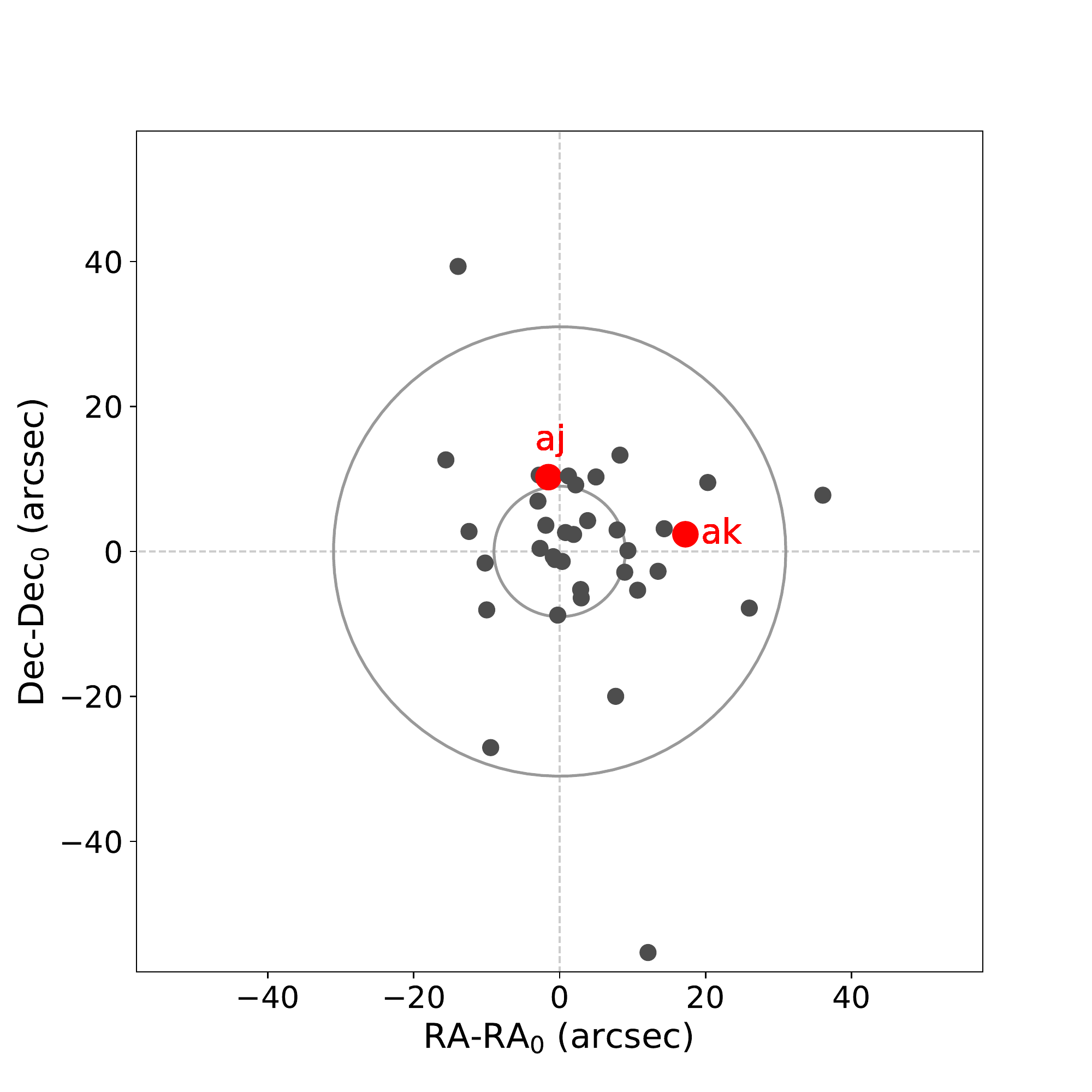}
\caption[Positions of Ter5aj and Ter5ak with respect to the cluster gravitational center and to the other MSPs in the cluster.]{Positions of Ter5aj and Ter5ak with respect to the cluster gravitational center and to the other MSPs in the cluster. The inner and outer circles are the cluster core and half mass radius, respectively \citep{lanzoni10}.}
\label{positions}
\end{figure}

Ter5aj is an isolated MSP with a spin period of 2.96 ms and a $\mathrm{ DM=238.63 \ pc \
cm^{-3}}$, very close to the cluster mean value. It is located $10.4\arcsec$  north
from the cluster gravitational center (\citealt{lanzoni10}, see Figure~\ref{positions}). Its spin period derivative is partially contaminated by the
effect of the MSP motion in the cluster potential field. We will analyze this in more
detail in Section~\ref{ter5_intrinsic}. 


{We roughly estimated the PSR mean flux density using again the radiometer equation on  daily detections. The values so obtained have been calibrated by comparison with those obtained applying the same method to other three isolated MSPs (namely Ter5R, Ter5S and Ter5T), for which measurements made referencing a flux calibrator are available (Ransom et al. 2017, in preparation). The average values of both the L-band and S-band flux densities are reported in Table~\ref{ajtiming}. The typical flux density of this MSP is of the order of that measured for the other faint isolated MSPs of this cluster (Ransom et al. 2017, in preparation)}.

\begin{table*}
\caption[Timing parameters for the new Terzan 5 MSPs.]{Timing parameters for the new Terzan 5 MSPs. Numbers in parentheses are uncertainties in the last digits
quoted; the time units are TDB and the adopted terrestrial time standard is UTC(NIST).}
\label{ajtiming}
\begin{center}{\scriptsize
\setlength{\tabcolsep}{4pt}
\renewcommand{\arraystretch}{1.3}
\begin{tabular}{l c c }
\hline
Pulsar  &   	Ter5aj                               &   Ter5ak           \\
\hline
\multicolumn{3}{c}{Timing Parameters}  \\
\hline
Right ascension, $\alpha$ (J2000)\dotfill  & 17:48:05.0119(2)  & 17:48:03.6860(2)    \\
Declination, $\delta$ (J2000)  \dotfill  & $-$24:46:34.85(7)   & $-$24:46:37.83(8)  \\
Spin frequency, $F$ (Hz)  \dotfill  &  337.96234149929(4) &  529.07066473956(4) \\
Spin frequency derivative, $\dot F$ $(10^{-14}$ Hz s$^{-1}$) \dotfill & $-$1.61313(7) & $-$2.4771(2)  \\
Spin frequency second derivative, $\ddot F$ $(10^{-25}$  Hz s$^{-2}$)  \dotfill & $-$1.8(4)  & - \\
Dispersion measure, DM (cm$^{-3}$ pc) \dotfill   & 238.633(6)   &  236.705(5)   \\
MJD range \dotfill  & 55423-57573 &  55423-57573  \\ 
Epoch (MJD) \dotfill & 56498 & 56498  \\
Data span (yr) \dotfill  & 5.9 &  5.9  \\
Number of TOAs \dotfill & 42  & 31   \\
Residuals RMS ($\mu$s) \dotfill  & 12.51 & 17.03  \\ 
Reduced $\chi^{2}$ value \dotfill & 1.00 & 1.04 \\
Solar system ephemeris model \dotfill & DE436 & DE436 \\
EFAC \dotfill & 1.14 & 1.15 \\
\hline
\\
\multicolumn{3}{c}{Derived Parameters} \\ \hline
Angular offset from cluster centre, $\theta_{\perp}$ (\arcsec) \dotfill & 10.4 & 17.4 \\
Spin period, $P$ (ms)  \dotfill  & 2.9589095505841(3) &   1.8901066845055(1) \\
Spin period derivative, $\dot P$ ($10^{-19}$) \dotfill & 1.41232(6) & 0.88495(6)  \\ 
Spin period second derivative, $\ddot P$ ($10^{-30}$ s$^{-1}$) \dotfill & 1.6(3) & - \\ 
Intrinsic spin period derivative, $\dot P_{\mathrm int}$ $(10^{-19})$\dotfill & $ <4.9$ &  $<2.1$\\
Characteristic age, $\tau_c$ (Myr) \dotfill & $>98$ & $>142$ \\
Surface magnetic field, $B_0$ ($10^8$ G) \dotfill &  $ <12.0$  & $<6.4$\\
Spin-down luminosity, $L_{SD}$ ($10^{35}$ erg s$^{-1}$) \dotfill & $<7.3$ & $<12.4$ \\
Flux density at 1.5 GHz, $S_{1.5}$ ($\mu$Jy)  \dotfill & 34 & 30 \\
Flux density at 2.0 GHz, $S_{2.0}$ ($\mu$Jy)  \dotfill & 18 & 16 \\
\hline
\end{tabular} }
\end{center} 
\end{table*}

\subsection{Ter5ak}

Ter5ak or J1748$-$2446ak has been discovered with a maximum stacked spectral power at $\mathrm{ DM=236.50
\ pc \ cm^{-3}}$. The PSR nature of this candidate has been confirmed by folding
the single observations, where it turned out to be visible in almost all of them. We managed to obtain a full timing solution for this object { using similar methods described for Ter5aj, except that we obtained the initial timing solution using the}
TEMPO based phase-connection routine available at \url{https://github.com/smearedink/phase-connect}, which will be presented in a forthcoming paper by Freire et al. (2017, in preparation). The timing solution is tabulated in
Table~\ref{ajtiming}, the post-fit residuals are reported in the bottom panel of
Figure~\ref{resid} {and the averaged pulse profile in the middle panel of Figure~\ref{sumprof}}.

Ter5ak is also an isolated MSP and it has a spin period of $\sim1.89$ ms, hence it is
the fourth fastest MSP in Terzan 5 and the fifth fastest among all the GC PSRs. { Its DM of $\mathrm{ 236.707 \ pc \ cm^{-3}}$ is well within the range covered by the other PSRs in the cluster}. Its position with respect to the other cluster
MSPs is reported in Figure~\ref{positions}, where it can be seen that it is located
at about $17.4\arcsec$ east from the cluster center. As for Ter5aj, its spin period
first derivative is clearly contaminated by its motion in the cluster potential field (see
Section~\ref{ter5_intrinsic}).
{The average flux density is also in this case of the order of that of the faintest isolated MSPs of this cluster.}

\subsection{Ter5al}

Ter5al is the last MSP identified in our analysis, with a spectral power peaked at
 $\mathrm{ DM=236.50 \ pc \ cm^{-3}}$. We have been able to reveal this
object in only $\sim20$ observations, being under the detection limit in all the
others. In the right panel of Figure~\ref{profili}, we report the detection plot of
the observation where Ter5al has the highest S/N. To date we have
not been able to obtain a timing solution for this system, likely because of the
insufficient number of  good detections. In the $\sim20$ detections we found no
evidence of acceleration, thus it is likely another isolated
system. Ter5al has a spin period of $\sim5.95$ ms. We determined its DM by measuring
the pulse TOAs in different sub-bands of the two brightest {L-band} observations
and we found $\mathrm{ DM=~236.48(3) \ pc \ cm^{-3}}$.

\begin{figure}
\centering
\includegraphics[width=11cm]{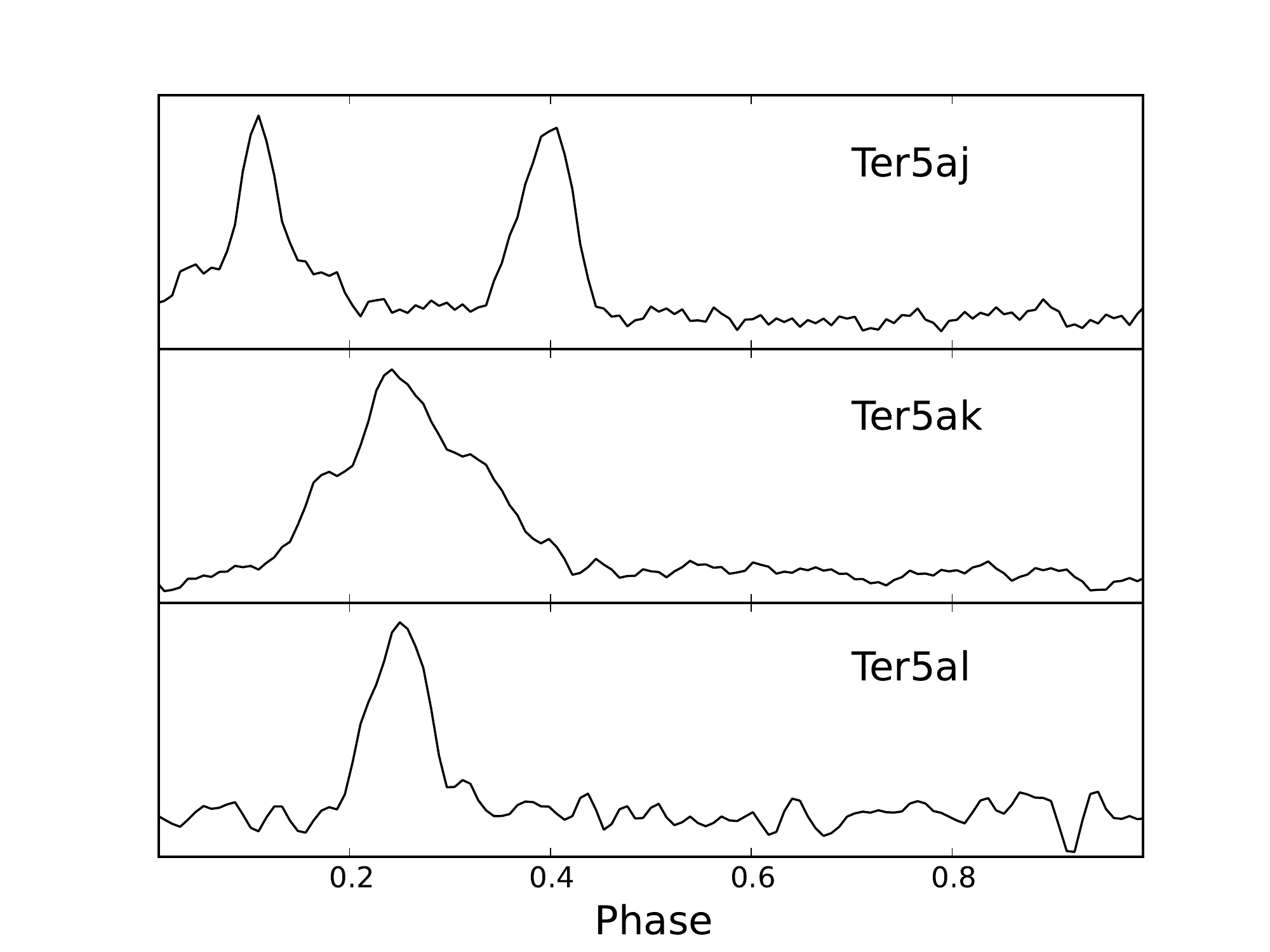}
\caption[Averaged pulse profile of Ter5aj, Ter5ak and Ter5al.]{Averaged pulse profile of Ter5aj (top panel), Ter5ak (middle panel) and Ter5al (bottom panel), obtained by coherently summing all the GUPPI detections of the MSPs.}
\label{sumprof}
\end{figure}

{The averaged pulse profile is reported in the bottom panel of Figure~\ref{sumprof}. This MSP turns out to be extremely faint. Indeed its average flux density is of only $\sim8 \; \mu$Jy in L-band and $\sim6\; \mu$Jy in S-band, making this PSR the faintest in the cluster and also explaining the small number of good detections}.

\section{Accelerations and physical parameters}
\label{ter5_intrinsic}
In this section we derive {some constraints on the accelerations and on the main} physical parameters of Ter5aj and
Ter5ak, the two new MSPs for which we have been able to obtain a timing solution.

For the case of MSPs in GCs, the measured spin period derivative ($\dot P_{meas}$),
derived through timing, does not represent a direct measurement 
of the MSP intrinsic spin-down, since it is the combination of
 different contributions \citep[see, e.g.,][]{phinney93}. Indeed, any motion of
a PSR with respect to the observer 
produces a change of the
observed spin period. In the case of GCs, the MSP motion in the cluster potential field induces a
change that can be large enough to match or even exceed the value due to the
intrinsic spin-down. Following \citet{phinney93}, $\dot P_{meas}$ can be written as
follows:
\begin{equation}
\label{accels}
\left( \frac{\dot P}{P}\right)_{meas} = \left( \frac{\dot P}{P}\right)_{int} + \frac{a_{z,GC}}{c} + \frac{a_g}{c} + \frac{a_s}{c}
\end{equation}
where $\left(\dot P/P\right)_{int}$  is the ratio between the intrinsic spin-down and the PSR
spin period, $a_c$ is the line of sight acceleration due to the GC potential field, $a_g$ is the
acceleration due to the Galactic potential, $a_s$ is an apparent centrifugal acceleration \citep[the
so-called Shklovskii effect;][]{shk70} and $c$ is the speed of light. The two latter terms are expected to be
negligible with respect to the former two, and following \citet[][and references within]{prager17}  
we know that
$a_g\approx 5.1\times10^{-10} \  \mathrm{ m \ s^{-2}}$ and $a_s\sim4.2\times10^{-12}  \  \mathrm{ m \
s^{-2}}$. According to \citet{freire05} and \citet{prager17}, the acceleration along our line of sight ($z$) due to the
cluster potential ($a_{z,GC}$) can be written as:
\begin{equation}
\label{ac}
a_{z,GC} (z,x) = -3.5\times10^{-7} \left( \frac{\rho_0}{10^{6} \; \Msun \ \mathrm{pc}^{-3}}\right)\left(\frac{z}{0.2 \ \mathrm{pc}} \right) \left(\sinh^{-1}(x) - \frac{x}{\sqrt{1+x^{2}}} \right) x^{-3} \ \mathrm{  m \ s^{-2}}
\end{equation}where $\rho_0$ is the cluster core density and $x\equiv r/r^{NS}_c$, where $r^{NS}_c$ is the
cluster core radius of the neutron star population and $r=\sqrt{r_{\perp}^2 +z^2}$ is distance of the PSR from
the cluster center. In the latter formula, $r_{\perp}=D\theta_{\perp}$ is the PSR
projected distance from the cluster center, where D is the distance of the cluster
from the Sun and $\theta_{\perp}$ the PSR angular offset from the cluster center. 

{\citet{prager17} used the ensemble of Terzan~5 MSPs, including Ter5aj and Ter5ak, to
derive the cluster physical properties. They found $\rho_0 = 1.58\pm0.13 \times 10^6 \; \Msun
\ \mathrm{pc^{-3}}$ and $r_c^{NS} = 0.16\pm0.01$ pc. 
Given the angular offsets of Ter5aj and Ter5ak from the cluster center (see Table~\ref{ajtiming}) and the cluster distance of 5.9
kpc \citep{lanzoni10}, we measured the possible line of sight accelerations of these two MSPs for different values of the line of sight distance $z$.  We found that the maximum allowed accelerations are $\pm 3.4 \times10^{-8} \ \mathrm{ m \ s^{-2}}$ and $\pm 2.0 \times10^{-8} \ \mathrm{ m \ s^{-2}}$ for Ter5aj and Ter5ak, respectively. We used these values to constrain, starting from Equation~\ref{accels}, the MSP intrinsic spin-down rates and, consequently, the characteristic ages,
surface magnetic fields and spin-down luminosities. All these values are tabulated in Table~\ref{ajtiming} and, for
both the MSPs, are in agreement with those typically expected for old and recycled MSPs.}

\cleardoublepage

\newcommand{\gras}[1]{\mbox{\boldmath $#1$}}
\def\etal{{et al.~}}
\def\ltsima{$\; \buildrel < \over \sim \;$}
\def\gtsima{$\; \buildrel > \over \sim \;$}
\def\lsim{\lower.5ex\hbox{\ltsima}}
\def\gsim{\lower.5ex\hbox{\gtsima}}
\def\lapp{\ifmmode\stackrel{<}{_{\sim}}\else$\stackrel{<}{_{\sim}}$\fi}
\def\gapp{\ifmmode\stackrel{>}{_{\sim}}\else$\stackrel{<}{_{\sim}}$\fi}
\def\mcom{M_{\rm COM}}
\def\mpsr{M_{\rm PSR}}
\def\apsr{A_{\rm PSR}}
\def\acom{A_{\rm COM}}
\def\dpsr{d_{\rm PSR}}
\def\rcom{R_{\rm COM}}
\def\rrl{R_{\rm RL}}
\def\Uc{\rm m_{F390W}}
\def\Vc{\rm m_{F606W}}
\def\Ic{\rm m_{F814W}}
\def\Hc{\rm m_{F656N}}
\def\apar{[}
\def\cpar{]}
\def\comA{COM-M71A}
\def\psrA{PSR J1953$+$1846A}
\newcommand{\msp}{J1953$+$1846A}
\def\Msun{\mathrm{M_{\odot}}}
\def\msun{$M_{\odot}$}

\def\Lsun{\mathrm{L_{\odot}}}
\def\Rsun{\mathrm{R_{\odot}}}

\chapter{The Black-Widow PSR J1953+1846A in the Globular~Cluster M71}
\label{cap_m71}
\begin{flushright}
\textit{Mainly based on \citealt{cadelano15a}, ApJ, 807:91}
\end{flushright}

\vspace{1cm}

\initial{T}his chapter is devoted to the identification and characterization of the companion star to PSR J1953+1846A, a ``black widow'' binary millisecond pulsar in the globular cluster M71. By using the accurate position and orbital
parameters obtained from radio timing, we identified the optical
companion in ACS/Hubble Space Telescope images. It turns out to be a faint (${ m_{F606W}}\approx 26$, ${ m_{F814W}}\approx 25$) and variable star 
located at only $\sim0.06"$ from the pulsar timing position. The
light curve shows a maximum at the pulsar inferior conjunction and a
minimum at the pulsar superior conjunction, thus confirming the
association with the system. The shape of the optical modulation suggests
that the companion star is heated, likely by the pulsar wind. The
comparison with the X-ray light curve possibly suggests the presence of an
intra-binary shock due to the interaction between the pulsar wind and the
material released by the companion. This is the second identification
(after COM-M5C) of an optical companion to a black widow pulsar in a
globular cluster. Interestingly, the two companions show a similar light curve
 and share the same position in the color magnitude diagram.

\clearpage

\section{Introduction}\label{intro}

PSR J1953+1846A (hereafter M71A)  is the only millisecond pulsar (MSP) known so far in M71 \citep{hessels07}. M71 is a low
density globular cluster \citep[GC; $\log \rho_{0}=2.83$ in units of $\mathrm{ L_{\odot}\; pc^{-3}}$;][2010 edition]{harris96}, 
in a disk-like orbit \citep{cadelano17a}, located at
$\sim4$ kpc from the Earth. It is one of the most metal-rich clusters
among GCs outside the bulge \citep{harris96} and its density profile shows
an extended core \citep[$r_{c}=56.2"$;][]{cadelano17a} and no
signatures of core collapse. M71A was discovered in a targeted survey
of all GCs visible with the 305-m Arecibo radio telescope
\citep{hessels07}. It is located at $\alpha {=19^{h}53^{m}46.42^{s} \ ; \
\delta=18^{\circ}47'04.84"}$, at a projected distance of only $20"$ (0.53 core radii) 
from the cluster center, it has a spin period of $\sim4.9$ ms and a low eccentricity orbit
of $\sim4.2$ hours. M71A is classified as a black-widow (BW). In fact, because of its
very low-mass function (${f\approx1.6\cdot10^{-5} \; \Msun}$), the companion is
expected to have a minimum mass of ${\sim0.032 \; \Msun}$. Moreover, as
commonly found for BW systems, the radio signal shows eclipses for about
20\% of the orbital period (at 1400 MHz observing frequency), likely due
to stripped material from an evaporating companion. A Chandra X-ray
observation of this cluster revealed a source in a position compatible
with the PSR location and a luminosity of about $10^{31}$ erg s$^{-1}$ in the $0.3-8.0$ keV spectral range \citep{elsner08}. The light
curve is consistent with a non-steady source and the photon index ($\Gamma
= 1.89 \pm 0.32$) suggests magnetospheric radiation and/or an emission
from intra-binary shocks. In the context of an optical study of the M71 X-ray sources, \citet{huang10}
suggested as possible optical companion to M71A a star located at $\sim0.1\arcsec$ from the radio PSR and lying along the red side of the cluster main sequence (MS), in a region commonly occupied by binary systems.
Nonetheless, its absolute magnitude
(${M_{V}\sim8.5}$) implies a mass of about ${0.5 \; \Msun}$, inconsistent
with radio-derived mass function (in fact such a large mass would be
compatible only with a nearly face-on orbit, where no radio eclipses are
expected). Hence, \citet{huang10} concluded that this object is unlikely
to be the real companion, which could be still below the detection
threshold or, alternatively, that M71A could be a hierarchical triple
system.

\section{Radio Timing}

Timing observations were carried out with the 305-m William E.\ Gordon telescope at the Arecibo Observatory in Puerto Rico, between MJDs 52420 (2002 May 26) and 53542 (2005 June 21), with the initial discovery observations on MJD 52082 (2001 June 22) incorporated into the timing solution.   The Gregorian L-band Wide receiver was used for the observations, sending dual-polarization data to the Wideband Arecibo Pulsar Processor \citep[WAPP; see][]{dsh00} autocorrelation spectrometers.  

All the data were then folded modulo the best-known PSR ephemeris.  A Gaussian profile was fit to the summed profile from several observations for use as a standard profile, and the {\tt FFTFIT} algorithm \citep{tay92} was used to determine pulse times of arrival (TOAs).  Time segments corresponding to eclipses and to the times when M71 transited at Arecibo (during which the telescope could not track the cluster) were not considered in the timing analysis.

\begin{table*}
\caption{Timing parameters for PSR~J1953+1847 (M71A). Numbers in parentheses are uncertainties in the last digits
quoted.}
\label{tab:M71A_params}
\begin{center}{\scriptsize
\setlength{\tabcolsep}{4pt}
\renewcommand{\arraystretch}{1.3}
\begin{tabular}{l c }
\hline
\multicolumn{2}{c}{Measured Parameters}  \\
\hline
Right ascension, $\alpha$ (J2000)\dotfill  & 
  19:53:46.41966(3)  \\
Declination, $\delta$ (J2000)  \dotfill  & 
  +18:47:04.8472(7) \\
Spin frequency, $F$ (Hz)  \dotfill  & 
204.57006473073(3)\\
Spin frequency derivative, $\dot F$ ($10^{-15}$ Hz s$^{-1}$) \dotfill & $-2.0299(3)$ \\
Spin frequency second derivative, $\ddot F$ $(10^{-25}$ Hz s$^{-2}$)  \dotfill & 5.4(3)\\
Epoch (MJD) \dotfill & 52812.0 \\
Dispersion measure, DM (cm$^{-3}$pc) \dotfill   & 
 117.3941(15) \\
DM derivative (cm$^{-3}\mbox{pc\,yr}^{-1}$)  \dotfill  & 
$-$0.0274(17) \\
Orbital period, $P_b$ (d) \dotfill  & 0.1767950297(2)\\ 
Projected semi-major axis, $x$ (s) \dotfill  & 0.0782246(12)\\
Epoch of Ascending Node, $T_0$ (MJD) \dotfill  & 52811.8761877(3)\\ \hline
\\
\multicolumn{2}{c}{Derived Parameters} \\ \hline
Spin period, $P$ (ms) \dotfill & 4.8883007458412(6) \\
Spin period derivative, $\dot P$ $(10^{-20})$ \dotfill & 4.8506(8)\\
Spin period second derivative, $\ddot P$ $(10^{-29}$ s$^{-1}$)  \dotfill & $-1.28(7)$\\
Angular offset from cluster centre, $\theta_{\perp}$ ($^\prime$) \dotfill & 0.33 \\
Intrinsic period derivative, $\dot P_{{ int}}$ $(10^{-20})$\dotfill & $4.3 <\dot P_{\mathrm{ int}}< 5.4$\\
Characteristic age, $\tau_c$ (Gyr) \dotfill & $1.4 < \tau_c < 1.8$ \\
Surface magnetic field, $B_0$ ($10^8$ G) \dotfill & $4.6 < B_0 < 5.2$\\
Mass function $f$ (M$_\odot$) \dotfill & 0.0000164427(8) \\
Minimum companion mass $m_c$ (M$_\odot$)\tablenotemark{b} \dotfill & 0.032\\
\hline
\end{tabular} }
\end{center} 
\end{table*}

\begin{figure}
  \begin{center}
\includegraphics[width=12.5cm]{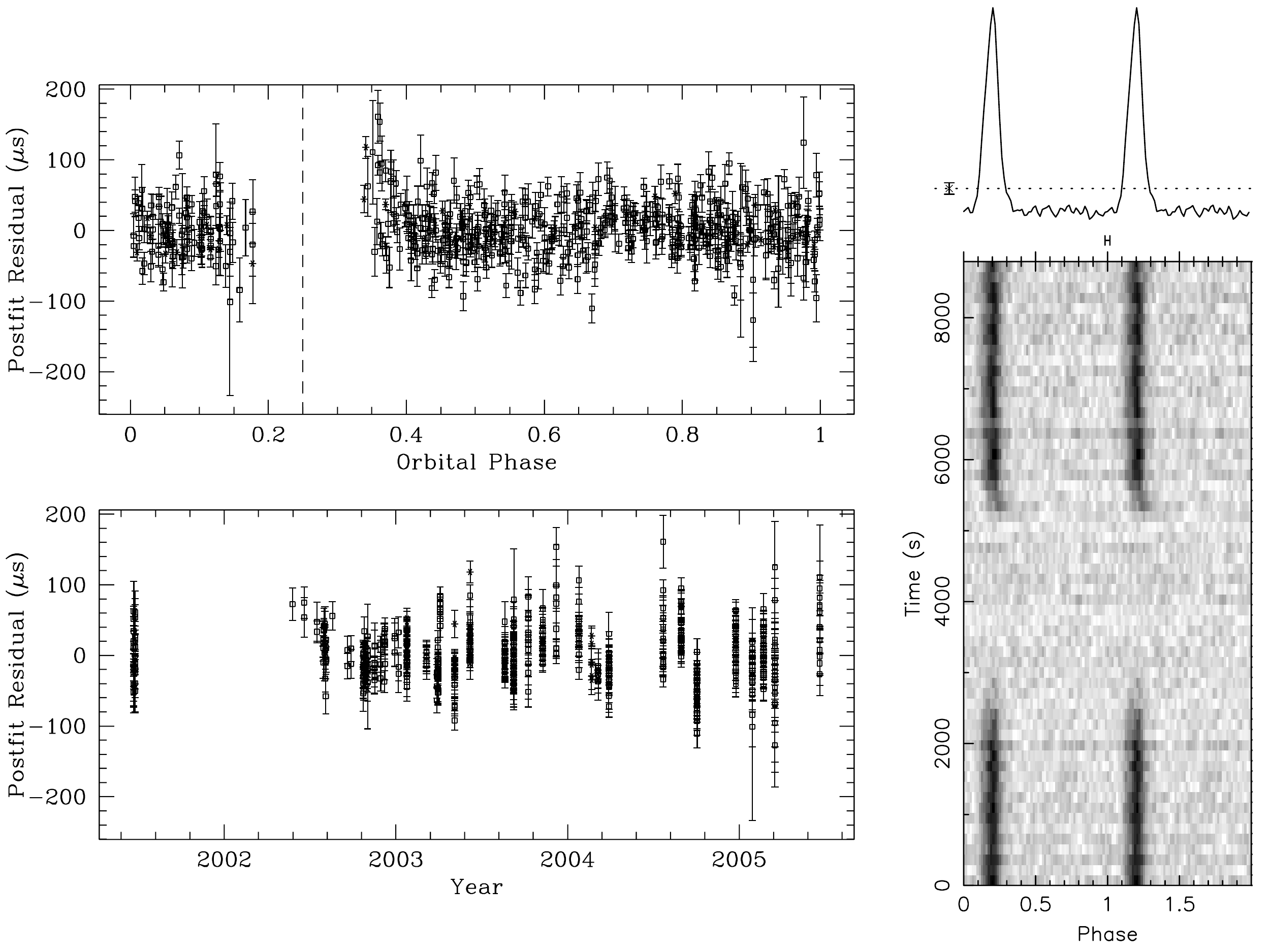}
\caption[Postfit timing residuals and pulse profile of M71A]{{\it Left panel:} Postfit timing residuals for M71A, as functions of orbital phase and date.  The dashed line indicates orbital phase 0.25, the PSR superior conjunction. {\it Right panel:} An observation of M71A on MJD 52798 (2003 June 8). The abrupt disappearance of the PSR at the start of eclipse, as well as the slight dispersive delay on reappearance, are clearly visible.  The cumulative pulse profile is plotted twice at the top of the figure.}
\label{fig:M71A_timing}
\end{center}
\end{figure}

\begin{figure}[!b]
  \begin{center}
\leavevmode 
\includegraphics[width=10cm]{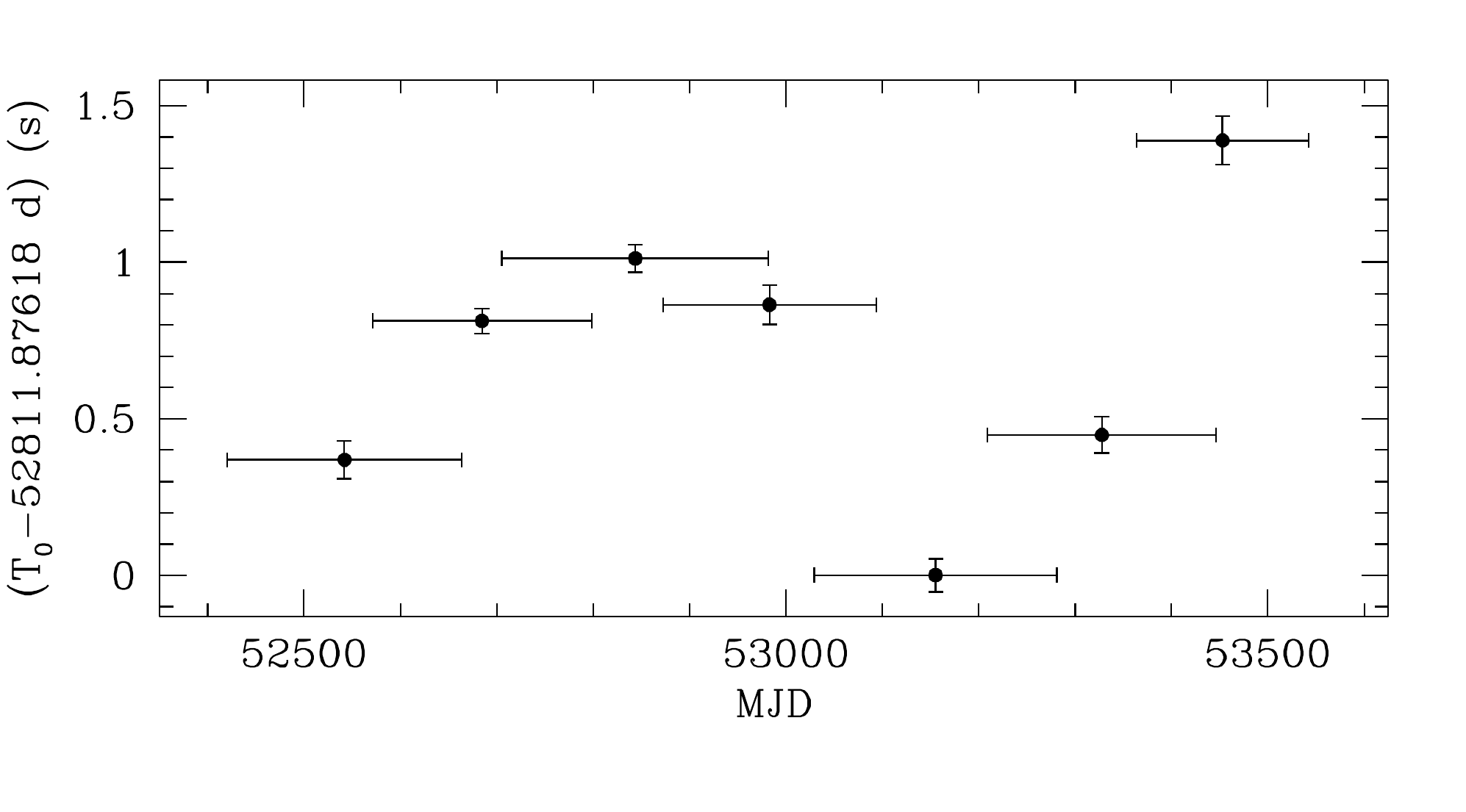}
\caption[Variation of the time of passage through ascending node]{Variation of the time of passage through ascending node (orbital phase 0), computed for overlapping segments of data and holding all other timing parameters fixed at their nominal values.}
\label{fig:t0seg}
\end{center}
\end{figure}

Timing analysis was performed with the {\tt TEMPO} software package\footnote{tempo.sourceforge.net} using the DE421 Solar System ephemeris and the TT(BIPM) clock standard.  The {\tt BT} timing model of \citet{bt76} was used, as the orbit has no significant eccentricity.  The timing parameters are listed in Table~\ref{tab:M71A_params} and residuals are presented in Figure~\ref{fig:M71A_timing}.  The root-mean-square postfit residual is 35$\,\mu$s.  The reduced-$\chi^2$ of the fit is 4.8; however, we list the parameter uncertainties as reported by {\tt TEMPO} without scaling, as the epoch-to-epoch wander in the residuals is likely due to the interactions between the two stars (see Figure~\ref{fig:t0seg}) rather than to any misestimation of the TOA uncertainties.   The high-precision radio timing position is slightly offset (0.06$^{\prime\prime}$) from the position of the optical counterpart (see Section~\ref{identification}), but agrees within the much larger uncertainty (0.2$^{\prime\prime}$) of the latter. Following the reasoning in \citet{fhn+05}, we find the maximum possible acceleration due to the { gravitational field} of the cluster for this line of sight to be $\pm 3.2 \times 10^{-10}$\,m\,s$^{-2}$.  This implies that most of { the} observed pulse period derivative (${\dot P}$) is intrinsic.  Further corrections due to the differential acceleration in the Galaxy \citep[e.g.][]{nt95,rmb+14} are small.  Given the small velocity dispersion in the core of the cluster \citep[$\mathrm{ 2.3 \ km \ s^{-1}}$;][]{harris96}, the velocity of the PSR relative to that of the cluster should be very small; therefore its proper motion should be very similar to the proper motion of the cluster \citep[$\mathrm{\sim3.5 \ mas \ yr^{-1}}$;][]{cadelano17a} as a whole, making the corresponding correction to ${\dot P}$ \citep{shk70} about half the size of that due to the Galactic acceleration.  The timing data do not allow us to derive a reliable proper motion for the PSR.  We use the range of allowed accelerations to constrain the intrinsic ${\dot P}$ as well as the characteristic age and surface magnetic field in Table~\ref{tab:M71A_params}.

The PSR is asymmetrically eclipsed between approximate orbital phases of 0.18 and 0.35, where orbital phase 0.25 represents superior conjunction.  The eclipses begin fairly abruptly but when the signal returns, it at first suffers excess dispersive delay due to ionized material within the orbit (Figure~\ref{fig:M71A_timing}). The mass loss from the companion star has a further manifestation in the variation of orbital parameters: Figure~\ref{fig:t0seg} shows the value of the time of ascending node passage for overlapping subsets of the data.  The variation is comparable to that seen in other black widow eclipsing systems \citep[e.g.,][]{aft94,nbb+14} and significantly less than what is typically present in the redback (RB) systems \citep[e.g.,][]{akh+13}, which have much more massive, likely non-degenerate companion stars.

\section{Optical Photometry of the Companion Star}\label{identification}
\subsection{Observations and data analysis}\label{Sec:dataan}

The identification of the companion to M71A has been performed through two datasets of high resolution
images acquired with the Wide Field Camera (WFC) of the Advanced Camera for Surveys (ACS) mounted on
the Hubble Space Telescope (HST). The primary dataset has been
obtained on 2013 August 20 (GO12932, P.I.: Ferraro) and consists of a set of ten images in the F606W
filter (with exposure times: $\mathrm{ 2\times459 \ s ;\ 3\times466 \ s ; \ 5\times500 \ s}$) and nine
images in the F814W filter ($\mathrm{ 5\times337 \ s; \ 3\times357 \ s; \ 1\times440 \ s}$). We also
analyzed an archival dataset, obtained on 2006 July 1 (GO1775, P.I.: Sarajedini) with the same
instrument and same filters. It consists of four F606W images with an exposure time of 75 s and four
F814W images with an exposure time of 80 s.\\

 The standard photometric analysis has been performed
on the ``flc'' images, which are corrected for flat field, bias, dark counts and charge transfer
efficiency. These images have been further corrected for ``Pixel-Area-Map''\footnote{For more
details see the ACS Data Handbook.} with standard IRAF procedures. By using the {\tt DAOPHOT II ALLSTAR}
and {\tt ALLFRAME} packages \citep{stetson87}, we performed an accurate photometric analysis of each
image, following the procedure described in Chapter~\ref{opticalcom}. The resulting catalog of stars contains instrumental magnitudes that have been calibrated to the VEGAMAG system
cross-correlating with {\tt CataXcorr} our catalog with that by \citet{anderson08},
using the $\sim7600$ stars in common.
Since the WFC images suffer heavily from geometric distortion, we corrected the instrumental
positions (x,y) by applying the equations reported by \citet{meurer03} and using the coefficients in
\citet{hack01}. Then we transformed instrumental positions into the absolute astrometric system
($\alpha, \delta$) using the stars in common with the \citet{anderson08} catalog. The resulting
astrometric solution has an accuracy of $\sim0.14''$ in $\alpha$ and of $\sim0.13''$ in $\delta$,
corresponding to a total position accuracy of $\sim0.2''$.

\subsection{The companion to M71A}

\begin{figure*}[!t]
\begin{center}
\leavevmode
\includegraphics[width=10cm]{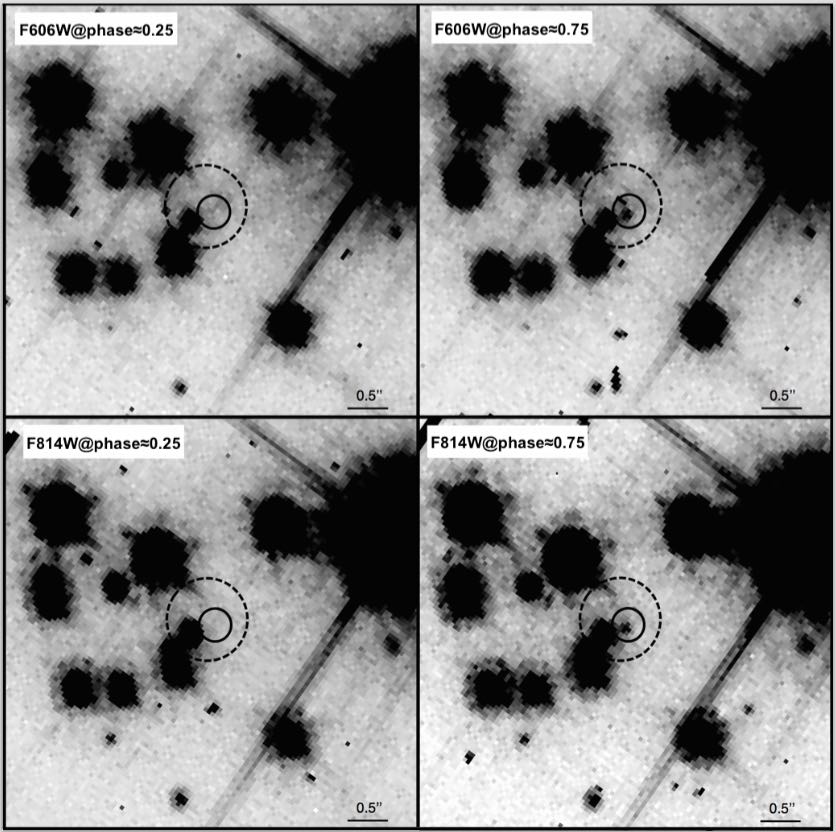}
  \caption[Primary dataset HST images of the $5" \times 5"$ region around the nominal position of M71A]{Primary dataset HST images of the $5" \times 5"$ region around the nominal position of M71A. The filters and the orbital phases are labeled in each panel. The solid circle is centered on the radio position and it has a radius of $0.2"$ (which is larger than the formal uncertainty from PSR timing). The dashed circle is centered on the X-ray counterpart and it has a radius of $0.5"$. The relatively bright star on the left border of the solid circle is the candidate optical companion proposed by \citet{huang10}. COM-M71A is clearly visible inside the solid circle in the right panels (corresponding to the inferior conjunction of the PSR, where the companion reaches maximum brightness), while in the left panels (at superior conjunction of the PSR) it is below the detection threshold.}
  \label{charts_M71A}
\end{center}
\end{figure*}

The search for the companion star to M71A was performed by means of an accurate photometric analysis
of all the detectable objects within a $5^{\prime\prime} \times5^{\prime\prime}$ wide region centered on the nominal
position of the MSP. Figure \ref{charts_M71A} shows the zoomed ($0.5^{\prime\prime} \times0.5^{\prime\prime}$) central
part of that region. As can be seen, a relatively bright object is found to have a position
compatible with the X-ray source (dashed circle) and the radio source (solid-line circle). This is
the star proposed by \citet{huang10} to be the optical counterpart to M71A. However, a much fainter
object, showing a strong variability, is visible in the figure. This is a quite promising object and
it is located at ${ \alpha=19^{h}53^{m}46.4062^{s} \ ; \ \delta=18^{\circ}47^{\prime}04.793^{\prime\prime}}$,
only $0.06^{\prime\prime}$ from the radio position and $0.13^{\prime\prime}$ from the X-ray source, thus in
perfect positional coincidence within our positional uncertainty ($\sim0.2^{\prime\prime}$). In
the primary dataset, it has been detected in 9 (out of 10) images in the F606W filter, with a
magnitude variation ranging from ${ m_{F606W}\approx24.3}$ to ${ m_{F606W}\approx27}$, while in
the F814W filter has been detected in 6 images (out of 9) and the magnitude varies from ${ m_{F814W}\approx23.4}$ to ${ m_{F814W}\approx24.9}$. Unfortunately, the images in the archival
dataset are too shallow to properly detect this faint object: in fact it turned out to be above the
detection threshold in only one exposure in the F814W filter.  For the four deep exposures of the
primary dataset in which the star is not visible, we estimated an upper magnitude limit by
simulating an artificial star of decreasing magnitude at the position of the candidate companion.
The derived detection threshold turned out to be ${ m_{F606W}\sim26.5}$ and  ${
m_{F814W}\sim25.9}$.\\

\begin{figure*}[!b]
\begin{center}
\leavevmode
\includegraphics[width=10cm]{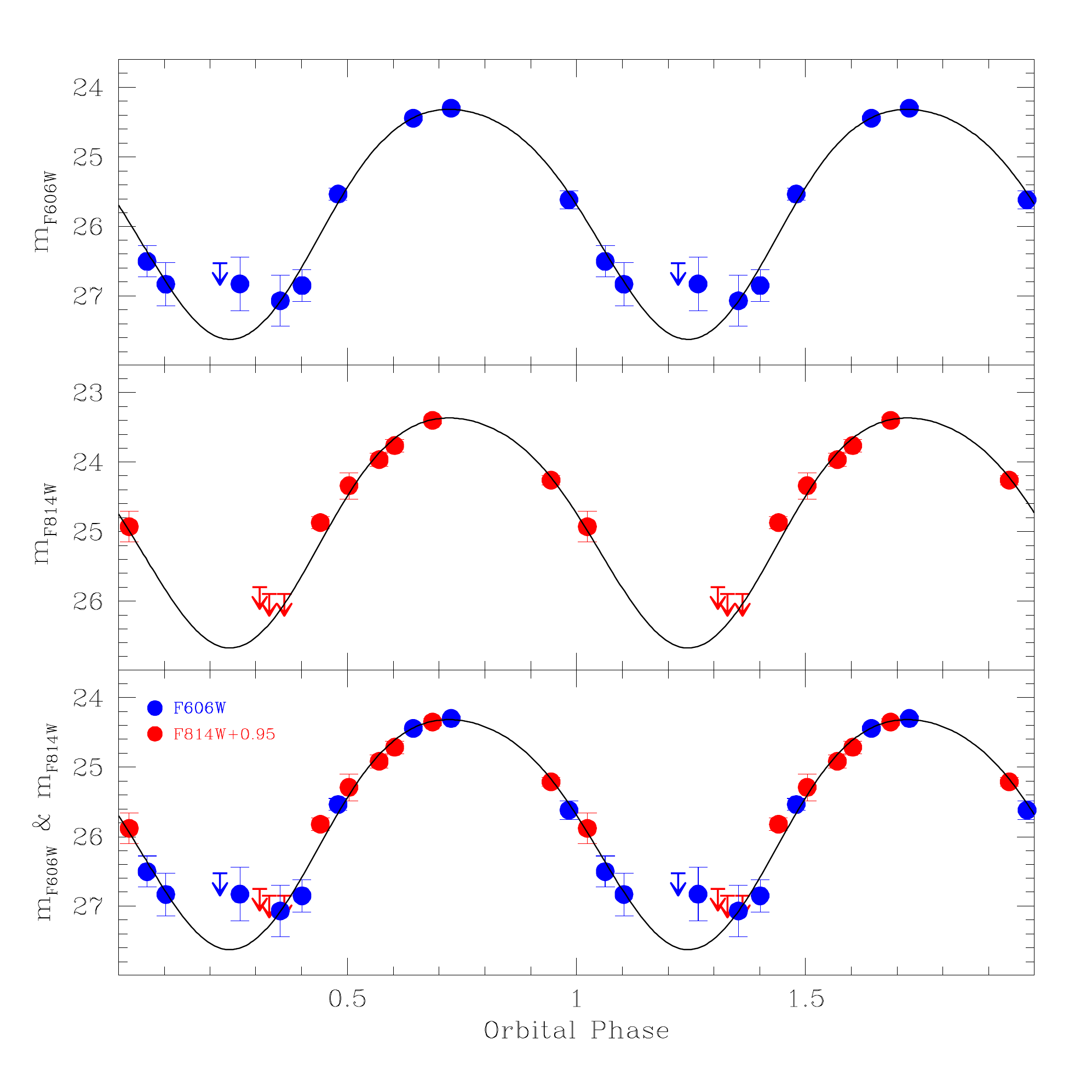}
  \caption[Light curves of COM-71A in the F606W and F814W filters]{Light curves of COM-71A in the F606W and F814W filters separately (upper and middle panels) and for the combination of the two (bottom panel), obtained after a 0.95 mag shift of the F814W magnitudes. All curves are folded with the radio parameters and two periods are shown for clarity. Circles mark the observed points, arrows are the magnitude upper-limits for the images where the star is below the detection threshold. The black curve in each panel is the best analytical model obtained from the combined light curve and then adapted to each filter.}
  \label{curve}
\end{center}
\end{figure*}

\begin{figure*}
\begin{center}
\includegraphics[width=10cm]{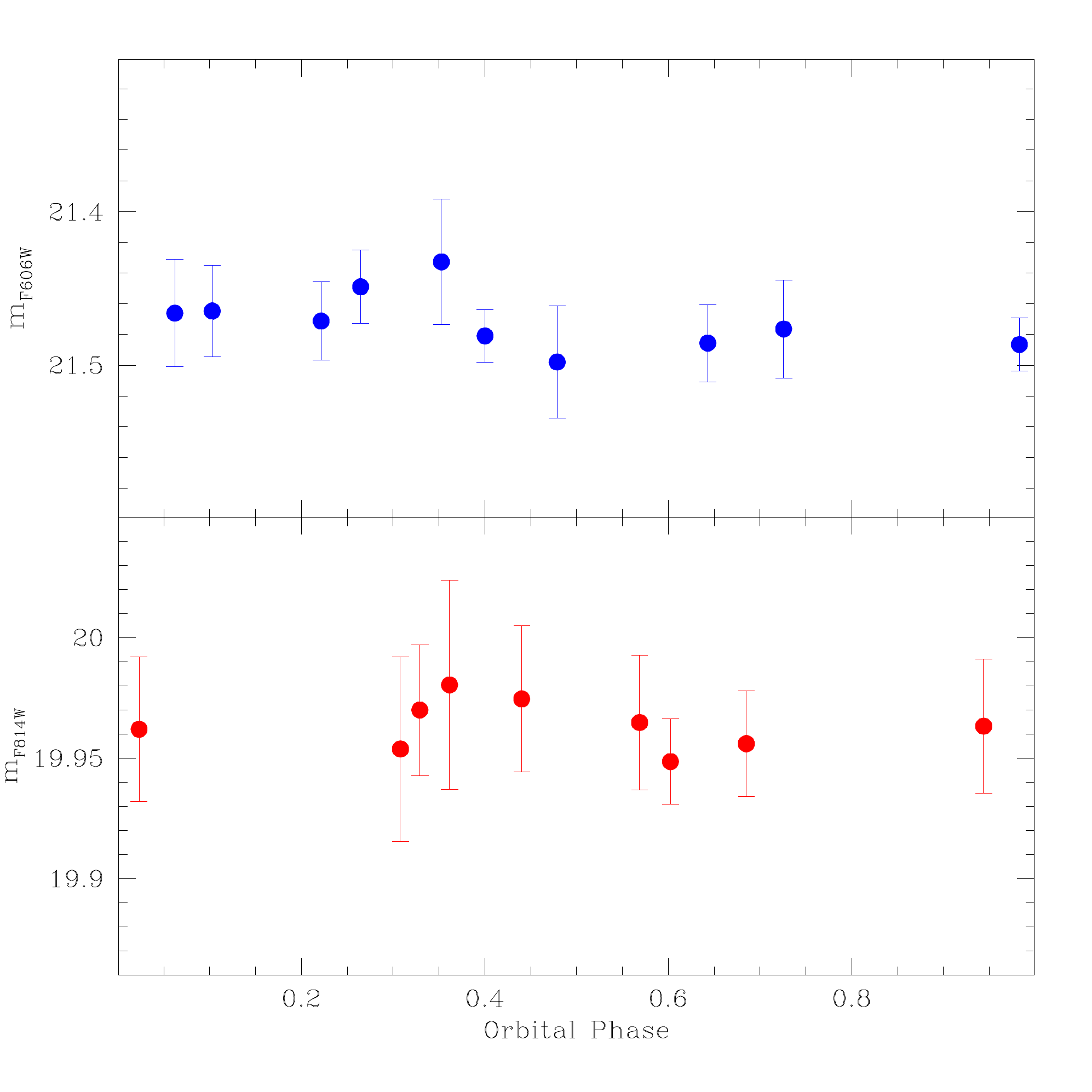}
  \caption[Light curves of the candidate companion proposed by \citet{huang10}]{Light curves of the candidate companion proposed by \citet{huang10}, folded with radio orbital parameters. The absence of any magnitude modulation as a function of the orbital phase is the definitive confirmation that this object is not connected to M71A.}
  \label{curvano}
\end{center}
\end{figure*}

In order to reliably establish that the detected star is the binary companion to M71A, we
built the light curve in both the available filters by folding the optical measurements with the
orbital period and the ascending node time of the PSR (see Table~\ref{tab:M71A_params}). The results are shown in Figure \ref{curve}, and in Table~\ref{tab1} we report the MJD of the images with their related orbital phases and magnitudes. As
can be seen, the light curves show a sinusoidal modulation spanning at least three magnitudes and it
is fully consistent with the orbital period of the binary system. This establishes the physical
connection between the variable star and the MSP. Indeed the exposures in which the star is not
detected nicely correspond to the light curve minima. The curves have a maximum at $\phi
\approx0.75$, corresponding to the PSR inferior conjunction (where we observe the companion side
facing the PSR) and a minimum at $\phi \approx0.25$, corresponding to the PSR superior conjunction
(where we observe the back side of the companion). This behavior is indicative of a strong heating
of the companion side exposed to the PSR emission and it is in good agreement with the observed
optical properties of other similar objects \citep[e.g.][]{stappers01, reynolds07,
pallanca12, breton13, pallanca14a, li14}. For the sake of comparison, in Figure
\ref{curvano} we plot the light curve (folded following the same procedure described above) of the
possible companion suggested by \citet{huang10}. As can be seen the star does not show any
significant flux variation. \\

\begin{table}[!t]
\centering
\scriptsize
\captionsetup{justification=centering}
 \caption{Optical observations of COM-M71A}
   \begin{tabular}{|c|c|c|c|}
     \hline
   $\phi$ & $t$ (MJD) & ${ m_{F606W}}$  &  ${ m_{F814W}}$ \\
         \hline
   0.02 &  56524.57602385  & $-$ & $24.9 \pm 0.2$ \\
   \hline
  0.06 &  56524.58290459  &  $26.5 \pm 0.2$ & $-$  \\
      \hline
   0.10 &   56524.59012681 &  $26.8 \pm 0.3$ & $-$ \\
      \hline 
   0.22 &  56524.43431532 &  $>26.5$ & $-$   \\
      \hline 
  0.26 &    56524.44193085 & $26.8 \pm 0.4$ & $-$ \\
      \hline
  0.31 &  56524.44960589 & $-$ & $>25.9$ \\
      \hline
 0.33 & 56524.63013255 & $-$ &  $>25.9$ \\
      \hline
 0.35 & 56524.45750982 &  $27.1 \pm 0.4$  & $-$ \\
      \hline
  0.36 &  56524.63586163 & $-$ & $>25.8$ \\
      \hline
 0.40 & 56524.64270200 &  $26.8 \pm 0.2$ & $-$ \\
      \hline
  0.44 & 56524.64977385 &  $-$ & $24.87 \pm 0.09$ \\
      \hline
   0.48 &   56524.65661404 &  $25.54 \pm 0.08$ & $-$ \\ 
   \hline
  0.50 & 53867.78512088$^{a}$ &  $-$ & $24.3 \pm 0.2$ \\
      \hline
   0.57 & 56524.49569959 & $-$ & $23.97 \pm 0.09$\\
      \hline
  0.60 & 56524.50166033 &  $-$ & $23.76 \pm 0.09$ \\
      \hline
 0.64 & 56524.50885348 &  $24.44 \pm 0.03$ & $-$ \\
      \hline
  0.68 & 56524.51627848 &  $-$ & $23.40 \pm 0.05$ \\
      \hline
   0.73 &  56524.52347163 & $24.30 \pm 0.03$ & $-$\\
      \hline 
  0.94 &  56524.56203070 & $-$ &  $24.26 \pm 0.06$ \\
      \hline
  0.98 &  56524.56891144  &  $25.6 \pm 0.1$ & $-$ \\
      \hline
         \end{tabular}    
         \tablecomments{Orbital phases ($\phi$), corresponding MJD ($t$) of the observations and observed magnitudes or upper-limits in both filters.\\  $^{a}$ This is the only image of the archival dataset where the companion star is above the detection threshold.  
}
       \label{tab1}      
   \end{table}

All these pieces of evidence suggest that the faint variable (which we name COM-M71A) is the optical
companion to the BW M71A. It is the tenth MSP optical companion and the second to a BW system in a
GC. In the color-magnitude diagram (CMD), COM-M71A is located at faint magnitudes in
a region between the MS and the white dwarf (WD) cooling sequences, where no normal GC stars are expected. This
position is indicative of a non degenerate or semi-degenerate, low-mass and swollen star.
Interestingly, the position of this object in the CMD is quite similar to that of COM-M5C, the only
companion to a BW system known in GCs up to now and recently identified by \citet{pallanca14a} in
the GC M5.\\

\section{Discussion}

Since the available data do not uniformly sample the orbital phases of
the system in either the F606W or the F814W filters (see
upper and middle panels of Figure~\ref{curve}), in order to accurately determine
the light curve of the companion star we combined the two datasets
together, by applying a 0.95 mag shift to the F814W magnitudes (bottom
panel of Figure~\ref{curve}). We then used the software {\tt GRATIS}\footnote{``Graphical Analyzer for TIme Series'' is a software aimed at studying stellar variability phenomena. Developed by Paolo Montegriffo at INAF-Osservatorio Astronomico di Bologna.} and a $\chi^{2}$ criterion to determine the two harmonics best fit
model\footnote{Note that a single harmonic model (i.e. a sinusoidal function) does not provide a good match of the observed light curve.}  to the curve. This is shown as a solid line in
the bottom panel of Figure~\ref{curve}.  Finally, we verified that the same
solution also provides a good fit to the light curves in each filter
separately. Indeed, the reduced $\chi^{2}$ turned out to be 1.50 for the
F606W filter, and 1.75 in F814W (see the solid curves in the upper and
middle panels of Figure~\ref{curve}).  This demonstrates that no significant
modulation of the stellar color (temperature) along the orbit is
measurable from the available dataset, and a much finer sampling of
the light curve is needed to provide additional clues on this
possibility. In Table~\ref{tab3} we report the maximum and minimum values
for both the magnitude and the flux in each filter, evaluated from the
best-fit model by following the procedure described in \citet{bolin12}
for the ACS. The uncertainties are calculated by using the mean
photometric errors for stars with similar magnitudes.  The magnitude
shift needed to superimpose the F814W light curve to that in the F606W
filter implies a color index of $0.95\pm0.12$ for the companion
star. By adopting a $0.54 \; \Msun$ WD cooling sequence from the BaSTI catalog\footnote{http://basti.oa-teramo.inaf.it} \citep{manzato08, salaris10}, this value can be converted into a
temperature of $5100\pm800$ K, which is in good agreement with those
evaluated for other BW systems \citep[e.g.][]{stappers01, pallanca12, breton13, pallanca14a, li14}.\\

\begin{table}[t]
\centering
\scriptsize
\captionsetup{justification=centering}
 \caption{Optical properties of COM-M71A}
  \begin{tabular}{|c|c|c|}
     \hline
       & F606W & F814W \\
   \hline
   ${m_{bright}}$ & $24.31\pm0.01$ & $23.37\pm0.02$ \\
   \hline
   ${m_{faint}}$ & $27.62\pm0.09$ & $26.7\pm0.1$ \\
   \hline
   ${ F_{bright}}$  ($10^{-17} \  \mathrm{ erg \ cm^{-2} \ s^{-1}}$) & $126\pm1$ & $128\pm2$ \\
   \hline
   ${ F_{faint}}$ ($10^{-17} \  \mathrm{ erg \ cm^{-2} \ s^{-1}}$)  & $5.9\pm0.5$ & $6.1\pm0.6$ \\
   \hline
   $\Delta$F ($10^{-17} \   \mathrm{ erg \ cm^{-2} \ s^{-1}}$) & $120\pm50$  & $120\pm60$ \\
   \hline
   ${ L_{bright}}$ ($10^{29} \mathrm{ erg \ s^{-1}}$) & $24.2\pm0.2$ & $24.5\pm0.4$ \\
   \hline
   ${ L_{faint}}$ ($10^{29} \mathrm{ erg \ s^{-1}}$) &   $1.15\pm0.09$ & $1.2\pm0.1$ \\
         \hline    
    \end{tabular}  
     \tablecomments{Maximum and minimum luminosities of COM-M71A as derived from the model light curve.}
       \label{tab3}
\end{table}

\begin{figure*}[!t]
\begin{center}
\includegraphics[width=10cm]{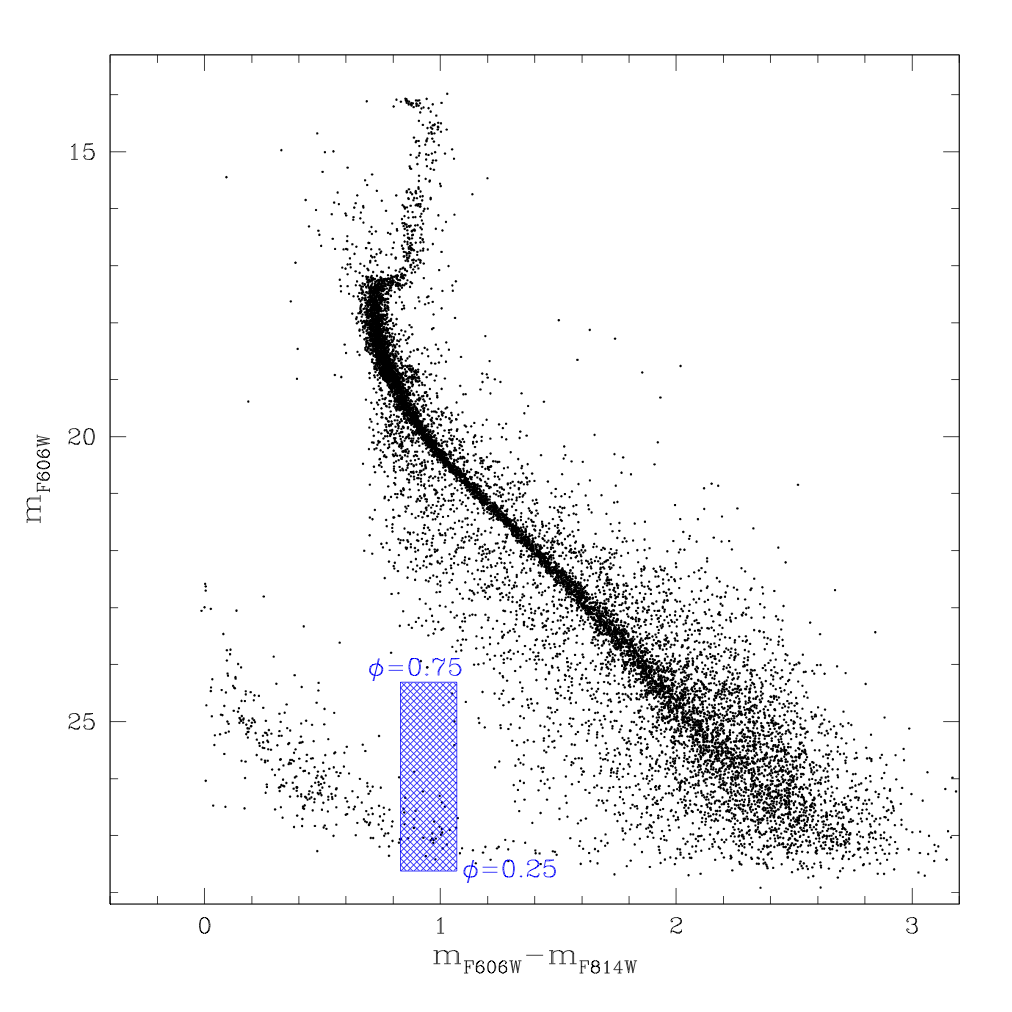}
  \caption[CMD of M71 with highlighted in blue the region occupied by COM-M71A]{CMD of M71 with highlighted in blue the region occupied by COM-M71A during the whole orbital period, as predicted by the light curve model (see text, Figure~\ref{curve} and Table~3).}
  \label{cmdmotion}
\end{center}
\end{figure*}

In Figure~\ref{cmdmotion} we show the CMD, with the shaded rectangle
marking the region occupied by COM-M71A during the orbital period. The height of the rectangle
corresponds to the maximum ${ m_{F606W}}$ magnitude difference expected from the best-fit model shown 
in Figure~\ref{curve}, while the width corresponds to the photometric error at the minimum 
luminosity. As already mentioned, the star is
located between the MS and the WD cooling sequence, and it spans a range
of about three magnitudes. Of particular interest is the predicted
star position during the PSR superior conjunction, where we expect to
see the stellar side not exposed to the PSR flux (clearly this is
exactly the case only for $i = 90^{\circ}$). The CMD position of COM-M71A
in this phase could be compatible with the He WD cooling sequence,
suggesting a semi-degenerate stellar structure. However, at these low
luminosities, our analytical model is not appropriately constrained by
data. Therefore, in order to confirm this possibility, further
observations are needed. In principle, the companion mass can be constrained from the
comparison of its CMD position and theoretical isochrones.  However,
in the case of strongly perturbed stars the mass inferred in this way
can be overestimated, or suggestive of inclination angles too small to
be consistent with the presence of radio eclipses \citep[see][]{fer03_msp, pallanca10, mucciarelli13}. In our
case, not only the companion position in the CMD is clearly out of the
canonical evolutionary sequences, but also its minimum luminosity is
not properly constrained by the observations.\\

Assuming that the companion optical emission is mostly due to black-body radiation, the luminosity and temperature of this star would be consistent with an object of radius $R_{BB}\leq0.02 \; R_{\odot}$. However, the companions to BWs usually suffer from strong tidal distortion due to the interaction with the PSR, therefore they are swollen up and possibly they can even fill their Roche Lobes. Furthermore, the presence of radio eclipses suggests that the simple $R_{BB}$ is a gross underestimate of the true stellar radius. Indeed, the Roche Lobe (RL) radius is far more appropriate to describe the size of the companion \citep[e.g.][]{stappers96, pallanca12, breton13, pallanca14a}. According to \citet{eggleton83}, the Roche Lobe (RL) radius can be computed as:
\begin{equation}
\label{RL}
{R_{RL}\simeq \frac{0.49q^{\frac{2}{3}}a}{0.6q^{\frac{2}{3}}+\ln\left(1+q^{\frac{1}{3}}\right)}},
\end{equation}
where q is the ratio between the companion and the PSR masses and a is the orbital separation. Combining this relation with the PSR mass function, assuming a neutron star mass ranging from $1.2\; \Msun$ to $2.5 \; \Msun$ \citep{ozel12} and an inclination angle ranging from $0^{\circ}$ to $90^{\circ}$, we find $0.22\; \Rsun<{R_{RL}}<1.24 \; \Rsun$.\\

Interestingly, the light curve shape presents a hint of asymmetric structure in both filters: the increase to the maximum seems to be smoother than the decrease to the minimum. Despite the low statistic, this behavior could be due to a slight asynchronous rotating companion, as in the case of PSR J2051$-$0827 \citep[see][]{stappers01}. This could be the result of a tidal torque from the wind of a magnetically active star, which can result in a companion angular velocity that differs from the orbital angular velocity. Moreover, in this case the angular velocity could be subject to a variation with time due to a secular time dependence of the orbital period, due itself to a variation of the companion quadrupole moment \citep[see e.g.][]{applegate94, doroshenko01, handbook}. However, in order to probe this intriguing possibility, an uniform sampling of the light curve from new observations is needed.\\

\subsection{Reprocessing efficiency and Roche Lobe filling factor}

\begin{figure*}[!h]
\begin{center}
\includegraphics[width=10cm]{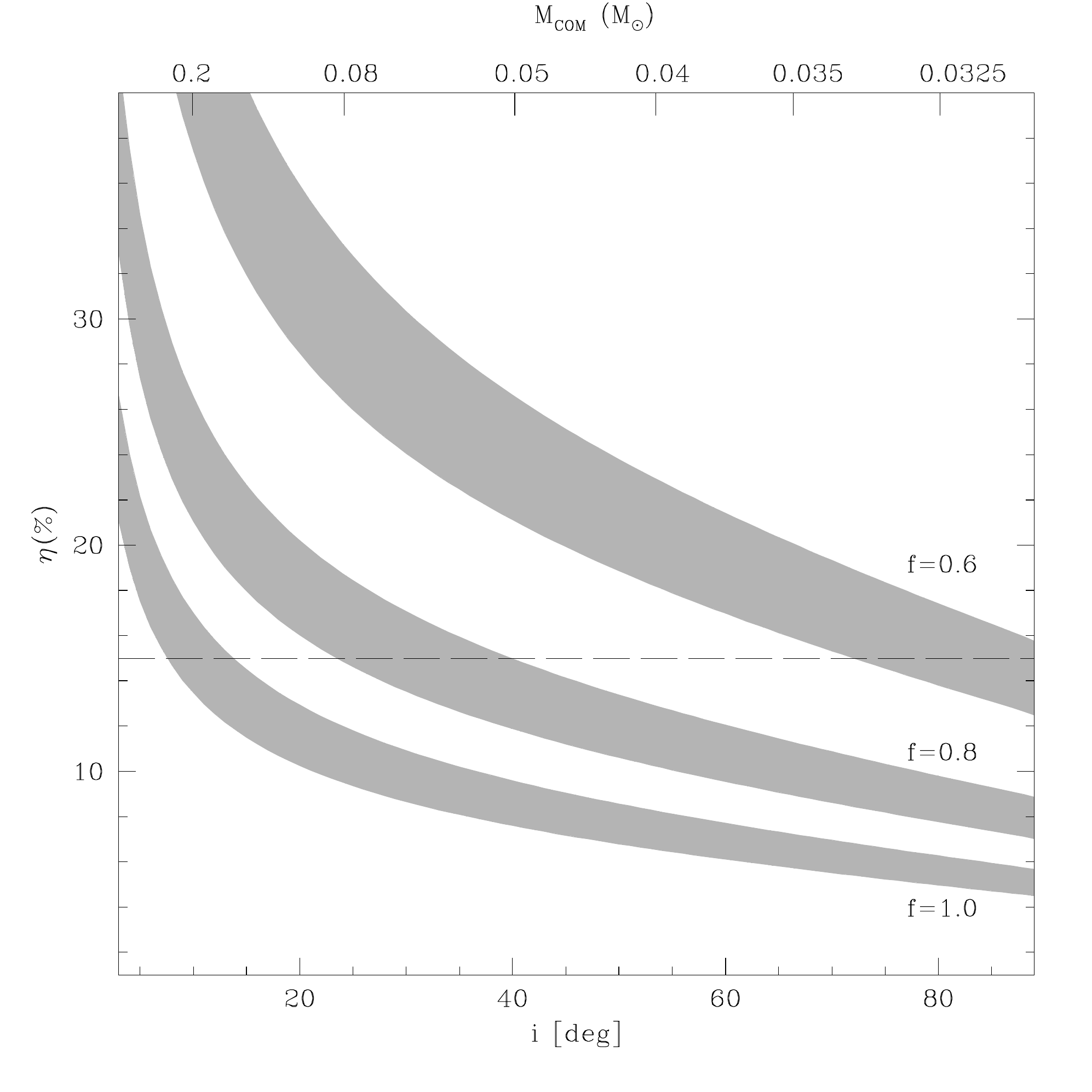}
  \caption[Reprocessing efficiency of the PSR emitted energy as a function of the inclination angle]{Reprocessing efficiency of the PSR emitted energy as a function of the inclination angle, assuming three different values of the RL filling factor and a PSR mass of $1.4 \; \Msun$. The thickness of each strip corresponds to the range of spin-down energies measured for this PSR (see text). The horizontal dashed line at $\eta=15\%$ is a typical reprocessing efficiency reported in \citet{breton13}. On the top axis, the companion masses for a PSR mass of $1.4 \; \Msun$ are reported.}
 \label{eff}
\end{center}
\end{figure*}

Under the assumption that the optical magnitude modulation is mainly due to the heating of the
companion surface by the PSR flux, we can compare the observed flux amplitude of the light curve with the expected one ($ {\Delta  F_{exp}}$) as a function of the
inclination angle ($i$), by the following relation \citep{pallanca14a}:
\begin{equation}
\label{delta}
{\Delta F_{exp}(i)=\eta \frac{\dot{E}}{a^{2}}R_{COM}^{2}(i)\frac{\epsilon(i)}{4 \pi d_{PSR}^{2}}}, 
\end{equation}
where $\eta$ is the reprocessing efficiency under the assumption of a PSR isotropic emission, ${R_{COM}(i)}=f{R_{RL}(i)}$ is the companion star radius, { where $f$ is the volume-averaged Roche Lobe filling factor}, ${d_{PSR}}$ is the MSP distance, assumed to be equal to the GC distance \citep[${d_{PSR}=4.0}$ kpc;][]{harris96} and ${\epsilon(i)}$ parametrizes the difference of the heated surface visible to the observer between maximum and minimum, as a function of the inclination angle. ${\dot{E}=4\pi I \frac{\dot{P}_{int}}{P^{3}}}$ is the PSR spin-down luminosity where $I$ is the momentum of inertia. Using the the spin period and its intrinsic first derivative obtained from radio timing (see Table 1) and assuming $I=10^{45}$ g cm$^{2}$, we found that $\dot E = 4.6-5.8 \times 10^{33}$ erg s$^{-1}$, typical values within the Galactic eclipsing MSP population. Setting ${ \Delta F_{exp}= \Delta F_{obs}}$ in the F606W filter (see Table 3), we evaluated the reprocessing efficiency as a function of the inclination angle for different values of the RL filling factor. Results are shown in Figure~\ref{eff}. As can be seen, for high inclination angles and a RL filling companion, the reprocessing efficiency is $\sim5\%$, while for filling factor $f=0.8$ is $\sim8\%$. A typical value of 15\% \citep{breton13} would be consistent with $f\sim0.6$. Values of $f<1$ would be plausible since some works showed that BW companions not always completely fill  their RL  \citep[e.g.][]{callanan95, stappers99, breton13}. Similar results hold for the F814W filter.\\ 

It is worth noting that by using ${R_{BB}}$ instead of ${R_{RL}}$ for the stellar radius, the efficiency
increases over 100\% for almost every meaningful configuration. This can be admitted only if an
anisotropic PSR emission is assumed. However, the presence of long radio eclipses and the
behavior of similar objects is a strong indication that ${R_{BB}}$ heavily underestimates the stellar
true radius.

\subsection{A comparison between M71A and M5C}

\begin{figure*}[!t]
\begin{center}
\includegraphics[width=7.8cm]{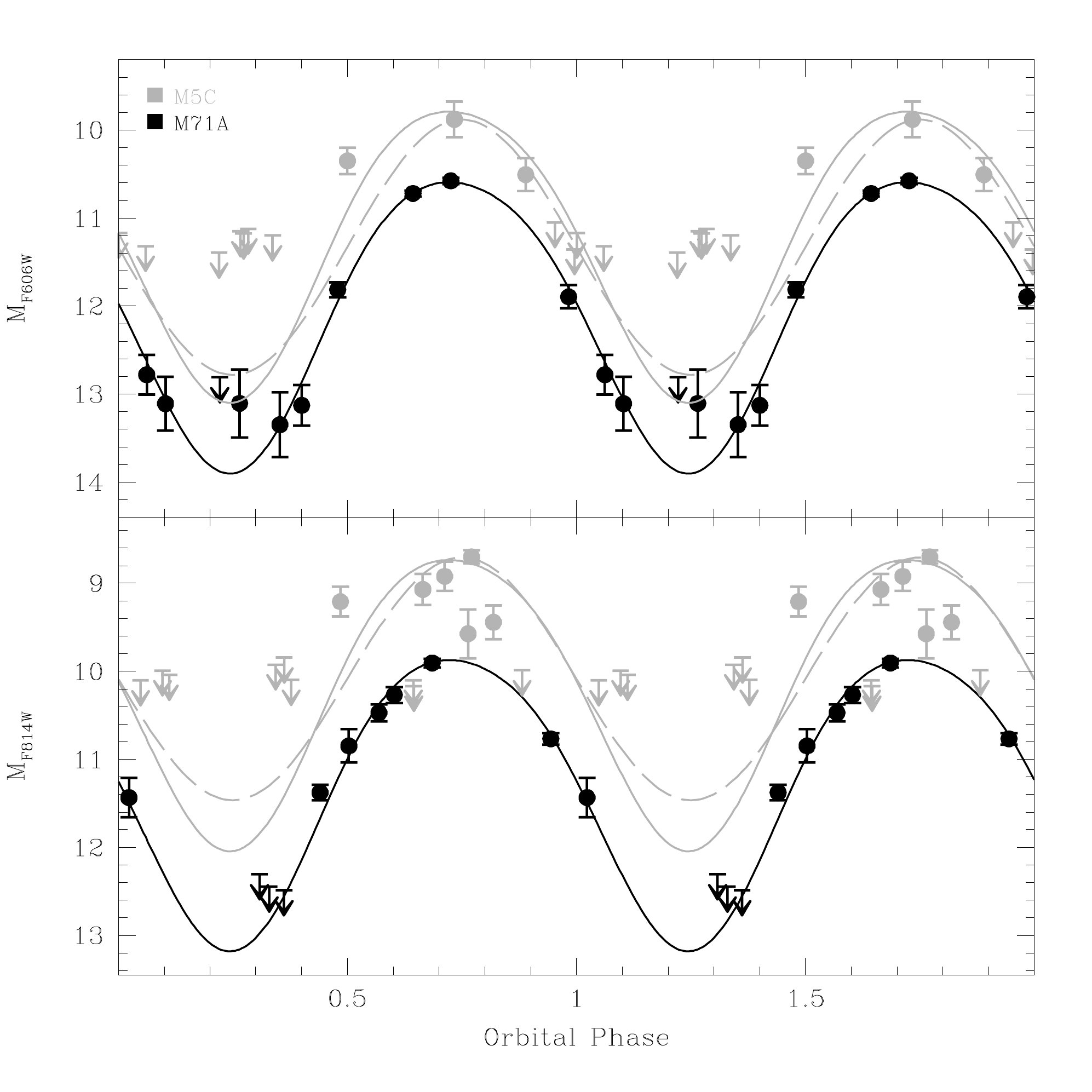}
\includegraphics[width=7.8cm]{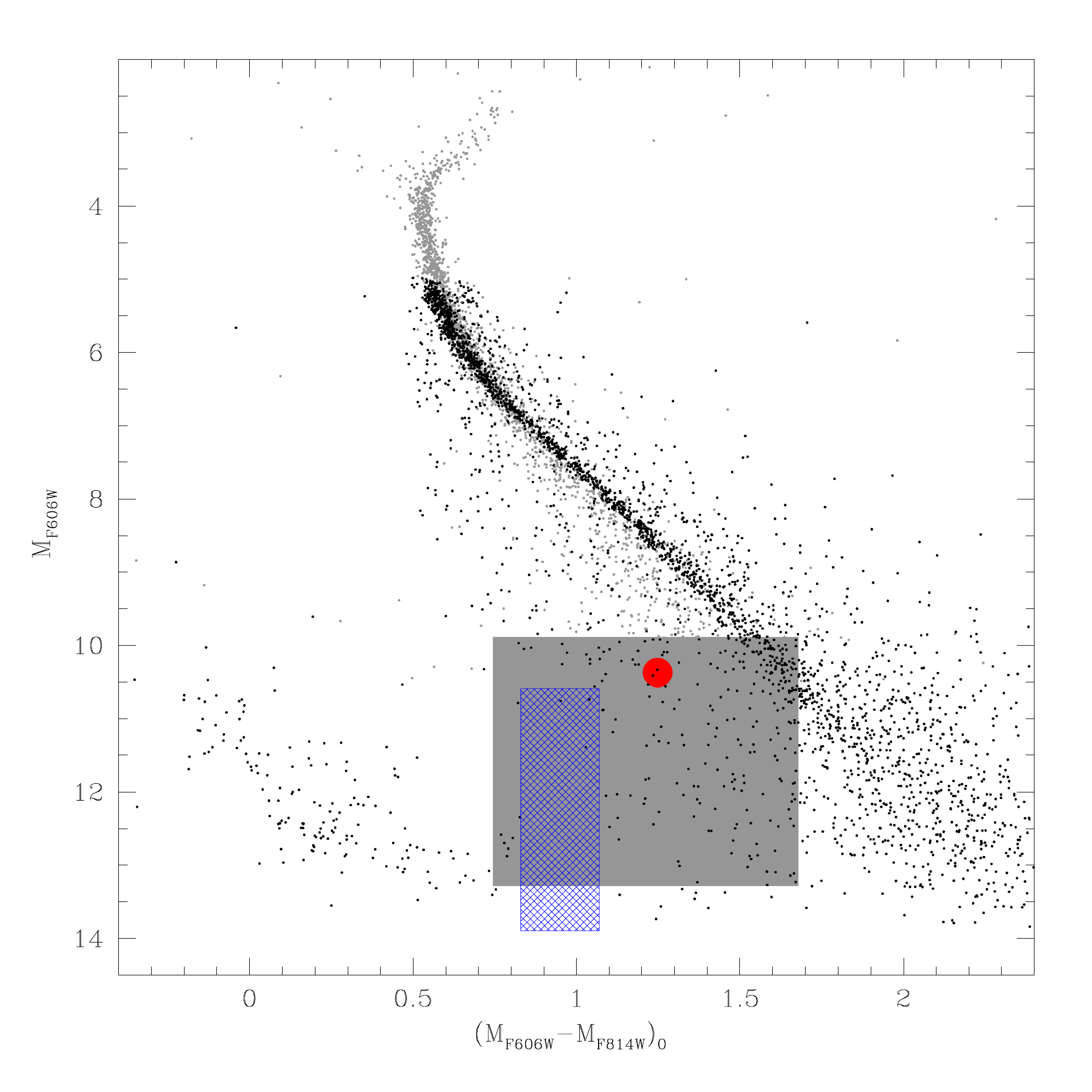}
  \caption[Light curves and CMD positions of COM-M71A and COM-M5C]{{\it Left panel:} Optical light curves of COM-M71A (this work; black points and lines) and COM-M5C \citep[from][gray points and dashed lines]{pallanca14a}, with magnitudes reported to absolute values. The gray solid lines are COM-M71A model adapted to COM-M5C to reproduce the observed points. {\it Right panel:} CMD of M71 (black dots) and M5 (gray dots). The blue shaded region is the position of COM-M71A along the whole orbital phase, as derived from the light curve model (see Figure \ref{cmdmotion}). The red point and the gray area are the indicative position of the COM-M5C \citep[see][for more details]{pallanca14a}. Both the objects are located between the MS and the WD cooling sequence, suggesting common properties of these two BW companions.}
  \label{confronto}
\end{center}
\end{figure*}

So far, the optical companion to PSR J1518+0204C (hereafter M5C) was the only BW companion known in
a GC. M5C is a 2.48 ms PSR with an orbital period of $\sim2.1$ hr located in the GC M5. Its radio timing and the optical photometry of the companion star (COM-M5C) is discussed in \citet{pallanca14a}. In section 3, we anticipated some
interesting analogies between this star and COM-M71A. In order to further investigate similarities
between these BW companions, we compared the optical properties of the two objects. Figure
\ref{confronto} shows their light curves, with the magnitudes reported to the absolute values.
Indeed, despite the low sampling of the COM-M5C light curves, these two objects seem to have a quite
similar optical behavior. As reference, we used COM-M71A analytical models (solid lines) for the
COM-M5C, from which we inferred that a similar light curve structure could hold even for COM-M5C,
being more appropriate than the simple sinusoid (dashed lines) used by \citet{pallanca14a}, given
the sparse number of measurements that prevent them to build an accurate model.  Figure
\ref{confronto} shows the position of the two objects in the CMD. Again, considering the
uncertainties in COM-M5C magnitudes and colors, we found that they are located in the same region,
suggesting a common evolutive path for these low-mass, possibly non degenerate, swollen and heated
companions. Interestingly, in the CMD the two BW companions are located in a region completely different from that usually occupied by RBs \citep{pallanca14b}. Of course, additional identifications of BW companions are needed to firmly characterize the evolution
of these objects. In addition, using equation (\ref{delta}) for COM-M5C, setting the filling factor $f=1$ and using the spin-down period from \citet{pallanca14a} to evaluate the spin-down luminosity ($\dot E = 0.7-3 \times 10^{34}$ erg s$^{-1}$), we found a reprocessing efficiency $\eta\sim5-20\%$, a value fully in agreement with what found for COM-M71A, thus further strengthening the analogies between these two systems.

\subsection{Comparing X-ray and optical light curves}

Usually, BWs with a high energy counterpart do not show any appreciable X-ray variability related to their orbital period \citep[see, e.g.,][respectively for the BWs in the GC 47 Tucanae and in the
Galactic field]{bogdanov06, gentile14}. However, this could be an observational bias, due to the
lack of deep enough and systematic surveys of BWs in the X-rays. On the other hand, it is worth noting that several RB systems clearly show orbital X-ray modulation likely due to the presence of intra-binary shocks \citep{bogdanov06, bogdanov11, bogdanov14}. M71A is an exception, since it has
been found to show periodic X-ray variability \citep{elsner08}. Very interestingly, the
determination of COM-M71A optical light curve offers the opportunity to perform a
comparison between the two. The most intriguing feature emerging from the comparison of the light  curves (both folded with the binary system parameters) is that the phase spanned by the radio eclipses ($0.18<\phi<0.35$) does not correspond to the phase of the X-ray minimum ($0<\phi<0.2$), but it nicely lines up with the optical minimum ($\phi \approx 0.25$). Thus we found that the X-ray minimum precedes the optical PSR superior conjunction. A similar effect was already observed for the RB 47TucW, a 2.35 ms binary MSP with an orbital period of $\sim3.2$ hr and a companion mass of $\sim0.15\; \Msun$ \citep{camilo00}, whose optical light curves indicate the presence of a strong heating \citep{edmonds02}, as usually observed for BW systems. For this object \citet{bogdanov06} argue that the X-ray variability can be attributed to the presence of an intra-binary shock that is
eclipsed by the companion star. The length of the X-ray eclipse suggests that this shock is located closer to the companion star than to the MSP. In particular this behavior could be due to a
swept-back shocked region, produced by the interaction between the PSR wind and the stream of gas issuing from the inner Lagrange point L1, elongated perpendicular to the semi-major axis of the binary \citep[see][for a detailed description]{bogdanov05}. Despite the low X-ray statistics, this
is likely to be the case also for M71A, where the intra-binary shock could be eclipsed just before
the PSR superior conjunction. Even for a companion that is under-filling its Roche Lobe, this shocked region can be created thanks to the stellar wind which can result in mass outflow through L1 \citep{bogdanov05}.

As discussed in \citet{bogdanov05}, the Accreting Millisecond X-ray Pulsar (AMXP) SAX J1808.4$-$3658,
during quiescent states, shows several analogies with the RB 47TucW, in terms of both the X-ray
spectrum and the optical variability. Based on the discussion above, these properties are also
similar to those observed for M71A and, very interestingly, even the companion mass is comparable in
these two cases: above $0.032\; \Msun$ for COM-M71A, and $\sim 0.05\; \Msun$ for the companion to
SAX J1808.4$-$365 \citep{campana04}.  This puts M71A in the middle of the riddle, supporting the
possibility that AMXPs could be the bridge between RB and BW systems \citep{roberts14}. Clearly,
multi-wavelength studies of these objects are urged to unveil connections between AMXPs and
eclipsing MSPs, and between BWs and RBs. Indeed, several important new connections between
AMXPs and RBs have been made in the last years, { especially with the discoveries of systems transitioning from one state to the other}  \citep[see][]{archibald09, papitto13, patruno14,
bassa14, stappers14}.

\cleardoublepage

\chapter{The He White Dwarfs Orbiting the Millisecond Pulsars in the Globular Cluster 47 Tucanae}
\label{cap_47tuc}
\begin{flushright}
\textit{Mainly based on \citealt{cadelano15b}, ApJ, 812:63}
\end{flushright}

\vspace{1cm}

\initial{I}n this chapter we show how we used ultra-deep UV observations obtained with the Hubble Space
Telescope to search for optical companions to the binary millisecond
pulsars in the globular cluster 47 Tucanae. The analysis allowed us to identify four
new counterparts (to 47TucQ, 47TucS, 47TucT and 47TucY) and
confirmed those already known (to 47TucU and 47TucW). In the
color-magnitude diagram, the detected companions are located in a
region between the main sequence and the CO white dwarf cooling
sequences, consistent with the cooling tracks of He white dwarfs of
mass between $\mathrm 0.15 \ \Msun$ and $\mathrm 0.20 \ \Msun$.  For each
identified companion, mass, cooling age, temperature and pulsar mass
(as a function of the inclination angle) have been derived and
discussed. For 47TucU we also found that the past accretion history
likely proceeded in a sub-Eddington rate. The companion to the redback
47TucW is confirmed to be a non degenerate star, with properties
particularly similar to those observed for black-widow systems.
  Two stars have been identified within the 2$\sigma$ astrometric
  uncertainty from the radio positions of 47TucH and 47TucI, but the
  available data prevent us from firmly assessing whether they are the
  true companions of these two MSPs.

\clearpage

\section{Introduction}\label{intro}

47 Tucanae (also known as NGC 104) is one of the most studied globular cluster (GC) in the Milky Way. It is located at a distance of about $4.7$ kpc from the Sun, its stellar population has an age of about 12 Gyrs  \citep[e.g.][]{brogaard17}, an intermediate metallicity among the Galaxy GCs \citep[$\mathrm{[Fe/H]=-0.7}$;][]{brogaard17} and its structural properties show no signature of core-collapse. It also has an high interaction rate \citep{bahramian13}, which can explain the largest population of X-ray sources (300 detected sources within the half-mass radius) ever identified, so far, in a GC \citep{heinke05}. 
Moreover, radio observations revealed the 47 Tucanae hosts the largest population of millisecond pulsars (MSPs) after Terzan 5. Indeed, 25 radio MSPs have been discovered
to date \citep{manchester90,manchester91,robinson95,camilo00,pan16}. All of them have spin periods shorter than 8 ms and 14 of them are located in binary systems\footnote{Please
  visit \url{http://www.naic.edu/~pfreire/GCpsr.html}, for a complete
  list of the main radio timing properties of MSPs in GCs.}. The discovery and timing of these systems \citep{freire01b,freire03,ridolfi16,freire17} opened the possibility to study stellar evolution and the cluster dynamics with unprecedented detail. Indeed it has been possible to detect for the very first time the ionized interstellar medium within a GC \citep{freire01a}, to measure the cluster proper motion as a whole with a better precision than that obtained from optical studies and to put constrains on the possible presence of an intermediate-mass black hole by measuring the pulsar (PSR) line of sight accelerations \citep{freire17}. All the PSRs with a precise timing position have also an X-ray counterpart, whose emission is usually dominated by the blackbody radiation of the extremely hot neutron star (NS) surfaces \citep{bogdanov06}. The optical counterparts to the canonical MSP 47TucU and to the redback (RB) system 47TucW have been already identified in the past \citep{edmonds01,edmonds02}.

 In this chapter we describe the identification and the properties of four new MSP
companions in 47 Tucanae and we present the follow-up study of two
previously known companions. In Table~\ref{tab1psr} we report the main
radio timing properties of the analyzed objects, which are useful in
the following discussions.

\begin{table*}
\caption{Radio timing ephemeris of the analyzed MSPs. From left to right: MSP name, position, offset from the
  GC center, orbital period, mass function and characteristic age. Numbers in parentheses are uncertainties in the last digits quoted. 
  Reference: \citet{ridolfi16,freire17}.}
\label{tab1psr}  
\begin{center}{\scriptsize
\setlength{\tabcolsep}{4pt}
\renewcommand{\arraystretch}{1.3}
\begin{tabular}{l c c c c c c c}
\hline
MSP  & ${\alpha \ (h \ m \ s)}$  & ${\delta \ (^{\circ} \ ' \ \arcsec)}$ & Offset (') &  ${ P_b}$ \ (d) &  $f \ (\Msun)$ &   ${M_{COM}^{min} \ (\Msun)}\tablenotemark{a}$  & ${\tau _{c}}$ (Gyrs) \\
\hline
47TucH  & 00 24 6.7032(2) & -72 04 6.8067(6) & 0.77 &  2.36 &   1.927$\times10^{-3}$  & 0.168 & $>1.9$ \\
\hline
47TucI & 00 24 7.9347(2) & -72 04 39.6815(7)  & 0.28 & 0.23 & 1.155$\times10^{-6}$ & 0.0132&  $0.6$ \\
\hline 
47TucQ  & 00 24 16.4909(1) & -72 04 25.1644(6) & 0.95 & 1.19  &  2.374$\times10^{-3}$  & 0.181 & $>5.0$ \\
\hline
47TucS  & 00 24 3.9794(1) & -72 04 42.3530(4) & 0.21  & 1.20  &   3.345$\times10^{-4}$  & 0.091 & $>1.3$  \\
\hline
47TucT  & 00 24 8.5491(5) & -72 04 38.932(3) & 0.32 & 1.13  &   2.030$\times10^{-3}$  & 0.171 & $>0.43$ \\
\hline 
47TucU & 00 24 9.8366(1) & -72 03 59.6882(4) & 0.94  & 0.43  &   8.532$\times10^{-4}$  & 0.126 & $3.8$ \\
\hline
47TucW & 00 24  6.058(1) & -72 04 49.088(2) & 0.09  & 0.13  &  8.764$\times10^{-4}$ & 0.123 & $>1.15$  \\
\hline
47TucY & 00 24 1.4026(1) & -72 04 41.8363(4) & 0.37  & 0.52 &  1.178$\times10^{-3}$  & 0.141 & $>3.1$  \\
\hline
\multicolumn{8}{p{.8\textwidth}}{$^{a}$ Computed assuming an orbital inclination angle of 90$^{\circ}$ and a PSR mass of 1.4\,M$_\odot$.}
\end{tabular} }
\end{center} 
\end{table*}

\section{Optical Photometry of the Star Cluster}
\label{identification}
\subsection{Observations and data analysis}
\label{Sec:dataan}
The identification of the MSP companions has been
performed through an ultra-deep, high resolution, photometric dataset
acquired under GO 12950 (P.I: Heinke) with the UVIS camera of the Wide
Field Camera 3 (WFC3) mounted on the Hubble Space Telescope (HST). The
dataset consists of 8 images in the F390W filter, with exposure times
of $567-590$~s, and 24 images in the LP F300X filter, with exposure
times of $604-609$~s.
 
The standard photometric analysis has been performed, following the prescription reported in Chapter~\ref{opticalcom}, on the ``flt'' images, which are
corrected for flat field, bias and dark counts\footnote{``flc'' images, corrected also for charge transfer efficiency, were not available at the time this work was performed.}.  These images have
been further corrected for ``Pixel-Area-Map''\footnote{For more
  details see the WFC3 Data Handbook.} with standard IRAF
procedures. The photometric analysis of the images resulted in the creation of a catalog of the cluster sources detectable dataset images, with instrumental positions and magnitudes. The latter have been calibrated to the VEGAMAG system by
using the zero points quoted in the WFC3 Data Handbook and by
performing aperture corrections (see Chapter~\ref{opticalcom}).

\subsection{Astrometry}

Since the WFC3 images suffer from geometric distortions, we corrected
the instrumental positions ($x$,$y$) following \citet{bellini11}. In order
to transform the instrumental positions into the absolute astrometric
system ($\alpha, \delta$), we used, first of all, the wide field
  catalog presented in \citet{ferraro04}. Its astrometric solution
has been improved by using {\tt CataXcorr} to cross-correlate it with the UCAC4 astrometric standard catalog (\citealp{zacharias13}; $\sim4600$ stars have been found in common
  between the two catalogs). The latter is based on the
International Celestial Reference System, thus allowing a more
appropriate comparison with the MSP positions derived from timing
using solar system ephemerides (which are referenced to the same
system). The newly astrometrized wide field catalog has then
been used as a secondary reference frame to astrometrize the WFC3
  data set, by means of $\sim22000$ stars in common. The resulting
  $1\sigma$ astrometric uncertainty is $0.10\arcsec$ and $0.11\arcsec$
  in $\alpha$ and $\delta$, respectively. Thus the final total
  astrometric uncertainty is $\sim0.15\arcsec$. Unfortunately, there
are only few stars in common between the WFC3 and the UCAC4 catalogs,
since the latter does not cover the cluster central regions. This
prevented a direct cross-correlation between the two catalogs and thus
we could not take into account the stellar proper motions between the
two observation epochs, which would have reduced the astrometric
uncertainty.

\subsection{Identification of the MSP companions}

\begin{figure*}
\begin{center}
\leavevmode
\includegraphics[width=5.1cm]{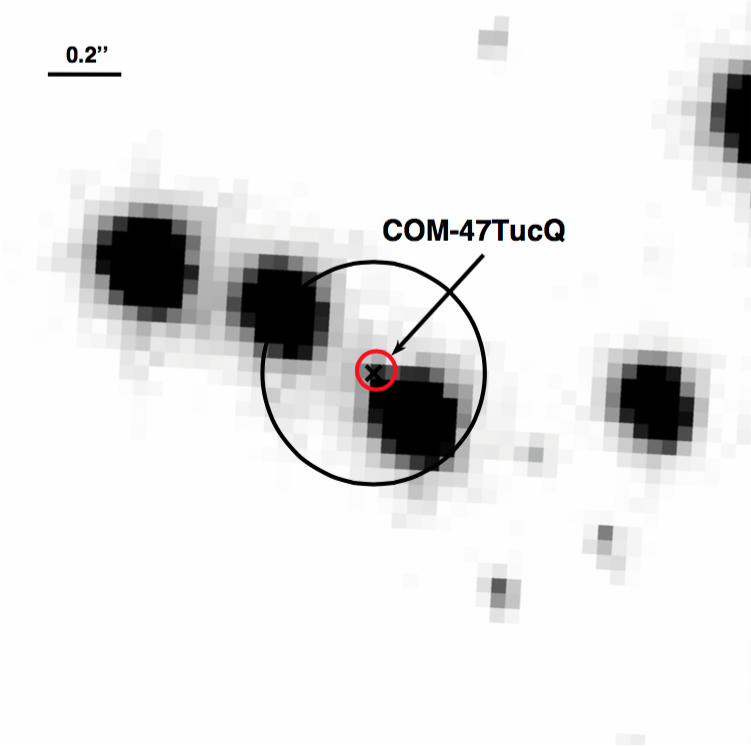}
\includegraphics[width=5.1cm]{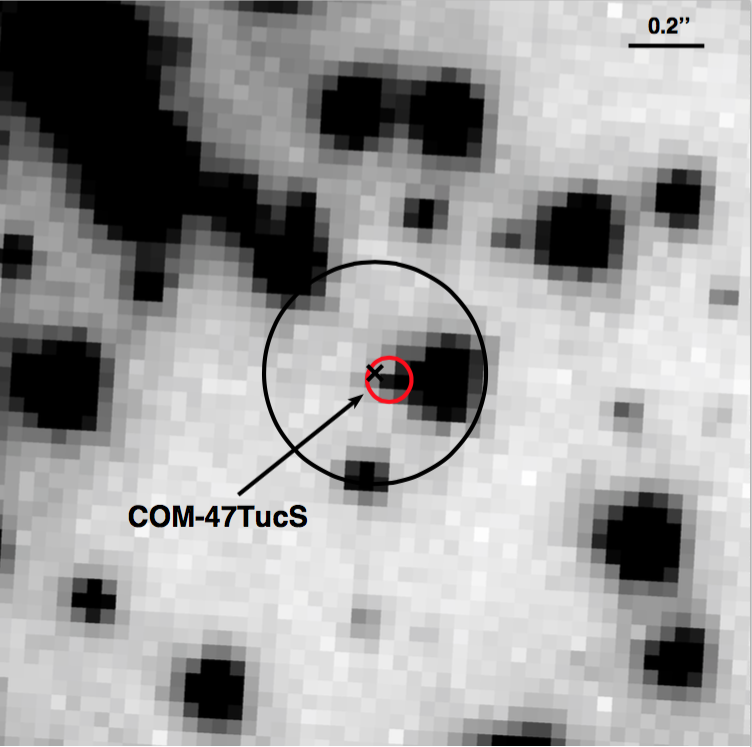}
\includegraphics[width=5.1cm]{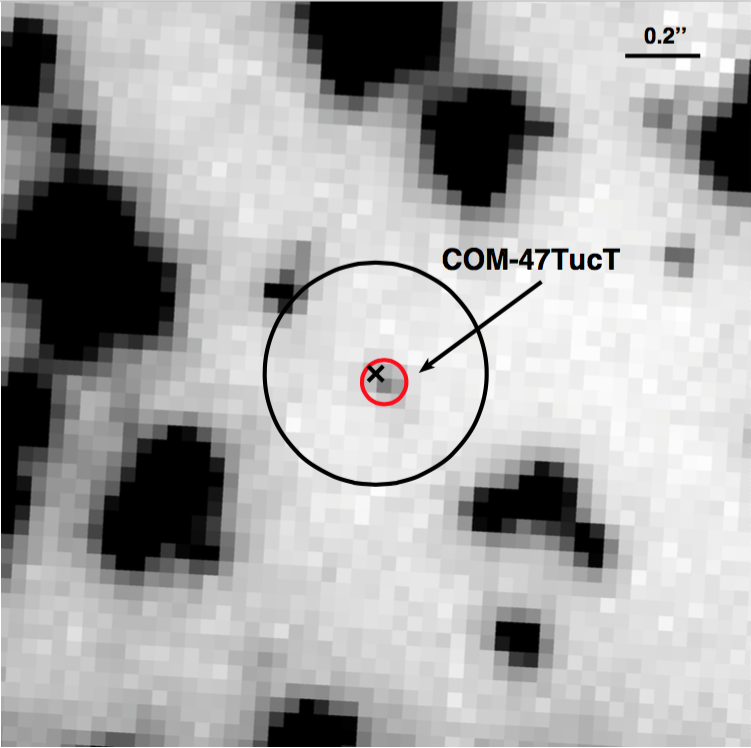}
\includegraphics[width=5.1cm]{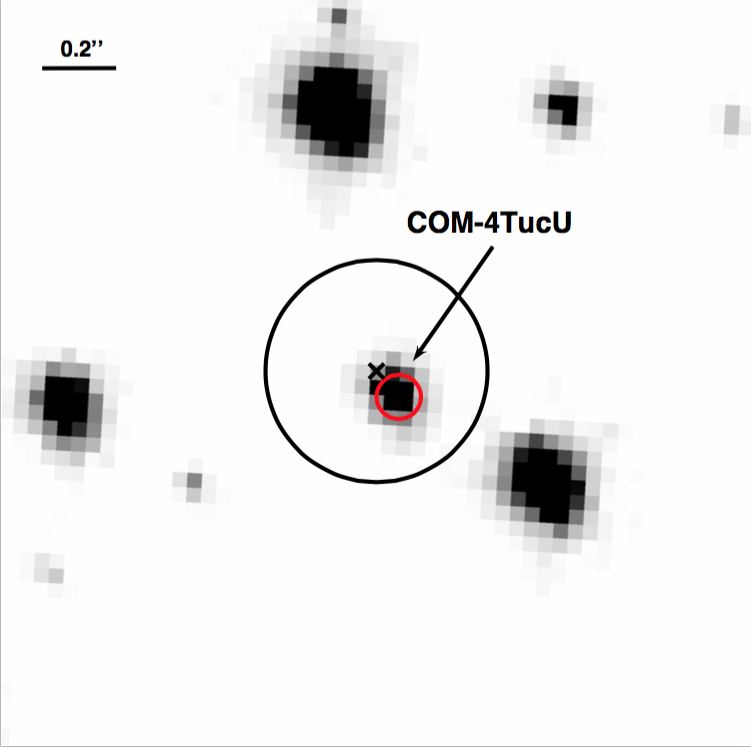}
\includegraphics[width=5.1cm]{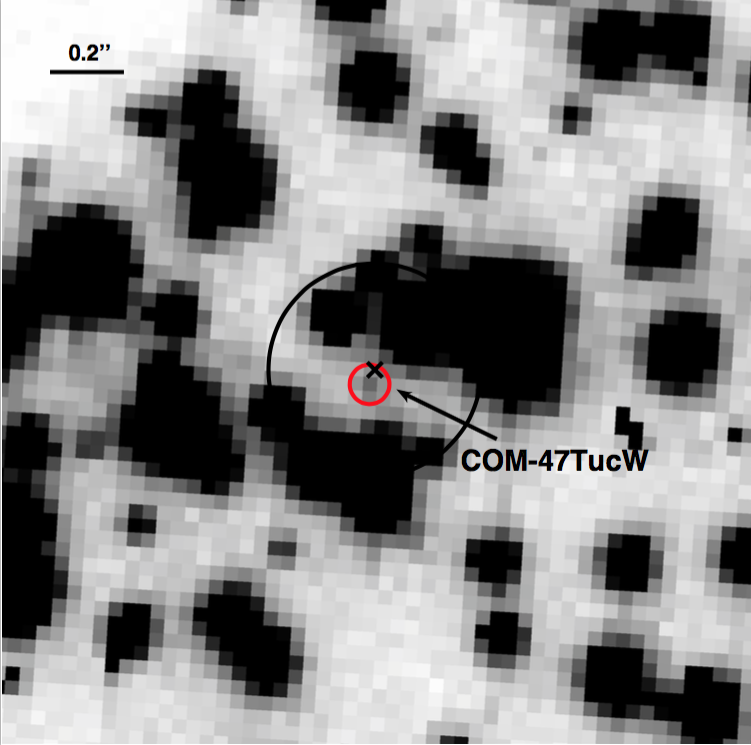}
\includegraphics[width=5.1cm]{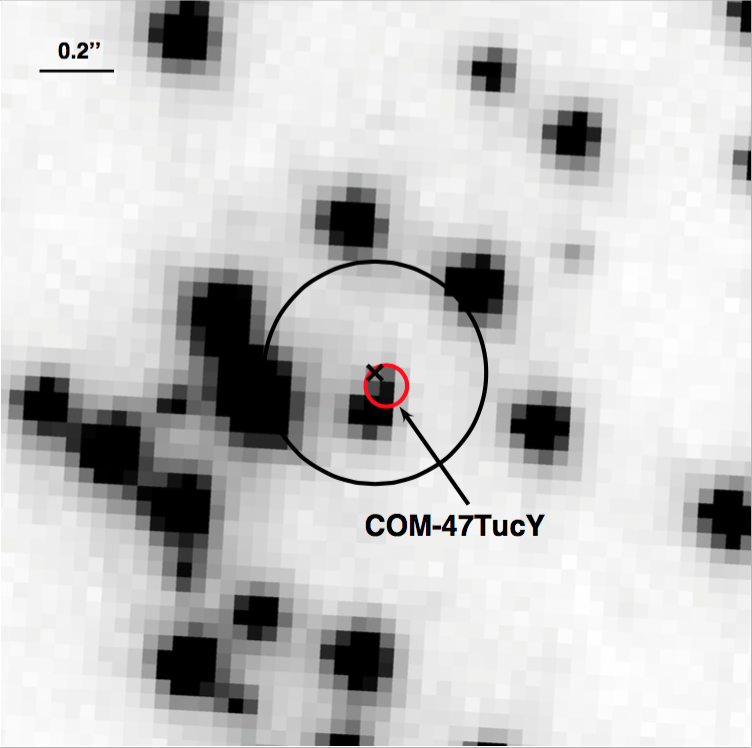}
\includegraphics[width=5.1cm]{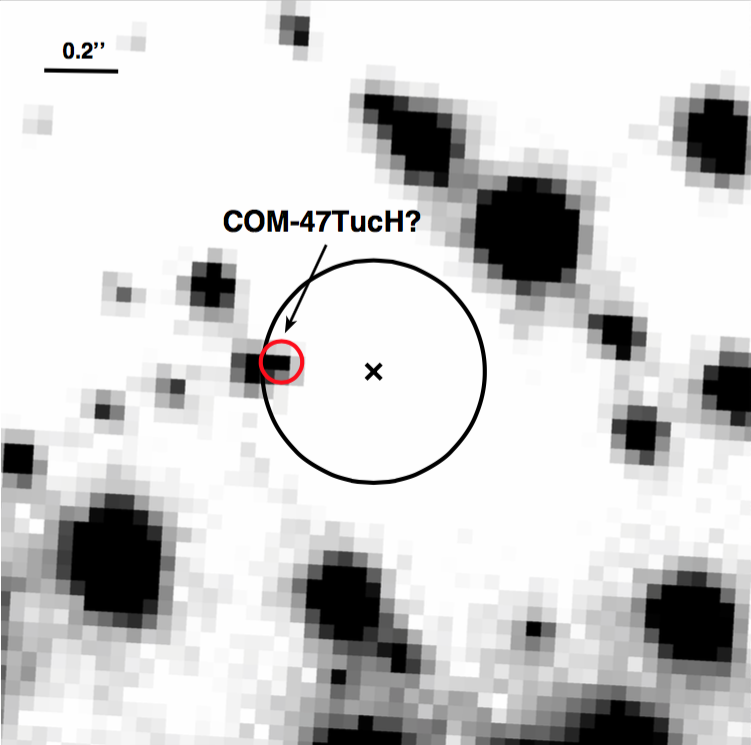}
\includegraphics[width=5.1cm]{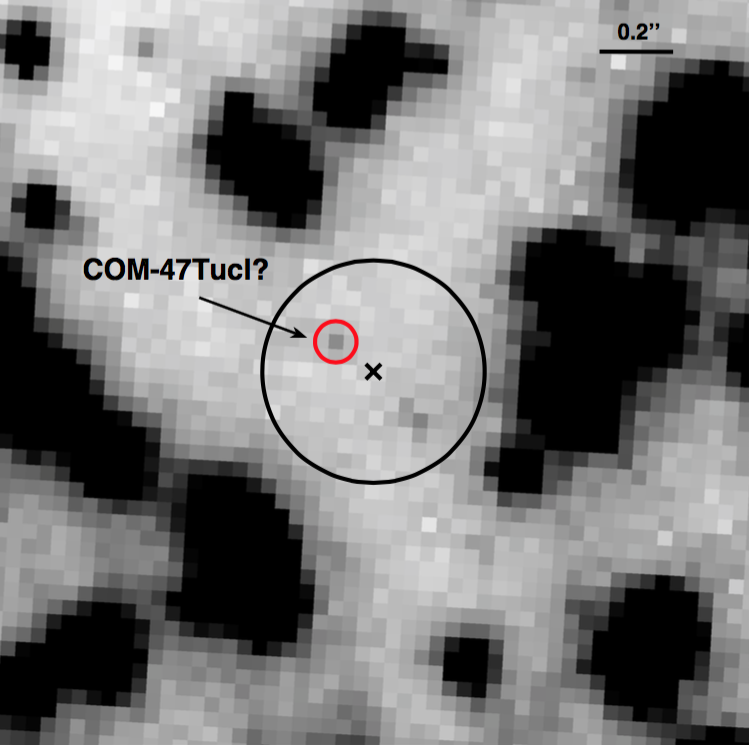}
\caption[HST images of the $2\arcsec \times 2\arcsec$ region around
  the position of the seven analyzed MSPs]{HST images of the $2\arcsec \times 2\arcsec$ region around
  the nominal position of the seven MSPs analyzed in this chapter. North
  is up and east is left. All the charts are obtained from a
  combinations of the available F300X images, with the exception of
  that of 47TucW that is from an image where the companion star is
  at its maximum luminosity. The black circles are centered on the
  radio PSR nominal position in the optical astrometric system and
  their radii are equal to our $2\sigma$ astrometric uncertainty
  ($0.30\arcsec$). The red circles mark the identified MSP companions.}
\label{fig1}
\end{center}
\end{figure*}

First of all, in order to search for the companions to the MSPs in 47
Tucanae, we checked the precision of our astrometric solution
re-identifying the two companion stars already known in the cluster
\citep[see][]{edmonds01,edmonds02}. To this aim, we performed a
detailed analysis of all the detectable objects within a $5\arcsec
\times 5\arcsec$ wide region centered on the nominal position of each
MSP. The companion to 47TucU (COM-47TucU; hereafter all the companions
will be named as COM-47Tuc followed by the letter of the respective
MSP) and 47TucW have been re-identified in stellar sources located at
$0.06\arcsec$ from the MSP nominal positions. Both the identifications
turn out to be largely within our astrometric uncertainty, thus
confirming the accuracy of the adopted astrometric solution. The
finding charts of these two reference objects are shown in Figure
\ref{fig1}.

\begin{figure*}[!t]
\begin{center}
\leavevmode
\includegraphics[width=10cm]{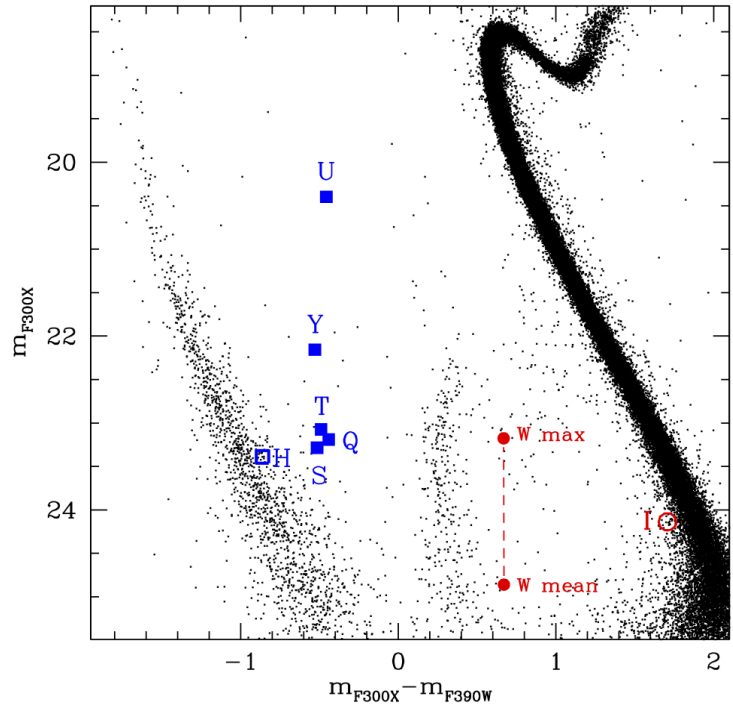}
  \caption[UV CMD of the GC 47 Tucanae]{UV CMD of the GC 47 Tucanae. Only stars with sharpness
    parameter $\mathrm |sh|\leq0.05$ are plotted. The blue solid squares
    mark the companions to the canonical MSPs. The possible
    counterparts to 47TucH and 47TucI are plotted as an open square and circle respectively. Since
    COM-47TucW is a strongly variable object, we report its position
    at the maximum and mean luminosities, as derived by the best-fit
    models (see Section \ref{w} and Figure~\ref{curva_w}).}
  \label{cmd_all}
\end{center}
\end{figure*}

\begin{figure*}[t]
\begin{center}
\leavevmode
\includegraphics[width=10cm]{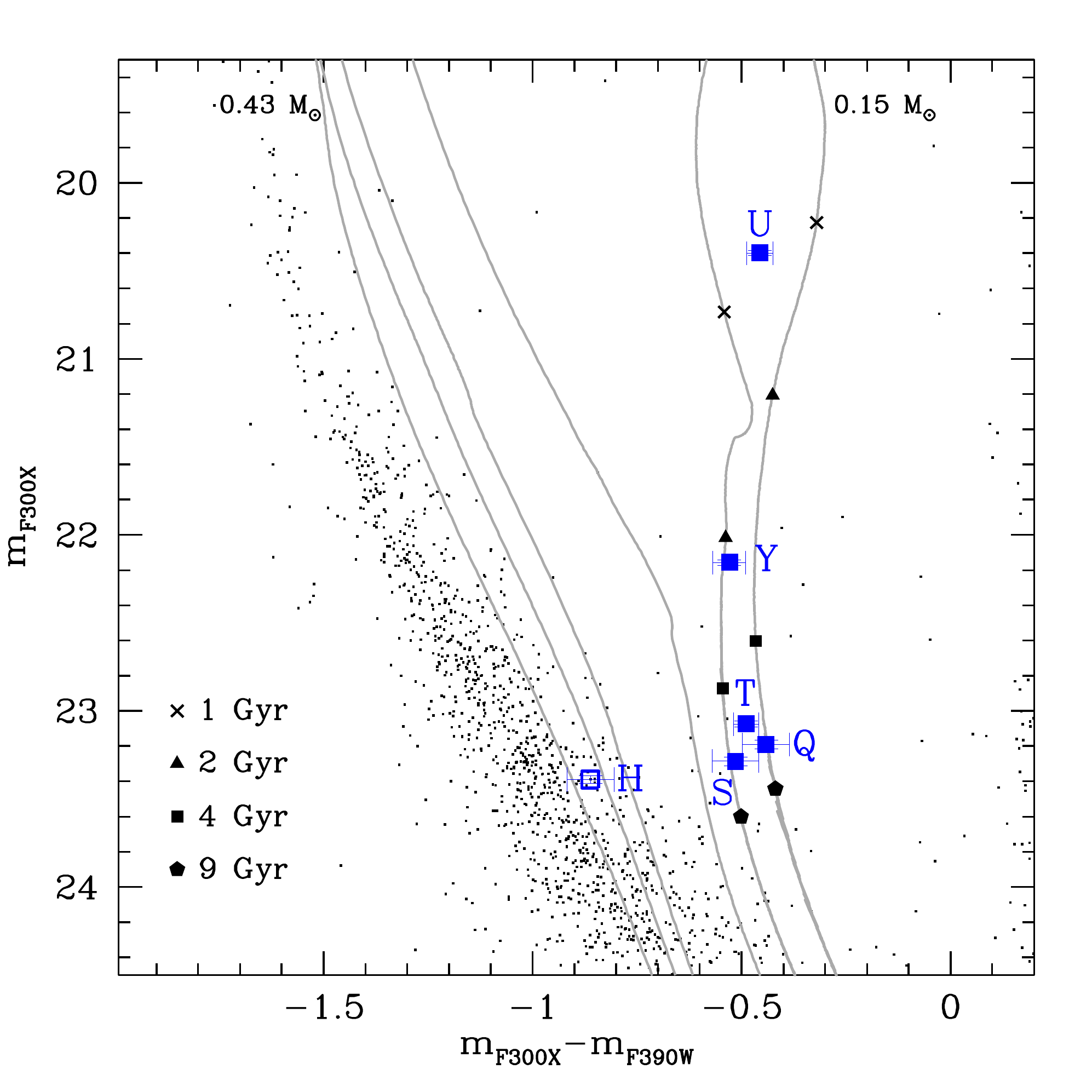}
  \caption[Same as in Figure~\ref{cmd_all}, but zoomed into the WD
    region]{Same as in Figure~\ref{cmd_all}, but zoomed into the WD
    region. The continuous curves are reference He WD cooling tracks
    for stars of $\mathrm 0.15 \ \Msun$, $\mathrm 0.17 \ \Msun$, $\mathrm 0.20
    \ \Msun$, $\mathrm 0.32 \ \Msun$, $\mathrm 0.36 \ \Msun$ and $\mathrm 0.43
    \ \Msun$ (from right lo left). For the two rightmost tracks, points at 1,2,4 and 9 Gyrs have been marked with different symbols. The photometric errors of the
    companion stars are also drawn. }
  \label{cmdwd}
\end{center}
\end{figure*}

Following the same procedure, we searched for the companions to all
the other MSPs with a known position \citep{ridolfi16,freire17}.  Stars located within the $2\sigma$
  uncertainty from the PSR position have been considered as
possibile counterparts. Four companions (to 47TucQ, 47TucS, 47TucT and
47TucY) have been identified on the basis of their positional
coincidence (all of them are located at a distance $\leq0.06\arcsec$
from the nominal radio position) and of their position in the color-magnitude diagram (CMD). Two faint stars have been detected also
  within the $2\sigma$ uncertainty circle from 47TucI and
  47TucH. However, their distances ($0.15\arcsec$ and $0.24\arcsec$
  respectively) from the PSR radio positions are significantly
  larger than in all the other cases, thus casting doubts about these
  objects being the true optical counterparts (see more discussion in
  Section~\ref {comH}). The finding charts of all these objects are
shown in Figure~\ref{fig1} and their main photometric properties are
reported in Table~\ref{tab_wdprop}. Their location in the cluster CMD
is shown in Figure~\ref{cmd_all}, where only the stars with a sharpness
parameter\footnote{The sharpness parameter is a {\textrm DAOPHOT II}
  output that quantifies the stellar-like structure of each object
  fitted with the PSF model. See the User Manual for more details.}
$\mathrm |sh|\leq0.05$ are plotted. As can be seen, with the
  exception of the candidate companion to 47TucI, all the newly
identified counterparts are located in the region where He white dwarfs (WDs) are
expected, although the candidate companion to 47TucH could be
compatible also with the CO WD cooling sequence (see Section
  \ref{comH}). Since the radio timing properties suggest that these
systems are the product of the canonical recycling scenario, their
location along the He WD cooling sequences guarantees their connection
with the MSPs. Note in fact that the probability of a chance
coincidence with another He WD is extremely low ($\mathrm
P\approx0.1\%$)\footnote{The chance coincidence probability has been
  evaluated predicting the number He WD expected within a radius equal
  to the $2\sigma$ astrometric uncertainty. We derived the He WD
  density by direct counting of the all objects located among the
  cooling tracks (see Section~\ref{prop} and Figure~\ref{cmdwd}) and
  dividing this number by the size of the WFC3 field of view. Please
  note that even including all the stars of the catalog with sharpness
  $\mathrm |sh|>0.05$, the chance probability remains $\lesssim0.5\%$.},
since these objects can only be the product of the late stage of the
evolution of exotic objects like, for example, MSPs and cataclysmic
variables. The candidate companion to 47TucI is instead a main
  sequence-like object, and its properties will be briefly discussed
  in Section \ref{comH}.  As concerns the previously known
companions, COM-47TucU is also located along the He WD sequence, while
the RB COM-47TucW is located in an anomalous region between the
main sequence and the WD cooling sequence (see Section~\ref{w}).

With the exception of the COM-47TucW, no significant variability
related to the orbital period has been detected. For 47TucU, 47TucY
and 47TucW (see Section~\ref{w}) the observations sample a significant
fraction of the orbital period. Instead, for the other systems (with
orbital periods longer than 1 day) the coverage is too poor to allow
any appropriate variability analysis.  However, a strong magnitude
modulation, as the one observed for non degenerate companions
\citep[see e.g.][]{stappers99,edmonds02,reynolds07, pallanca10,
  romani11, pallanca14a,cadelano15a}, is not expected and usually not
observed for degenerate objects, since the flux enhancement due to
re-heating of the companion star by the PSR emitted energy is
negligible.

\begin{table*}
\caption{Optical properties of the companion stars. From left to right: MSP companion name, position, distance from
  the radio MSP nominal position, F300X and F390W magnitudes and the
  relatives uncertainties.}
\label{tab_wdprop} 
\begin{center}{\scriptsize
\setlength{\tabcolsep}{4pt}
\renewcommand{\arraystretch}{1.3}
\begin{tabular}{l c c c c c}
\hline
Name & $\alpha$ (h m s) & $\delta$ ($^{\circ}$ $'$ $\arcsec$) & dist ($\arcsec$) & ${ m_{F300X}}$  & ${ m_{F390W}}$ \\
\hline
COM-47TucQ  & 00 24 16.489 &  -72 04 25.209   &   0.04   &$23.19\pm0.02$ & $23.63\pm0.05$ \\
\hline
COM-47TucS  & 00 24 3.977 & -72 04 42.385    &   0.03    &$23.29\pm0.02$ & $23.80\pm0.05$\\
\hline
COM-47TucT  & 00 24 8.549 &  -72 04 38.965   &   0.04    &$23.07\pm0.02$ & $23.56\pm0.03$\\
\hline 
COM-47TucU & 00 24 9.835 & -72 03 59.746   &   0.06   &$20.40\pm0.01$ & $20.85\pm0.03$\\
\hline
COM-47TucW & 00 24 6.063 & -72 04 49.133    &   0.06    & 24.28$^{a}$ & 23.62$^{a}$  \\
\hline
COM-47TucY & 00 24 1.401 & -72 04 41.875    &   0.04    & $22.16\pm0.02$ & $22.69\pm0.04$\\
\hline
COM-47TucH?  & 00 24 6.755    & -72 04 6.781    &   0.24    & $23.39\pm0.02$ & $24.25\pm0.05$ \\
\hline    
COM-47TucI? & 00 24 7.953 & -72 04 39.559 & 0.15 & $24.14\pm0.04$ & $22.43\pm0.03$\\
\hline
\multicolumn{6}{p{.7\textwidth}}{$^{a}$ The values for COM-47TucW correspond to the mean
  magnitudes of the best-fit models (see Figure~\ref{curva_w}).}
\end{tabular} }
\end{center} 
\end{table*}

\section{Discussion}

\subsection{The physical properties of the He WD companions}
\label{prop}
\begin{figure}
\begin{center}
\leavevmode
\includegraphics[width=5.1cm]{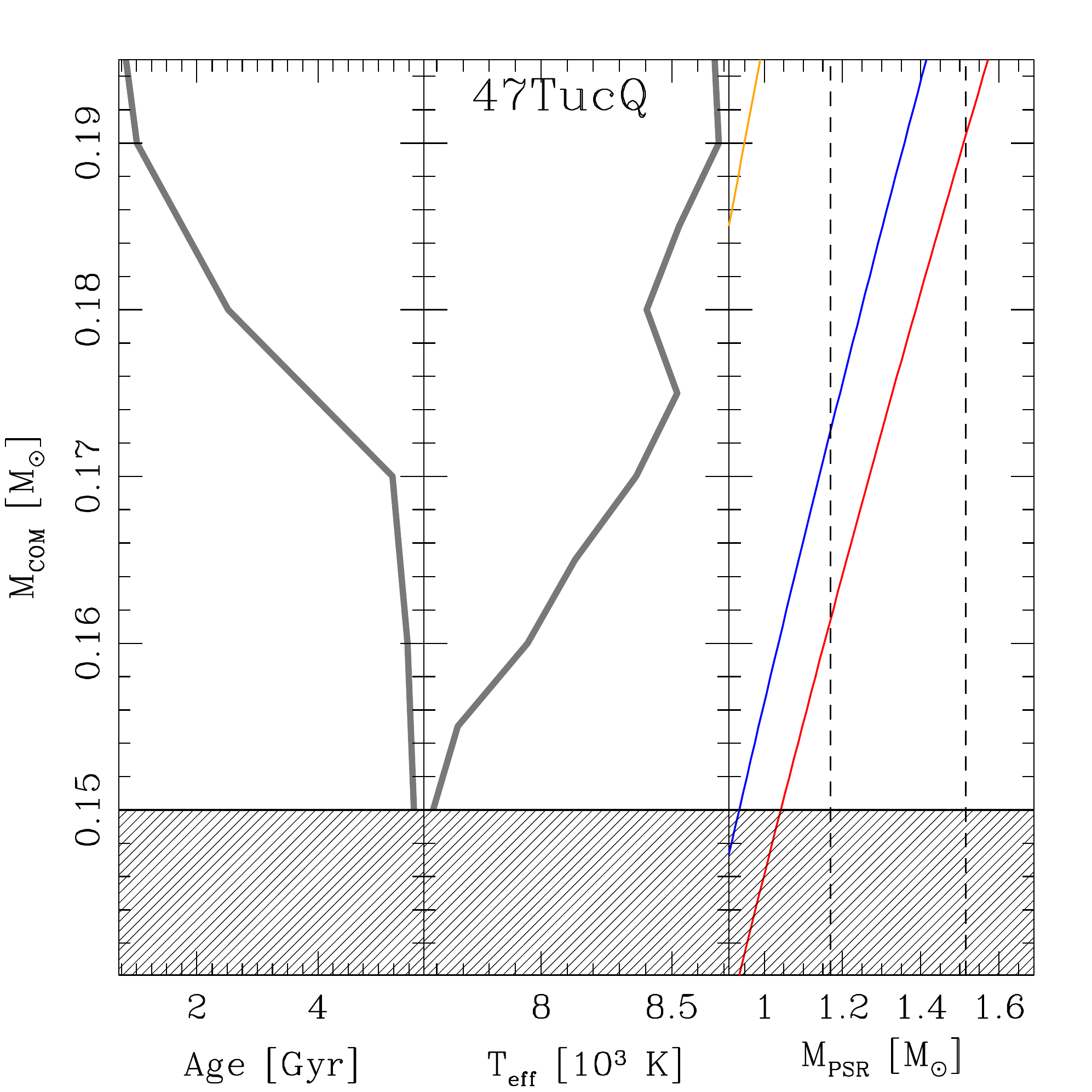}
\includegraphics[width=5.1cm]{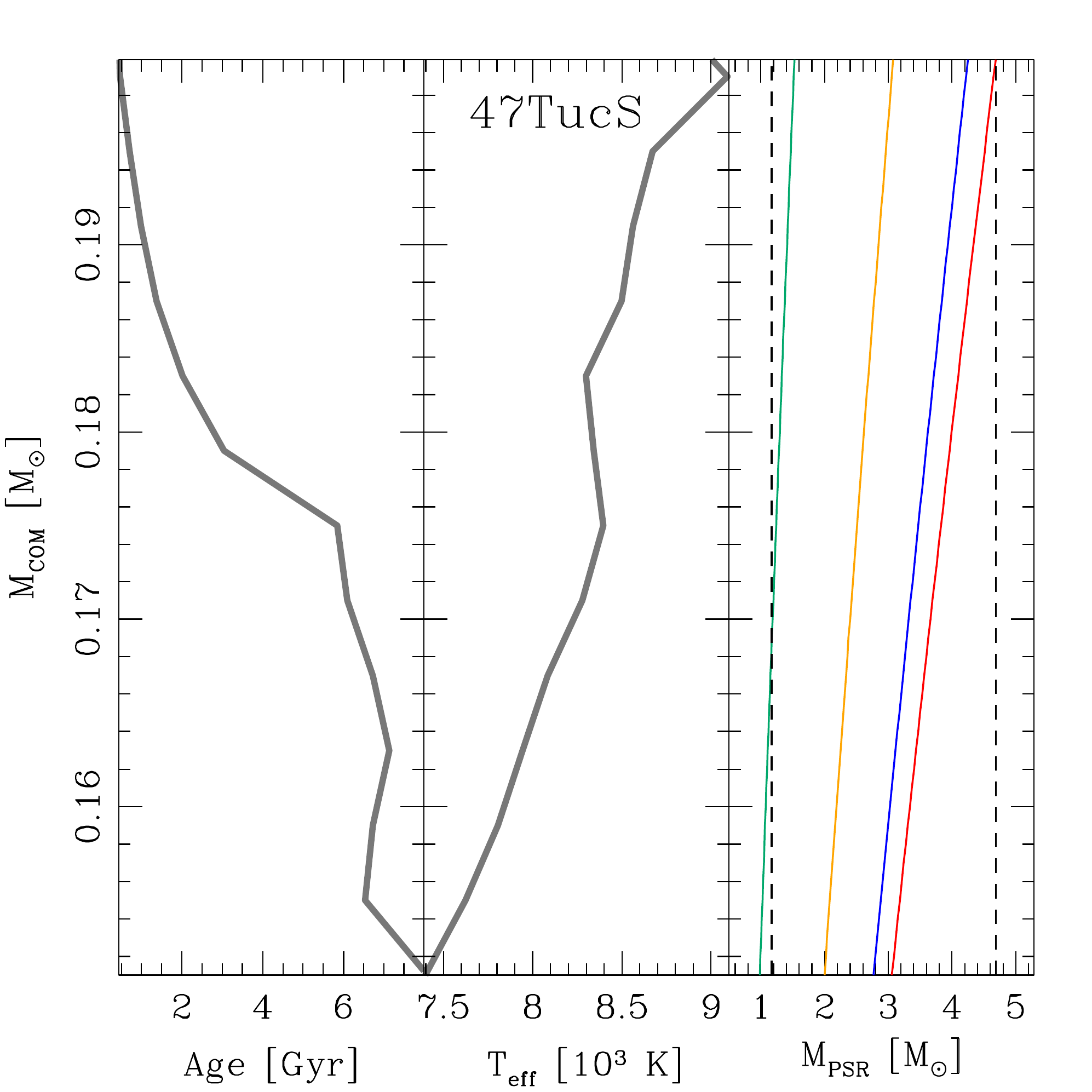}
\includegraphics[width=5.1cm]{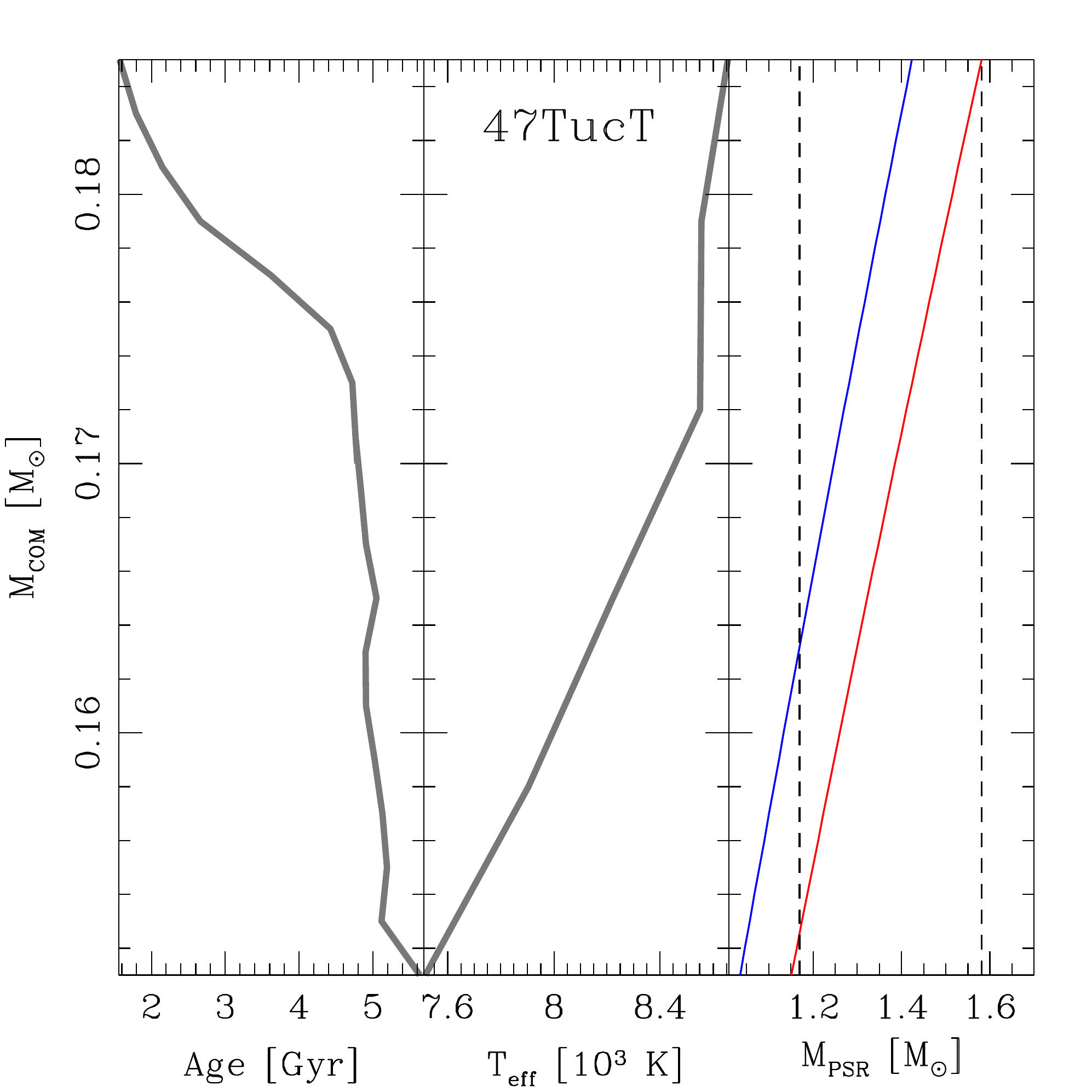}
\includegraphics[width=5.1cm]{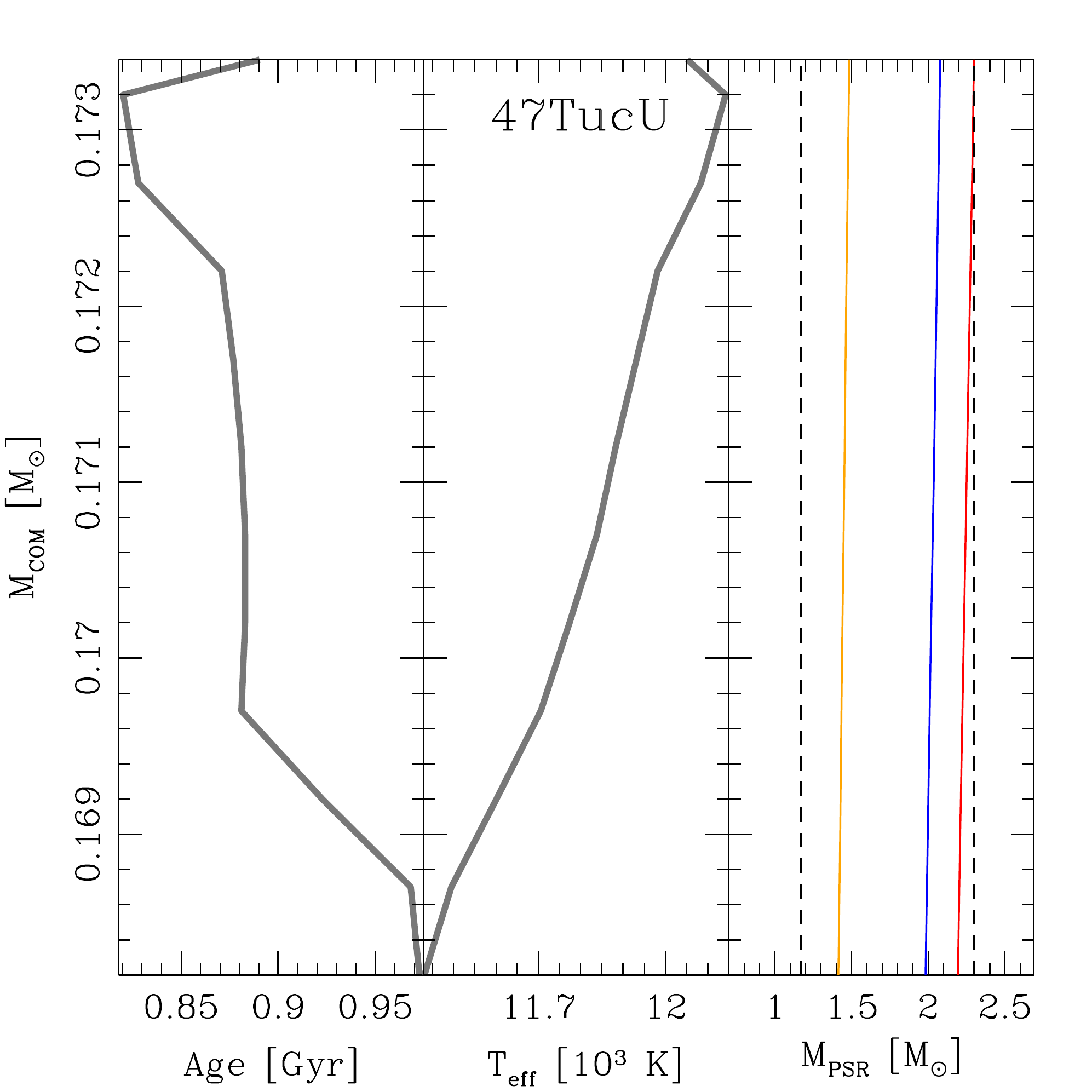}
\includegraphics[width=5.1cm]{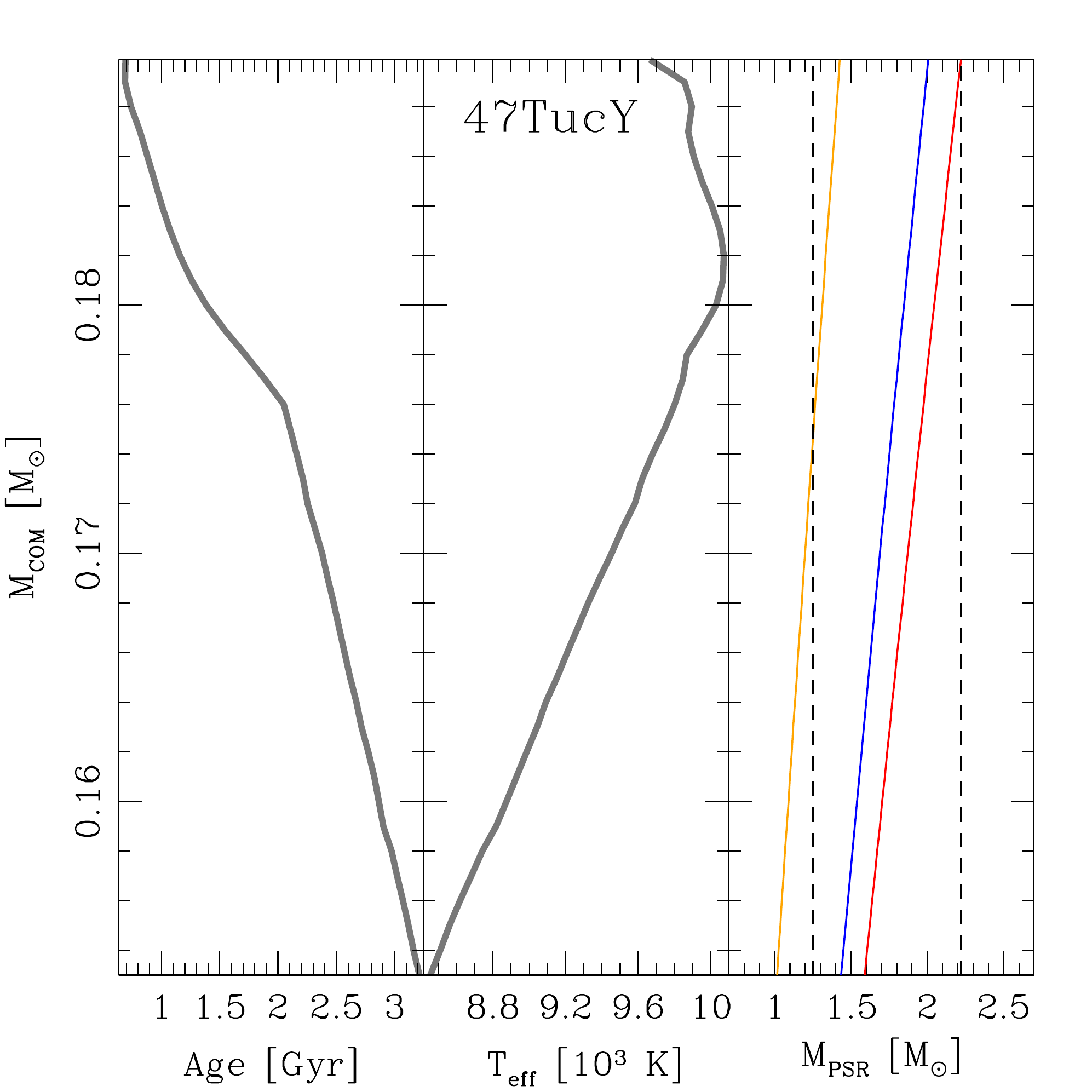}
  \caption[Physical properties of 47TucQ, 47TucS, 47TucT, 47TucU and 
    47TucY]{Physical properties of 47TucQ, 47TucS, 47TucT, 47TucU and 
    47TucY (see labels), as derived from the comparison between the
    photometric characteristics of each companion and the WD cooling
    track models. In each plot, the gray lines drawn in the left and
    central panels show the allowed combinations between companion
    mass and cooling age or temperature (see text). In the case of COM-47TucQ
    the shaded area marka the region not
    allowed by the theoretical cooling tracks. In the rightmost panel
    of each plot, the solid curves represent the combination of values
    allowed by the PSR mass function for different inclination angles
    ($i=90^{\circ}$ in red, $i=70^{\circ}$ in blue, $i=50^{\circ}$ in
    orange, and $i=30^{\circ}$ in green). The blue dashed lines
    correspond to the assumed minimum NS mass \citep[$\sim 1.17\; 
      \Msun$;][]{janssen08} and the largest NS mass value obtained for
    $i=90^{\circ}$.}
\label{wdprop}    
\end{center}
\end{figure}

In order to constrain the main properties of the He WD companions, we
have compared the position of each candidate in the CMD with a set of
He WD cooling tracks computed by \citet{althaus13}. These models span
a mass range from $\mathrm 0.15 \ \Msun$ to $\mathrm 0.43 \ \Msun$, spaced at
about $0.005 \ \Msun$ for masses between $\mathrm 0.15 \ \Msun$ to $\mathrm
0.19 \ \Msun$ and up to $\mathrm 0.07 \ \Msun$ for larger masses. We
transformed the theoretical luminosities and temperatures into the
absolute F300X and F390W magnitudes, by applying the bolometric
corrections kindly provided by P. Bergeron
\citep[see][]{holberg06,bergeron11}.  Then, the model absolute
magnitudes have been transformed into the apparent ones by using the
distance modulus $(m-M)_{0}=13.32\pm0.10$ \citep{ferraro99}\footnote{
  Many literature works reported on different values of 47 Tucanae
  distance modulus \citep[see, e.g.][and references
      therein]{woodley12}. However, all these possibile values have
  only a minimal influence on our derived companion properties
    (e.g. the derived companion masses would vary of less than
    $\sim7\%$ for all the companion stars)}. and the color excess
$E(B-V)=0.04\pm0.02$ \citep{ferraro99,zoccali01,salaris07} and
extinction coefficients $A_{F300X}/A_{V}=1.77309$,
$A_{F390W}/A_{V}=1.42879$
\citep[][]{cardelli89,odonnel94}. Figure~\ref{cmdwd} shows the zoomed
portion of the CMD in the WD region with a sample of cooling tracks
for different masses overplotted. As can be seen, the range in mass of
the models is large enough to properly sample the portion of the CMD
where all the companions are located. Therefore we used this set of
models to derive the combinations of parameters (mass, cooling age and
temperature) that simultaneously satisfy the observed photometric
magnitudes in both the filters, also taking into account the
uncertainties on the companion magnitudes, distance modulus and
reddening. The best values have been evaluated with a simple
$\chi^{2}$ statistic. In doing this, linear interpolations (for
different masses but equal ages) among the tracks have been performed
in order to have a tighter mass sampling. We assumed that each
companion is located at the distance of 47 Tucanae\footnote{Even
  though \citet{freire01a} measured distance offsets between the
  cluster MSPs, such differences are very small and can be neglected
  for our goals.} and it is affected by the same extinction\footnote{
  The effects of differential reddening are negligible for our goals
  \citep[see][]{milone12a,milone12b}.}. Figure \ref{wdprop} shows, for
each system, the combination of cooling age (left panel), temperature
(central panel) and PSR mass (right panel) appropriate for the derived
value of the companion mass. In particular, in each plot the right
panel shows the results obtained for different values of the
inclination angle and interesting constraints on each system can be
drawn. For instance, by setting the inclination angle to $90^{\circ}$,
the maximum PSR mass allowed from the inferred companion mass can be
evaluated. Conversely, by assuming the minimum PSR mass equal to $\mathrm
1.17 \ \Msun$ \citep[the lowest mass ever measured for a
  NS;][]{janssen08}, a conservative lower limit to the inclination
angle can be derived. All these results are also summarized in
Table~\ref{tab3}\footnote{The reader should be aware that the WD parameters
  should not be assumed at face value as perfectly correct but as
  estimations, since they are model dependent and could also suffer
  from some hardly quantifiable uncertainty linked to the bolometric
  corrections.}. Note that we are not analyzing here the cases of
  47TucH and 47TucI, which we will discuss in Section~\ref{comH}.

\begin{table*}
\caption{Derived properties of the five MSPs with He WD companions. From top to bottom: companion mass, age, temperature,
  luminosity, PSR mass and inclination angle.}
\label{tab3}
\begin{center}{\scriptsize
\setlength{\tabcolsep}{4pt}
\renewcommand{\arraystretch}{1.3}
\begin{tabular}{l c c c c c}
\hline
Parameter & 47TucQ & 47TucS & 47TucT & 47TucU & 47TucY \\
\hline
${ M_{COM} \ (\Msun)}$ & $0.150-0.195$ & $0.150-0.200$ & $0.150-0.185$ & $0.168-0.173$ & $0.150-0.190$  \\ \hline
Age (Gyr) & $0.7-5.5$ & $0.4-8.1$ & $1.6-6.0$ & $0.82-0.93$ & $0.6-3.2$  \\ \hline
$T$ ($10^{3}$ K) &  $7.5-8.7$ & $7.4-9.1$ & $7.5-8.6$ & $11.4-12.1$ & $8.4-10.1$  \\ \hline
$L$ ($\mathrm{ 10^{-3} \ L_{\odot}}$)  &  $8.6-9.7$ & $7.8-9.1$ & $9.4-10.5$ & $141-165$ & $22.5-23.5$  \\ \hline
${ M_{PSR} \ (\Msun)}$ & $<1.57$ & $<4.69$ & $<1.58$ & $<2.30$ & $<2.22$  \\ \hline
$i \ (^{\circ})$ & $>58$ & $>26$ & $>57$ & $>42$ & $>45$   \\   
\hline
\end{tabular} }
\end{center} 
\end{table*}

\begin{figure*}
\begin{center}
\leavevmode \includegraphics[width=10cm]{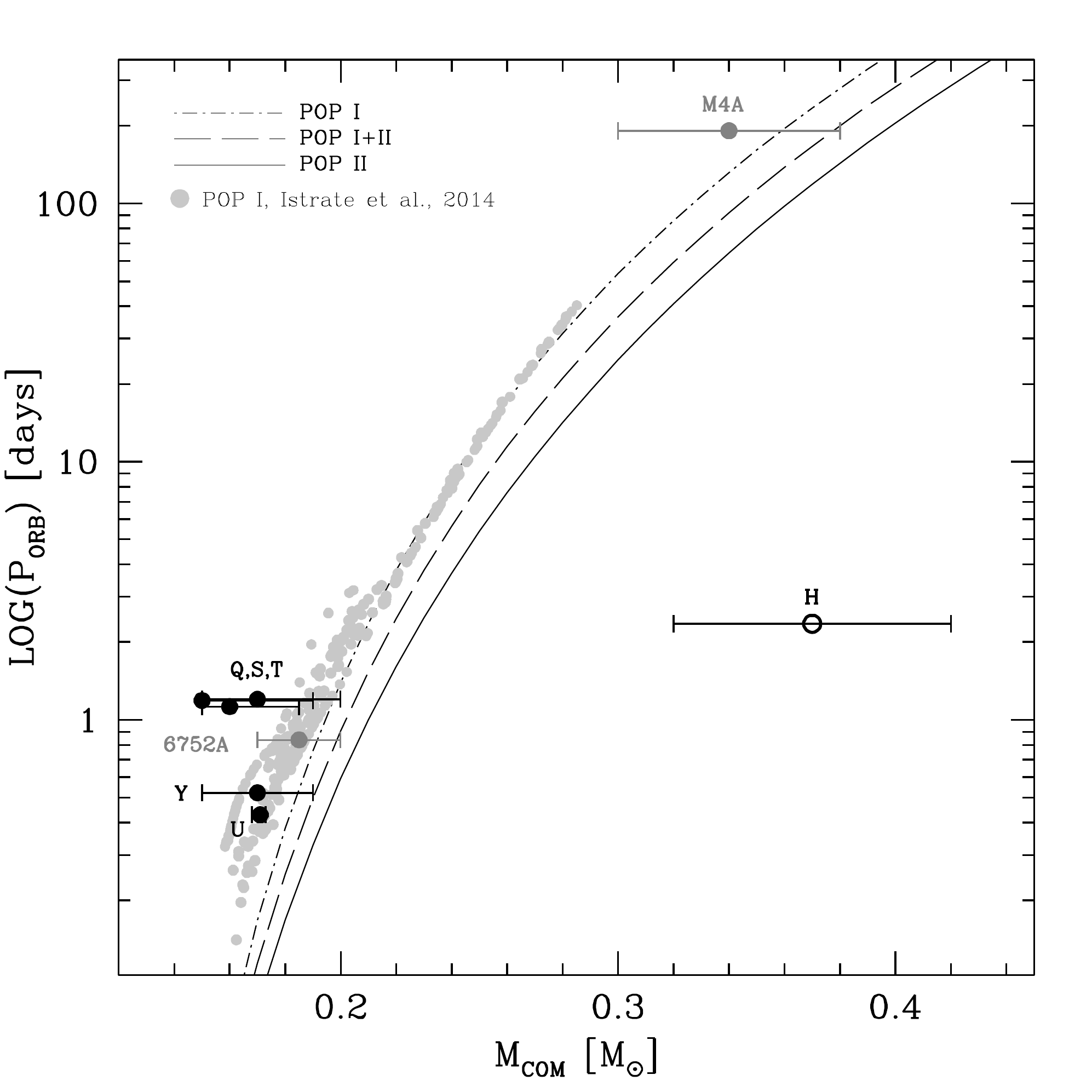}
  \caption[MSP orbital periods as a function of the best-fit
    companion masses]{MSP orbital periods plotted as a function of the best-fit
    companion masses, for each identified object, plus
    the ones detected in NGC 6752 and M4 \citep[dark gray points;
      see][]{fer03_msp,sigurdsson03}. The three curves correspond to
    the theoretical predictions of \citet{tauris99} for three
    different stellar population progenitors, as reported in the
    top-left legend. The light gray points correspond to the
    theoretical results obtained by \citet{istrate14}.}
  \label{mp}
\end{center}
\end{figure*}

As can be seen, all the companions have masses between $\mathrm \sim0.15
\ \Msun$ and~$\mathrm \sim0.2 \ \Msun$. The derived ranges of ages are in
agreement with the lower limits to the PSR characteristic ages
reported in Table~\ref{tab1}. The only exception is COM-47TucU, which
is discussed below. In principle the mass of COM-47TucQ could be
smaller than our best-fit value ($0.15\; \Msun$), since not
theoretical tracks for masses below this value are available. However,
already a $0.15\; \Msun$ companion would imply an extremely low value
of the PSR mass ($ \lesssim 1 \; \Msun$). This puzzling result could
be partially explained with the difficulty of accurately determine the
color of the optical counterpart, because of the presence of a very
close bright object (see Figure~\ref{fig1}).

Our results rule out a massive NS in the case of 47TucQ and
47TucT, while the possibility of a $ \sim2 \; \Msun$ NS remains
opened in the cases of 47TucS, 47TucU and 47TucY. However,
Figure~\ref{wdprop} shows that the PSR mass can be significantly reduced
by assuming an intermediate-low inclination angle of the orbital
plane. In any case these systems are worthy of future, especially spectroscopical,
investigations. 

We also compared our results with the theoretical predictions on the
behavior of the orbital period as a function of the companion mass
discussed in \citet{tauris99}. Such a model has been already
empirically verified by \citet{corongiu12} and \citet{bassa06}.  As shown in
Figure~\ref{mp}, where we also added the two He WD companions
identified in NGC 6752 and M4 \citep[see][]{fer03_msp,sigurdsson03},
our results are in reasonable agreement with the model. The analytical
prediction seems to slightly overestimate the companion mass or to
underestimate the system orbital period. However, this model is valid
for binary systems with $\mathrm 0.18\ \Msun<M_{WD}<0.45\ \Msun$, thus
only marginally representative of our sample, where most of the
companions appear to be less massive than $\mathrm 0.18 \ \Msun$. More
updated models (from \citealt{istrate14}; gray points in
Figure~\ref{mp}) are in better agreement, although they are the
results of simulations of donor stars with metallicity $\mathrm Z=0.02$,
larger than that of 47 Tucanae {\citep[i.e. $\mathrm Z=0.008$,][]{lapenna15}.

The brightness of COM-47TucU allowed us to put tighter constraints to
the system parameters with respect to the other objects. Both its mass
and temperature are in excellent agreement with those reported in
\citet{edmonds01}, while our derived age ($\approx 0.9$ Gyr) turns out
to be 0.3 Gyr larger than their estimate. Such a discrepancy could be
due to the different theoretical models used. However, as already
noticed by \citet{edmonds01}, the cooling age is significantly lower
than the characteristic age of 3.8 Gyr (see Table~\ref{tab1psr}). This discrepancy should not alarm, since the PSR
characteristic ages are based on many assumptions and large deviations
from the companion cooling ages are commonly observed \citep[see
  e.g.][]{handbook,tauris12a,tauris12b}.  Using the WD age together
with the intrinsic spin period derivative ${
\dot{P}=2.7\pm0.5\times10^{-20}}$ \citep{freire17} and
the actual spin period $\mathrm P\approx4.343$ ms, we evaluated a MSP
birth spin period (the so-called equilibrium spin period) of $\mathrm
P_{0}\approx3.576$ ms. This value, combined with the surface magnetic
field $\mathrm B\approx3.145\times10^{8}$ G and assuming a NS with a
radius of 10 km and a canonical mass of $\mathrm 1.4 \ \Msun$, can be used
to infer the typical accretion rate that reaccelerated the NS during
the low-mass X-ray binary phase. By using equation (8) of
\citet{vandenheuvel09}, we find that the system past accretion history
likely proceeded at a sub-Eddington rate
$(\dot{M}/\dot{M}_{EDD}\sim0.02)$, as expected from the typical
  evolution of close binary systems with light donor stars
  \citep{tauris99,istrate14}. Although the mass accretion rate
strongly depends on the NS radius, the general result does not change
assuming different radii or even different NS masses.
\subsection{Possible candidate companion stars}
\label{comH}

As can be seen from Figures \ref{cmd_all}, \ref{cmdwd} and \ref{mp}, the
possible companion to 47TucH appears to have properties quite
different from those observed for the other companions, first of
  all its much larger distance from the MSP nominal position
  ($0.24\arcsec$), which corresponds to almost twice our astrometric
  uncertainty. Moreover, following the procedure adopted in the
previous section, we derived for this object a mass of $\mathrm
0.37\pm0.05 \ \Msun$. This value, combined with the binary system
total mass of $\mathrm 1.66 \ \Msun$ \citep[][]{freire03}, would imply a
PSR mass of $\mathrm \sim1.29 \ \Msun$, a value slightly lower than
expected for a recycled PSR, although still acceptable within the
uncertainties. Its position in the CMD is compatible also with
  the CO WD cooling sequence, which would increase the probability of
  a chance coincidence to $\sim 2-3\%$.  Furthermore, at odds with the
  others objects, this candidate counterpart occupies an anomalous
  region in the orbital period companion mass plane shown in
  Figure~\ref{mp}. Although this anomaly could be real (since 47TucH
  has a large eccentricity, probably due to some kind of dynamical
  interaction), all these pieces of evidence suggest that the observed
  object is probably an isolated WD and the true companion star is
  still under the detection threshold (see
    Section~\ref{nondec}).

A possible candidate companion to MSP 47TucI has been also
  detected (see Figure~\ref{fig1} for the finding chart). This is a
  binary system with a short orbital period ($\sim0.23$ days) and a
  very small eccentricity. From the PSR mass function, the companion
  is expected to be a very low-mass star (${ M_{COM}\sim0.015\; \Msun}$). The absence of radio eclipses, probably due to a low
  inclination angle, prevents its characterization as a black-widow (BW)
  system. At $0.15\arcsec$ from the PSR position, we identified a star
  located at the faint-end of the cluster main sequence (see
  Figure~\ref{cmd_all} and Table~\ref{tab_wdprop}). If we assume that the
  companion is a bloated star seen in a binary system with a low
  inclination angle, such a CMD position could be reasonable.
  However, the lack of any significant variability related to the
  orbital period prevents us from firmly associating this candidate to
  the MSP. In fact, the orbital period coverage of the F390W images is
  too poor, while the signal to noise ratio of the F300X data allows
  us to only infer that, in case of photometric variability, the
  maximum variation amplitude must be smaller than $\sim0.8$ mag.  We
  therefore conclude that it is more likely that the real companion
  star is still under the detection threshold. Indeed, the
probability of a chance coincidence with a main sequence star is non
negligible ($\sim45-50\%$). We finally note that another object
lies within the astrometric uncertainty circle, but its association
with the MSP can be excluded, since it is a common CO WD, with
properties incompatible with the MSP timing ephemeris.}

\subsection{Non detections}
\label{nondec}
No interesting counterparts have been identified for all the other
known binary MSPs. These non-detections are likely due to companion
stars still under the detection threshold (as in the case, e.g., of
47TucE and the BW 47TucJ), or to the severe crowding
conditions of the area surrounding the MSP positions (as in the case
of the BWs 47TucO and 47TucR). No search could be performed
for 47TucX since its position is outside the field of view.

 Considering that the companion to 47TucJ should be a
  non-degenerate object, its non-detection in UV passbands cannot be
  used to get useful information on its properties. Instead, the
  counterpart to 47TucE is expected to be He WD, which remains
  undetected down to our limiting magnitudes ($\sim25$ in the F300X
  filter and $\sim25.5$ in the F390W filter).  Hence, taking into
  account that the cooling age of a $\sim 0.17\; \Msun$ WD at these
  detection thresholds is larger than the cluster age ($\sim10-11$
  Gyr; \citealp{gratton03,hansen13}), it is unlikely that this star
  has a mass similar to that estimated for the other companions.  It
  is more probable that it is more massive than $0.2\; \Msun$
  (corresponding to a faster cooling) and its cooling age is larger
  than 1 Gyr.  The same should apply also to the case of 47TucH if its
  true companion is still under our detection limits (as suggested
  above).  Interestingly, according to the theoretical relation of
  \citet{tauris99} and the orbital periods of 47TucE and 47TucH
  ($\sim2.3$ and $\sim2.4$ days, respectively), the companions to both
  these MPSs are indeed expected to have masses $\gtrsim0.2\; \Msun$.

\subsection{The companion to the RB 47TucW}
\label{w}
\begin{figure*}[!t]
\begin{center}
\leavevmode
\includegraphics[width=10cm]{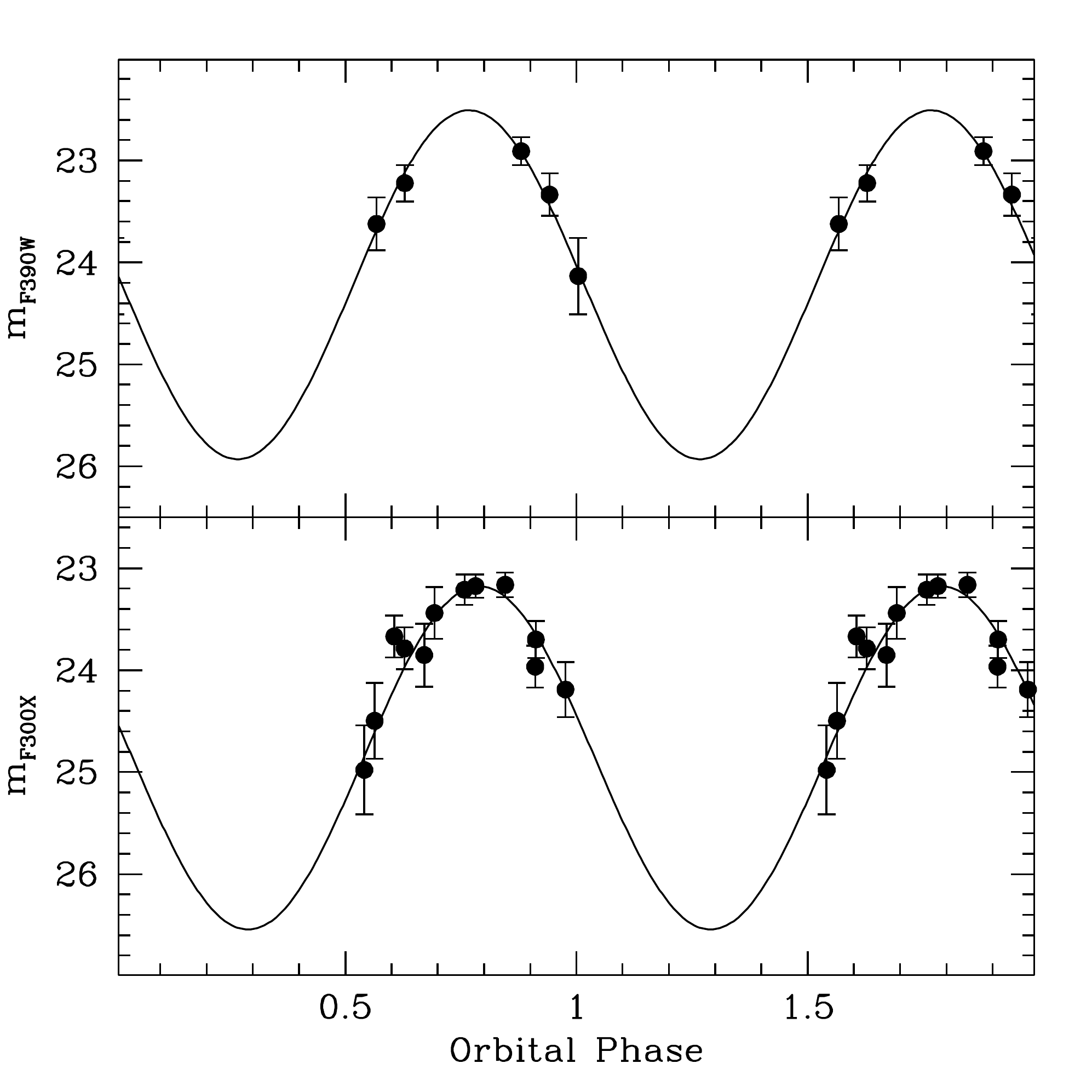}\\
  \caption[Light curves of COM-47TucW]{Light curves of COM-47TucW in the F390W (upper panel) and
    F300X (lower panel). The two curves are folded with the radio
    parameters and two periods are shown for clarity.  The black curve
    in each panel is the best analytical model obtained independently
    for each filter.}
  \label{curva_w}
\end{center}
\end{figure*}

47TucW is the only RB in 47 Tucanae with an available timing solution. It
is a binary MSP with a spin period of 2.35 ms, an orbital period of
$\sim3.2$ hr and a companion mass of $\mathrm \sim0.15
\ \Msun$. The first optical identification of this system was
presented in \citet{edmonds02}, who suggested that the companion is a
perturbed and non degenerate star with a light curve structure
indicating a strong heating by the PSR flux. In Figure~\ref{curva_w} we
show, for both the filters, the light curves we obtained by folding
our photometric measurements with the most updated radio timing
ephemeris \citep{freire17}. The zero orbital phase has been set at
the PSR ascending node time\footnote{Please note that we are using a
  different formalism with respect to \citet{edmonds02}.}. As can be
seen, in agreement with previous works \citep{edmonds02, bogdanov05}, the light curves present a
single maximum-minimum structure, likely due to the heating by the PSR
flux. Unfortunately, the star has been measured above the detection
threshold only near its maximum luminosity. Nonetheless,
modeling\footnote{We used the ``Graphical Analyzer for TIme Series'',
  a software aimed at studying stellar variability phenomena,
  developed by Paolo Montegriffo at INAF-Osservatorio Astronomico di
  Bologna.}  the sinusoidal light curve, we found that, in both the
filters, the companion spans $\sim3.5$ magnitudes between the maximum
and the derived minimum, in agreement with previous observations. The
best fit-model is shown as a solid curve in
Figure~\ref{curva_w}. Interestingly, the light curve structure is more
similar to the ones observed for BWs than for RBs
companions, which usually, but not always, show a double minimum-maximum structure due
to tidal deformation \citep[see
  e.g.][]{fer03_msp,cocozza08,pallanca10,li14}. The CMD position of
the companion during the maximum and at a mean phase (as derived by
the adopted model) is shown in Figure~\ref{cmd_all}. The system is located
between the main sequence and the WD cooling sequence, where no normal
stars are expected and thus suggesting a perturbed and strongly heated
companion star. Again, at odds with other RB systems, this lies
in a region more similar to that occupied by the two BW
companions identified so far in GCs \citep[][Cadelano et al.,
  2015]{pallanca14b}.  The X-ray counterpart to 47TucW shows a
variability which is likely due to an intra-binary shock between the
PSR wind and the matter lost by the companion
\citep{bogdanov05}. Interestingly, as discussed by \citet{bogdanov06}, the minimum of the X-ray light
curve is displaced with respect to the optical one. Such a behavior
has been also noticed for the BW M71A \citep{cadelano15a},
thus further strengthening the connection of this MSP with BW
systems. All this allows us to speculate that a scenario where 47TucW will evolve
into a canonical MSP with a He WD companion \citep[as in the case of
  MSP-A in NGC 6397; see][]{burderi02} is somewhat unlikely, opening the possibility to an
evolution toward the BW stages. Indeed
such an evolutionary path has been already suggested by the
simulations of \citet{benvenuto14}. The identification of new RB
companions will shed light on this possibility.

\chapter{The Optical Counterpart to the Accreting Millisecond X-ray Pulsar SAX J1748.9-2021 in the Globular~Cluster~NGC~6440}
\label{cap_6440}
\begin{flushright}
\textit{Mainly based on \citealt{cadelano17b}, ApJ, 844:53C}
\end{flushright}
\vspace{1cm}
\initial{W}e used a combination of deep optical and $\mathrm{H\alpha}$ images of the Galactic globular cluster NGC~6440, acquired with the Hubble Space Telescope,
to identify the optical counterpart to the accreting millisecond X-ray pulsar \sax during quiescence.
A strong $\mathrm{H\alpha}$ emission  has been detected from a main sequence star (hereafter COM-SAX J1748.9-2021)  
located at only $0.15"$ from the nominal position of the X-ray source. The position of the star  also  agrees with the optical counterpart found by \citet{verbunt00} during an outburst.  We propose this star as the most likely optical counterpart to the binary system. By direct comparison  with  isochrones,   we estimated that \com$\,$  has a mass of $0.70\; \Msun - 0.83\; \Msun$, a radius of $0.88\pm0.02\; \Rsun$ and a superficial temperature of $5250\pm80$ K. These parameters combined with the orbital characteristics of the binary suggest that the system is observed at a very low inclination angle  ($\sim8^{\circ}-14^{\circ}$) and that the star is filling or even overflowing its Roche-Lobe. This, together with the equivalent width of the $\mathrm{H\alpha}$ emission ($\sim20$ \AA), suggest on-going mass transfer. The possible presence of such an on-going mass transfer during a quiescence state also suggests that the radio pulsar is not active yet and thus this system, {\ despite its similarity with the class of redback millisecond pulsars, is not a transitional millisecond pulsar}. 

\clearpage
\section{Introduction}\label{intro_SAX}
As we described in Chapter~\ref{accreting}, Accreting Millisecond X-ray Pulsars (AMXPs) are a sub-group of transient low-mass X-ray binaries that show, during outbursts, X-ray pulsations from a rapidly rotating neutron star. During these outbursts, the matter lost from the companion star via Roche-Lobe overflow is channeled { down} from a truncated accretion disk onto the neutron star magnetic poles, producing X-ray pulsations at frequencies $\nu \geq 100$ Hz \citep[see][and references therein]{patruno12}. Two of these systems are located in the globular cluster (GC) NGC 6440, the only cluster, together with NGC 2808 and M28, known to host AMXPs. NGC 6440 is located in the Galactic bulge, above the Galactic plane, at $8.5$ kpc from the Sun \citep{valenti07}. It is a metal-rich system ($\mathrm{[Fe/H]\sim-0.5}$, \citealt{origlia97, origlia08a}) affected by a quite large and differential extinction, with a mean color-excess ${E(B-V)=1.15}$ \citep{valenti04}. The cluster also hosts six (classic) radio millisecond pulsars (MSPs) \citep{freire08}.

\sax was discovered with the {\it Beppo}SAX/WFC satellite in 1998 as a part of a program aimed at monitoring the X-ray activity around the Galactic center \citep{intzand99}. Since its discovery, it has experienced four outbursts, approximately one each five years: in 2001 \citep{intzand01}, 2005 \citep{markwardt05}, 2010 \citep{patruno10} and finally in 2015 \citep{bozzo15}. The X-ray pulsar (PSR) has been observed pulsating at a spinning frequency of $\sim 442$ Hz and these pulsations have been used to obtain a phase-coherent timing solution \citep{gavriil07,altamirano08,patruno09,sanna16} which revealed that \sax is a binary system with an orbital period of $\sim8.76$ hours, a projected semi-major axis of $\sim0.4$ light-seconds and a companion mass of at least $0.1\; \Msun$. \citet{altamirano08} suggested that the companion star is more likely a $0.85\; \Msun-1.1\; \Msun$ star, i.e. a bright main sequence or a slightly evolved star, thus implying a binary system seen at a low orbital inclination angle.  \citet{verbunt00} identified  the optical counterpart during the 1998 outburst as a blue star with $ B\simeq22.7$. This outburst counterpart was identified through images obtained with a ground-based telescope and during nonoptimal seeing conditions. 

The spin and orbital properties of \sax are similar to those observed in the emerging class of ``transitional millisecond pulsars'' (tMSPs, see Chapter~\ref{accreting}): binary systems that alternate between stages of classical rotation-powered emission, where  radio emission is detected as in a common eclipsing MSPs of the redback (RB) class, and stages of accretion-powered emission where the radio emission is off and X-ray pulsations are detected like in AMXPs. The similarity between \sax and the class of tMSPs suggests that this system might be a tMSP whose radio PSR emission during quiescence has not been revealed yet. In fact, no radio pulsed emission has been detected from this object, in spite of  the radio searches devoted to this aim \citep[see][]{patruno09}. The identification of the optical counterpart during quiescence  can provide  crucial information to understand the properties and the nature of the binary system.

\section{Observations and Data Reduction}

This work is based on two different datasets of images obtained with the {\it Hubble Space Telescope} (HST) using
the UVIS camera of the Wide Field Camera 3 (WFC3). The first dataset (GO12517, P.I.: Ferraro) has been obtained on July 2012 and it consists of 27 images in the F606W filter with an exposure time of 392 s each and 27 images in the F814W filter with an exposure time of 348 s each. The second dataset (GO13410, P.I.: Pallanca) has been acquired during three different epochs: October 2013, May 2014 and September 2014. Each epoch consists of 5 images in the F606W filter and exposure time of 382 s, 5 images in the F814W filter and exposure time of 222 s and 10 images in the F656N filter and exposure time of 934~s. 
 
{ We used the images processed, flat-fielded and bias subtracted (``flt'' images\footnote{``flc'' images, corrected also for charge transfer efficiency, were not available at the time this work was performed.}) by the standard HST pipeline}.  The photometric analysis has been performed using standard procedures, after correcting the images for ``Pixel-Area-Map"\footnote{For more details see the WFC3 Data Handbook.} and following the prescription described in Chapter~\ref{opticalcom}. The resulting instrumental catalog, containing instrumental magnitudes, has been calibrated on the VEGAMAG system using the WFC3 zeropoints publicly available at \url{http://www.stsci.edu/hst/wfc3/phot_zp_lbn} \citep{ryan16}.

The instrumental positions have been corrected for geometric distortions by applying the equations reported in \citet{bellini11}. In order to transform the instrumental positions to the absolute coordinate system ($\alpha, \delta$), we  used the Pan-STARRS1 catalog of stars \citep{flewelling16} reported, by means of $\sim120$ in common, to the UCAC4 astrometric standard catalog \citep{zacharias13}. Then, this catalog has been used as reference frame to astrometrize the HST dataset, by means of $~1200$ stars in common. The resulting $1\sigma$ astrometric uncertainty is $\sim 0.15\arcsec$ both in $\alpha$ and $\delta$, thus providing a total uncertainty of about $0.21\arcsec$.

 NGC 6440 is affected by a substantial differential reddening. In order to estimate the extinction variation  within the observed field of view, we adopted a method similar to that already applied  to other clusters (see \citealt{massari12}). The detailed procedure and the reddening map will be published in a forthcoming paper (Pallanca et al. 2018, in preparation). Here we just briefly describe the procedure. We first selected a sample of reference stars  with  small photometric errors and values of the sharpness parameter (``well-fitted" stars), located along the cluster evolutionary sequences in the color-magnitude diagram (CMD). These stars have been used to built a ``reference" mean ridge line. Then, for each 
star a mean ridge line has been constructed by using the fifty ``well-fitted'' stars spatially located close to it. Finally, the shift ${\delta E(B-V)}$  needed to register this ridge line to the ``reference" mean ridge line is computed. The derived extinction map  shows absorption clouds with a patchy structure,  and extinction variations as large as ${\delta(B-V)=1.0}$ mag have been measured. The map has been used to build the differential reddening-corrected CMD used in the following analysis.

\section{Results}
\label{results}

\begin{figure*}[b]
\begin{center}
\includegraphics[width=5.1cm]{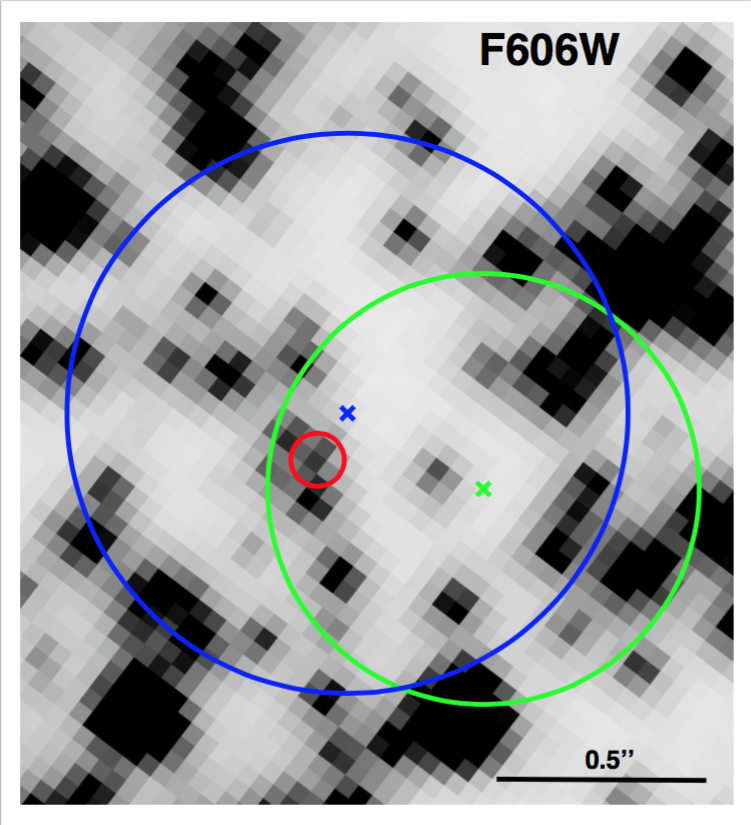}
\includegraphics[width=5.1cm]{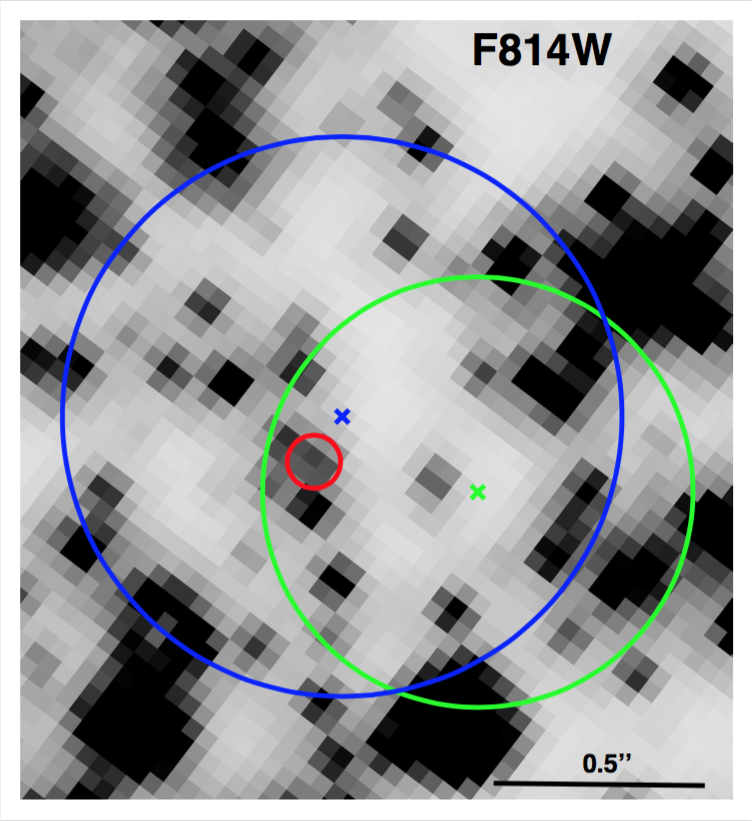}
\includegraphics[width=5.1cm]{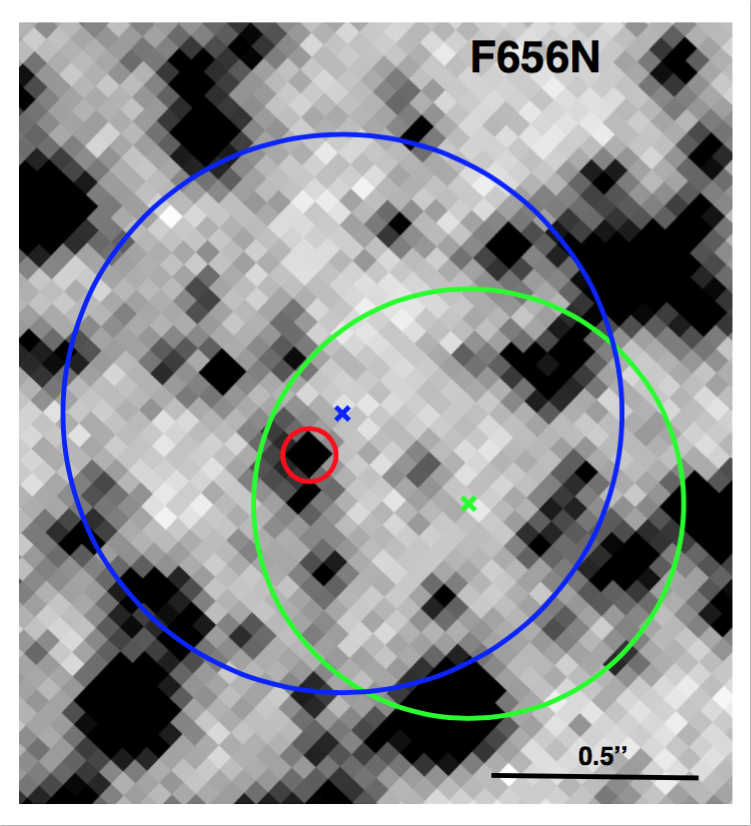}
  \caption[Finding charts of the counterpart to \com]{Finding charts of the counterpart to \com (north is up and east is left). The left, central and right panels are   combined images from the F606W, F814W and F656N expositions, respectively. In all the panels, the blue cross indicates the X-ray nominal position, while the blue circle, centered on the blue cross, has a radius equal to the combined X-ray and optical astrometric uncertainty. The green cross is centered on the outburst counterpart reported by \citet{verbunt00} and the circle has a radius equal to their astrometric uncertainty. The solid red circle marks the candidate optical counterpart. }
  \label{chart_SAX}
\end{center}
\end{figure*}

\begin{figure*}[h]
\begin{center}
\includegraphics[width=7.8cm]{cmd.png}
\includegraphics[width=7.8cm]{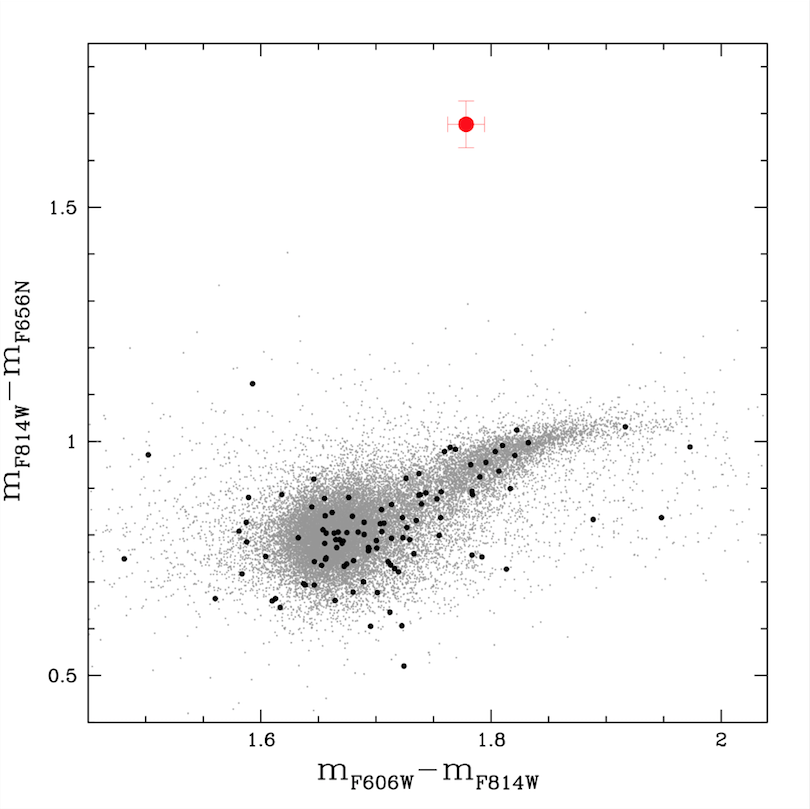}
  \caption[CMD and color-color diagrams of NGC 6440]{{\it Left Panel:} (${ m_{F606W}, m_{F606W}-m_{F814W}}$) differential reddening corrected CMD of NGC~6440. The stars within $2\arcsec$ from the nominal position of the X-ray source are shown as big black dots. Cluster stars detected in the WFC3 field of view are plotted in grey. The dashed blue curve is the best fit isochrone (see Section~\ref{results}) and the red point is the mean position of \com in the four epochs. {\it Right Panel:} (${ m_{F814W}-m_{F656N}, m_{F606W}-m_{F656N}}$) cluster color-color diagram. The  symbols are as in the upper panel. }
  \label{cmd_SAX}
\end{center}
\end{figure*}

In order to search for the optical counterpart to SAX J1748.9-2021, we analyzed all the objects located within a $3\sigma$ radius ($\sim2\arcsec$) from the X-ray position reported by \citet{pooley02}, where $\sigma$ is the combined X-ray ($\sim0.6\arcsec$) and optical ($\sim0.21\arcsec$) astrometric uncertainty.   The finding charts of the region around  the X-ray nominal position are shown, for all the filters,  in Figure~\ref{chart_SAX}. We emphasize that in all the epochs sampled by the observations discussed in this thesis,  \sax was in a quiescence state. Thus we did not expect to find a bright star like that reported in \citet{verbunt00}. From the analysis  of the ${H\alpha}$ images, we found a quite promising candidate. In the right panel of Figure~\ref{cmd_SAX}, we show the ${(m_{F814W}-m_{F656N}, m_{F606W}-m_{F814W}}$) color-color diagram of the entire cluster of stars (small grey dots), with all the stars detected in the region around the X-ray source position highlighted as large black dots. This diagram  has been found to be particularly powerful in pinpointing  ${H\alpha}$ emitters  \citep[e.g.]{beccari14}. 
All the stars detected in the X-ray source region appear to be standard stars, with the only exception of one object (the red dot) that shows  a quite anomalous ${H\alpha}$ color (${m_{F814W}-m_{F656N})\sim 1.7}$,  thus indicating a  strong ${H\alpha}$ excess (see Section~\ref{discussion}).   This object occupies instead a standard position in the optical CMD (left panel of Figure~\ref{cmd_SAX}), being located along the cluster main sequence, about $\sim2$ magnitudes below the turn-off. 
Interestingly enough, this object is located at ${ \alpha=17^{h}48^{m}52.161^{s}}$ and ${ \delta=-20^{\circ}21'32.406\arcsec}$, at only $\sim0.15\arcsec$ from the nominal position of the X-ray source and it is the closest star to the X-ray position. Its location is also consistent with that of the burst counterpart proposed by \citet{verbunt00}: in fact, the distance between our and their counterpart is $\sim0.35\arcsec$, smaller than the quoted uncertainty of the latter ($\sim0.5\arcsec$). 
 \begin{figure*}[!t]
\begin{center}
\includegraphics[width=10cm]{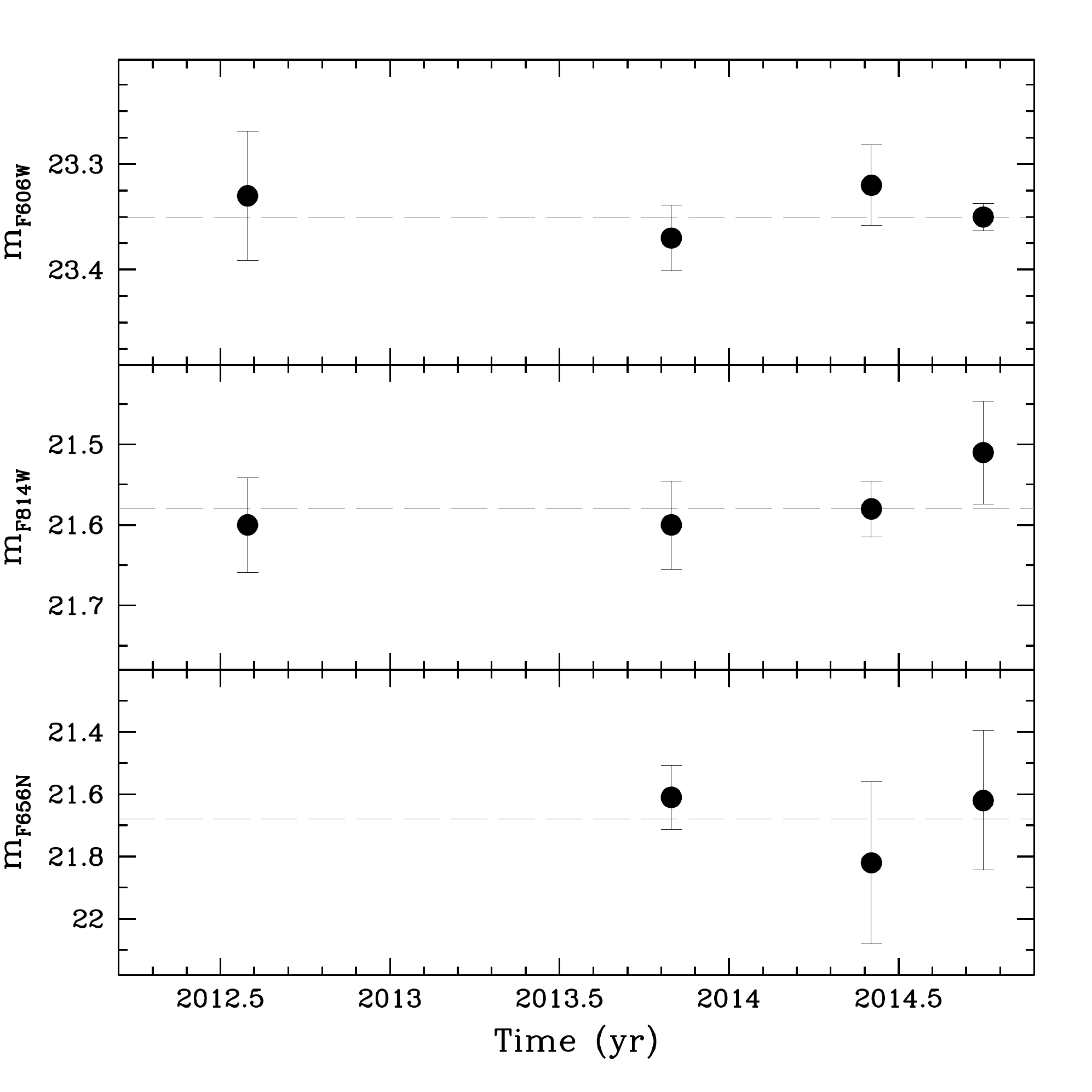}
  \caption[Magnitude of \com in the three different 
 filters and epochs]{Magnitude of \com in the three different 
 filters measured at the epochs at which observations are available.   
The mean magnitudes are indicated with the dashed horizontal lines.  No significant luminosity variations in all the photometric bands  across the different epochs can be detected.  }
  \label{magtime}
\end{center}
\end{figure*}

Therefore,  from both the positional agreement and the presence of ${H\alpha}$ emission we can conclude that this object is likely the companion star to the neutron star in the binary system \sax observed during quiescence (hereafter \com).   Figure~\ref{magtime} shows the measured  magnitudes of \com in different 
 filters at the four  epochs available.   As it can be seen, no significant variation is detected across the different epochs.
The mean magnitudes in each photometric band are: ${m_{F606W}=23.35\pm0.01}$, ${m_{F814W}=21.58\pm0.01}$ and ${m_{F656N}=21.68\pm0.05}$.  From the bottom panel of Figure~\ref{magtime} we can conclude that in the three epochs for which ${H\alpha}$ observations are available, a persistent  ${H\alpha}$ emission was present. Thus, these observations  indicate an ongoing  mass transfer activity from the companion toward  the neutron star when the AMXP is in a quiescence state. This is in line with what commonly observed in different classes of interactive binaries (e.g. low-mass X-ray binaries, cataclysmic variables) that are experiencing accretion phenomena also during quiescence  \citep[e.g.][]{ferraro00,torres08,beccari14,torres14}.

The physical properties of \com can be derived from the comparison of its position in the optical CMD with appropriate   
isochrone models. We used the  isochrone set from the Dartmouth Stellar Evolution Database \citep{dotter08}, for a 12 Gyr-old cluster \citep{origlia08b} with reddening, distance modulus and metallicity as reported in Section~\ref{intro_SAX}. The isochrone (reported as a blue dashed curve in the left panel of Figure~\ref{cmd_SAX}) nicely reproduces the cluster evolutionary sequences. By projecting the magnitude and color of \com onto the isochrone,  we found a stellar mass $M=0.73\pm0.01\; \Msun$, an effective temperature of ${T_e=5250\pm80}$ K and a bolometric luminosity of ${0.53\pm0.01\; \Lsun}$, the latter two corresponding to a radius of ${R=0.88\pm0.02\; \Rsun}$ (see the discussion in Section~\ref{discussion}). From the isochrone we can also infer that the expected $B$ magnitude of the object in quiescence should be ${B\simeq25.7}$. This value is 3 mag fainter than that measured by \citet{verbunt00} during the 1998 outburst. Such a large variation is similar to what observed between the outburst and the quiescence states of other AMXPs \citep[see e.g.][and references therein]{patruno12} and, more generally, of transient low-mass X-ray binaries \citep[e.g.][]{ferraro15a}. Since different isochrone models can lead to slightly different results, we re-made the computations by using isochrones from the $BaSTI$ database \citep{pietrinferni04} and isochrones from the Padova Stellar Evolution Database \citep{girardi00}, finding similar results.

It is worth mentioning that NGC 6440 hosts another AMXP: NGC6440X-2 \citep[][]{altamirano10}.  This is an ultracompact system with an orbital period of only $\sim0.96$ hours and a X-ray PSR pulsating at $\sim206$ Hz. From the binary system mass function, the companion mass is expected to be $\geq 0.007\; \Msun$. 
Despite a careful search for the optical counterpart to this system in the available set of images, we did not find any reasonable candidate. Likely, the optical counterpart of this AMXP is still under the detection threshold, given the extremely low-mass expected for this companion star. We can therefore provide only lower limits in luminosity for this system: ${m_{F606W}>25.0}$, ${m_{F814W}>23.5}$ and ${m_{F656N}>23.0}$.

\section{Discussion}
\label{discussion}
\begin{figure*}[t]
\begin{center}
\includegraphics[width=10cm]{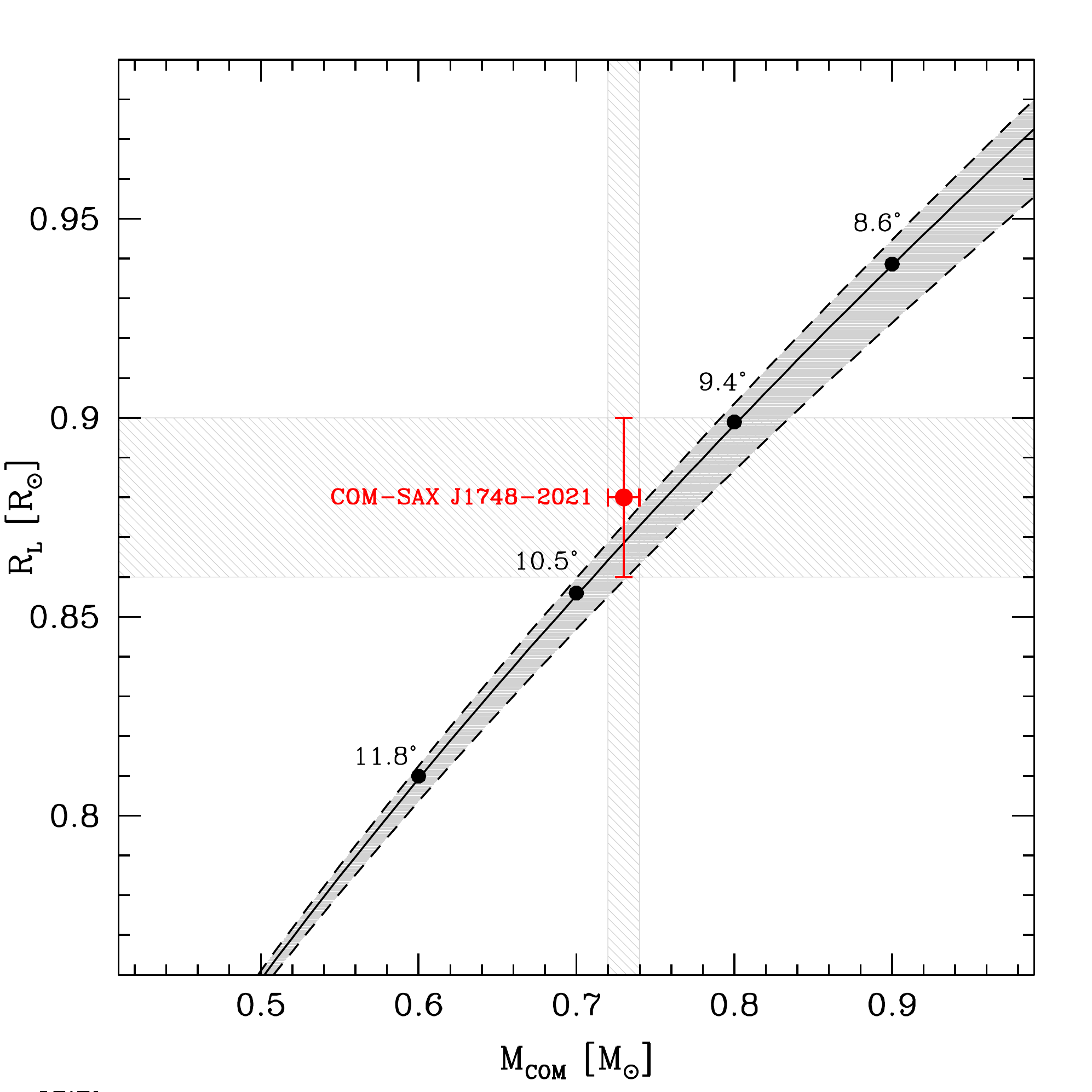}
  \caption[Roche Lobe radius as function of the \com mass]{Roche Lobe radius as function of the \com mass. The solid line represents the analytic prediction for a PSR mass of $1.4\; \Msun$, while the shaded gray area surrounded by the dashed curves correspond to the predictions for a PSR mass ranging from $1.2\; \Msun$ to  $2.4\; \Msun$. The labelled black dots indicate the inclination angle of the binary system as predicted by its mass function.   The red circle and gray striped area mark the derived \com radius and mass.}
  \label{raggio}
\end{center}
\end{figure*}

{\ We can now compare the radius of \com  obtained in
Section~\ref{results}  with the expected dimension of the Roche-Lobe (RL) radius}.  The latter quantity can be estimated according to the following relation  \citep{eggleton83}:
$$
{R_{L}=\frac{0.24M_{PSR}^{1/3}q^{2/3}\left(1+q\right)^{1/3}P_{b, hr}^{2/3}}{0.6q^{2/3}+\log(1+q^{1/3})}}
$$
where $\mathrm q$ is the ratio between the companion and the PSR mass and $\mathrm{P_{b, hr}}$ is the orbital period in hours. 
In Figure~\ref{raggio} (solid line) we plot the RL radius, computed by
assuming a PSR mass in the range $1.2\; \Msun-2.4\; \Msun$, as a function of the companion mass. {\ The position of \com  in this diagram (large filled dot) indicates that it is at least completely filling its RL and possibly even overflowing it}. Indeed, the  RL radius corresponding to the  estimated mass of \com   is $0.86\; \Rsun-0.87\; \Rsun$, implying a filling factor of $0.99-1.05$. {The derived filling factor is in agreement with what expected from such a system, where the presence of on-going mass transfer implies that the companion star is most likely in a RL overflow state. However, it is worth noticing that the mass derived for \com from standard stellar isochrones can be biased. In fact, the companion star could have suffered a strong mass loss if it is the same object that has recycled, via mass transfer, the neutron star. Such an effect is not accounted for by the stellar evolutionary models used to create isochrones, thus introducing a bias in the derivation of the companion physical properties \citep[see the notorious cases of PSR~J1740$-$5340A and PSR~J1824$-$2452H:][]{ferraro01a,pallanca10,mucciarelli13}. On the other hand, we can assume the derived photospheric radius more reliable, since it exclusively depends on the companion luminosity and temperature. Setting this measured radius equal to the RL radius, we found that the companion star is filling its Roche-Lobe in the mass range of $0.70\; \Msun-0.83\; \Msun$. However, the possible presence of heating of the companion star due to the neutron star emitted flux could affect the observed luminosity and temperature of the companion star, introducing an additional bias, which is difficult to quantify. Nevertheless, the position of \com, compatible with the cluster main sequence, suggests that this effect might not be very relevant for this system, at odds with what observed for strongly heated companion stars \citep[see, e.g.,][]{edmonds02,pallanca14a,cadelano15a}.} We can therefore conclude that the observed properties of  \com are likely compatible with that of a binary system where the secondary star has a mass of $0.70\;  \Msun -0.83\; \Msun$ and it is filling and possibly overflowing its RL, whose radius is $0.88\pm0.02\; \Rsun$. These quantities can be used to constrain the inclination angle of the system. Using the orbital solution reported by \citet{patruno09} and \citet{sanna16}, we found that, for a PSR mass in the range $1.2\; \Msun-2.4\; \Msun$, the binary inclination angle should be very low, between $8^{\circ}$ and $14^{\circ}$. Interestingly, a  low orbital inclination angle was independently suggested by the absence of dips and eclipses in the X-ray light curve \citep[e.g.][]{sanna16} and by the expected properties of the companion star discussed by \citet{altamirano08}.

The evidence of  ${H\alpha}$ emission previously discussed and shown in Figure~\ref{cmd_SAX} can be used to estimate the equivalent width (EW) of the emission line of main sequence stars. In doing this, we followed the method reported by \citet{demarchi10} and already used in previous papers \citep[]{pallanca13, beccari14}. Briefly, the  excess in the  de-reddened ${H\alpha}$ ${(m_{F606W}-m_{F656N})_0}$ can be expressed in terms of the equivalent width of the ${H\alpha}$ emission by using equation (4) in \citet{demarchi10}: ${EW=RW\times[1-10^{(-0.4\times \Delta H\alpha)}]}$, where :

\begin{itemize}
\item ${RW}$  is the ``rectangular width" of the adopted ${H\alpha}$ filter, its definition being similar to that of equivalent width used to measure the intensity of an emission/absorption line. According to Table 4 in \citet{demarchi10}, ${RW=17.48\AA}$ for the HST-WFC3 ${H\alpha}$ filter we adopted here.  

\item ${\Delta H\alpha}$ is the difference in the de-reddened ${H\alpha}$ color ${(m_{F606W}-m_{F656N})_0}$ between \com and the value expected from a star with the same optical color\\ ${(m_{F606W}-m_{F814W})_0}$ but showing no ${H\alpha}$ emission. 
\end{itemize}

\begin{figure*}[t]
\begin{center}
\includegraphics[width=10cm]{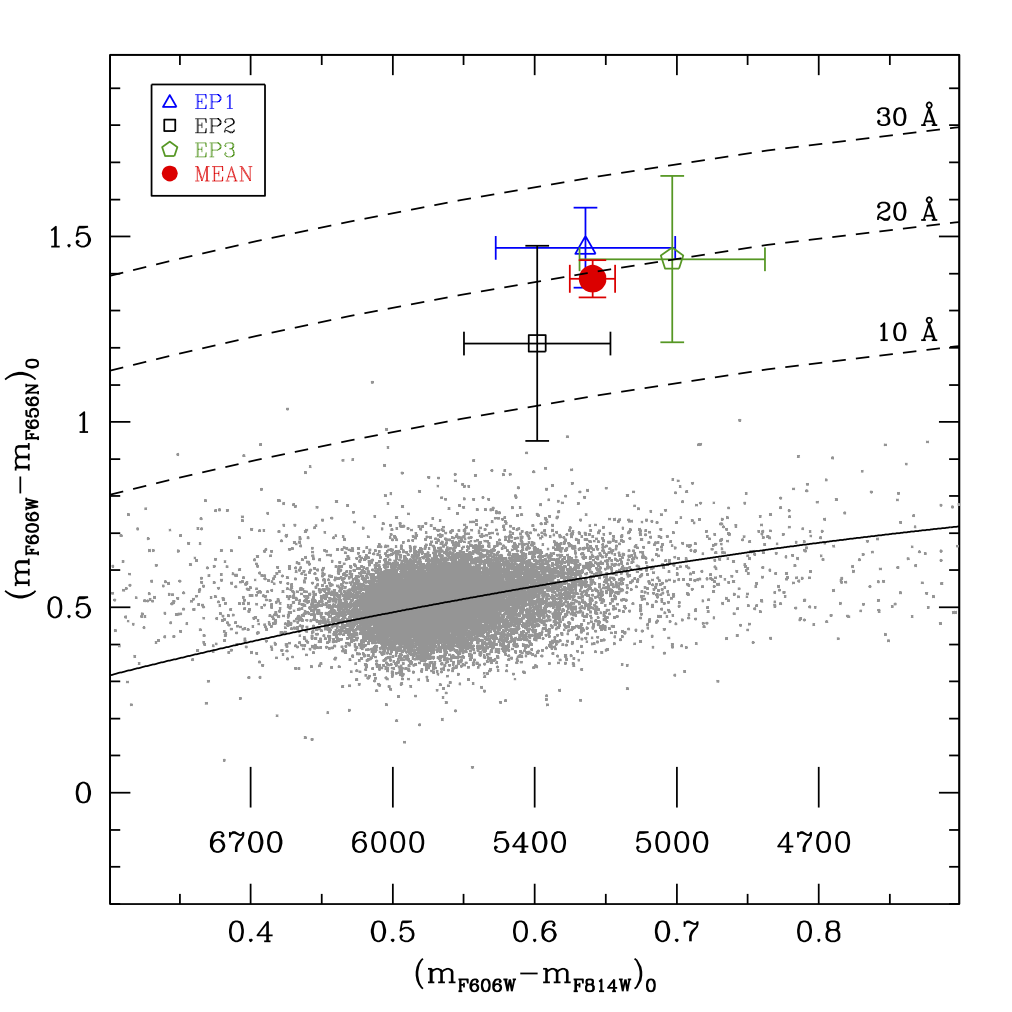}
  \caption[Color-color diagram of NGC 6440 used to determine the $H\alpha$ EW]{Reddening corrected ${ (m_{F606W}-m_{F814W})_0}$ vs ${ (m_{F606W}-m_{F656N})_0}$ color-color diagram of NGC 6440. The solid line marks the region occupied by main sequence stars with no ${H\alpha}$ emission, while the dashed ones show, respectively, the regions where stars with ${H\alpha}$ emission and an EW of ${10 \AA}$, ${20 \AA}$ and ${30 \AA}$ are located. The colored points are the positions of \com in the different epochs, as reported in the legend. Effective temperatures (in Kelvin) related to the corresponding colors are also marked.}
  \label{colcol}
\end{center}
\end{figure*}

On the basis of this relation, different curves at increasing ${H\alpha}$ EW can be computed and they are plotted in Figure~\ref{colcol}. When main sequence stars are plotted in this diagram, the vast majority of them are located around the ``no ${H\alpha}$ emission'' curve (solid curve), as expected by canonical cluster stars. \com   is instead located significantly above this line, showing a mean systematic excess of  ${\Delta H\alpha = 0.80\pm0.07}$ in all the sampled epochs.
As it can be seen from  Figure~\ref{colcol} such an excess corresponds to an EW of the ${H\alpha}$ emission  of $19\pm1$ \AA.
Such a value is too large to be attributed to chromospheric activity \citep{beccari14}. It is instead
 a typical value for system with a low mass accretion rate. In fact it turns out to be in agreement with the ${H\alpha}$ EWs measured in the majority of quiescent low mass X-ray binaries with a neutron star accretor \citep[${ EW = 20\; \mathrm \AA}-\mathrm{50\; \AA}$, see][and references therein]{heinke14}. The evidence of such a low-level accretion rate in this system was already suggested by the X-ray studies of \citet{bahramian15}, performed two years after and one year before a burst. The value of ${\Delta H\alpha}$ just measured can be used to directly  estimate the ${H\alpha}$ luminosity due to the accretion processes ${L(H\alpha)}$, by using the photometric zeropoints and the values of the inverse sensitivity (PHOTFLAM parameter)  publicly available for all the WFC3 filters (see \url{http://www.stsci.edu/hst/wfc3/phot_zp_lbn}). At the cluster distance, we found ${L(H\alpha)=1.27\pm0.08\times10^{-4}\; \Lsun} $. This value, together with the \com mass and radius quoted in Section~\ref{results}, can be inserted in equation (7) of \citet{demarchi10} to estimate the mean mass transfer rate of the binary system, that turns out to be $\dot{m}\sim3 \times 10^{-10}\; \Msun$ yr$^{-1}$. 
Note, however, that the derived value of ${\dot{m}}$ must be taken with extreme caution, since the \citet{demarchi10}  method has been calibrated for accretion processes in pre main sequence stars, hence its applicability to the different cases (as low-mass X-ray binaries) could be risky. 

{\ \citet{sanna16} measured for this binary system a large orbital period derivative of $1.1\times10^{-10}$. This large value is interpreted as the result of a non conservative mass transfer driven by the emission of gravitational waves. In this model, the large orbital period derivative implies a large time-averaged mass transfer rate ($\mathrm{\sim10^{-8}\; \Msun}$ yr$^{-1}$) that can be explained by a companion star with a low-mass of $\sim0.12\; \Msun$, where only the $3\%$ of its lost mass is accreted by the neutron star (see their Figure 9). Our findings are not in agreement with such a scenario, since the companion mass is significantly more massive. If we assume that the large orbital period derivative is indeed the result of a strong mass transfer, our and their results could be reconciled by assuming that the fraction of lost mass that is accreted by the neutron star is even smaller than $3\%$. However, in the previous paragraph we roughly estimated that the mass transfer rate during quiescence is ${\dot{m}\sim3 \times 10^{-10}\; \Msun}$ yr$^{-1}$. This value, although extremely uncertain, is $\sim100$ times smaller than their predicted value. This could suggest that the mass transfer rate, estimated by \citet{sanna16} on the basis of the orbital period derivative, is overestimated. More generally, the disagreement between our and \citet{sanna16} results could suggests that the large orbital period derivative is due to a different effect such as, for example, a variable quadrupole moment of the companion star \citep{applegate92,applegate94,hartman08,patruno12b}, a phenomenon commonly invoked to explain the time evolution of the orbits of black-widows, RBs and tMSPs \citep{applegate94,archibald13,pallanca14a,pletsch15}. The similarity of SAX J1748.9-2021 to the RB class might corroborate this hypothesis, although other alternatives exist \citep[see for example][for a discussion]{patruno16}.}\\

\begin{figure*}[h]
\begin{center}
\includegraphics[width=10cm]{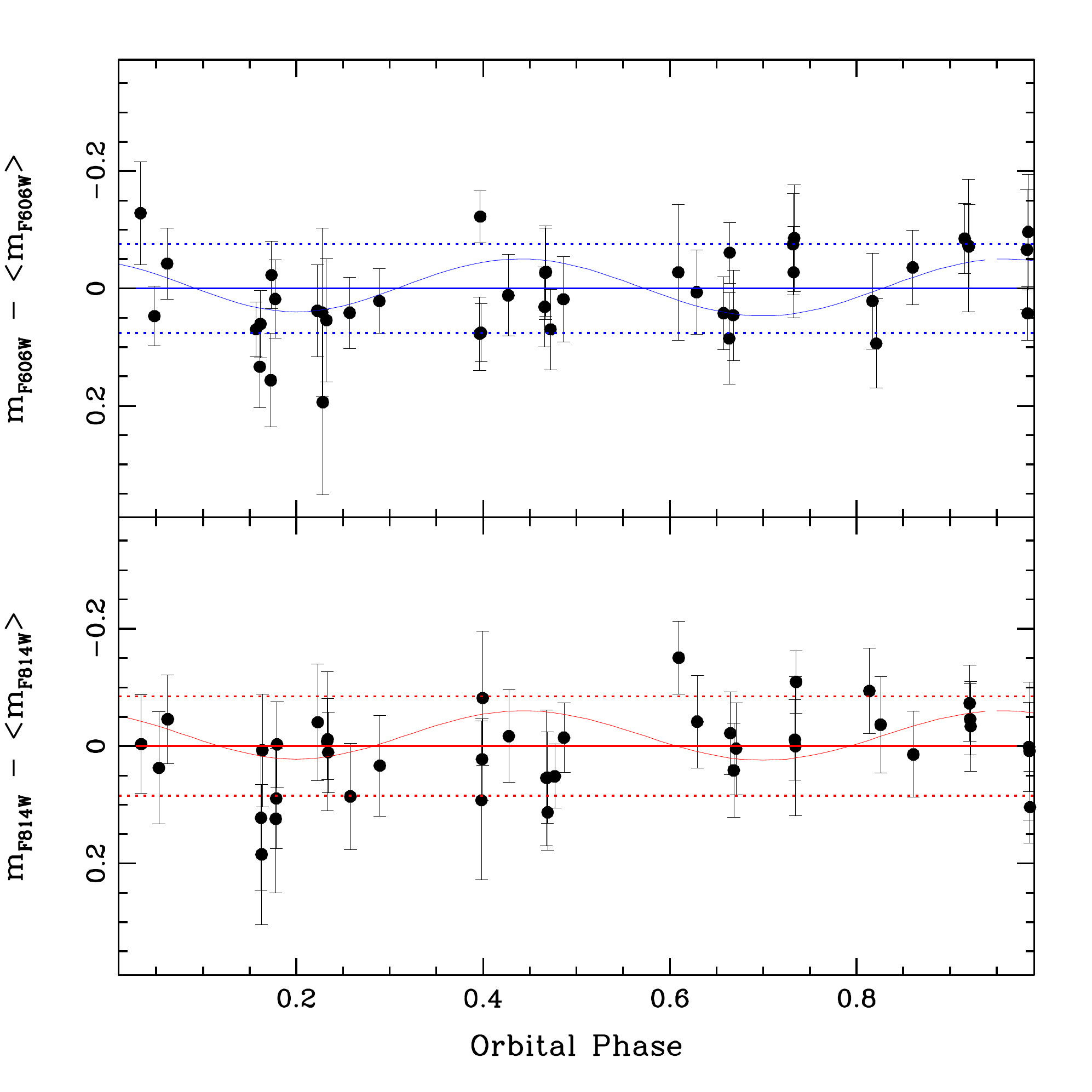}
  \caption[Light curve of \com obtained by folding the optical measurements]{{ Light curve of \com (black circles and error bars) obtained by folding the F606W measurements (top panel) and F814W measurements (bottom panel) with the binary system orbital parameters. The solid and dashed horizontal lines are respectively the mean and the standard deviations of the measurements. No evidence of variability associated to the orbital period is visible with the photometric errors on the single exposures. For illustrative purposes, we also show the simulated light curves obtained with {\tt NIGHTFALL} (blue and red curves) for an orbital inclination of $30^{\circ}$, which is, however, excluded by our analysis.}}
  \label{curva}
\end{center}
\end{figure*}

Since the analyzed dataset samples almost homogeneously the entire orbital period of the system,
an additional  aspect that we can investigate is the possible presence of light modulations.
Indeed, sinusoidal variations  due to irradiation processes or ellipsoidal deformation of the star, are expected to be observed in these systems \citep[see e.g.][]{homer01,davanzo09}, although the amplitude of the modulation strongly depends on the system inclination angle. In order to determine  the amplitude of the light curve expected from the system, we 
constructed a very basic model of \sax by using the software {\tt NIGHTFALL}\footnote{This software is publicly available at \url{http://www.hs.uni-hamburg.de/ DE/Ins/Per/Wichmann/Nightfall.html}.}. We simulated a set of light curve models (in the F606W and F814W filters)\footnote{Note that since the software does not allow to evaluate light curves for the specific  WFC3 photometric filters,  we  used the  V and I Johnson filter to simulate respectively the F606W and F814W light curves.} with a point-like primary star of $1.4\; \Msun$ and a RL filling companion star with masses in the range of $0.1\; \Msun - 1\; \Msun$ (compatible with both the binary mass function and the cluster stellar population)\footnote{The binary system mass function predicts, for a system with very low inclination angles ($i<5^\circ$), companion masses larger than $2\; \Msun$, incompatible with the old population of stars in GCs.}. { We found that amplitudes $\lesssim0.01$ mags are expected for an inclination angle of $\sim10^{\circ}$, corresponding to the derived companion mass ($\sim0.7\; \Msun$). Since the typical photometric uncertainty on the single measurements is $\sim0.08$ mags, such a small magnitude modulation cannot be detected with our dataset. Magnitude modulations comparable to or larger than our typical photometric uncertainty are expected only for $i\geq30^{\circ}$, corresponding to companion masses $\leq0.2\; \Msun$ (see an example in Figure~\ref{curva}), which have been excluded by our analysis. This seems to further support the conclusion that the system is seen at a small inclination angle. However,  this is a very basic model and the addition of processes like irradiation from the primary star, truncated disk,  etc.. can modify the light curve shape. Hence, deeper observations are needed before drawing solid conclusions about the optical variability of the system.}\\

The spin and orbital properties of \sax are quite similar to those generally observed in RB systems. This, combined with the periodical occurrence of outbursts, might suggest that this system is a tMSP whose radio pulsed emission has not been revealed yet (see Section~\ref{intro_SAX}). However, here we presented some observational evidence suggesting {\ an on-going mass transfer during the quiescence state. This is not expected in the radio PSR state of tMSPs \citep[see e.g.][]{archibald09,pallanca13} and suggests that in the case of \sax the radio emission mechanism is not active and thus that the system is not a tMSP. The ${H\alpha}$ emission detected in \sax clearly indicates that this system behaves as a typical low-mass X-ray binaries in quiescence, with mass transfer currently on-going and a possible residual accretion disk still present around the neutron star. This shows that not all accreting neutron stars with main sequence companions and orbital parameters similar to RB behave as tMSPs.}

\cleardoublepage

\clearpage


\def\ltsima{$\; \buildrel < \over \sim \;$}
\def\gtsima{$\; \buildrel > \over \sim \;$}
\def\lsim{\lower.5ex\hbox{\ltsima}}
\def\gsim{\lower.5ex\hbox{\gtsima}}
\def\lapp{\ifmmode\stackrel{<}{_{\sim}}\else$\stackrel{<}{_{\sim}}$\fi}
\def\gapp{\ifmmode\stackrel{>}{_{\sim}}\else$\stackrel{<}{_{\sim}}$\fi}

\def\msol{\,\mathrm{M}_\odot}
\def\media#1{\langle #1\rangle}
\newcommand{\masyr}{${\mathrm mas\, yr^{-1}}$}
\newcommand{\masyrarc}{${\mathrm mas\, yr^{-1}\, arcsec^{-1}}$}

\chapter{The Optical Identification of the X-ray Burster EXO~1745-248 in Terzan 5}
\label{Ter5burst}
\begin{flushright}
\textit{Mainly based on \citealt{ferraro15a}, ApJ, 807L:1F}
\end{flushright}
\vspace{1cm}
\initial{W}e report on the optical identification of the neutron star burster
EXO 1745-248 in Terzan 5.  The identification was performed by
exploiting HST/ACS images acquired in Director's Discretionary
Time shortly after (approximately 1 month) the Swift detection of the
X-ray burst.  The comparison between these images and previous 
  archival data revealed the presence of a star that currently
brightened by $\sim 3$ magnitudes,  consistent with expectations
  during an X-ray outburst.   The centroid of this object well
  agrees with the position, in the archival images, of a star located
  in the Turn-Off/Sub Giant Branch region of Terzan 5.  This supports
  the scenario that the companion should has recently filled its Roche
  Lobe. Such a system represents the pre-natal stage of a
millisecond pulsar, an evolutionary phase during which heavy mass
accretion on the compact object occurs, thus producing X-ray outbursts
and re-accelerating the neutron star.

\clearpage
\section{Introduction}
As discussed in Chapter~\ref{CapIntro}, low-mass X-ray binaries (LMXBs) and radio millisecond pulsars (MSPs)
are thought to be, respectively, the starting and the ending stages of
a common evolutionary path, where a neutron star accretes matter (and
angular momentum) from a companion (e.g., \citealp{alpar82,batta91}).  The early phases of this evolutionary
path are characterized by active mass accretion accompanied by intense
X-ray emission (larger than $\sim 10^{35}$ erg s$^{-1}$). These
systems are observed as LMXBs characterized by a few outburst in
  the X-ray due to accretion disk instabilities.  These objects are
usually called ``X-ray transients'' \citep{white84} and are commonly found both in globular clusters and in the Galactic field. \\

X-ray transient outbursts are easily and promptly detectable in the X-ray energy bands thanks to the constant sky monitoring of X-ray satellites such as INTEGRAL and Swift. Indeed, on March 13, 2015, Swift/BAT observations detected an X-ray burst in
Terzan 5 \citep{altamirano15}.  The Swift/XRT observations promptly
following the Swift/BAT detection localized the transient source at
RA(J2000)$=267.0207\deg$, DEC(J2000)$=-24.779\deg$, with a 90\%
uncertainty of $3.5\arcsec$ \citep{bahramian15}.  The
measured spectrum turned out to be consistent with a relatively hard
photon index of $1.0 \pm 0.2$ and a hydrogen column density $N_{\mathrm H}
=(4\pm 0.8)\times 10^{22}$ cm$^{-2}$. The latter is larger than the
typical value measured in Terzan 5 \citep{bahramian14} and well in
agreement with the hydrogen column density of the previously known
transient EXO~1745-248 \citep{kuulkers03}.  Indeed, the subsequent
position refinement by \citet{linares15} centered the system around
EXO~1745-248 with a $2.2\arcsec$ error circle. These data therefore
strongly suggest that the new Swift/BAT outburst coincides with EXO
1745-248, an X-ray neutron star transient that already showed
outbursts in 2000 and 2011 \citep{degenaar12}.  Such an identification
has been also confirmed by radio VLA observations \citep{tremou15},
which locate the source position within $0.4\arcsec$ of the published
coordinates of EXO 1745-248 obtained from \emph{Chandra} data (source
CX3 in \citealp{heinke06}).  The most recent Swift/XRT observations
indicate that the source is probably on the way to transit to a soft
state \citep{yan15}.\\

Such an intriguing object is not uncommon in Terzan 5. In fact, as already discussed in Chapter~\ref{CapIntro} and \ref{cap_t5}, this
stellar system is known to harbor several X-ray sources (see, e.g.,
\citealp{heinke06}) and to be the most efficient furnace of MSPs in
the Milky Way \citep{ransom05a, cadelano18}.
\citet{ferraro09} recently demonstrated that, at odds with what is
commonly thought, Terzan 5 is not a globular cluster, but a system
hosting stellar populations characterized by significantly different
iron abundances, spanning a total metallicity range of 1 dex (see also
\citealp{origlia11, origlia13, massari14}) and a very large collision rate
\citep{lanzoni10}, the largest among all Galactic globular clusters
\citep[see also][]{verbunt87}, which can explain the production of the large
population of MSPs and low-mass X-ray binaries (LMXBs) now observed in the system.

Since  the X-ray outburst detected by SWIFT is expected to also
produce a significant enhancement of the optical luminosity
  \citep[see][]{shahbaz98,charles06,testa12, pallanca13}, we
successfully applied for HST Director Discretionary Time to urgently
survey the central region of Terzan 5 and thus provide new insights
into this still unexplored phase of the LMXB-to-MSP path. Here we
report on the identification of the optical counterpart to EXO
1745-248 obtained from the analysis of these images.

\section{Observations and Data Reduction}
\label{obs}
To search for the expected optical emission from EXO 1745-248 during
its X-ray bursting phase, we submitted a HST Director Discretionary
Time proposal (GO 14061, PI: Ferraro) asking for two orbits with the
Advanced Camera for survey (ACS/WFC).  The observations have been
promptly performed on April 20, 2015, about one month  into the
  X-ray outburst (continuing at time of writing).  The dataset
(hereafter Epoch 3, EP3) consists of $5\times398$ s images in F606W,
$5\times371$ s images in F814W, and one short exposure per filter
($50$ s and $10$ s, respectively; these latter have not been used in
the present work).

Previous optical images of Terzan 5 acquired with the same instrument
in the same filters were already present in the HST Archive: GO 12933
(PI: Ferraro) performed on August 18th, 2013 (hereafter EP2), and
GO 9799 (PI: Rich) performed on September 9th, 2003 (hereafter EP1).
We already used these data to construct the deepest optical
color-magnitude diagram (CMD) of Terzan 5 (see \citealp{ferraro09, lanzoni10,
  massari12}).

For the present study, all the datasets have been homogeneously
analyzed, following the prescription reported in Chapter~\ref{opticalcom} on the (flc) images corrected for Charge Transfer
Efficiency. For each image we modeled a point spread
function (PSF) by using 150-200 bright and nearly isolated stars.
Afterwards, we performed source detection in each  image
imposing a 3-$\sigma$ threshold over the background level.
By using all the sources detected and PSF fitted in at least 1 out of
2 images in EP1, and 3 out of 10 images in EP2 and EP3, we then
created a catalog for every epoch.  In spite of including sources
detected in only one filter, such an approach allowed us to avoid
losing very faint (but possibly real) objects, while safely discarding
spurious detections, as cosmic rays and detector artifacts.  The
obtained master lists (one for every epoch) have been then used to
identify the stellar sources in each single frame and the PSF model
has been applied to derive the final magnitudes. As a final step, to
build the EP2 and EP3 catalogs we considered all the stars with
magnitude measured in both filters in at least 3 out of 5 images,
while the EP1 catalog obviously consists of the objects detected in
both the available images.  For each star, the magnitudes estimated in
different images of the same filter have been homogenized
\citep[see][]{ferraro92} and their weighted mean and standard deviation
have been finally adopted as the star magnitude and photometric
error.
The magnitude calibration to the VEGAMAG system has been performed by
using the catalog by \citet{massari12} as reference.

To precisely determine the star coordinates, we first applied the
equations reported by \citet{meurer03} and corrected the instrumental
positions for the known geometric distortions affecting the ACS
images. Through cross-correlation with the catalog of
\citet{massari12}, which had been placed onto the 2MASS system, we
then obtained the absolute coordinates for each star, with a final
astrometric accuracy of $\sim 0.2\arcsec$ in both right ascension and
declination.

\begin{figure}[!htb]
\centering%
\includegraphics[scale=0.7]{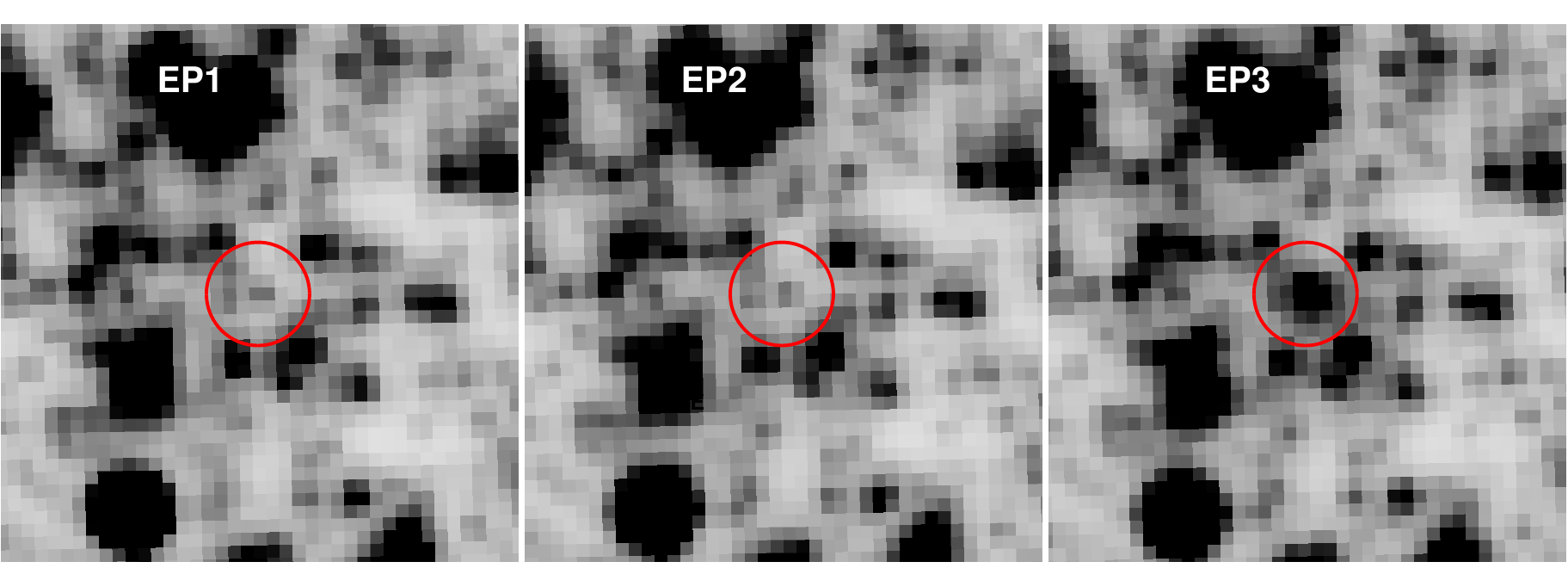}
\caption[Images of the region around EXO 1745-248.]{HST/ACS drz combined images of the $2\arcsec\times 2\arcsec$
  region around EXO 1745-248, in the F814W filter, for the three
  epochs (EP1, EP2, EP3, from left to right, respectively). The source
  (highlighted with a red circle) is visible as a faint star during
  the quiescence epochs EP1 and EP2, while it is observed in an
  outburst stage during EP3. North is up, east is to the left.}
\label{ident1}
\end{figure}

\begin{figure}[!htb]
\centering%
\includegraphics[scale=0.6]{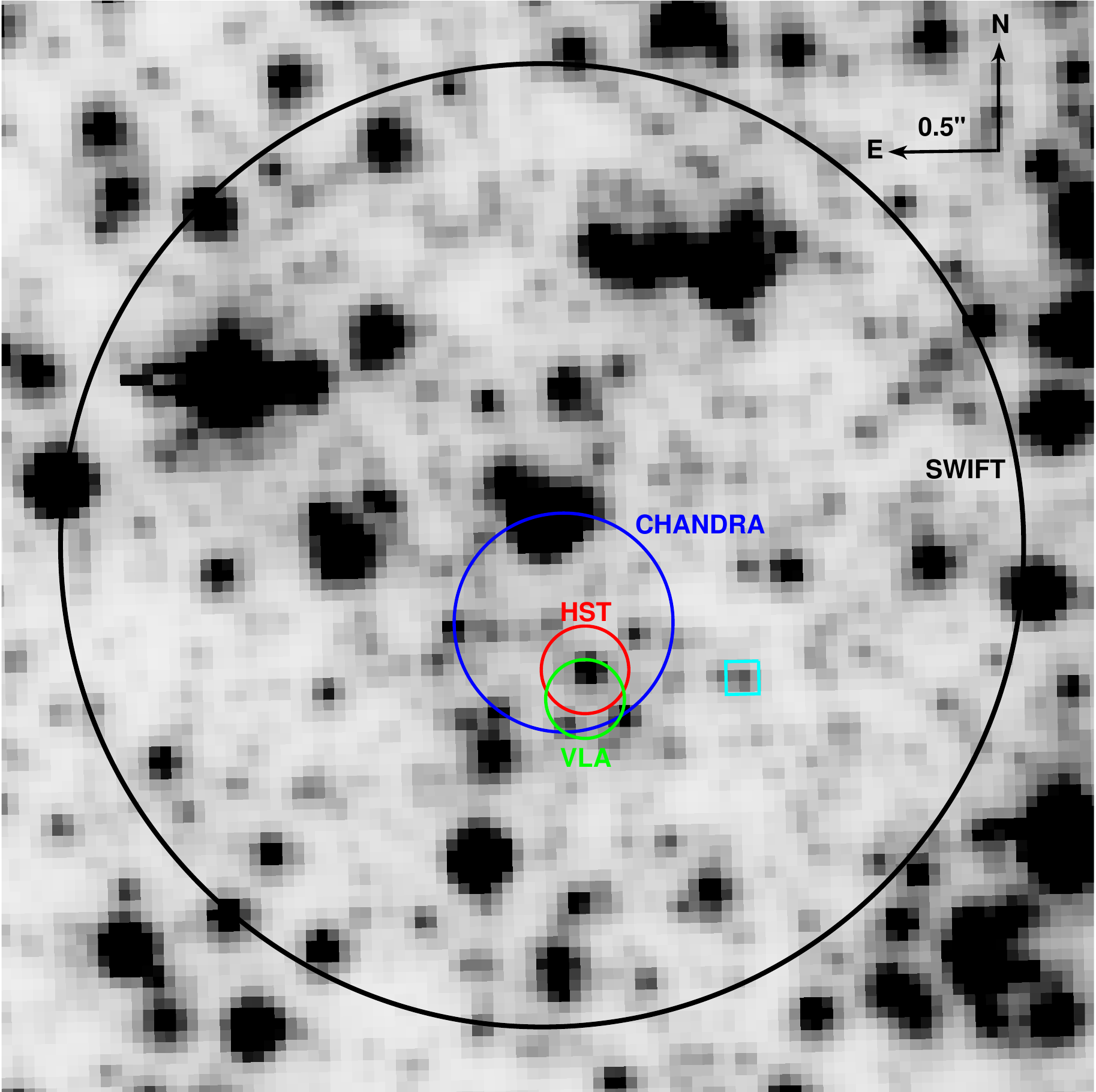}
\caption[F814W-band drz combined image of the region around EXO 1745-248.]{F814W-band drz combined image of the $5\arcsec\times
  5\arcsec$ region around EXO 1745-248 in the EP3 exposure. The source
  positions and uncertainties obtained from the various observational
  campaigns are marked: the Swift/XRT $2.2\arcsec$ radius error circle
  is shown in black, the Chandra error circle in blue, the VLA measure
  in green, and the HST optical determination in red. The cyan square
  marks the star previously proposed \citep{heinke03} as the possible
  optical counterpart to EXO 1745-248.}
\label{ident2}
\end{figure}

\begin{figure}[!htb]
\centering%
\includegraphics[scale=0.4]{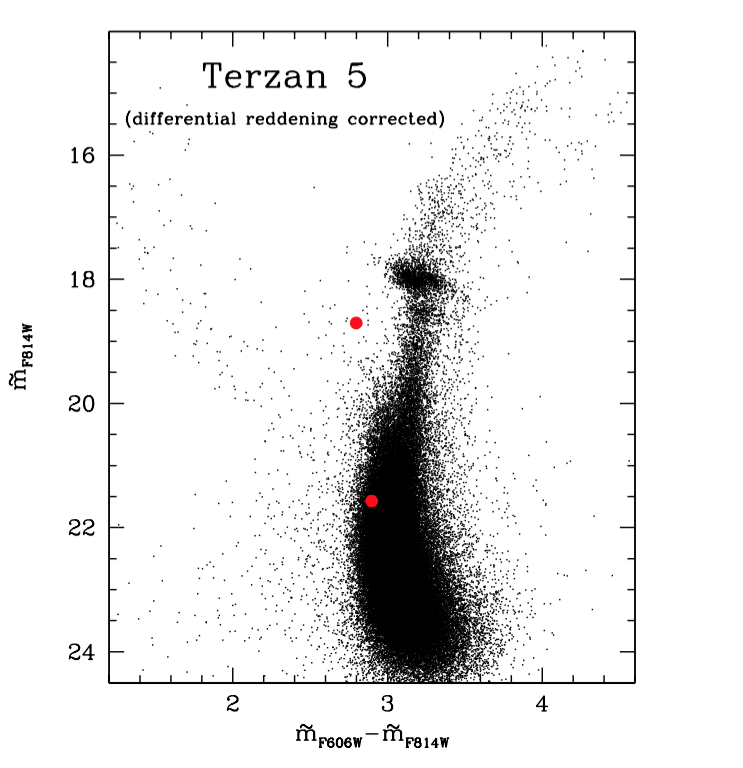}
\caption[CMD of Terzan 5 corrected for differential reddening.]{($m_{ F814W}$, $m_{ F606W}-m_{ F814W}$) CMD of
  Terzan 5 corrected for differential reddening (according to
  \citealp{massari12}).  The position of the optical counterpart to
  EXO 1745-248, in the outburst and in quiescence states, is marked
  with large red circles.}
\label{cmd1}
\end{figure}

\begin{figure}[!htb]
\centering%
\includegraphics[scale=0.6]{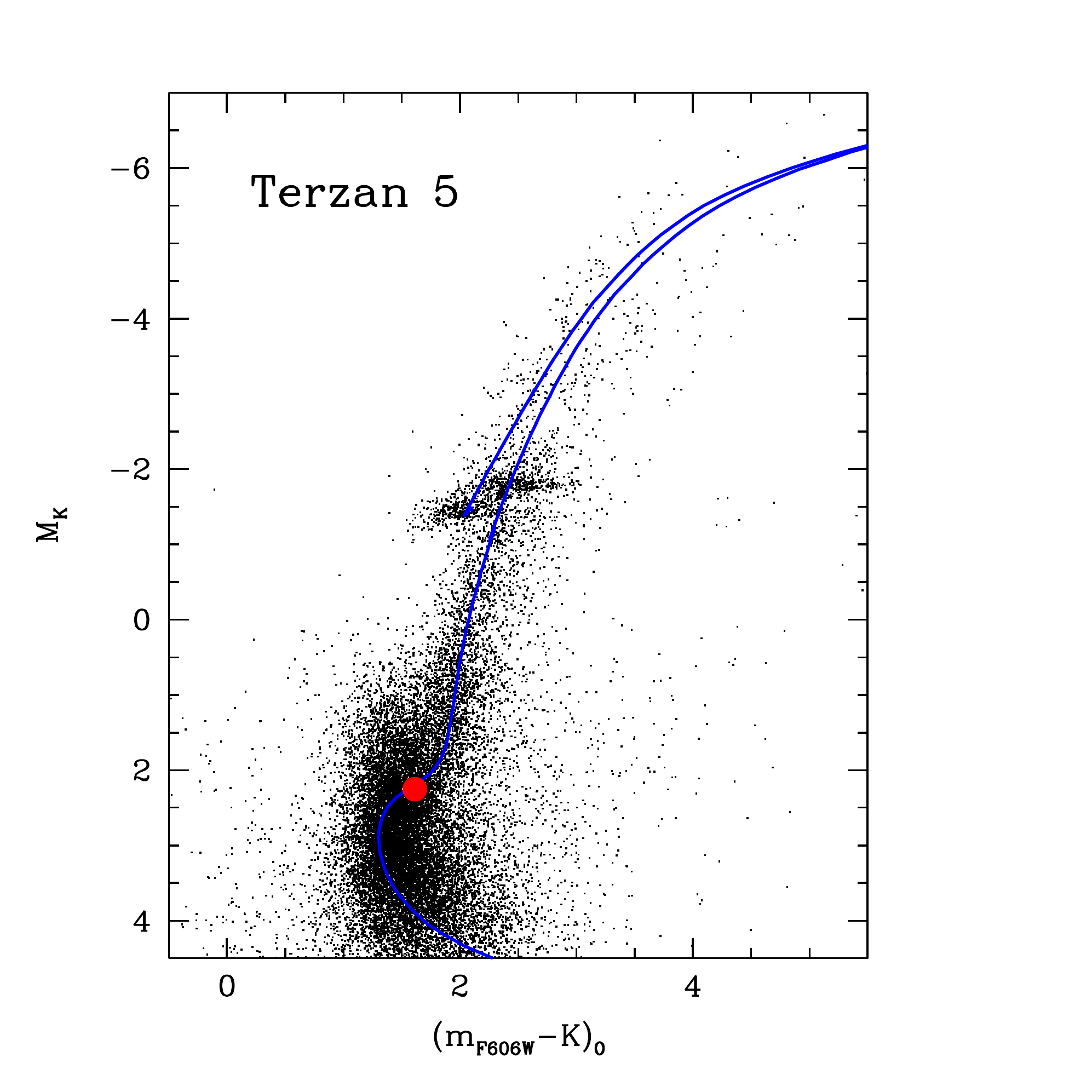}
\caption[CMD of Terzan 5 obtained from HST/ACS and ESO/MAD observations.]{Absolute $(M_K$, $m_{ F606W}- K)_0$ CMD of Terzan 5
  obtained from a combination of HST/ACS and ESO/MAD observations. The
  position of COM-EXO 1745-248 in the quiescent state is marked with
  the large red circle. The blue line corresponds to a 12 Gyr
  isochrone with [Fe/H]$=-0.3$ (from \citealp{girardi10}), well
  reproducing the main metal poor sub-population of Terzan 5.
}
\label{cmdmad}
\end{figure}
 
\section{Results}
\label{results}
The photometric analysis of our dataset in a region around the
position of EXO 1745-248 immediately revealed, in EP3, the presence of
 a bright star that was not visible in EP1 and EP2 images (see
  Figure \ref{ident1}).  The comparison of the three epochs
  unequivocally identifies the bright object (hereafter COM-EXO
1745-248) as the optical counterpart to EXO 1745-248.  The absolute
position of the optical source is $\alpha=17^h 48^m 05.23^s$,
$\delta=-24^o 46\arcmin 47.6\arcsec$.  This is consistent at
1-$\sigma$ with the VLA position quoted by \citet[][see the red and
  the green circles in Figure \ref{ident2}]{tremou15}.  Instead, the
star previously suggested as the possible optical counterpart to this
X-ray transient \citep{heinke03} is located $\sim 0.7\arcsec$ to the
west (cyan square in the figure).

 After the astrometric transformations, the centroid position of
  the bright object in EP3 is within 0.05 pixels from the centroid of
  a fainter star clearly detected in EP1 and EP2 (see Figure
  \ref{cmd1}). This could be the optical counterpart caught in
  quiescence. The probability that the true counterpart is a fainter,
  non detected star aligned (within 0.05 pixels) along the line of
  sight is very low ($P \sim 0.4\%$).\footnote{To estimate the
    probability of a chance superposition with a star fainter than the
    proposed counterpart, the number of stars down to 5 magnitudes
    below the Turn Off level at the same distance ($\sim 5\arcsec$)
    from the cluster center is needed. Since no data-set available for
    Terzan 5 reaches such a faint magnitude limit, we adopted as
    reference the luminosity function of 47 Tucanae derived from deep
    HST observations \citep{sarajedini07}. Star counts have been
    normalized to the number of Terzan 5 stars counted between the
    Turn Off level and two magnitudes above, in a ring of $2\arcsec$
    width, centered at $5\arcsec$ from the center.} In addition, the
  brightness profile of this star does not show any significant
  deviation from symmetry, thus supporting the hypothesis that it is a
  single object. Hence, the most natural conclusion is that identified
  star is indeed the counterpart in quiescence.  The identified
object passed from observed magnitudes $m_{ F606W}=24.74$ and
$m_{ F814W}=21.74$ during quiescence (EP1 and
EP2)\footnote{The magnitudes of the star in the EP1 ($m_{\mathrm
      F606W}=24.7\pm0.1$; $m_{ F814W}=21.6\pm0.1$) and EP2 ($m_{
      F606W}=24.74\pm0.04$; $m_{\mathrm F814W}=21.7\pm0.1$) quiescent
    stages are fully consistent within the errors.}, to $m_{
  F606W}=21.77$ and $m_{ F814W}=18.88$ in the outburst state (EP3),
thus experiencing a brightening of 3 magnitudes (corresponding to a
factor 16 in luminosity).  Because of its location in the inner
Galactic bulge, Terzan 5 is affected by a large extinction, with an
average color excess $E(B-V)=2.38$ \citep{valenti07},
showing strong variations, up to $\delta E(B-V)=0.67$ mag, within the
ACS field of view \citep{massari12}. We therefore applied the
high-resolution differential reddening map obtained by
\citet{massari12} to correct the observed magnitudes  (in the
  following, the notation $\widetilde { m}$ indicates magnitudes
  corrected for differential reddening). Figure \ref{cmd1} shows the
position of COM-EXO 1745-248 in the differential reddening corrected
CMD during the two states.  We found $\widetilde { m}_{
  F606W}=24.47$ and $\widetilde { m}_{ F814W}=21.57$ during EP1
and EP2, while $\widetilde { m}_{ F606W}=21.50$ and $\widetilde
{ m}_{ F814W}=18.70$ in EP3,  corresponding to a small (0.1
  mag) color variation, which is however within the errors.  No
variability has been detected over the period of $\sim 50$ min covered
by each HST orbit in EP3 and EP2. It is worth mentioning that EP3 data
were acquired on April 20, 2015, almost simultaneously to the X-ray
observations \citep{yan15} suggesting that the system is transiting
from a hard to a soft state.\\

An estimate of the orbital period of the system can be obtained
  by following \citep{shahbaz98}, who reporte a relation between the
  orbital period and the $V$-band luminosity variation. Since we
  observe $\Delta V \sim 3$ mag in the case of EXO 1745-248, the
  orbital period turns out to be $P\sim 1.3$ days. On the other hand,
  for LMXBs \citet{vanpa94} proposed an empirical relation between the
  absolute $V$ magnitude in outburst and the parameter $\Sigma$, which
  depends on the ratio between the X-ray and the Eddington
  luminosities ($L_{\mathrm X}/L_{\mathrm Edd}$) and the orbital period.  By
  assuming $L_{\mathrm X}/L_{\mathrm Edd}\sim 0.5$ \citep{yan15} and $M_V
  =1.37$ (in Johnson $V$ magnitude) for EXO 1745-248, we obtain
  $P\sim0.1$ days. From these {estimates}, the orbital period of the
  system {is likely to be} between 1.3 and 0.1 days.\\

In order to more deeply investigate the nature of COM-EXO 1745-248 in
the quiescence state under the assumption that the disk
  contribution to the observed magnitude is negligible, we identified
the star in the K-band adaptive optics images obtained with ESO/MAD,
used by \citet{ferraro09} to discover the two main multi-iron populations
hidden in this system.  We have first corrected the combined $(K,
m_{ F606W}-K$) CMD for differential reddening. Then we transformed
it into the absolute plane by assuming the average color excess quoted
above, and the distance modulus $(m-M)_0=13.87$ corresponding to a
distance of 5.9 kpc \citep{valenti07}.  The result is shown in Figure
\ref{cmdmad}, where the position of COM-EXO 1745-248 in the quiescent
state is marked.  A more detailed characterization of the nature
  of COM-EXO 1745-248 is strongly hampered by the complexity of the
  stellar populations harbored in Terzan5. The comparison with a 12
  Gyr old isochrone \citep{girardi10} well reproducing the main
  metal-poor sub-population of Terzan 5, at [Fe/H]$=-0.3$
  dex\footnote{As discussed in \citet{massari14}, this population
    consists of $\sim 62\%$ of the total, while a super-solar
    component at [Fe/H]$=+0.3$ dex accounts for $\sim 29\%$, and an
    even metal poorer component, at [Fe/H]$=-0.8$ dex, recently
    detected by \citet{origlia13}, corresponds to $\sim 5\%$ of the
    total. The CMD plotted in Fig. \ref{cmdmad} nicely shows two
    distinct red clumps at $M_K=-1.5$ and $M_K=-1.8$1, corresponding
    to the two major sub-populations first discovered in the system by
    \citet{ferraro09}.} suggests that COM-EXO 1745-248 could be a sub-giant
  branch star. On the other hand the metal-rich sub-population
  could be significantly (a few Gyr) younger than the main metal poor
  component \citep[see][]{ferraro09}. Thus, if COM-EXO 1745-248 belongs to
  the metal-rich component, it would be located below the sub-giant branch, in a
  position where companions to redback MSPs have been found (see,
  e.g., the case of COM-6397A in \citealp{ferraro01a}). Since no
  spectroscopic information on the metallicity of this star is
  available, both possibilities are equivalently valid. While in the
  case of redbacks any prediction on the stellar parameters based on
  the observed photometric properties can be difficult (see the case
  of COM-6397A), this is possible for a sub-giant branch star belonging to the
  metal poor population. In {this case}, the following stellar parameters are
  obtained: mass $M=0.9 \; \Msun$, effective temperature $T_{ eff}=
5440$ K, surface gravity $\log g=3.9$, and luminosity $\log
L/L_\odot=0.35$. The corresponding stellar radius therefore is
$R\sim1.7 \; \Rsun$. Hence, by assuming that the star has completely
filled its Roche-Lobe and adopting a canonical value for the neutron
star mass ($\sim 1.4 \; \Msun$), we derive an orbital separation $a
\approx 5.2 \; \Rsun$ and a period $P_{ orb}\sim 0.9$ d for the
binary system, fully in agreement with the range estimated above.
We estimate that the radial velocity
variations of such a binary system should have an amplitude of $\sim
170 \; \sin i$  km  s$^{-1}$ ($i$ being the system inclination angle),
which could be detectable through a dedicated spectroscopic follow-up.

\chapter*{Summary and Conclusions}
\addcontentsline{toc}{chapter}{Summary and Conclusions} 
\sectionmark{Summary and Conclusions}
\chaptermark{Summary and Conclusions}
\markboth{Summary and Conclusions}{Summary and Conclusions}

\initial{I}n this thesis work we presented the results obtained from the analysis of multi-wavelength observations of millisecond pulsars (MSPs) in different Galactic globular clusters (GCs). Observations obtained at radio, optical and near-UV wavelengths have been exploited to search for, timing and identify the optical counterparts to these exotic systems. In this chapter we summarize the main results obtained and discuss future perspectives.\\

\subsection*{Radio Observations}

In the first part of the thesis (Chapter~\ref{cap_t5}), we presented the analysis of radio observations obtained with the 100-m Robert Bird Green Bank Telescope. 

In Chapter~\ref{cap_t5} we discussed how to use archival observations to search for new pulsars (PSRs) and we applied this method to the case of the stellar system
Terzan 5. Instead of using classical search routines based on the analysis of
single observations, we developed an alternative method which combines multiple observations. In this method, 
 stacked power spectra at different dispersion measures (DMs) are created by summing the spectral
powers obtained in the single observations.  In order to remove radio frequency interference (RFI), a control DM spectrum has been subtracted from all the stacked power spectra.
All the candidates selected in these ``corrected'' stacked power spectra have been
processed with a KD-Tree algorithm, in order to select those that are likely
PSR candidates and discard those that are more likely persistent RFI. 

The application of the method to Terzan 5
 led us to discover three new MSPs in this stellar system.
For two of them, we have been able to obtain a phase connected timing solution that confirmed their
association with the GC, having a DM close to the cluster mean value and being
located within $\sim17\arcsec$ ($\sim1.9$ core radii) from the cluster gravitational center. 
These discoveries  bring the total
number of known MSPs in this system to 37, which corresponds to $\sim 25\%$ of the entire
PSR population identified so far in GCs.  

Indeed Terzan~5 turns out to be the most efficient factory of MSPs in
the Milky Way and the large number of  X-ray sources (see, e.g.,
\citealp{heinke06}) and  X-ray bursters (see the recent case of 
EXO 1745$-$248, \citealt{altamirano15,ferraro15}) suggests that this furnace is
currently very active. \citet{ferraro09} first pointed out that, at odds with what is
commonly thought, Terzan 5 probably is not a genuine GC, 
but a much more complex stellar system, since it hosts different stellar 
populations characterized by significantly different
iron abundances (see also \citealp{origlia11, origlia13, massari14}). 
Recently \citet{ferraro16} measured the ages of the two main 
sub-populations, (finding 12 and 4.5 Gyr for the sub-solar and super-solar metallicity
component, respectively), thus identifying  Terzan 5 as a site of
multiple bursts of star formation in the Galactic bulge.
Indeed, the measured chemical patterns
and the large age difference between 
the two main sub-populations could be naturally explained 
in a self-enrichment scenario where
Terzan 5 was originally much more massive ($\sim 10^8 \; 
\Msun$) than today ($\sim 10^6 \; 
\Msun$; \citealp{lanzoni10}), and therefore able to retain the
iron-enriched gas ejected by violent supernova explosions from which the second generation of stars formed. In particular, the chemical patterns measured in Terzan~5 require a large
number of type II supernovae. In turns, these should have also produced a large population of
neutron stars (NSs), mostly retained into the deep potential well of the
massive {\it proto}-Terzan 5 and likely forming binary systems through
tidal capture interactions.  This, together with its large collision rate (the largest among all Galactic GCs; \citealt{lanzoni10}) could have highly promoted PSR
recycling processes, which can naturally explain the large
population of MSPs and other stellar exotica now observed in the system. 

Population synthesis models predict that Terzan~5 could host from few hundreds up to one thousand MSPs \citep[e.g.,][]{bagchi11}, making this system a top priority target for future deep radio searches with the new generation of radio telescopes, which is expected to plausibly double to triple the current MSP population \citep{hessels15}. The relative simplicity and effectiveness of the method here proposed can be applied to observations obtained with the new generation of instruments. Together with the classical routine
searches on single observations, it could be a complementary method to discover
extremely faint PSRs by using the large amount of GC data that the new generation
of radio telescopes (such as ``MeerKAT'') is going to produce.

While the new generation of radio telescope is still under development, this method can be applied to the currently available archival data of other clusters. Indeed, being quite a powerful tool to search for very faint
PSRs, it opens the possibility to identify a significant
number of still unknown objects by simply using the large amount of observations that have been
devoted to GCs in the past decades. Special attention should be reserved to those clusters expected to harbor a rich population of exotic systems like isolated and very long period binaries, typical of dynamically evolved cluster such as NGC 6752, NGC 6624 and NGC 6522. Furthermore, this method can be improved by implementing techniques such as the ``phase-modulation search'' (see \ref{search}). This would make this method sensitive also to MSPs in very compact binaries.\\

\subsection*{Optical Observations}
The second part of the thesis was focused on the identification and characterization of optical counterparts to MSPs at different evolutionary stages. Six new companion have been identified.

In Chapter~\ref{cap_m71} we presented the identification of the companion to the black-widow (BW) PSR J1953+1846A (M71A) in the GC M71. Taking advantage of the precise measure of the PSR position from the radio timing analysis, we have used a set of high resolution ACS/HST images to search for its companion star in the optical bands. We identified a faint and strongly variable star (COM-M71A), showing a modulation of at least three magnitudes in both the filters used (F606W and F814W). In the color-magnitude diagram (CMD), COM-M71A lies in the region between the cluster main sequence and the WD cooling sequences, thus suggesting that it is a low-mass, non-degenerate or at least semi-degenerate star, with a temperature of about 5100~K. Unfortunately, because of its faintness, it
was detectable only in 16 out of 27 images, mostly during the PSR
inferior conjunction. The light curve shows a sinusoidal shape with a period fully
consistent with that of the binary MSP. The maximum, during the PSR inferior conjunction, and
the minimum, during the PSR superior conjunction, suggest a strong heating of the
companion star side exposed to the PSR flux. Such a behavior is comparable
with that observed for similar objects in the Galactic field. By modeling the light
curve, we showed that the companion reprocessing efficiency of the PSR energy is $\sim5\%$ 
for a Roche-Lobe filling companion, while a typical value of $15\%$ is found by assuming a filling factor of 0.6. The comparison between the optical and X-ray light curves suggests the possible presence of an
intra-binary shock, similar to that observed for the redback (RB) 47TucW. A
X-ray and optical follow-up will highlight the presence of this shock and, possibly,
will allow to characterize its property and structure. Unfortunately, the star
is too faint to allow a spectroscopic follow-up with the available instruments.
However, an optimized photometric follow-up would provide the opportunity to better
constrain the system properties, and by using, for example, phase-resolved
observations with a narrow ${H\alpha}$ filter we could constrain the presence of
ionized material, possibly related to the intra-binary shock.\\ 
COM-M71A is, so far, the second BW companion identified in a GC, after COM-M5C in M5 \citep{pallanca14a}. Interestingly, both the light curve shape and the position in the CMD are quite similar in the two systems. This suggests that probably the two objects underwent a similar evolutionary path. Even though the statistic is by far too limited to draw any solid conclusion, at the moment no significant differences between BWs in GCs and those in the Galactic field are observed, both in the radio and in the optical band. This, together with the increasing number of eclipsing MSPs identified in the Galactic field \citep[e.g.][Sanpa-arsa et al. 2018, in preparation]{breton13,li14}, confirms that no dynamical interactions are strictly needed for forming these systems, at odds with what usually believed in the past \citep[e.g.][]{king03}. \\

In Chapter~\ref{cap_47tuc} we showed that, thanks to ultra-deep, high resolution near-UV WFC3/HST observations, we identified the companions to four binary MSPs in 47~Tucanae (47TucQ,
47TucS, 47TucT and 47TucY) and confirmed the two already known objects
(COM-47TucU and COM-47TucW). The optical counterparts have coordinates
compatible, within the errors, with the PSR nominal positions. In the
CMD, all the objects are located along the He WD cooling sequences, as
expected from the MSP canonical evolutionary scenario. The only
exception is the companion to the RB system 47TucW, which is
located in an anomalous region between the main sequence and the WD
cooling sequences, suggesting that it is a low-mass main sequence star
highly perturbed and heated by the PSR flux. We compared the observed
CMD positions of the detected He WD companions with a set of cooling
tracks and derived the companion main properties (as masses, cooling
ages, temperatures) and also some constraints on the PSR masses. All
the companion stars have masses between $\sim 0.15 \ \Msun$ and
$\sim 0.20 \ \Msun$, and all the derived cooling ages are
  smaller than the cluster stellar population age ($\sim10-11$ Gyr). The orbital
periods vs companion masses are in fair agreement with the
evolutionary models of \citet{tauris99} and \citet{istrate14}.  By
combining the cooling age with the PSR spin down rate we found that
the accretion history of 47TucU likely proceeded at a sub-Eddington
rate.\\


In Chapter 7 we presented the optical identification and characterization of the AMXP \sax during quiescence. We identified a possible counterpart (\com) in a star located at only $\sim0.15\arcsec$ from the X-ray nominal position. This star, although being located along the cluster main sequence, shows an excess in the F656N filter, thus implying the presence of ${H\alpha}$ emission.  {\  We discussed the physical properties of the companion star and showed that it has a mass of $0.70\; \Msun - 0.83\; \Msun$, an effective temperature of 5250 K and it is filling, or even overflowing, its Roche-Lobe radius of $0.88\pm0.02 \; \Rsun$. This mass, combined with the binary system mass function and assuming a canonical range of NS masses, implies that the binary system is observed at a very low inclination angle ($\sim8^{\circ}-14^{\circ}$)}. This can also explain the absence of any significant magnitude variability and also the absence of dips and eclipses in the X-ray light curve. The EW of the ${H\alpha}$ emission has been evaluated to be of about $\mathrm{20 \; \AA}$, which corresponds to a mean mass transfer rate during quiescence of $\mathrm{\sim10^{-10} \, \, M_{\odot} \, \, yr^{-1}}$. {\ The possibility of on-going mass transfer and residual accretion disk around the NS during quiescence implies that the radio PSR is not reactivated yet. Hence \sax is probably not a tMSP and its behavior during quiescence is comparable with that commonly observed in classical quiescent low-mass X-ray binaries, even though its orbital and spin parameters are very similar to those observed for RBs. This directly implies that not all the RB-like AMXP with main sequence companions are tMSPs. For some reasons, the PSR in \sax is not able to turn-on the radio emission during the quiescence state, at odds with what happens for tMSPs. The reasons behind this are still obscure. Intriguingly, it is worth noticing that \com has a mass larger than that measured for the companion stars of RBs and tMSPs \citep[$0.2\; \Msun-0.4\; \Msun$, see, e.g.,][]{breton13,mucciarelli13,bellm16}. Therefore the companion mass could be one of the ingredients to understand why this RB-like AMXP is not behaving like a tMSP.} \\

In Chapter 8 we presented the identification of the optical counterpart to the
NS transient EXO 1745-248 in Terzan 5. This has been performed by exploiting a data-set of high resolution images obtained during three different epochs. During the third epoch, this object is detected in an outburst state and shows a brightening of $\sim
3$ magnitudes.  In the quiescence state it is a sub-giant branch star,
i.e., an object that is experiencing its first envelope expansion,
while evolving toward the red giant branch stage. The analysis of this source revealed that the companion star has a mass of about $0.9 \; \Msun$ and effective temperature of about 5500 K, while the binary orbital period has been constrained to be in the range 0.1- 1.3 days. Very interestingly, the X-ray emission of EXO 1745-248 during
quiescence was found to be highly variable both on short and long
time-scales \citep{degenaar12}, and \citet{linares14} recently
underlined that these properties are similar to those
observed for  the tMSP J1824$-$2452I \citep{papitto13}.
This evidence suggests that EXO~1745-248 could be another system
  belonging the rare class of objects caught shortly before the
  formation of a radio MSP, possibly in a stage preceding
  the swinging phase in which tMSPs are observed.  \\

\begin{figure}
\centering
\includegraphics[width=10cm]{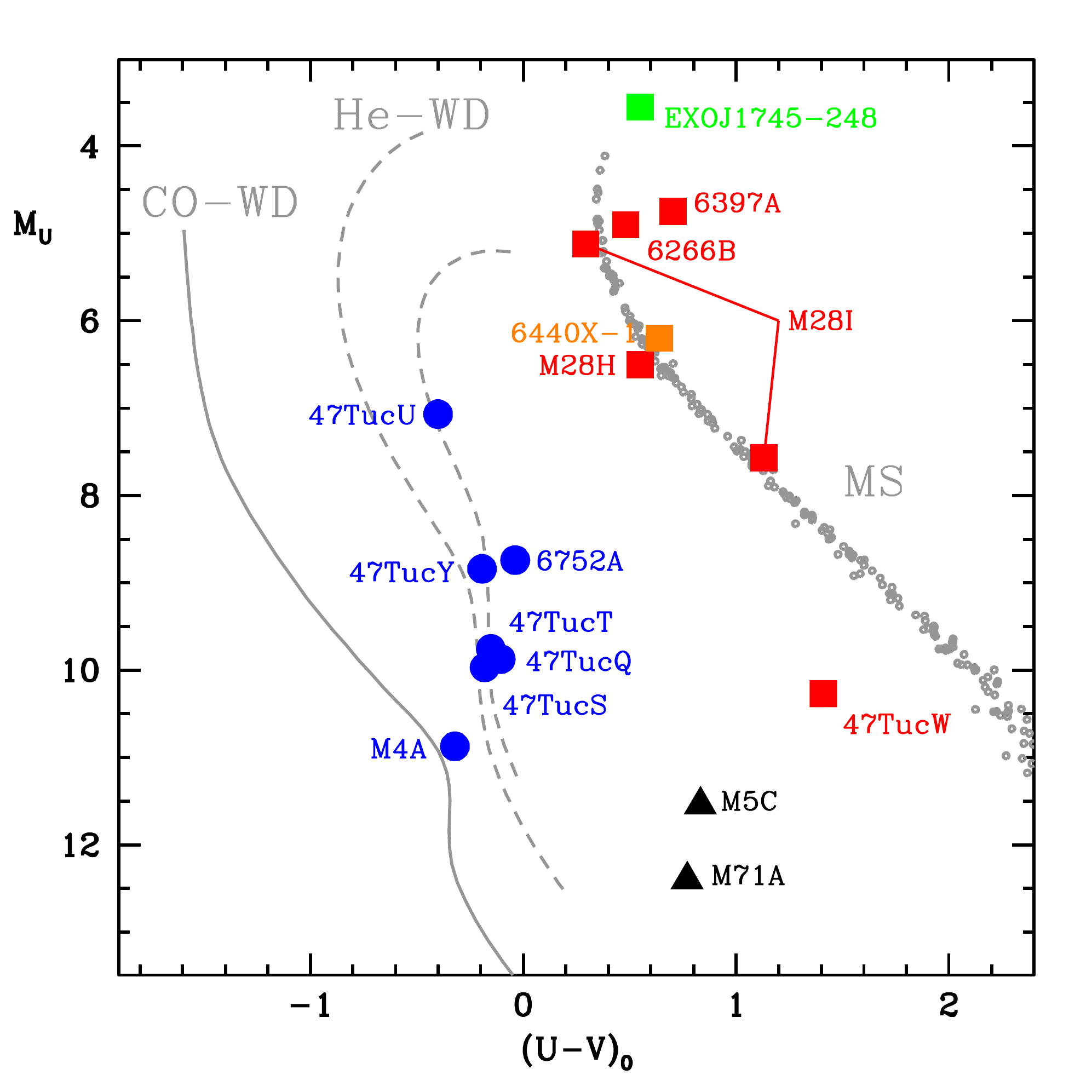}
\caption[Position of the MSP optical counterparts in an absolute CMD after this thesis work.]{Position of the MSP optical counterparts in an absolute CMD. This plot is the same as in Figure~\ref{figcom1} but includes all the companion stars identified in this thesis work. Blue dots mark  the position of  canonical MSPs, while red squares and black triangles the position of RBs and BWs, respectively. The AMXP SAX J1748.9$-$2021 is in orange and labelled as 6440X-1 for clarity, while the companion to EXO J1745$-$248 is marked with a green square.}
\label{figcom_post}
\end{figure}

Summarizing, in this thesis work we have identified six new optical counterparts to MSPs in GCs. This corresponds to have increased the total sample of such objects by $40\%$, reaching the number of 15. Moreover we have identified the optical counterpart to a X-ray burster in Terzan~5. It is now possible to clarify the evolutionary mechanism of MSPs, identifying different possible paths. Figure~\ref{figcom_post} reports the CMD position of all the companions known after this thesis work (see Figure~\ref{figcom1} for the state of the art before this thesis). The first inspection of this updated plot clearly reveals that different typologies of companion stars correspond to different typologies of MSPs. In other words, the classification of the companion stars alone appear to be sufficient to classify the binary MSPs. Even not accounting for the radio properties of the MSPs, the positions of the companions in an absolute CMD allow a clear classification of the typology of each binary. Indeed, as can be seen, canonical MSPs have exclusively low-mass He WD companion stars, the vast majority with masses $\lesssim0.2 \; \Msun$. Only one system, M4A, has a companion with a slightly larger mass of $0.35\; \Msun$. As shown in Figure~\ref{mp}, this mass range is exactly what expected from binary evolution models \citep{tauris99,istrate14}. This suggests that the cluster environment does not significantly affect the evolution of canonical MSPs. While it is well known and assessed that the cluster environment is able to create a large number of new MSPs, their general properties appear to remain unaltered during the typical timescale of their formation and evolution ($\sim 1$~Gyr). Signatures of environmental effects can be found in those systems with eccentric orbits, mildly recycled PSRs or very long orbital periods, which have been, so far, almost exclusively found in the densest GCs \citep{verbunt14}. Unfortunately, no optical counterpart to these ``perturbed'' systems has been identified so far, and therefore it has been impossible to compare the properties of their companion stars with those of standard canonical low-mass He WDs.

At odds with canonical MSP companions, both the spider MSP companions (COM-M71A and COM-47TucW) are, as expected, non degenerate objects. The companions to RB systems are all closely located along the cluster main sequence. The vast majority of them are packed in the upper main sequence, few magnitudes below the turn-off point. While this could be a selection effect, it is interesting to note that the only exception is the companion to 47TucW, located on the blue side of the main sequence, in a low-luminosity region $\sim7$ magnitudes below the turn-off. BW systems have, on the other hand, very peculiar companion stars located in a very low-luminosity region, between the main sequence and the WD cooling sequence. This is a region where no standard evolving stars are expected, thus confirming the high degree of perturbation that these systems suffer during their evolution. 

Aside from the position in the CMD, another key difference between canonical and spider MSP companions is the presence of flux variability, usually showing a periodicity compatible with the orbital period of the binary. This is usually not observed and not expected in canonical MSPs, since the PSR flux intercepted by their degenerate (thus very compact) companions is negligible (but see the case discussed in \citealt{edmonds01}). Instead, the flux variability of spider MSPs comes in two flavors: RB systems show mildly variable light curves, with typical magnitude differences of few tenths and with a double minimum and double maximum structure that is due to the tidal distortions of the companion star caused by the NS \citep{ferraro01a,cocozza06,pallanca10,breton13,li14}. Heating effects are sometimes observed but do not dominate the light curve shapes. On the other hand, all the BW systems and some RB systems (such as 47TucW, \citealt{edmonds02}) present light curves with strong variability, usually reaching magnitude differences larger than 3, and with a single minimum and a single maximum structure, which is the result of the heating of the stellar side exposed to the PSR injected flux \citep{pallanca12,pallanca14a,li14,cadelano15a}. This effect dominates the light curves, while other effects, such as tidal distortions and spin asynchronicity of the companion stars, are secondary. 

The scenario explained above confirms that the efficiency of the irradiation of the companion star by the PSR injected flux is a key ingredient to understand the evolution and the differences among RBs and BWs, as already pointed out by theoretical works \citep[e.g.][]{chen13,benvenuto14}. In spite of the new discoveries, however, the formation and evolution of these systems is still difficult to underline. \citet{benvenuto14} proposed that RBs represent somehow an early stage of MSP evolution: some of them evolve toward the canonical MSP stage, where the companion star forms a He degenerate nucleus and loses its remaining thin envelope \citep[see also][]{burderi02}; others evolve toward the stage of BWs, where the companion star is no able to form (or only partially forms) a He degenerate nucleus and a strong evaporation due to the primary injected flux leads to the creation of a very low-mass companion with a mass of few hundredths of solar masses. However, ad odds with these results, \citet{chen13} stated that the evolution of canonical, RB and BW systems is independent. According to their results, RBs do not evolve neither into canonical MSPs, nor in BWs. From an observational point of view, starting from the results obtained by the identification of these objects in GCs and in the Galactic field \citep[e.g.][]{reynolds07,kaplan12,pallanca12,bellm13,breton13,li14}, we can see that the majority of RB companions are not subject to strong heating. Therefore their evolution toward the BW stage would appear unlikely. Their evolution toward the He WD stage still cannot be probed, especially because of the lack of identifications of objects in an intermediate stage between the two. However, it is interesting to note that there are some RB systems, both in the Galactic field and in GCs, such as 47TucW, whose companion light curves are dominated by the strong irradiation due to the PSR flux, as commonly observed for all the BW companions. Moreover, as can be seen from Figure~\ref{figcom_post}, in the CMD, 47TucW is located on the blue side of the main sequence, in a region close to that occupied by the two BW systems. This suggests that RB systems whose companions are subject to strong heating and thus ablation can indeed evolve to the BW stage, and 47TucW would be therefore a prototype of a RB evolving into a BW. As concerns the final destiny of BWs, it is possible that the progressive evaporation of the companion star could in the end lead to its total destruction, thus creating an isolated MSP. However, the evaporation time-scale appear to be too large to account for the observed number of isolated MSPs \citep[e.g.][]{chen13}, and since the latter are mostly found in dense GCs, it is reasonable to think that their companion stars have been unbound through a dynamical interaction with another star or a binary in the cluster. However, the total disruption of the BW companions can still explain the small number of isolated MSPs in the Galactic field. \\

The CMD position of the companion to the AMXP SAX J1748.9-2021 is compatible with that of RBs, and it closely resembles that of COM-M28H. This is not surprising, since this system is a RB-like AMXP and therefore can be considered, within an evolutionary scenario, as the precursor of a RB system or also of a tMSP. This system will evolve toward the tMSP/RB stage as soon as the residual accretion halts and the PSR emission will be finally reactivated. 

Finally, the CMD position of the NS burster EXO 1745$-$248 reveals that the companion is a sub-giant branch star. Therefore, we can speculate that EXO1745-248 is experiencing the very early phase of the
mass accretion stage, when an expanding star (a sub giant branch
object) is filling its Roche-Lobe and transferring material that
eventually spins-up the NS.  Indeed, the few outbursts in
the X-ray occurring during this stage unambiguously indicate that
heavy mass accretion on the NS is taking place.  As time
passes, the mass accretion rate will decrease and the system will
 enter in a later stage of the evolution, possibly characterized by a cyclic alternation between accretion and rotation-powered emission and finally by the re-activation of the re-accelerated NS. 

\subsection*{Future Developments}

The study of MSPs in GCs will receive a major boost in the near future thanks to the performances of the new generation of telescopes. Radio observations with ``MeerKAT'' and then with the ``Square Kilometer Array'' (SKA) should be able to identify most of the expected MSPs, especially those in nearby clusters and in those systems (such as Omega Centauri, M80, NGC 6388, etc...) expected to host rich populations of MSPs not revealed yet. The new discoveries will open the possibility to perform systematic studies of the different typologies of MSP populations in GC characterized by different structural and dynamical properties. This will improve our understanding of the cluster itself. New large sample of exotic binaries will be discovered and will be an important ground where to test fundamental physics. The number of identified companion stars will be easily enhanced thanks to the infrared capabilities of the ``James Webb Space Telescope'' (JWST), whose sensitivity and angular resolution will provide an unique view on the faintest companion stars, especially those in exotic systems like BWs and especially those in heavily obscured bulge GCs like Terzan~5, NGC~6440, NGC~6441, etc..., which are known or expected to host very rich populations of MSPs. Ground-based new facilities like the ``European Extremely Large Telescope'' (E-ELT) will complement the observation of these clusters in the optical bands, also providing the opportunity of perform spectroscopic studies that will give us new clues on the chemical properties of these objects, and thus on their evolutionary processes.

\clearemptydoublepage
%
%

\appendix
\def\rmxaa{Rev. Mexicana Astron. Astrofis.}

\chapter{Proper Motions and Structural Parameters of the Galactic Globular Cluster M71}
\label{appm71}
\begin{flushright}
\textit{Mainly based on \citealt{cadelano17a}, ApJ, 836:170}
\end{flushright}
\vspace{1cm}
\initial{I}n this appendix we show how we have exploited the two ACS/HST datasets used to the analysis presented in Chapter~\ref{cap_m71} to additionally study the cluster kinematic. Since the two dataset are separated by a temporal baseline of
$\sim7$ years, we used them to determine the relative stellar proper motions
(providing membership) and the absolute proper motion of the Galactic
globular cluster M71.  The absolute proper motion has been used to
reconstruct the cluster orbit within a Galactic, three-component,
axisymmetric potential. M71 turns out to be in a low latitude
disk-like orbit inside the Galactic disk, further supporting the
scenario in which it lost a significant fraction of its initial mass.
Since large differential reddening is known to affect this system, we
took advantage of near-infrared, ground-based observations to
re-determine the cluster center and density profile from direct star
counts.  The new structural parameters turn out to be significantly
different from the ones quoted in the literature. In particular, M71
has a core and a half-mass radii almost 50\% larger than previously
thought.  Finally we estimate that the initial mass of M71 was likely
one order of magnitude larger than its current value, thus helping to
solve the discrepancy with the observed number of X-ray sources.

\clearpage
\section{Introduction}
\label{intro}
As discussed in Chapter~\ref{CapIntro}, the high central densities of globular clusters provide
the ideal ground to the formation of exotic objects like blue
straggler stars, cataclysmic variables, low-mass X-ray binaries and
millisecond pulsars \citep[e.g.][]{ferraro97,ferraro03,pooley03, ransom05a, heinke05}.
In this respect, remarkable is the case of M71, which is a low-density
GC located at a distance of about 4 kpc from Earth. It has a quite
high metallicity ([Fe/H]$=-0.73$), a color excess ${E(B-V)=0.25}$
\citep[][2010 edition]{harris96} and a total mass of about $2 \times
10^4 \; \Msun$ \citep{kimmig15}. X-ray observations revealed that it
hosts a large population of X-ray sources, most likely consisting of
stellar exotica.  Surprisingly, as discussed in \citet{elsner08,
  huang10}, the number of X-ray detections in M71 is significantly
larger than what is expected from its present-day mass and its
collisional parameter (which is a characteristic indicator of the
frequency of dynamical interactions and thus of the number of stellar
exotica in a GC; e.g. \citealp{bahramian13}).  However, it is worth
noticing that M71 is located at a low Galactic latitude
(${l=56.75^{\circ},b=-4.56^{\circ}}$), likely on a disk-like orbit
\citep{geffert00}. Hence, it could have have lost a substantial
fraction of its initial mass, due to heavy interactions with the
Galactic field and to shocks caused by encounters with molecular
clouds and/or spiral arms.  { Moreover the structural parameters 
of this cluster have been estimated from shallow optical images \citep{peterson97}, 
and therefore need to be re-determined more accurately.}
Hence, the value of the collisional parameter,
which directly depends on the cluster structural parameters
\citep{verbunt87}, could be biased.

By taking advantage of two epoch of observations { obtained with the {\it Hubble Space Telescope} (HST) }and wide-field
near-infrared and optical datasets for M71, in this appendix we present the
determination of: \emph{(i)} the stellar proper motions (which allow
us to distinguish cluster members from Galactic contaminants),
\emph{(ii)} the absolute PM of the system (from which we estimate its
orbit within the Galaxy during the last 3 Gyr), and $(iii)$ the
cluster gravitational center and structural parameters. 

In Section \ref{obs} we describe the procedures adopted for the data
reduction and analysis. Sections \ref{pm} and \ref{abspm} are devoted
to the determination of relative stellar proper motions (PMs), and of
the cluster absolute PM and orbit, respectively. In Section
\ref{clust} we present the new determination of cluster gravity
center, density profile and structural parameters from near-infrared
data and we study how the latter change if optical observations are used
instead. We also provide an estimate of the initial mass of the
system. Finally, in Section \ref{conclu} we summarize the results and discusse the
X-ray source abundance discrepancy in light of the new values of the
cluster structural parameters and the initial mass estimate.

\section{Observations and Data Reduction}
\label{obs}  
The work presented in this Appendix is based on two different datasets. Their
characteristics and the adopted data reduction procedures are
described in the following.

\begin{figure*}[h]
\centering
\includegraphics[width=7.8cm]{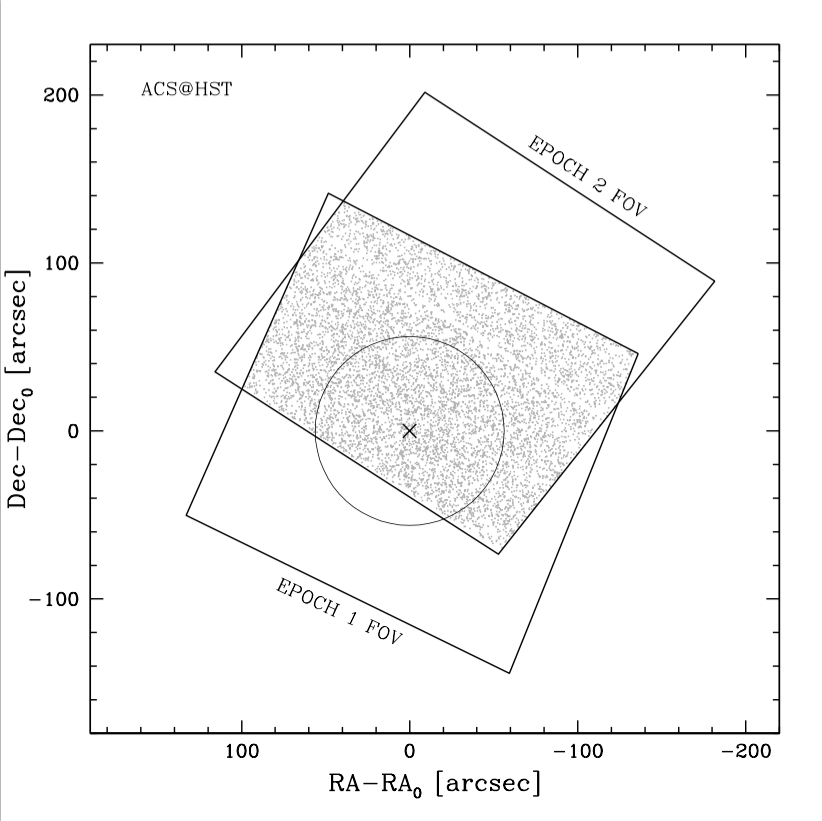}
\includegraphics[width=7.8cm]{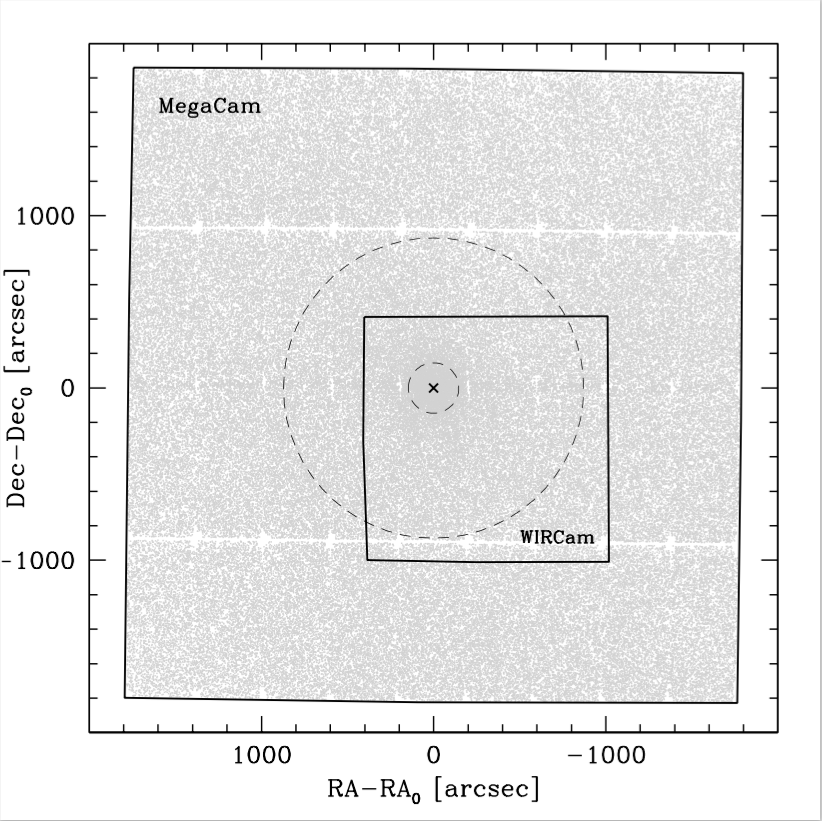}
\caption[FOVs of the ACS, MegaCam and WIRCam datasets]{{\it Left:} FOVs of the ACS first and second epoch datasets, centered on
  the newly estimated gravity center of M71 (black cross; see Section
  \ref{center}). The grey dots highlight the stars in common are between
  the two datasets, which has been used to measure the stellar proper
  motions. The solid circle marks the core radius of the cluster as
  derived in this work (${r_c=56.2\arcsec}$; see Section
  \ref{density}). {\it Right:} FOVs of the MegaCam and WIRCam datasets, centered on the
cluster gravity center (black cross). The small and large dashed
 circles mark, respectivey, the half-mass and the tidal radii derived
 in this work.}
\label{mappafov}
\end{figure*}

\begin{figure*}[ht]
\centering
\includegraphics[width=9.2cm]{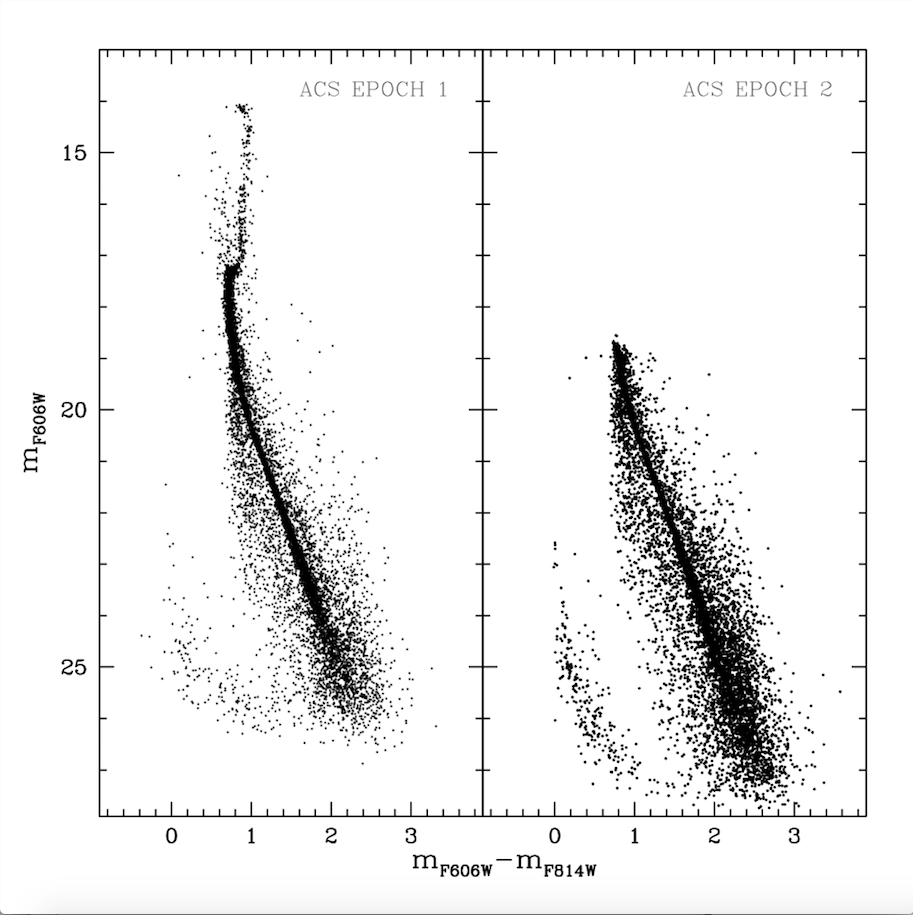}
\caption[Optical CMD of M71]{Optical CMD of M71 obtained from the first and second epoch
  ACS datasets (left and right panels, respectively).}
  \label{cmd}
\end{figure*}

{\it High-Resolution Dataset --} This has been used to determine the
stellar PMs. It consists of two sets of images acquired { with the Wide Field Camera (WFC) of the} 
Advanced Camera for Surveys (ACS) mounted on HST (see the left panel of Figure \ref{mappafov} for a map of the fields of view - FOVs - covered by these observations). This camera provides a FOV of  $202\arcsec\times202\arcsec$ with a pixel scale { of ${ 0.05\arcsec \ \mathrm{pixel}^{-1}}$}. The first epoch data have been
collected under GO10775 (P.I.: Sarajedini) on 2006 July 1, and consist
of a set of ten dithered images, five in the F606W filter (with
exposure times: $1 \times 4$ s; $4 \times 75$ s) and five in the F814W
filter ($1 \times 4$ s; $4 \times 80$ s).  The second epoch is
composed of proprietary data obtained under GO12932 (P.I.: Ferraro) on
2013, August 20. It consists of a set of ten deep images acquired
through the F606W filter ($2 \times 459$ s; $3 \times 466$ s; $5
\times 500$ s) and nine images in the F814W filter ($5 \times 337$ s;
$3 \times 357$ s; $1 \times 440$ s). The photometric analysis has been
performed on the {\textrm -flc} images (which are corrected for flat
field, bias, dark counts and charge transfer efficiency) following the
procedures described in detail in \citet{anderson06}. Briefly, both
the epochs have been analyzed with the publicly available program
{\textrm img2xym\_WFC.09x10}, which uses a pre-determined model of a
spatially varying point spread function (PSF) plus a single
time-dependent perturbation PSF (to account for focus changes or
spacecraft breathing). { The final output of this process are two catalogs (one for each epoch)
with instrumental magnitudes and { positions for all the sources above a given threshold}.
Star positions were corrected in each catalog for geometric distortion by adopting the solution provided by \citet{anderson07}.}
By using the stars in common with the public catalog of
\citet[][see also \citealp{anderson08}]{sarajedini07}, instrumental
magnitudes have been calibrated on the VEGAMAG system and instrumental
positions have been reported on the absolute right ascension and declination coordinate
reference system ($\alpha$ and $\delta$, respectively). The final
color-magnitude diagrams (CMDs) are shown in Figure \ref{cmd} for the
two different epochs.

{\it Wide-field Dataset --} To determine the cluster gravitational
center and structural parameters, we used ground-based near-infrared
images (Prop ID: 11AD90; PI: Thanjavur) obtained with the wide field
imagers WIRCam mounted at the Canada-France-Hawaii Telescope
(CFHT). To study the effect of differential reddening, we also made
use of optical wide-field images (Prop ID: 04AC03, 03AC16; PI: Clem)
acquired with MegaCam at the same telescope.  The WIRCam camera
consists of a mosaic of four chips of 2040$\times$2040 pixels each,
with a pixel scale of ${ 0.31\arcsec \ \mathrm{pixel}^{-1}}$, providing a
total FOV of $\sim21.5\arcmin \times 21.5\arcmin$. We analyzed seven
images obtained with the $J$ and $K_s$ filters, with exposure times of 5 s
and 24 s, respectively. A dither pattern of few arcseconds was applied
to fill the gaps among the detector chips.  The MegaCam camera
consists of a mosaic of 36 chips of 2048$\times$4612 pixels each, with
a pixel scale of ${ 0.185\arcsec \ \mathrm{pixel}^{-1}}$ providing a FOV of
$\sim1^{\circ}\times1^{\circ}$. A total of 50 images have been
acquired, both in the g' and in the r' bands, with exposure times of
250 s each. A dither pattern of few arcseconds was adopted for each
pointing, thus allowing the filling of most of the interchip gaps,
with the exception of the most prominent, horizontal ones. The right panel of Figure~\ref{mappafov} shows the map of the Wide-field dataset.

\begin{figure*}[hb]
\centering
 \includegraphics[width=9.2cm]{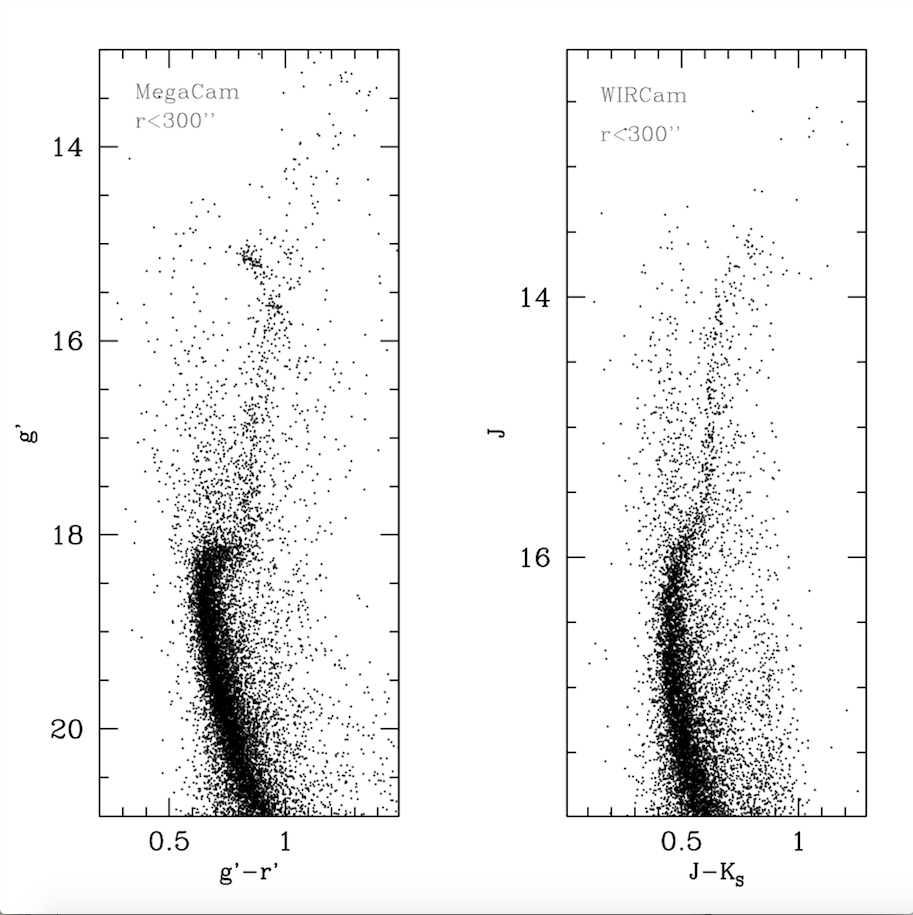}
\caption[Optical and near-infrared CMDs of M71]{Optical and near-infrared CMDs of M71, obtained from the
  MegaCam (left panel) and the WIRCam datasets (right panel),
  respectively.  In both cases, only stars within $300\arcsec$ from
  the center are plotted. }
\label{cmd2}
\end{figure*}

For both these sets of observations, the images were pre-processed
(i.e. bias and flat-field corrected) by means of the Elixir pipeline
developed by the CFHT team and the photometric analysis has been
performed independently on each chip by following the procedures
described in Chapter~\ref{opticalcom}. The obtained star instrumental magnitudes have been
reported to the SDSS photometric system\footnote{See
  \url{http://www.cfht.hawaii.edu/Science/CFHTLS-DATA/megaprimecalibration.html\#P2.}}
for the MegaCam catalog, and to the 2MASS system for the WIRCam
catalog. Finally the instrumental positions have been reported to
the absolute coordinate reference frame by using the stars in common
with the 2MASS catalog\footnote{Publicly available at
  \url{http://vizier.u-strasbg.fr}}. The CMDs for these datasets are
shown in Figure \ref{cmd2} for stars located at less than $300\arcsec$
from the center.

As can be seen from both Fig. \ref{cmd} and Fig. \ref{cmd2}, the
standard evolutionary sequences are well defined. However, they are
also heavily contaminated by foreground objects, as expected from the
location of M71 close to the Galactic disk.

\clearpage
\section{Relative Proper Motions}
\label{pm}
\begin{figure*}[h]
\centering
\includegraphics[width=10cm]{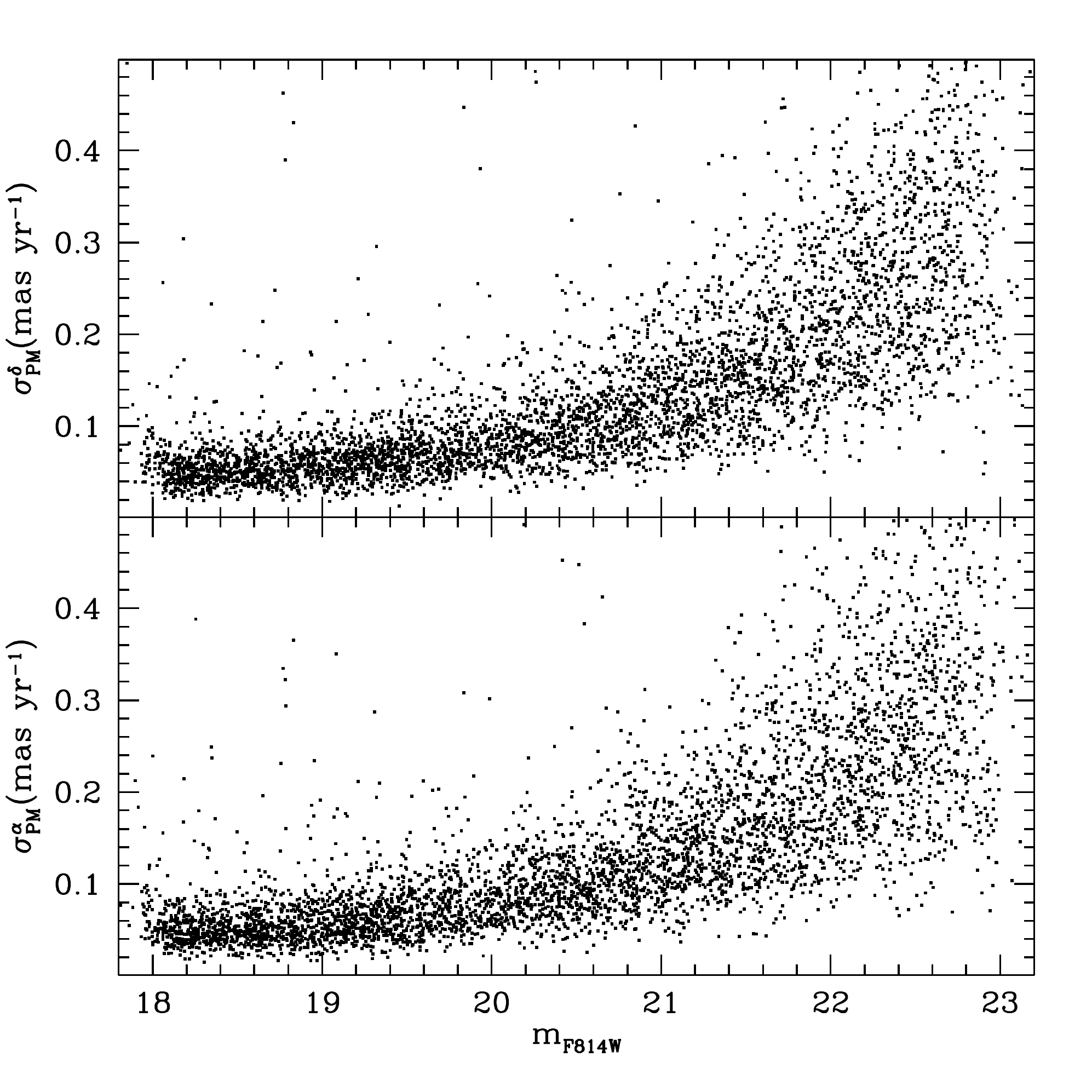}
\caption[Estimated uncertainties of the derived proper motions]{Estimated uncertainties of the derived proper motions as a
  function of the $m_{F814W}$ magnitude of the measured stars. The
  upper and the lower panels show, respectively, the uncertainties in
  the $\alpha$ and in the $\delta$ directions. For stars with ${m_{F814W}\lesssim 21}$ the typical error is smaller than ${0.07
  \ \mathrm{mas \ yr}^{-1}}$. }
\label{errmoti}
\end{figure*}

To study the PMs of M71, we used the high resolution datasets. These
are separated by a temporal baseline of 7.274 years and { because of
their different orientation, pointing and magnitude limit, the PM analysis could be performed only on the $\sim5000$ stars in common, located in the overlapping FOV} (see Figure~\ref{mappafov}) and
having magnitudes ${18<m_{F814W}<24}$ (corresponding to magnitudes ${19<m_{F606W}<25}$).
 We adopted the procedure described in \citet[][see also \citealp{dalessandro13,
    bellini14, massari15}]{massari13}. { Briefly, we used six parameters linear transformations\footnote{{ To do this we applied six parameters linear transformations using {\tt CataXcorr}.}} to report the coordinates of the stars in each exposure to the distortion-free reference catalog of
\citet{sarajedini07}. Since we are interested in the stellar PMs relative to the cluster frame, these transformations have been determined by using a sample of $\sim6600$ stars that, in the reference catalog, are likely cluster members on the basis of their CMD position (i.e. stars located along the main sequence). Moreover, the transformations have been determined independently on each detector chip in order to maximize the accuracy. At the end of the procedure, for each of the $\sim5000$ stars we have up to ten position measurements in the first epoch catalog and up to nineteen in the second epoch catalog.}  To determine the relative PMs, we computed the mean X and Y positions of each star in
each epoch, adopting a $3\sigma$ clipping algorithm. The star PMs are
thus the difference between the mean X,Y positions evaluated in the
two epochs, divided by ${\Delta t=7.274}$ years. The resulting PMs
are in units of pixels $\mathrm{years}^{-1}$. Since the master frame is
already oriented according to the equatorial coordinate system, the
X-component of the PM corresponds to a projected PM along the
(negative) RA and the Y-component corresponds to a PM along the
Dec. Therefore, we converted our PMs in units of mas years$^{-1}$
by multiplying the previous values for the ACS pixel scale (${0.05\arcsec \; \mathrm{pixel}^{-1}}$), and we named ${\mu_\alpha \cos(\delta)}$ and ${\mu_\delta}$
the PMs along the RA and Dec directions, respectively.  To maximize
the quality of our results, we built a final PM catalog by taking
into account only stars for which at least three position measurements
are available in each epoch. At the end of the procedure we counted
4938 stars with measured PMs. The errors in the position of the stars
in each epoch ($\sigma_{1,2}^{\alpha,\delta}$) have been
calculated as the standard deviation of the measured positions around
the mean value. Then the errors in each component of the PM have been
assumed as the sum in quadrature between the error in the first and
second epoch: ${\sigma_{PM}^{\alpha,\delta} = \sqrt{
  (\sigma_1^{\alpha,\delta})^2 + (\sigma_2^{\alpha,\delta})^2 }/
\Delta t}$. The errors as a function of the star magnitudes are shown
in Figure \ref{errmoti}. For both the PM directions, the typical
uncertainty for stars with ${m_{F814W}<21}$ is less of ${\sim
0.07 \ \mathrm{mas \ yr}^{-1}}$, demonstrating the good quality of our
measurements.  Following \citet{bellini14}, we also verified that our PM measurements are not affected by chromatic effects, i.e., there is no dependence of ${\mu_\alpha \cos(\delta)}$ and ${\mu_\delta}$ on the (F606W-F814W) color.
Finally, our PM measurements are not even affected by positional
effects, i.e., there is no dependence of the derived PMs on the
instrumental (X,Y) positions.

\begin{figure*}[!h]
\centering
\includegraphics[width=10cm]{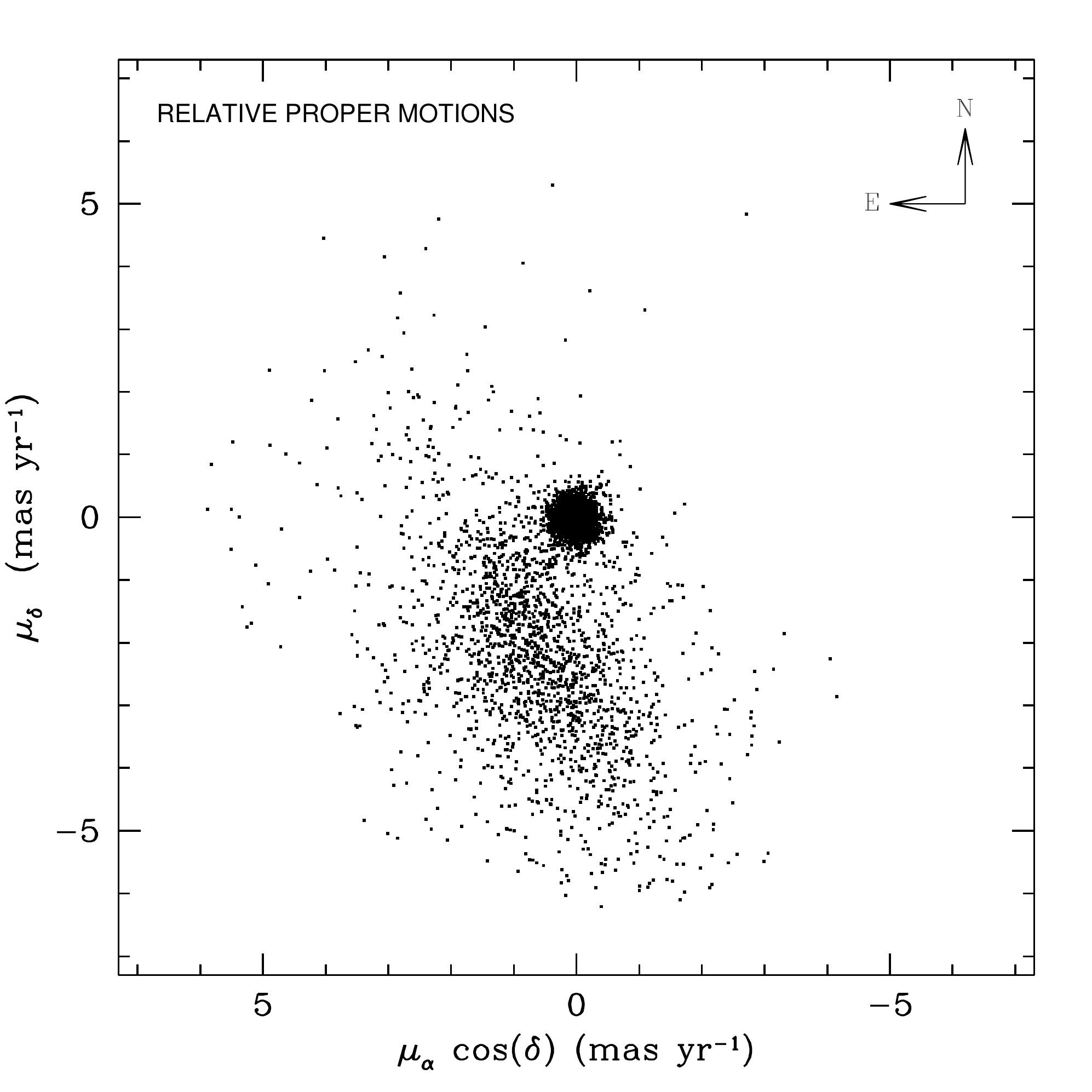}
\caption[VPD of the relative PMs]{VPD of the relative PMs. The clump in the (0,0) $\mathrm{mas
  \ yr}^{-1}$ position is dominated by the cluster population. The
  elongated region beyond this clump is instead due to contaminating
  stars, mostly from the Galactic disk.}
\label{motirel}
\end{figure*}

\begin{figure*}[h]
\centering
\includegraphics[width=10cm]{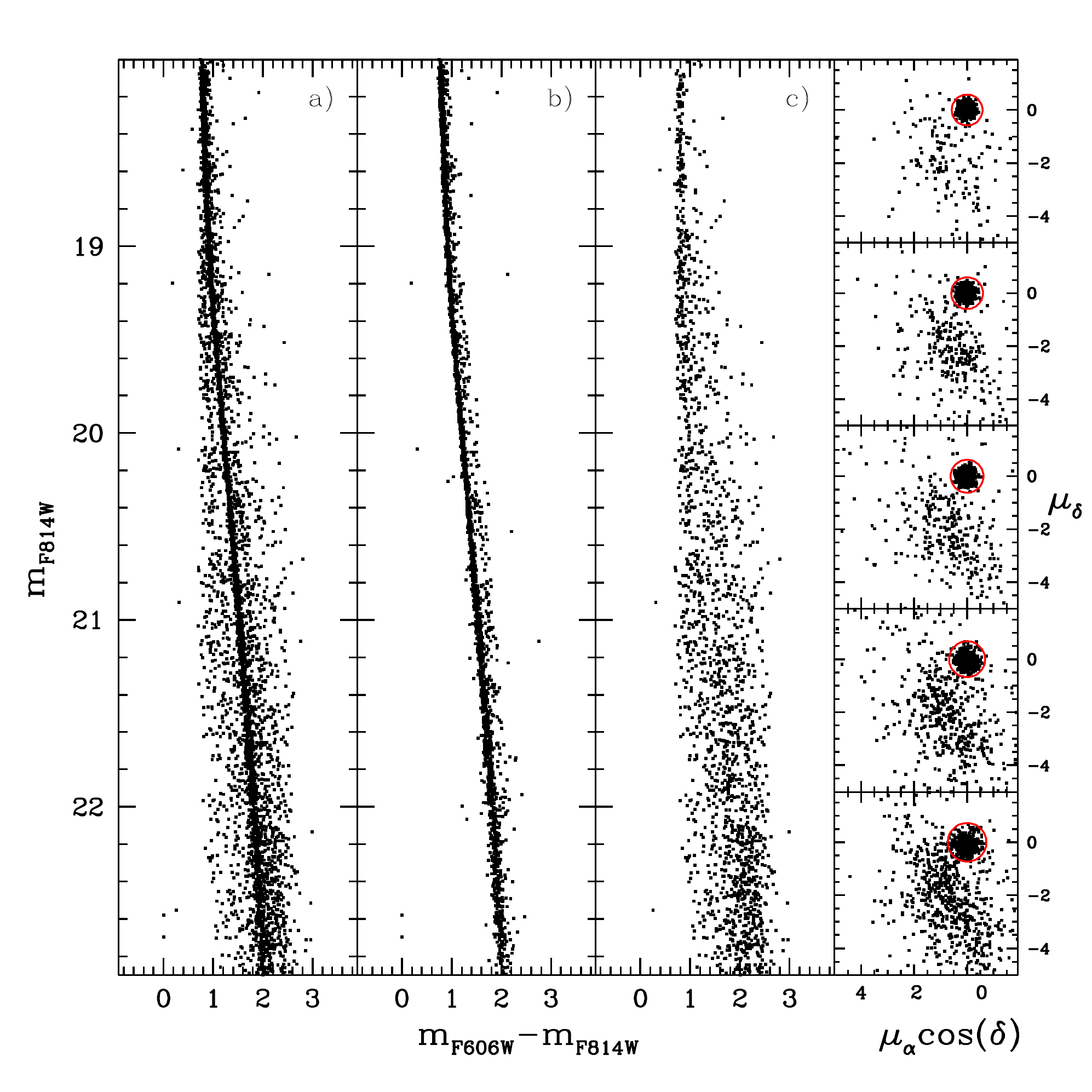}
\caption[Optical CMD decontaminated using the derived proper motions]{{\it Panel a:} Optical CMD of all the stars in common
  between the two observation epochs. {\it Panel b:} Decontaminated
  CMD obtained by using only the likely cluster members selected from
  the VPDs shown in the rightmost column. As can be seen, a sharper
  and more delineated main sequence and the binary sequence are now
  appreciable. {\it Panel c:} CMD made of all the contaminating
  objects, selected from the VPDs as those with PMs not compatible
  with the that of the GC. The Galactic sequence can be appreciate
  from this plot. {\it Rightmost column:} VPDs of the measured stars
  divided in bins of magnitudes. The solid red circles contain all the
  objects selected as likely cluster members.}
\label{cmdmoti}
\end{figure*}

In Figure \ref{motirel} we show the PM distribution in the vector
points diagram (VPD). As can be seen, the VPD is dominated by two
prominent features: the clump in the center with zero relative PM is,
by definition, dominated by the cluster population, while the
elongated sparse distribution of points extending beyond this clump is
dominated by contaminating field stars, mostly from the Galactic
disk. { At first inspection of the VPD we can see that only $\lesssim
60\%$ of the $\sim5000$ analyzed stars are likely cluster members}. A high
percentage of field contamination is indeed expected in the case of
M71, since it has a quite low stellar density and is located in a
crowded field at low Galactic latitudes. By selecting in the VPD the
likely cluster members (i.e. the stars with relative PM around 0 both
in $\alpha$ and in $\delta$) we find that the mean motion is ${0.01 \ \mathrm{mas
\ yr}^{-1}}$ with a standard deviation of ${0.1 \ \mathrm{mas \ yr}^{-1}}$ both
in $\alpha$ and in $\delta$, thus, as expected, consistent with zero.

The effect of decontaminating the CMD from field stars is shown in
Figure \ref{cmdmoti}, where we have separated the objects with PM$\lesssim {
0.6 \ \mathrm{mas \ yr}^{-1}}$ (likely cluster members), from those with larger
PMs.  The selection of stars in the VPD is shown, per bins of one
magnitude, in the left-hand column of the figure, with the objects
having PM$\lesssim {0.6 \ \mathrm{mas \ yr}^{-1}}$ encircled in red. The effect on
the CMD is shown in the other three columns: from left to right, the
observed CMD, the CMD of cluster members only, and the CMD of field
stars.  In the latter, it is well appreciable the main sequence of the
Galactic field.  Instead, the decontaminated CMD clearly shows a sharp
and well defined main sequence, also revealing the binary sequence.
The few stars on the blue side of the main sequence could be cluster
exotic objects, such as cataclysmic variables, X-ray binaries or
millisecond pulsars \citep[e.g.][]{ferraro01b,pallanca10,cohn10,cadelano15a},
where a main sequence companion star is heated by a compact object.  Nonetheless, we cannot completely rule out the
possibility that some of these stars are field objects with PMs
compatible with those of the cluster members.

\section{Absolute Proper Motions}
\label{abspm}
\begin{figure*}[h] 
\centering
\includegraphics[width=10cm,angle=-90]{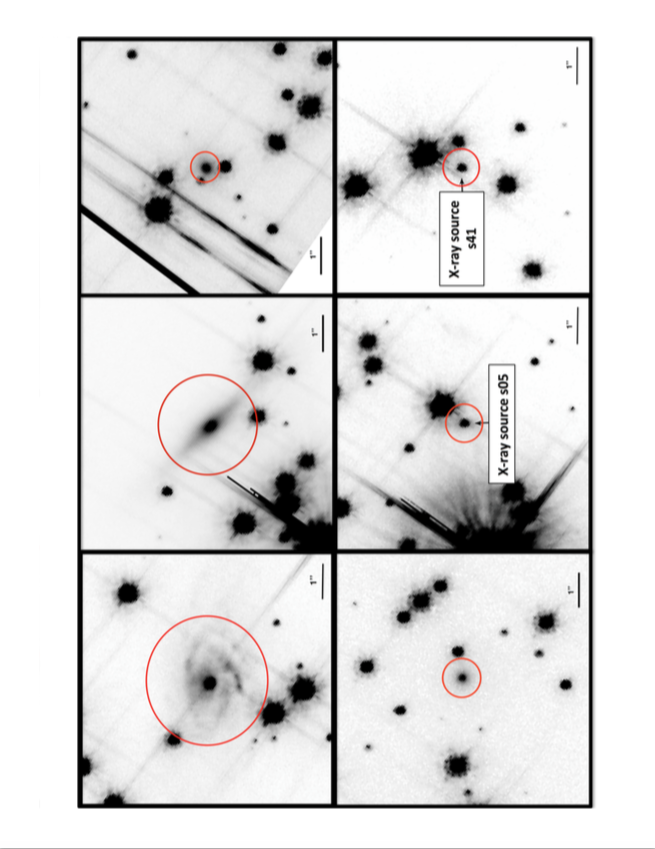}
\caption[Finding charts of six selected extra-Galactic objects]{Finding charts of the six selected extra-Galactic objects
  used to determine the absolute reference frame zero point. The
  charts are taken from an ACS image acquired in the F814W filter. The
  three upper panels and the bottom-left panel show the four galaxies
  found through the visual inspection of the images.  The central and
  right bottom panels show the optical counterparts to the X-ray
  sources s05 and s41, classified as AGNs. The point-like structure of
  these sources allowed a precise determination of the centroid
  position and of their PM.}
   \label{charts}
\end{figure*}

\begin{figure*}[!t]
\centering
\includegraphics[width=10cm]{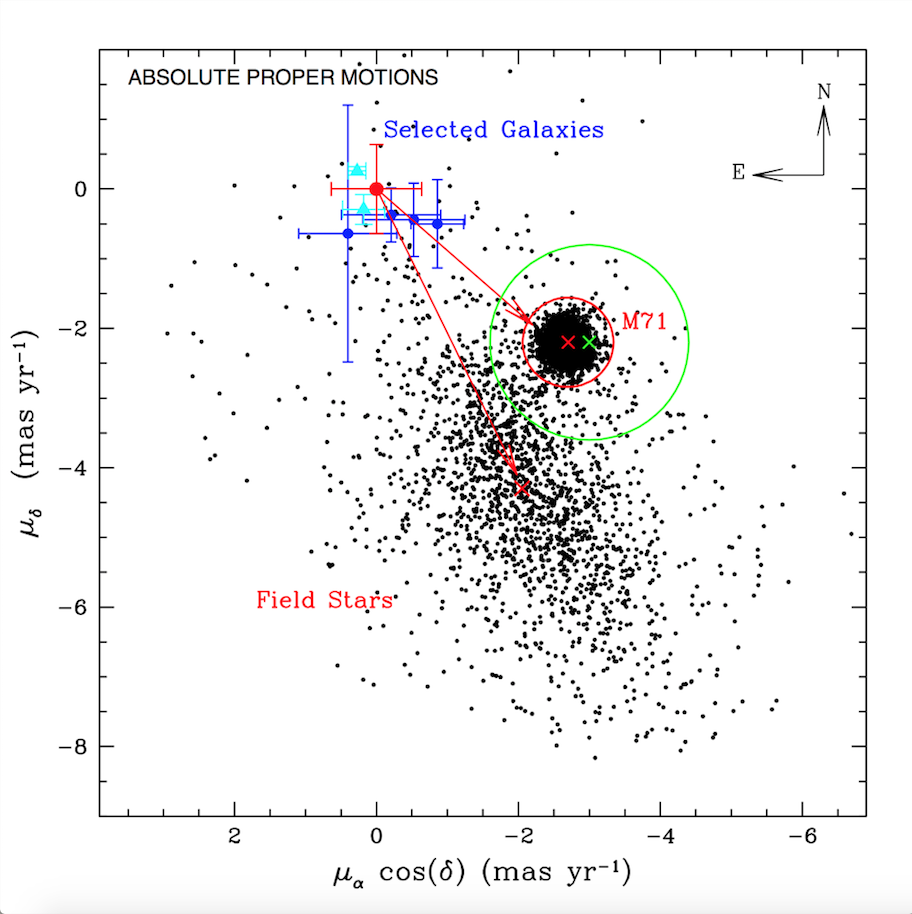}
\caption[VPD of absolute PMs]{VPD of absolute PMs. The extra-Galactic objects used to
  determine the absolute reference frame zero point are marked as blue
  dots (galaxies) and cyan triangles (AGNs), and their mean motion
  is marked a large red dot.  The red arrows indicate the absolute
  PM vectors of M71 and of the Galactic field population.  The
  absolute PM value of M71 estimated in this work is marked with a red
  cross and the red circle represents its $\sim1 \sigma$ confidence
  region. The values previously determined in the literature
  \citep{geffert00} are marked with a green cross and a green
  circle. The red cross centered on the elongated structure is centered on the
  field mean PM (see Section~\ref{systematics}).}
  \label{motiass}
\end{figure*}

To transform the relative PMs into absolute ones, we used
background galaxies as reference, since they have negligible PMs due
to their large distances. This method has been successfully used in
several previous works \citep[e.g.][]{dinescu99, bellini10,
  massari13}. Unfortunately, the NASA Extragalactic Database report no
sources in the FOV used for the PM estimate. Thus, we carefully
inspected our images in order to search for diffuse galaxy-like
objects. We found four galaxies with central point-like structure and
relative high brightness, which allowed us to precisely determine
their centroid position.  Although many other galaxies are present in
the FOV, they have no point-like structure or are too faint to allow
the determination of a reasonable PM value. Moreover, as part of a
project aimed at searching for optical counterparts to X-ray sources,
we identified two promising  active galactic nuclei (AGN) candidates. Two Chandra X-ray sources,
named s05 and s41 in \citet{elsner08}, have high energy and optical
properties that can be attributed either to AGNs or to cataclysmic
variables \citep[see][for more details]{huang10}. In order to
distinguish between these two possibilities, we analyzed their PMs. We
reported our relative PM reference frame to the absolute cluster PM
(${ \mu_{\alpha}\cos \delta, \mu_{\delta}=-3.0\pm1.4, -2.2\pm1.4
\ \mathrm{mas \ yr}^{-1}}$) previously determined by \citet{geffert00} and found
that these two sources have an absolute PM significantly different
from the cluster motion and compatible with zero. We therefore
conclude that these two objects are likely background AGNs\footnote{Of
  course, these sources could be foreground cataclysmic variables with
  PMs almost perfectly aligned with our line of sight, but this
  possibility seems to be quite unlikely.} and add them to the list of
objects used to determine our reference absolute zero point.  The six
selected objects are located very close to each other in the VPD, as
expected for extragalactic objects, and their finding charts are shown
in Figure \ref{charts}. We defined the absolute zero point as the
weighted mean of their relative PMs and assumed as error the
uncertainty on the calculated mean. By anchoring this mean position to
the (0,0) mas yr$^{-1}$ value, we find that the absolute PM of M71
is:
\begin{equation}
{ (\mu_{\alpha}\cos \delta, \mu_{\delta})=(-2.7\pm0.5, -2.2\pm0.4) \ \mathrm{mas \ yr}^{-1}}.
\end{equation}
This value is in good agreement with (but more accurate than) the
previous determination \citep{geffert00}, and it remains unchanged
within the errors even if the two candidate AGNs are excluded from the
analysis: in that case we get: ${(\mu_{\alpha}\cos \delta,
\mu_{\delta})=(-2.4\pm0.6, -1.9\pm0.1) \ \mathrm{mas \ yr}^{-1}}$, still
in agreement with the previous results.  The VPD in the absolute frame
is plotted in Figure \ref{motiass}, with the red and green crosses and
circles marking, respectively, the absolute PM and its uncertainty as
determined in this study and as quoted in \citet{geffert00}.

\label{systematics}
Since every absolute PM measurement is strictly dependent on the
accuracy of the absolute reference frame, we need to verify the
possible presence of systematic errors in its determination. One of
the possible source of systematic errors is the rotation of the GC on
the plane of the sky. Indeed, since we used only cluster stars to
define the relative reference frame, if the GC is rotating, then our
frame will be rotating too. This would introduce an artificial
rotation to background and foreground objects around the cluster
center. { To quantify this possible effect we followed the procedure described in \citet{massari13}. We selected a sample of field stars as those that in the VPD of Figure~\ref{motirel} have relative PMs larger than ${0.8 \ \mathrm{mas \ yr}^{-1}}$. Then we decomposed their PM vectors into a 
radial and tangential component with respect to the cluster center. If the GC is rotating,
we would expect to find a clear dependence of the PM tangential component on the distance from the cluster center. Such a dependence is however excluded by our results, thus that the internal regions of M71 are not rotating, in agreement with the recent findings by \citet{kimmig15}}.

{ We also compared} the field star motion to that expected
from theoretical Galactic models in the analyzed FOV. To evaluate the
field mean motion, we followed the procedure described in
\citet{anderson10}. First, we excluded the stars within ${0.8 \ \mathrm{mas
\ yr}^{-1}}$ from the cluster mean motion. Then we iteratively removed
field stars in a symmetric position with respect to the GC exclusion
region and evaluated the weighted mean motion by applying a $3\sigma$
algorithm. We found ${(\mu_{\alpha}\cos \delta,
\mu_{\delta})=(-2.0\pm0.2, -4.3\pm0.2) \ \mathrm{mas \ yr}^{-1}}$. We
compared these values with those predicted for the same region of the
sky in the Besan\c{c}on Galactic model \citep{robin03}, simulating a
sample of $\sim2000$ artificial stars distant up to 15 kpc from the
Galactic center, in a FOV centered on M71, covering a solid angle of
$\sim11\arcmin$, and having V magnitudes ranging 12 from to 25. The
predicted field mean motion is ${(\mu_{\alpha}\cos \delta,
\mu_{\delta})=(-2.4, -4.7) \ \mathrm{mas \ yr}^{-1}}$, in good agreement
with our results.


\subsection{The cluster orbit}
\label{orbitaround}
\begin{figure*}[h]
\centering
\includegraphics[width=7.8cm]{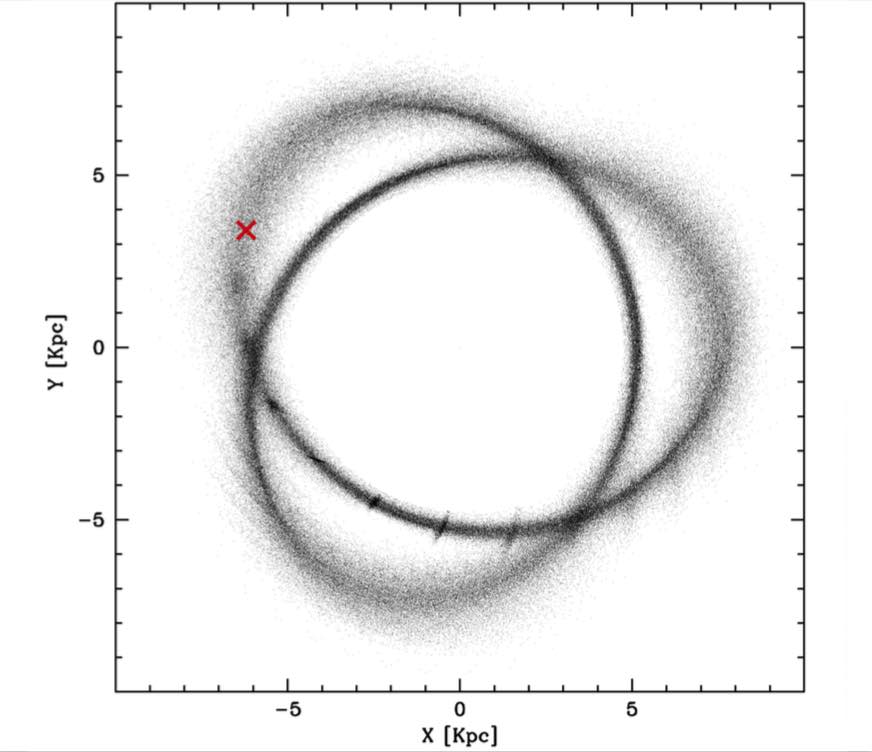}
\includegraphics[width=7.8cm]{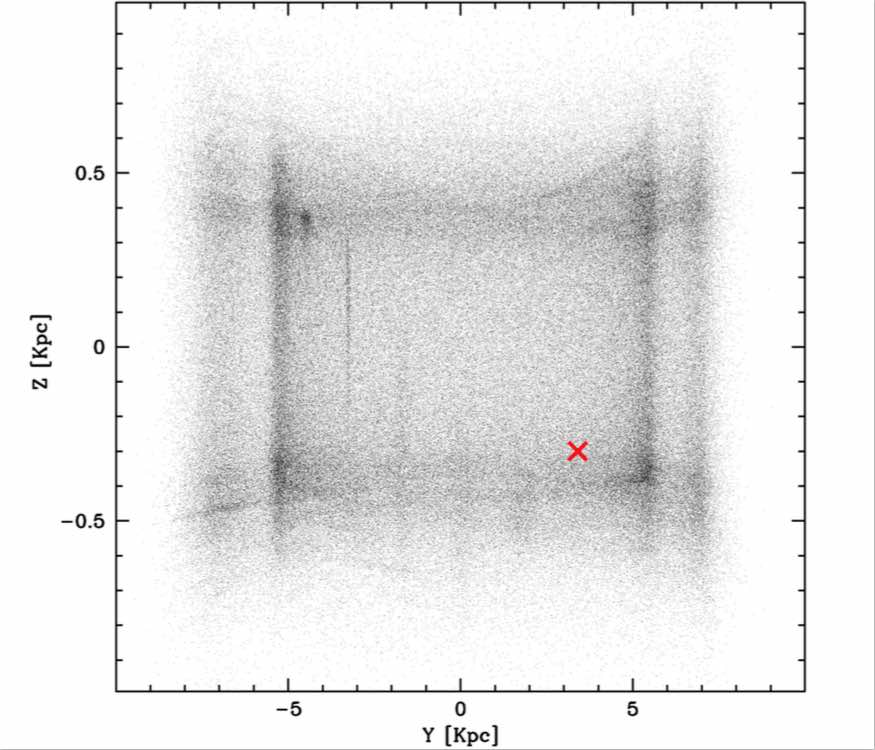}
\caption[Simulation of M71 orbit in the Galaxy]{{\it Left:} Simulated positions occupied by M71 during the
  last 3 Gyr along its orbit in the equatorial Galactic plane. Each
  point represents the position of one (out of 1000) cluster in one of
  the 32000 snapshots obtained during the numerical integration. The
  red cross marks che current cluster position. {\it Right:} Same as
  in the left panel, but for the orbit in the meridional Galactic
  plane.}
  \label{orbit}
\end{figure*}

The GC absolute PM, combined with the radial velocity
${v_r=-23.1\pm0.3}$ km s$^{-1}$ from \citet{kimmig15}, can be used to
determine 3D space velocity of the cluster in a Cartesian
Galactocentric rest frame. Using the formalism described in
\citet{johnson87}, assuming the Local Standard of Rest  velocity equal
to ${ 256 \ \mathrm{km \ s}^{-1}}$ \citep{reid09} and using the value of the Sun velocity
with respect to it from \citet{schonrich10}, we obtained ${(v_x,v_y,v_z )=(52\pm10,204\pm6,31\pm12) \ \mathrm{km
\ s^{-1}}}$, where the major source of uncertainty is the GC absolute PM
error.
We then used the 3D velocity of the cluster and its current
Galactocentric position\footnote{We adopted the Galactic coordinates
  quoted in \citet{harris96} and the convention in which the X axis
  points opposite to the Sun, i.e., the Sun position is $(-8.4,0,0)$, { where the distance of the Sun from the Galactic center is from \citet{reid09}}.}
${(x,y,z )=(-6.2\pm0.6,3.4\pm0.3,-0.32\pm0.03
)}$ kpc, to reconstruct its orbit in the axisymmetric potential
discussed in \citet{allen91}, which has been extensively used to study
the kinematics of Galactic stellar systems
\citep[e.g.][]{ortolani11,moreno14,massari15}.  The orbit was
time-integrated backwards, starting from the current conditions and
using a second-order leapfrog integrator \citep[e.g.][]{hockney88}
with a small time step of $\sim100$ kyr. Since the adopted Galactic
potential is static, we choose to back-integrate the orbit only for 3
Gyr, since longer backward integrations become uncertain due to their
dependence on the Galactic potential variations as a function of
time. This numerical integration required about 32000 steps and
reproduced $\sim20$ complete cluster orbits. The errors on the
conservation of the energy and the Z-component of the angular momentum
never exceeded one part over $10^9$ and $10^{16}$, respectively. We
generated a set of 1000 clusters starting from the phase-space initial
conditions normally distributed within the uncertainties. For all of
these clusters we repeated the backward time
integration. Figure~\ref{orbit} shows the resulting cluster orbits in
the equatorial and meridional Galactocentric plane. As can be seen,
the cluster has a low-latitude disk-like orbit within the Galactic
disk. Indeed in the equatorial plane it reaches a maximum distance of
$\sim8$ kpc from the Galactic center and a minimum distance of $\sim5$
kpc. Thus, it orbits around the assumed spheroidal bulge, never
crossing it. Moreover, it persists on a low-latitude orbit, with a
typical height from the Galactic plane of about $\pm0.4$ kpc, thus
again confined within the disk. The estimated orbits indicate that, at
least during the last 3 Gyr, M71 tightly interacted with the inner
Galactic disk. { With respect to the large majority of Galactic GCs, which are on large orbits across the (low-density) halo, these interactions likely induced heavy mass-loss  \citep{vesperini97} in M71, thus supporting the possibility that it lost a significant fraction of its initial mass,} as already suggested by its flat mass function
\citep{demarchi07}. Moreover, such a heavy mass-loss could finally
explain why M71 harbors a large population of X-ray sources, in spite
of its present low mass \citep{elsner08}.

\section{Gravitational Center, Structural Parameters and Initial Mass}
\label{clust}
In this section we present the determination of the gravitational
center and of the new structural parameters of M71.
 
\subsection{Gravitational center}
\label{center}
To avoid biases due to the strong differential reddening affecting the
system \citep[e.g.][]{schlegel98}, for the determination of the
cluster center of gravity $C_{grav}$ we used the near-infrared
WIRCam catalog, which has the same level of completeness of the ACS
one in the magnitude range ${14 < K_s < 16.8}$.  ${C_{grav}}$ has been determined following an iterative procedure that, starting from a first-guess center, selects a sample of stars within a circle of radius r and re-determine the center as the average of the star coordinates ($\alpha$ and $\delta$). The procedure stops when convergence is reached, i.e., when the newly-determined center coincides with the previous ones within an adopted tolerance limit \citep[][see also \citealt{montegriffo95,lanzoni07}]{lanzoni10}.
For M71,
which is a relative loose GC
\citep{harris96}, we repeated the procedure eighteen times, using
different values of $r$ and selecting stars in different magnitude
ranges, chosen as a compromise between having high enough statistics
and avoiding spurious effects due to incompleteness and saturation.
In particular, the radius $r$ has been chosen in the range
$140\arcsec-160\arcsec$ with a step of $10\arcsec$, thus guaranteeing
that it is always larger than the literature core radius
${r_c=37.8\arcsec}$ \citep{harris96}. For each radius $r$, we have
explored six magnitude ranges, from ${K_s>14}$ (in order to exclude stars
close to the saturation limit), down to ${K_s=16.3-16.8}$, in steps of 0.1
magnitudes. As first-guess center we used that quoted by
\citet{goldsbury10}.  The final value adopted as ${C_{grav}}$ is the
mean of the different values of $\alpha$ and $\delta$ obtained in the eighteen
explorations, and its uncertainty is their standard deviation. We
found ${\alpha=19^{h}53^{m}46.106^{s}}$ and
${\delta=+18^{\circ}46\arcmin43.38\arcsec}$, with an uncertainty of about
$1.7\arcsec$.  The newly determined center of M71 is $\sim 5.7\arcsec$
west and $\sim 0.3.\arcsec$ north from the one measured from optical
ACS data by \citet{goldsbury10}. Such a discrepancy is likely
ascribable to an effect of differential reddening impacting the
optical determination. 

\subsection{Stellar density profile}
\label{density}
\begin{figure*}[!b]
\centering
\includegraphics[width=10cm]{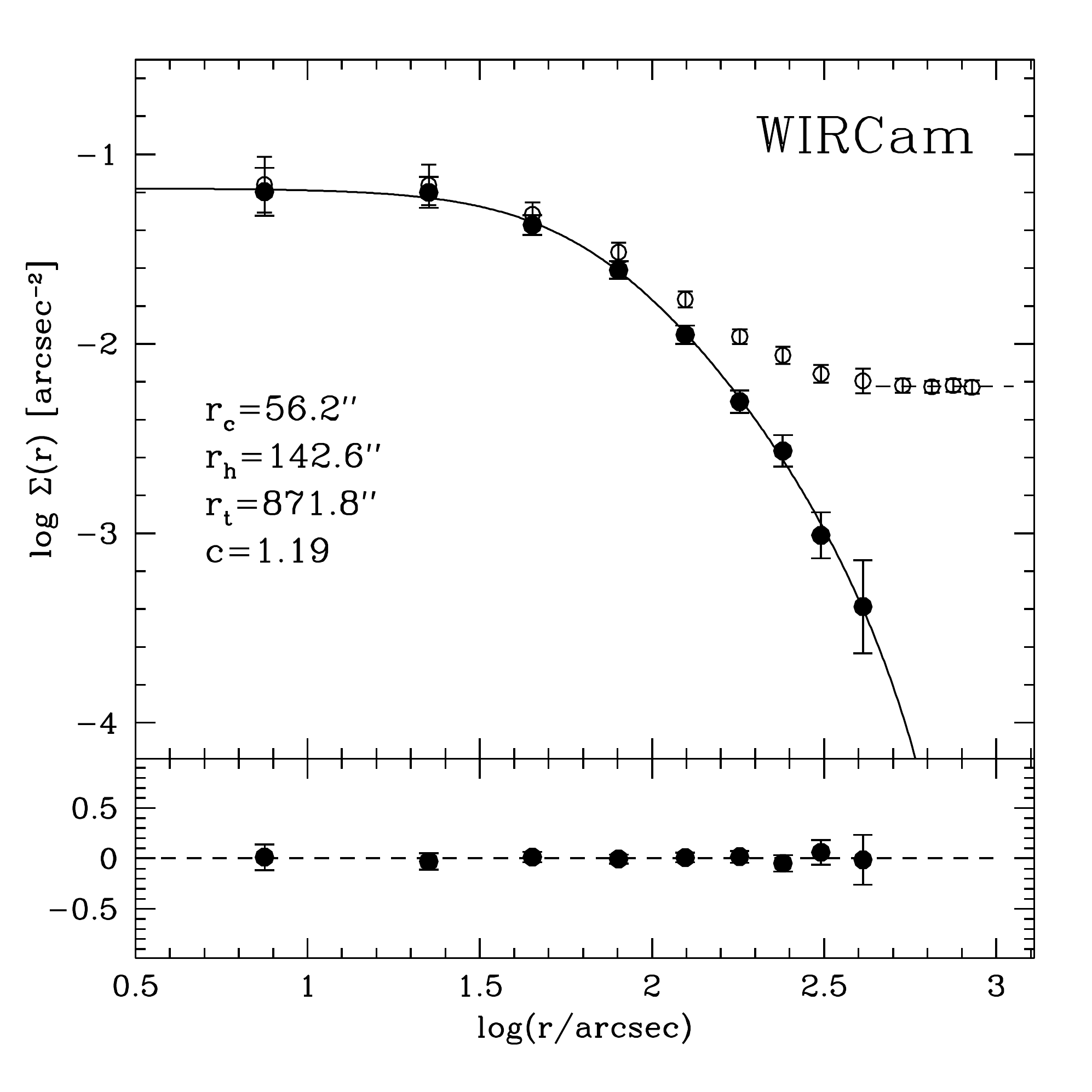}
\caption[Observed density profile of M71 obtained
  near-infrared dataset]{Observed density profile of M71 obtained from the WIRCam
  near-infrared dataset (open circles). The dashed line marks the
  density value of the Galactic field background, obtained by
  averaging the four outermost points. The black filled circles are
  the density values obtained after background subtraction (see
  Section \ref{density}). The best-fit King model (solid line) is
  overplotted to the observations and the residuals of the fit are
  reported in the bottom panel. The best-fit structural parameters are
  also labelled in the figure.}
\label{profwircam}
\end{figure*}

Since the surface brightness profile can suffer from strong biases and
fluctuations due to the presence of few bright stars \citep[see, e.g.,
  the case of M2 in][]{dalessandro09}, in order to re-evaluate the
structural parameters of M71 we used direct star counts.  The
determination of the stellar density profile (number of stars per unit
area, in a series of concentric annuli around ${C_{grav}}$) has been
performed following the procedure fully described in
\citet{miocchi13}. Also in this case, in order to minimize the
differential reddening effect we used the near-infrared WIRCam data,
which covers distances out to $\sim 1000\arcsec$ from ${C_{grav}}$
in the south-west portion of the cluster (see the right panel Figure~\ref{mappafov}).
To build the density profile we considered 13 concentric annuli around
${C_{grav}}$, each one divided into four sub-sectors. We then
counted the number of stars with ${14<K_s<16.5}$ in each sub-sector and
divided it by the sub-sector area.  The projected stellar density in
each annulus is the mean of the values measured in each sub-sector and
the uncertainty has been estimated from the variance among the
sub-sectors. The stellar background has been estimated by averaging
the outermost values, where the profile flattens, and it has been
subtracted to the observed distribution to obtain decontaminated
density profiles.  The result is shown in Figure \ref{profwircam}.

The cluster structural parameters has been derived by fitting the
observed density profiles with a spherical, isotropic, single-mass
\citet{king66} model.\footnote{These models can be generated and
  freely downloaded from the Cosmic-Lab web site:
  \url{http://www.cosmic-lab.eu/Cosmic-Lab/Products.html}. The fitting
  procedure is fully described in \citet{miocchi13}}. The single-mass
approximation is justified by the fact that the magnitude range chosen
to build the profile includes cluster stars with negligible mass
differences.  The best-fit model results in a cluster with a King
dimensionless potential ${W_0=5.55\pm 0.35}$, a core radius
${r_c=56.2^{+4.5}_{-4.0}}$ arcsec, a half-mass radius ${r_h=146.2^{+11.5}_{-10.0}}$ arcsec, a truncation radius ${r_t=871.8^{+247}_{-164}}$ arcsec and, thus, a concentration parameter,
defined as the logarithm of the truncation to the core radius,
${c=\log (r_t/r_c)=1.19}$.

{ There is a significant difference between these parameters and those quoted in the \citet{harris96} catalog, originally estimated by \citet{peterson97} from a surface brightness profile obtained from shallow optical images: ${ r_c = 37.8\arcsec}$, ${r_h = 100.2\arcsec}$ and ${r_t = 533.9\arcsec}$ (the latter being derived from the quoted value of the concentration parameter: ${c = 1.15}$). To further investigate this discrepancy, we built the cluster surface brightness profile using a K-band 2MASS image, and we found that it is in agreement with the number density profile shown in Figure~\ref{profwircam}, thus further reinforcing the reliability of the newly-determined parameters. On the other hand, if we take into account only the brightest pixels of the K-band image, we find a surface brightness profile consistent with the literature one. This implies that the structural parameters quoted in the literature (which are determined from the light of the most luminous giants only) are not representative of the overall cluster profile.}

\begin{figure*}[h]
\centering
\includegraphics[width=7.6cm]{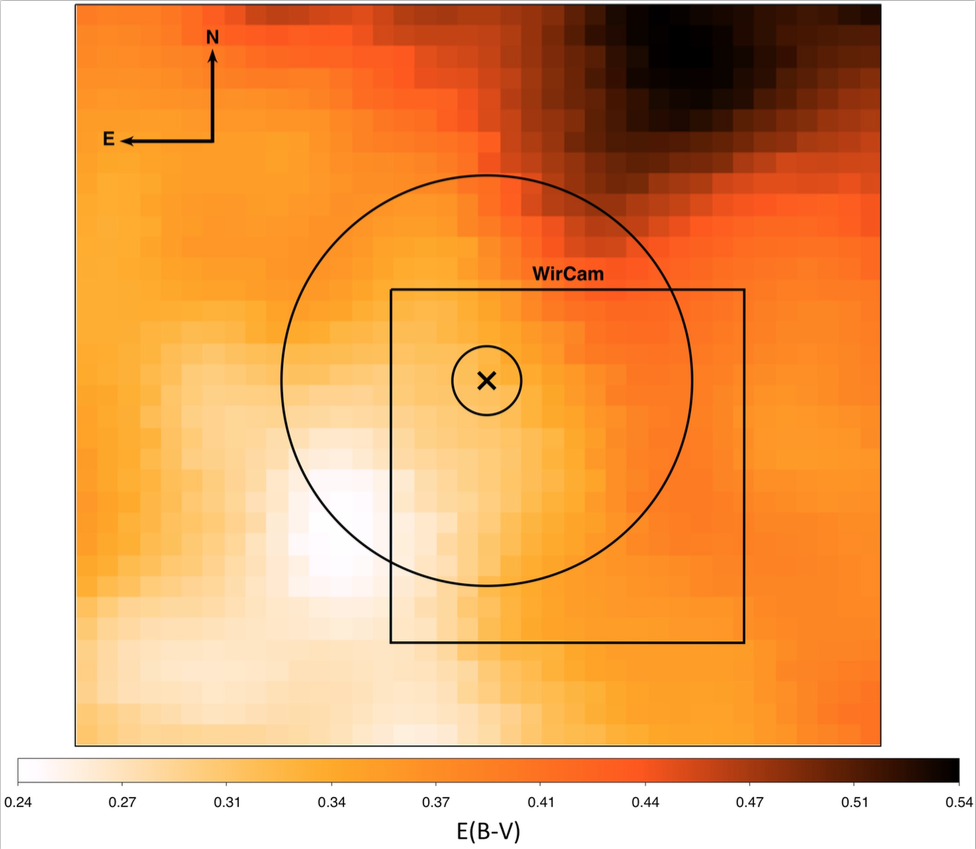}
\includegraphics[width=7.7cm]{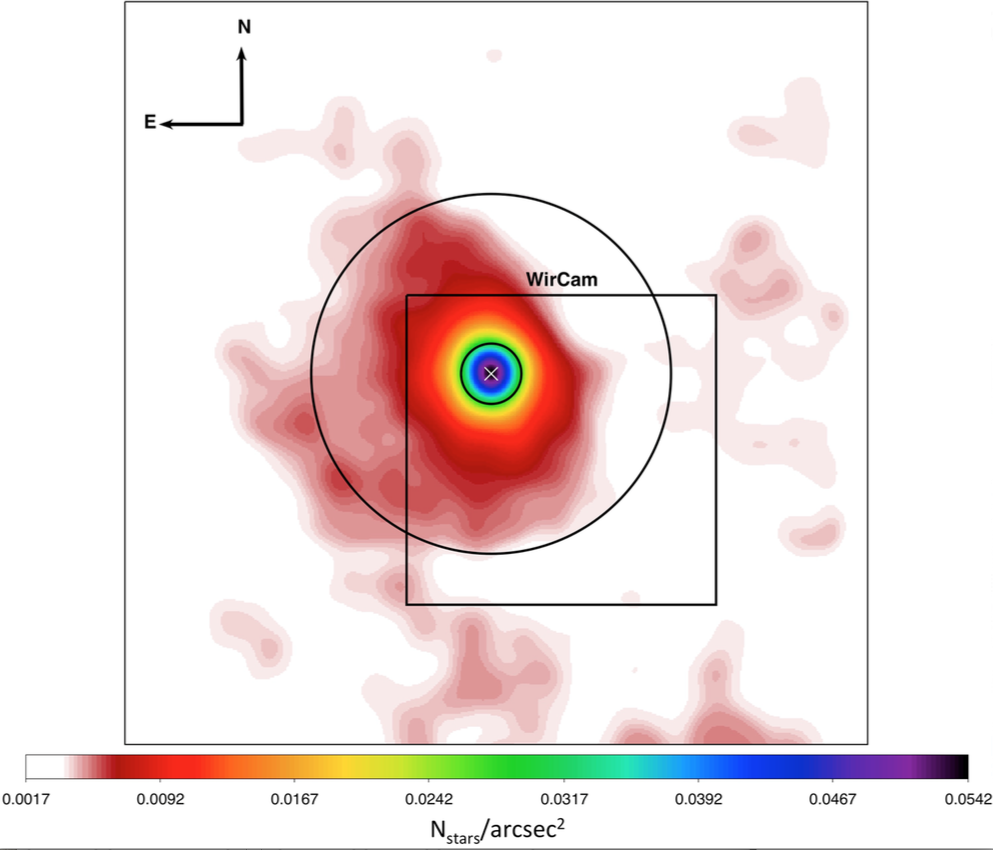}
\caption[Reddening map of the $1500\arcsec\times 1500\arcsec$ region centered on M71]{{\it Left Panel:} Reddening map (from \citealp{schlegel98})
  of the $1500\arcsec\times 1500\arcsec$ region centered on M71. The
  small and large circles mark, respectively, the half-mass and
  truncation radii of the cluster. The square marks the FOV of the
  WIRCam dataset. The presence of severe differential reddening across
  the system, especially in the north-west sector, is apparent.  {\it
    Right Panel:} 2D stellar density map in the same region of the left panel, obtained by counting all the stars with
  $13< g'<19$ in the optical (MegaCam) dataset. The comparison with the
  reddening map on the left makes the effect of extinction well
  visible: the significant decline in the stellar density observed in
  the external regions of the map (especially in the north-east
  sector) is clearly due to the presence of thick dust clouds.}
\label{mapred}
\end{figure*}

The availability of a very wide ($\sim1^{\circ}\times1^{\circ}$)
sample at optical wavelengths (the MegaCam dataset) with an analogous
level of completeness (comparable to the ACS one for $13< g'<19$)
allowed us to investigate how the derivation of the cluster stellar number 
density profile from optical observations can be affected by the
presence of large differential extinction.  Figure \ref{mapred}
compares the extinction map and the 2D density map of the
$1500\arcsec\times 1500\arcsec$ region of the sky centered on M71. The
former is obtained from \citet{schlegel98} and shows that the color
excess ${E(B-V)}$ varies from $\sim 0.24$ to $\sim 0.54$, with several ``spots'' and a clear gradient across
the field. 
The density map in the right-hand panel shows the number of
stars with $13<g'<19$, per unit area, detected in the MegaCam
sample. As expected, at large scales it reveals a direct
correspondence with the extinction map: in particular, the stellar
density manifestly drops in the north-west sector, where the color
excess is the highest, while the opposite is true in the south-east
part of the cluster. Obviously, this is expected to significantly
impact the density profile obtained from star counts in the optical
bands.
\begin{figure*}[!b]
\centering
\includegraphics[width=10cm]{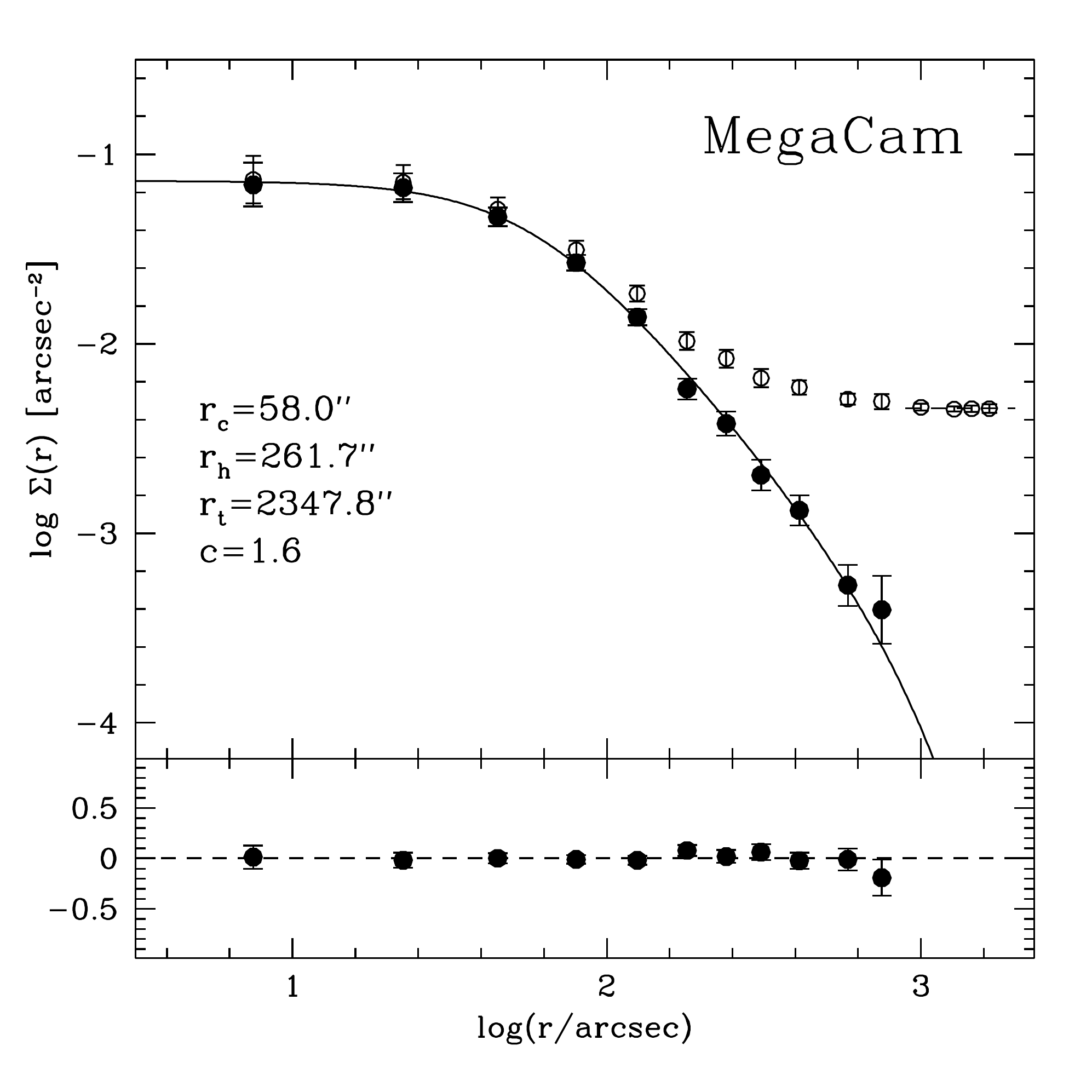}
\caption[Density profile of M71 obtained from the optical
  dataset]{Density profile of M71 obtained from the optical (MegaCam)
  dataset. The meaning of all symbols and lines is as in
  Fig. \ref{profwircam}. The best-fit King model parameters
  (especially the truncation radius and the concentration) are
  significantly different from those obtained from near-infrared
  observations, mainly because the Galactic background is
  underestimated at optical wavelengths, due to the large differential
  reddening affecting the external regions of the covered FOV (compare
  to Fig. \ref{mapred}).}
\label{profmega}
\end{figure*}

To quantitatively test this effect, we determined the cluster density
profile by using the MegaCam (optical) data. The result is plotted in
Figure~\ref{profmega} and shows that, indeed, the structural parameters
of the best-fit King model turn out to be very different from those
obtained from the near-infrared (almost reddening-unaffected) dataset
(compare with Fig. \ref{profwircam}). In particular, the concentration
parameter is much larger (${c=1.6}$), as a consequence of a comparable
core radius (${r_c=58\arcsec}$ \emph{versus} $56.2\arcsec$), but a more
than doubled truncation radius (${r_t=2347.8\arcsec}$ \emph{versus}
$871.8\arcsec$).  Such a severe over-estimate of ${r_t}$ is due to the
high extinction affecting the external portions of the MegaCam sample,
where the Galactic field background is evaluated, and it clearly
demonstrates how important is to take differential reddening under
control for the determination of a cluster density profile.

\subsection{Cluster initial mass}
\label{in_mass}
In Sect. \ref{orbitaround} we have argued that M71 likely lost a
significant fraction of its original mass, mostly due to environmental
effects. In this section, we attempt to estimate the total cluster
initial mass. Although many recipes can be used to this aim
\citep[e.g.][]{vesperini13}, we adopted the simple analytical approach
described in \citet{lamers05,lamers06}. It describes the way a cluster
loses its mass due to the effects of both stellar and dynamical
evolution (including processes such as interactions with the Galactic
tidal field and shocks due to encounters with giant molecular clouds
or spiral arms).  Although this method has been developed specifically
for open clusters, it can be used also in the case of M71, since its
current mass (${M=2.0^{+1.6}_{-0.9}\times10^4 \; \Msun}$; from
\citealp{kimmig15}) and orbit are consistent with those typical of
open clusters \citep[see also][for a similar implementation of this
  procedure]{dalessandro15}. The initial mass ${M_{ini}}$ of the
cluster can be expressed as follows:
\begin{equation}
{ M_{ini}\simeq \left[ \left( \frac{M}{M_\odot} \right) ^\gamma +  \frac{\gamma t}{t_0}  \right]   ^{\frac{1}{\gamma}} [1-q_{ev}(t)]^{-1}}, 
\end{equation}
where $M$ is the cluster current mass, $t=12\pm1$ Gyr is the cluster
age \citep{dicecco15}, ${t_0}$ is the dissolution time-scale parameter,
$\gamma$ is a dimensionless index and ${q_{ev} (t)}$ is a function
describing the mass-loss due to stellar evolution. The dissolution
time-scale parameter is a constant describing the mass-loss process,
which depends on the strength of the tidal field.  Small values of
${t_0}$ are typically associated with encounters with molecular clouds
and spiral arms, while larger values are used to describe the effect
of the Galactic tidal field \citep[see][]{lamers05}.  Since M71 has an
orbit and a structure quite similar to those of open clusters, we
assumed ${t_0}$ in the same range of values (${2.3<t_0<4.7}$ Myr)
constrained in \citet{lamers05}.  The parameter $\gamma$ depends on
the cluster initial density distribution and is usually constrained by
the value of the King dimensionless potential ${W_0}$. We adopted ${\gamma = 0.62}$,
corresponding to ${W_0 = 5}$, a typical value for an averagely concentrated cluster.
The function ${q_{ev} (t)}$, which describes the
mass-loss process due to stellar evolution, can be approximated by the
following analytical expression:
\begin{equation}
{\log q_{ev}(t) = (\log t-a)^{b} +c, \ \ for \ t > 12.5 \ \mathrm{Myr}},
\end{equation}
where $a$, $b$ and $c$ are coefficients that depend on the cluster
metallicity.  The iron abundance ratio of M71 is [Fe/H]=$-0.73$
\citep{harris96}, which corresponds to $a=7.03$, $b=0.26$ and
$c=-1.80$ \citep{lamers05}.

\begin{figure*}[h]
\centering
\includegraphics[width=10cm]{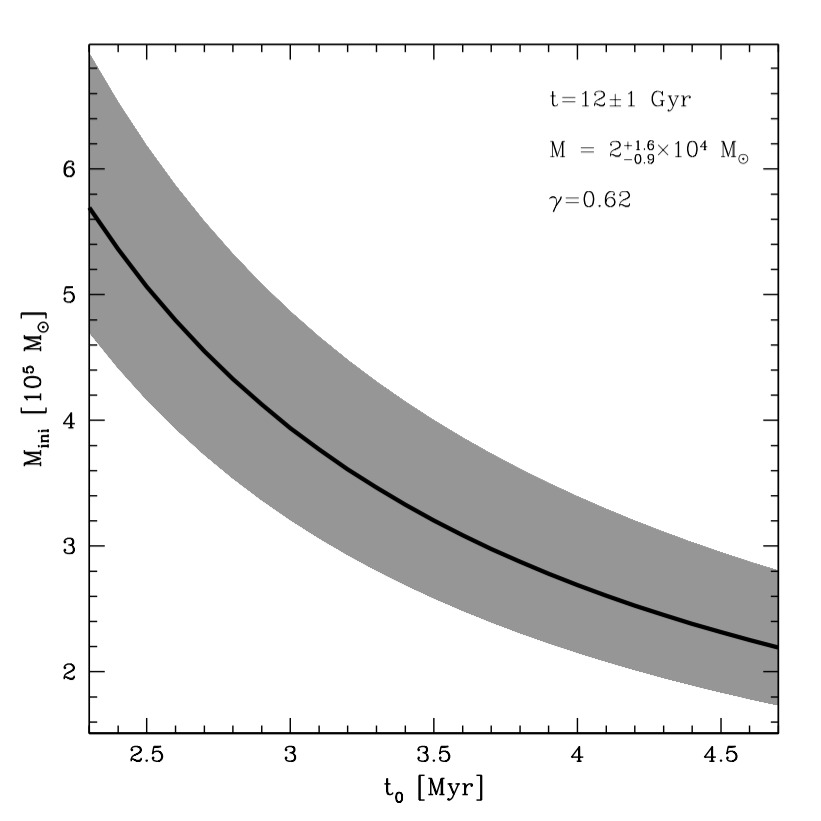}
\caption[Initial mass of M71 as a function of the dissolution
  time-scale parameter]{Initial mass of M71 as a function of the dissolution
  time-scale parameter ${t_0}$, estimated as described in
  Sect. \ref{in_mass}. The black curve shows the values obtained by
  assuming a cluster current mass of $2\times10^4 \ \Msun$ and an
  age of 12 Gyr. The gray shaded area includes all the values
  determined by considering the uncertainties in these quantities (see
  labels). }
  \label{mini}
\end{figure*}

The resulting initial mass of the cluster is shown in Figure~\ref{mini} as a function of the explored range of values of ${t_0}$.  It
varies between 1.8 and $6.8\times10^5 \; \Msun$, which are all values
typical of the mass of Galactic halo GCs, and is one order of
magnitude (or more) larger than the current mass.  Also considering
the largest possible value of ${t_0}$ ($\sim30$ Myr; see
\citealp{lamers05}), we find that the cluster initial mass is at least
twice its current value.  Clearly, this estimate is based on a
simplified approach and on parameters derived by the average behaviors
of open clusters, and different assumptions may lead to different
results. { However, it is interesting to note that, while such a high mass loss would be unlikely for a halo GC, it can be reasonable for a system moving along an orbit confined within the disk (see Sect.~\ref{orbitaround}).}

\section{Summary}
\label{conclu}
{ By using two high-resolution ACS datasets separated by a temporal baseline of $\sim7$ years, we determined the relative PMs of $\sim5000$ individual stars in the direction of the low-mass GC M71, finding that only $\sim60\%$ of them have PMs consistent with being members of the cluster.} The identification of four galaxies and two AGNs within the
sampled FOV, allowed us to also constrain the absolute PM of M71. This
has been used to infer the orbit of the cluster within the Galactic
potential well, which has been modeled by using a three-component
axisymmetric analytic model. { It turned out that, at least during the last 3 Gyr, M71 has been in a disk-like orbit confined within the Galactic disk. It therefore seems reasonable to suppose that M71 suffered a number of dynamical processes (e.g., with the dense surrounding environment, the Milky Way spiral arms, various molecular clouds) that made it lose an amount of mass significantly larger than what expected for the majority of Galactic GCs, which are on halo-like orbits.}
We re-determined the gravitational center
and density profile of M71 by using resolved star counts from a
wide-field near-infrared catalog obtained with WIRCam at the
CFHT. This allowed us to minimize the impact of the large and
differential reddening affecting the system. With respect to the
values quoted in the literature (which have been determined from
optical data), we found the cluster centre to be located almost
$6\arcsec$ to the west, a $\sim 50\%$ larger core and half-mass radii.
Finally, we used a simplified analytical approach to take into account
mass-loss processes due to stellar and dynamical evolution, and thus
estimate the initial cluster mass, finding that the system likely was
one order of magnitude more massive than its current value.

As discussed in Sect. \ref{intro}, M71 is known to harbor a rich
population of X-ray sources \citep{elsner08}, in a number that exceeds
the predictions based on the values of its mass and its collision
parameter $\Gamma$ \citep{huang10}. Since this latter depends on the
cluster central luminosity density and core radius (${\Gamma
\propto\rho_0^{1.5} r_c^2}$; \citealp{verbunt87, huang10}), we have
re-evaluated it by using the newly determined structural parameters.
By adopting the central surface brightness quoted in \citet{harris96}
and equation (4) in \citet{djorg93}, we found ${\log\rho_0=2.60}$ (in
units of $\mathrm{L_\odot}$/pc$^{-3}$).  From this quantity and the value of
${r_c}$ here determined, the resulting value of $\Gamma$ is about half
the one quoted in \citet{huang10}, and the discrepancy in terms of the
expected number of X-ray sources aggravates.  Instead, the much larger
initial mass here estimated for the system would be able to naturally
account for the currently observed X-ray population, thus reinforcing
the hypothesis that M71 lost a large fraction of stars during its
orbit.  An accurate investigation of the possible presence of tidal
tails around the cluster would be important to confirm such a
significant mass-loss from the system. However, this is currently
hampered by the large differential reddening affecting this region of
the sky, and a wide-field infrared observations are urged to shed
light on this issue.

\clearpage


\clearemptydoublepage

%
\backmatter
\bibliographystyle{plainnat} 
\refstepcounter{chapter}
\bibliography{biblio}
\clearemptydoublepage
%
%
\chapter*{Acknowledgements}
\addcontentsline{toc}{chapter}{Acknowledgements} 
\sectionmark{Acknowledgements}
\chaptermark{Acknowledgements}
\markboth{Acknowledgements}{Acknowledgements}

First of all, I want to thank my supervisor Francesco R. Ferraro for giving me the opportunity to carry out this thesis project. \\

A big warm thank to Cristina Pallanca, Barbara Lanzoni and Emanuele Dalessandro who daily helped me throughout these three years.\\

Another big warm thank to Scott Ransom for all his teachings and his hospitality during my stay in Charlottesville. Those three months have been awesome! \\

Finally, I want to acknowledge all the other people that somehow contributed to this thesis work. In alphabetical order: Marta Burgay, Paulo Freire, Jason Hessels, Alessio Mucciarelli, Alessandro Patruno, Andrea Possenti, Ingrid Stairs. \\

and above all: my Family, my Walden, my Home. \\

\vspace{3cm}

\flushright{{\it ``...from so simple a beginning \\ endless forms most beautiful and most wonderful \\ have been, and are being, evolved.''} \\ \vspace{0.3cm} Charles Darwin - On the origin of species}
   
\end{document}